%% file: thesis.tex
\documentclass[english,fontsize=12pt, a4paper,BCOR=16mm]{scrbook} 
\usepackage[utf8x]{inputenc}
\usepackage[T1]{fontenc} 
\usepackage[linktocpage]{hyperref} 
\usepackage[
twoside,
a4paper,
inner=32mm,
outer=32mm,
marginparwidth=25mm,
top=30mm,
bottom=40mm,
]
{geometry}

\usepackage{float}
\usepackage[singlespacing]{setspace} 
\usepackage{sectsty} 
\usepackage{listings}
\usepackage{tabularx} 
\usepackage{csquotes} 
\newcolumntype{C}{>{\centering\arraybackslash} m{6cm} } 
\newcommand{\heading}[1]{\multicolumn{2}{c}{\textbf{#1}}} 
\newcommand{\headingtext}[1]{\multicolumn{2}{l}{\parbox{.96\textwidth}{#1}}} 
\usepackage{cite} 
\usepackage{algorithm,algpseudocode} 
\newcommand{\DQNNNISQ}{DQNN\textsubscript{NISQ }}
\newcommand{\DQNNsNISQ}{DQNNs\textsubscript{NISQ }}
\newcommand{\DQNNNISQP}{DQNN\textsuperscript{+}\textsubscript{NISQ }}
\newcommand{\DQGANNISQ}{DQGAN\textsubscript{NISQ }}
\newcommand{\DQGANsNISQ}{DQGANs\textsubscript{NISQ }}
\usepackage[letterspace=-80]{microtype}

\usepackage{color,xcolor}
\definecolor{color1}{HTML}{4e9773}
\definecolor{color1L}{HTML}{eef6f2}
\definecolor{color2}{HTML}{fab73d}
\definecolor{color2L}{HTML}{fffbf4}
\definecolor{color3}{HTML}{d2561a}
\definecolor{color3L}{HTML}{faede6}
\definecolor{color0}{HTML}{949494}
\definecolor{color0M}{HTML}{DCDCDC}
\definecolor{color0L}{HTML}{F8F8F8}
\subsectionfont{\normalsize}  
\usepackage{enumitem}
\usepackage[color=color2]{todonotes}
\hypersetup{colorlinks,filecolor=black, linkcolor=color1, citecolor=color1, urlcolor=color1}

\usepackage{fancyhdr} 
\usepackage{titlesec, blindtext}
\newcommand{\hsp}{\hspace{20pt}}
\titleformat
{\chapter} 
[display] 
{\sffamily\centering\bfseries\fontsize{90pt}{90}\selectfont\color{color1}} 
{\thechapter} 
{-4ex}
{\textcolor{color1}{\rule{\textwidth}{2pt}}	\vspace{2ex}
\Large\color{color1}
}
[
\vspace{-3.0ex}%
\textcolor{color1}{\rule{\textwidth}{2pt}}\vspace{-3ex}
]

\titleformat{\section}[hang]{\sffamily\centering\large\bfseries\color{color1}}{\thesection\hsp}{0pt}{}
\titleformat{\subsection}[hang]{\sffamily\centering\bfseries}{\thesubsection\hsp}{0pt}{}
\usepackage{tocloft} 
\usepackage{bibunits} 
\usepackage[titletoc]{appendix}
\usepackage{tocloft}

\usepackage{scalerel,stackengine,amsmath} 
\usepackage{amssymb} 
\usepackage{braket  }
\usepackage{amsthm} 
\usepackage{bbm} 
\usepackage{bbold} 
\DeclareFixedFont{\ttb}{T1}{txtt}{bx}{n}{12} 
\DeclareFixedFont{\ttm}{T1}{txtt}{m}{n}{12}  
\DeclareMathOperator{\tr}{tr}
\DeclareMathOperator{\notgate}{\textls{\textit{NOT}}}
\DeclareMathOperator{\cnot}{\textls{\textit{CNOT}}}
\DeclareMathOperator{\swap}{\textls{\textit{SWAP}}}

\DeclareMathOperator{\can}{\textls{\textit{CAN}}}
\DeclareMathOperator{\rxx}{\textls{\textit{RXX}}}
\DeclareMathOperator{\ryy}{\textls{\textit{RYY}}}
\DeclareMathOperator{\rzz}{\textls{\textit{RZZ}}}
\allowdisplaybreaks
\newcommand{\mathcolorbox}[1]{\fcolorbox{color2}{color2L}{$\displaystyle #1$}}
\makeatletter
\renewcommand*\env@matrix[1][\arraystretch]{%
\edef\arraystretch{#1}%
\hskip -\arraycolsep
\let\@ifnextchar\new@ifnextchar
\array{*\c@MaxMatrixCols c}}
\makeatother
\makeatletter
\patchcmd{\env@cases}{1.2}{0.8}{}{}
\makeatother

\usepackage{nameref,cleveref}
\newtheorem{defi}{Definition}[chapter]

\newtheorem{prop}[defi]{Proposition}
\crefname{chapter}{Chapter}{Chapter}
\Crefname{chapter}{Chapter}{Chapter}
\crefname{section}{Section}{Sections}
\Crefname{section}{Section}{Sections}
\crefname{algorithm}{Algorithm}{Algorithm}
\Crefname{algorithm}{Algorithm}{Algorithm}
\crefname{equation}{Equation}{Equation}
\Crefname{equation}{Equation}{Equation}
\crefname{figure}{Figure}{Figure}
\Crefname{figure}{Figure}{Figure}
\crefname{appendix}{Appendix}{Appendices}
\Crefname{appendix}{Appendix}{Appendices}
\crefname{prop}{Proposition}{Propositions}
\Crefname{prop}{Proposition}{Propositions}
\crefname{defi}{Definition}{Definitions}
\Crefname{defi}{Definition}{Definitions}

\usepackage{graphicx}
\usepackage{tikz}
\usepackage{pgfplots}
\usepackage{subcaption}
\usepackage{adjustbox} 		
\usepackage[section]{placeins} 
\usetikzlibrary{matrix,fit,calc,shapes.geometric,backgrounds,positioning,decorations.markings,decorations.pathreplacing,arrows,knots,hobby,angles,quotes,shapes,snakes,automata,petri}
\usepackage{wrapfig} 
\usepackage{hhline,colortbl} 
\tikzset{
perceptron0/.style = {circle,draw=color0,line width=1pt,fill=color0L,minimum width=0.7cm},
perceptronS0/.style = {circle,draw=color0,line width=1pt, fill=color0L,minimum width=0.2cm},
perceptron1/.style = {circle,draw=color1,line width=1pt,fill=color1L,minimum width=0.7cm},
perceptronS1/.style = {circle,draw=color1,line width=1pt,fill=color1L,minimum width=0.2cm},
perceptron2/.style = {circle,draw=color2,line width=1pt,fill=color2L,minimum width=0.7cm},
perceptronS2/.style = {circle,draw=color2,line width=1pt,fill=color2L,minimum width=0.2cm},
line0/.style = {draw=white,line width=3pt},
lineD/.style = {line width=1pt},
line1/.style = {draw=color1,line width=1pt},
line2/.style = {draw=color2,line width=1pt},
line3/.style = {draw=color3,line width=1pt},
operator0/.style = {text=black,draw=color0,line width=1pt,fill=color0L,minimum width=1cm,minimum size=1.5em},
operator1/.style = {draw, text=black, draw=color1,line width=1pt, fill=color1L,minimum width=1cm,minimum size=1.5em},
operator2/.style = {draw, text=black, draw=color2,line width=1pt, fill=color2L,minimum width=1cm,minimum size=1.5em},
brace0/.style = {decorate,decoration={brace,amplitude=5pt},color0},
dot/.style = {draw,fill,shape=circle,minimum size=5pt,inner sep=0pt},
dotwhite/.style = {draw,fill=white,shape=circle,minimum size=5pt,inner sep=0pt},
cross/.style={path picture={ \draw[thick,black](path picture bounding box.north) -- (path picture bounding box.south) (path picture bounding box.west) -- (path picture bounding box.east);	}},
circlewc/.style={draw,circle,cross,minimum width=0.3 cm},
dcross/.style={path picture={ \draw[thick, black](path picture bounding box.north west) -- (path picture bounding box.south east) (path picture bounding box.south west) -- (path picture bounding box.north east);}},
halfcross/.style={path picture={ \draw[thick, black](path picture bounding box.south west) -- (path picture bounding box.north east);}},
meter/.append style={draw, fill=white, inner sep=10, rectangle, font=\vphantom{A}, minimum width=30, line width=.7,
path picture={\draw[black] ([shift={(.1,.3)}]path picture bounding box.south west) to[bend left=50] ([shift={(-.1,.3)}]path picture bounding box.south east);\draw[black,-latex] ([shift={(0,.1)}]path picture bounding box.south) -- ([shift={(.3,-.1)}]path picture bounding box.north);}},
networkcircle1/.style = {line1,circle,fill=white, text width=2mm, minimum height=0.7cm},
networkellipse1/.style = {line1, ellipse,fill=white,minimum height=1.5cm,minimum width=0.7cm},
networkellipseM/.style = {line1, ellipse,fill=white,minimum height=2cm,minimum width=0.7cm},
networkellipseX/.style = {line1, ellipse,fill=white,minimum height=3cm,minimum width=0.9cm},
vertex0/.style = {regular polygon,regular polygon sides=7,draw=color0,line width=1pt,fill=color0L,minimum width=1cm},
vertex1/.style = {regular polygon,regular polygon sides=7,draw=color1,line width=1pt,fill=color1L,minimum width=1cm},
vertex2/.style = {regular polygon,regular polygon sides=7,draw=color2,line width=1pt,fill=color2L,minimum width=1cm},
}
\newcommand\equalhat{\mathrel{\stackon[1.5pt]{=}{\stretchto{%
\scalerel*[\widthof{=}]{\wedge}{\rule{1ex}{3ex}}}{0.5ex}}}}

\newcommand{\oneoneone}{\raisebox{-1.4pt}{\tikz[yscale=0.6,xscale=0.4]{
\node(1) [circle,draw,inner sep=0pt,minimum size=4.5pt] at (-1,0) {};
\node(2) [circle,draw,inner sep=0pt,minimum size=4.5pt] at (0,0) {};
\node(3) [circle,draw,inner sep=0pt,minimum size=4.5pt] at (1,0) {};
\node(e) [circle,draw=white,inner sep=0pt,minimum size=0.5pt] at (1.4,-0.2) {};
\draw (1)--(2) -- (3);
}}}
\newcommand{\oneothreeone}{\raisebox{-1.4pt}{\tikz[yscale=0.9,xscale=0.6]{
\node(1) [circle,draw,inner sep=0pt,minimum size=4.5pt] at (-1,0.15) {};
\node(3) [circle,draw,inner sep=0pt,minimum size=4.5pt] at (0,-.08) {};
\node(4) [circle,draw,inner sep=0pt,minimum size=4.5pt] at (0,0.15) {};		
\node(5) [circle,draw,inner sep=0pt,minimum size=4.5pt] at (0,0.38) {};			
\node(6) [circle,draw,inner sep=0pt,minimum size=4.5pt] at (1,0.15) {};
\node(e) [circle,draw=white,inner sep=0pt,minimum size=0.5pt] at (1.2,0.3) {};
\draw (1)--(3);
\draw (1)--(4);
\draw (1)--(5);
\draw (6)--(3);
\draw (6)--(4);
\draw (6)--(5);
}}}
\newcommand{\twothreetwo}{\raisebox{-1.4pt}{\tikz[yscale=0.9,xscale=0.6]{
\node(1) [circle,draw,inner sep=0pt,minimum size=4.5pt] at (-1,0) {};
\node(2) [circle,draw,inner sep=0pt,minimum size=4.5pt] at (-1,0.3) {};
\node(3) [circle,draw,inner sep=0pt,minimum size=4.5pt] at (0,-.08) {};
\node(4) [circle,draw,inner sep=0pt,minimum size=4.5pt] at (0,0.15) {};		
\node(5) [circle,draw,inner sep=0pt,minimum size=4.5pt] at (0,0.38) {};			
\node(6) [circle,draw,inner sep=0pt,minimum size=4.5pt] at (1,0) {};
\node(7) [circle,draw,inner sep=0pt,minimum size=4.5pt] at (1,0.3) {};
\node(e) [circle,draw=white,inner sep=0pt,minimum size=0.5pt] at (1.2,0.3) {};
\draw (1)--(3) -- (2);
\draw (1)--(4) -- (2);
\draw (1)--(5) -- (2);
\draw (6)--(3) -- (7);
\draw (6)--(4) -- (7);
\draw (6)--(5) -- (7);
}}}
\newcommand{\twothree}{\raisebox{-1.4pt}{\tikz[yscale=0.9,xscale=0.6]{
\node(1) [circle,draw,inner sep=0pt,minimum size=4.5pt] at (-1,0) {};
\node(2) [circle,draw,inner sep=0pt,minimum size=4.5pt] at (-1,0.3) {};
\node(3) [circle,draw,inner sep=0pt,minimum size=4.5pt] at (0,-.08) {};
\node(4) [circle,draw,inner sep=0pt,minimum size=4.5pt] at (0,0.15) {};		
\node(5) [circle,draw,inner sep=0pt,minimum size=4.5pt] at (0,0.38) {};			
\node(e) [circle,draw=white,inner sep=0pt,minimum size=0.5pt] at (0.2,0.3) {};
\draw (1)--(3) -- (2);
\draw (1)--(4) -- (2);
\draw (1)--(5) -- (2);
}}}

\newcommand{\twotwo}{\raisebox{-1.4pt}{\tikz[yscale=0.9,xscale=0.6]{
\node(e) [circle,draw=white,inner sep=0pt,minimum size=0.5pt] at (0.3,0.3) {};
\node(1) [circle,draw,inner sep=0pt,minimum size=4.5pt] at (-1,0) {};
\node(2) [circle,draw,inner sep=0pt,minimum size=4.5pt] at (-1,0.3) {};
\node(4) [circle,draw,inner sep=0pt,minimum size=4.5pt] at (0,0) {};
\node(5) [circle,draw,inner sep=0pt,minimum size=4.5pt] at (0,0.3) {};
\draw (1)--(4);
\draw (2)--(4);
\draw (1)--(5);
\draw (2)--(5);
}}}

\newcommand{\threeone}{\raisebox{-1.4pt}{\tikz[yscale=0.9,xscale=0.6]{
\node(1) [circle,draw,inner sep=0pt,minimum size=4.5pt] at (1,0.22) {};
\node(3) [circle,draw,inner sep=0pt,minimum size=4.5pt] at (0,0) {};
\node(4) [circle,draw,inner sep=0pt,minimum size=4.5pt] at (0,0.22) {};		
\node(5) [circle,draw,inner sep=0pt,minimum size=4.5pt] at (0,0.44) {};			
\node(e) [circle,draw=white,inner sep=0pt,minimum size=0.5pt] at (1.5,0.3) {};
\draw (1)--(3);
\draw (1)--(4);
\draw (1)--(5);
}}}

\pgfplotsset{
/pgfplots/bar shift auto/.style={
/pgf/bar shift={%
-0.5*(\numplotsofactualtype/2*\pgfplotbarwidth + ((\numplotsofactualtype/2)-1)*(#1)) +
(.5+round((\plotnumofactualtype+1)/2)-1)*\pgfplotbarwidth + (round((\plotnumofactualtype+1)/2)-1)*(#1)
},
},
}

\usepackage{ifthen} 
\newlength{\hintscolumnwidth}
\newlength{\separatorcolumnwidth}
\newlength{\maincolumnwidth}
\newcommand*{\hintstyle}[1]{{\textbf{\textcolor{color1}{#1}}}}
\newcommand*{\cventry}[7][.25em]{%
\cvitem[#1]{#2}{%
{\bfseries#3}%
\ifthenelse{\equal{#4}{}}{}{, {\slshape#4}}%
\ifthenelse{\equal{#5}{}}{}{, #5}%
\ifthenelse{\equal{#6}{}}{}{, #6}%
.\strut%
\ifx&#7&%
\else{\newline{}\begin{minipage}[t]{\linewidth}\small#7\end{minipage}}\fi}}
\newcommand*{\cvitem}[3][.25em]{%
\begin{tabular}{@{}p{\hintscolumnwidth}@{\hspace{\separatorcolumnwidth}}p{\maincolumnwidth}@{}}%
\raggedleft\hintstyle{#2} &{#3}%
\end{tabular}%
\par\addvspace{#1}}

\begin{document}
\doublespacing
\include{text/cover}

\onehalfspacing
\include{text/quote}

\singlespacing  
\addtocontents{toc}{\cftpagenumbersoff{chapter}}
\addtocontents{toc}{\protect\vspace{0.2cm}} 
\addtocontents{toc}{\cftpagenumberson{chapter}}
\markboth{ }{ } 
\include{text/abstract}
\markboth{ }{ } 
\include{text/zusammenfassung}

\newpage
\markboth{ }{ } 
\include{text/acknowledgements}
\newpage
\markboth{\quad}{\quad} 
\include{text/publications}
\markboth{\quad}{\quad} 
\addtocontents{toc}{\cftpagenumbersoff{chapter}}
\addtocontents{toc}{\protect\vspace{1cm}} 
\addtocontents{toc}{\cftpagenumberson{chapter}}

\onehalfspacing
\markboth{}{} 
\tableofcontents
\markboth{}{} 
\singlespacing  

\input{text/introduction}
\input{text/ML}

\input{text/QI}
\input{text/DQNN} 
\input{text/NFL}

\input{text/graphs}
\input{text/gan}

\input{text/conclusion}

\newpage
\thispagestyle{empty}
\quad \newpage
\vspace*{\fill} 
\centerline{\sffamily\bfseries\Huge\color{color1} Appendix}
\vspace*{\fill} 
\newpage \thispagestyle{empty} 
\addtocontents{toc}{\cftpagenumbersoff{chapter}}
\addtocontents{toc}{\protect\vspace{1cm}} 
\addtocontents{toc}{\cftpagenumberson{chapter}}
\begin{appendices}
\input{text/appendixDQNN}

\input{text/appendixgraph}
\input{text/appendixGAN}
\end{appendices}

\addtocontents{toc}{\cftpagenumbersoff{chapter}}
\addtocontents{toc}{\protect\vspace{1cm}} 
\addtocontents{toc}{\cftpagenumberson{chapter}}

\newpage
\small
\bibliographystyle{nosort_habbrv}
\addcontentsline{toc}{section}{Bibliography}
\bibliography{literatur}

\input{text/CV}
\end{document}

%% file: text/cover.tex
\thispagestyle{empty}
\pagenumbering{Roman}

\vspace*{\fill} 
\begin{figure}[h!]
\begin{center}
\begin{tikzpicture}[scale=1.4]
\foreach \x in {-.5,.5} {
\draw[line0] (0,\x) -- (2,-1);
\draw[line3] (0,\x) -- (2,-1);
\draw[line0] (0,\x) -- (2,0);
\draw[line2] (0,\x) -- (2,0);
\draw[line0] (0,\x) -- (2,1);
\draw[line1] (0,\x) -- (2,1);
}
\foreach \x in {-1.5,-0.5, ..., 1.5} {
\draw[line0] (2,-1) -- (4,\x);
\draw (2,-1) -- (4,\x);
\draw[line0] (2,0) -- (4,\x);
\draw (2,0) -- (4,\x);
\draw[line0] (2,1) -- (4,\x);
\draw (2,1) -- (4,\x);
}
\foreach \x in {-1.5,-0.5, ..., 1.5} {
\draw[line0] (4,\x) -- (6,-0.5);
\draw (4,\x) -- (6,-0.5);
\draw[line0] (4,\x) -- (6,0.5);
\draw (4,\x) -- (6,0.5);
}
\foreach \x in {-1,0,1} {
\node[perceptron0] at (2,\x) {};
}
\foreach \x in {-1.5,-0.5, ..., 1.5} {
\node[perceptron0] at (4,\x) {};
}
\node[perceptron0] at (0,-0.5) {};
\node[perceptron0] at (0,0.5) {};
\node[perceptron0] at (6,-0.5) {};
\node[perceptron0] at (6,0.5) {};
\end{tikzpicture}
\end{center}
\end{figure}

\begin{center}
\textbf{\textcolor{color1}{\sffamily{\Huge{Quantum neural networks}}}}\\ \vspace{1cm}


Von der Fakultät für Mathematik und Physik\\ der Gottfried Wilhelm Leibniz Universität Hannover\\
\vspace{1cm}
zur Erlangung des akademischen Grades \\ Doctor rerum naturalium \\ Dr.\ rer.\ nat.\\
\vspace{1cm}
genehmigte Dissertation von \\
\vspace{1cm}
\textbf{M.\ Sc.\ Kerstin Beer}\\ 
\vspace{1cm}2022
\end{center}
\vspace*{\fill} 

\newpage 
\begin{flushleft}
\textbf{Examination board:}\\
Prof. Dr. Mich\'ele Heurs (chair)\\
Prof. Dr. Tobias J. Osborne (supervisor)\\
Prof. Dr. Avishek Anand\\
\vspace{1cm}
\textbf{Referees:}\\
Prof. Dr. Tobias J. Osborne\\
Prof. Dr. Avishek Anand\\
Prof. Dr. Chiara Macchiavello\\
\vspace{1cm}
Day of the defence: 22.02.2022\\
\end{flushleft}

%% file: text/quote.tex


\newpage 
\vspace*{\fill} 
\begin{quote} 
	\centering 
	\textcolor{black}{
	\textit{\enquote{I have never tried that before, \\so I think I should definitely be able to do that.}}\\
	Pippi Longstocking (Astrid Lindgren)}
\label{quote}
\end{quote}
\vspace*{\fill}

%% file: text/abstract.tex
\newgeometry{twoside,
	a4paper,
	inner=32mm,
	outer=32mm,
	marginparwidth=25mm,
	top=30mm,
	bottom=37mm}

\phantomsection\addcontentsline{toc}{section}{Abstract}\section*{Abstract}
Quantum computing is one of the most exciting research areas of the last decades. At the same time, methods of machine learning have started to dominate science, industry and our everyday life. In this thesis we combine these two essential research topics of the 21st century and introduce \emph{dissipative quantum neural networks} (DQNNs), which are designed for fully quantum learning tasks, are capable of universal quantum computation and have low memory requirements while training. 

We start the discussion of this interdisciplinary topic by introducing artificial neural networks, which are a very common tool in classical machine learning. Next, we give an overview on quantum information. Here we focus on quantum algorithms and circuits, which are used to implement quantum neural networks. Moreover, we explain the opportunities and challenges arising with today's quantum computers.  

The discussion of the architecture and training algorithm of the DQNNs forms the core of this work. These networks are optimised with training data pairs in form of input and desired output states and therefore can be used for characterising unknown or untrusted quantum devices. We not only demonstrate the generalisation behaviour of these quantum neural networks using classical simulations, but also implement them successfully on actual quantum computers. 

To understand the ultimate limits for such quantum machine learning methods, we discuss the \emph{quantum no free lunch} theorem, which describes a bound on the probability that a quantum device, which can be modelled as a unitary process and is optimised with quantum examples, gives an incorrect output for a random input. This gives us a tool to review the learning behaviour of quantum neural networks in general and the DQNNs in particular.

Moreover we expand the area of applications of DQNNs in two directions. In the first case, we include additional information beyond just the training data pairs: since quantum devices are always structured, the resulting data is always structured as well. We modify the DQNN's training algorithm such that knowledge about the graph-structure of the training data pairs is included in the training process and show that this can lead to better generalisation behaviour.

Both the original DQNN and the DQNN including graph structure are trained with data pairs in order to characterise an underlying relation. However, in the second extension of the algorithm we aim to learn characteristics of a set of quantum states in order to extend it to quantum states which have similar properties. Therefore we build a generative adversarial model where two DQNNs, called the generator and discriminator, are trained in a competitive way. 

Overall, we observe that DQNNs can not only be trained efficiently but also, similar to their classical counterparts, modified to suit different applications.
\begin{flushleft}
	\emph{Keywords:} quantum computing, neural network, machine learning
\end{flushleft}

%% file: text/zusammenfassung.tex
\phantomsection\addcontentsline{toc}{section}{Kurzzusammenfassung}\section*{Kurzzusammenfassung}

\begin{sloppypar}Quantencomputer bilden eines der spannendsten Forschungsgebiete der letzten Jahrzehnte. Zur gleichen Zeit haben Methoden des maschinellen Lernens begonnen die Wissenschaft, Industrie und unseren Alltag zu dominieren. In dieser Arbeit kombinieren wir diese beiden wichtigen Forschungsthemen des 21. Jahrhunderts und stellen \emph{dissipative quantenneuronale Netze} (DQNNs) vor, die für Quantenlernaufgaben konzipiert sind, universelle Quantenberechnungen durchführen können und wenig Speicherbedarf beim Training benötigen.\end{sloppypar}

Wir beginnen die Diskussion dieses interdisziplinären Themas mit der Einführung künstlicher neuronaler Netze, die beim klassischen maschinellen Lernen weit verbreitet sind. Dann geben wir einen Überblick über die Quanteninformationstheorie. Hier fokussieren wir uns auf die zur Implementierung von quantenneuronalen Netzen nötigen Quantenalgorithmen und -schaltungen. Außerdem erläutern wir die Chancen und Herausforderungen der heutigen Quantencomputer.

Die Diskussion der Architektur und des Trainingsalgorithmus der DQNNs bildet den Mittelpunkt dieser Arbeit. Diese Netzwerke werden mit Trainingsdatenpaaren in Form von Eingangs- und gewünschten Ausgangszuständen optimiert und können daher zur Charakterisierung unbekannter oder nicht vertrauenswürdiger Quantenbauelemente verwendet werden. Wir demonstrieren nicht nur das Generalisierungsverhalten dieser Netze anhand klassischer Simulationen, sondern konstruieren auch eine erfolgreiche Implementierung für Quantencomputer.

Um die ultimativen Grenzen solcher Methoden zum maschinellen Lernen von Quantendaten zu verstehen, führen wir das \emph{quantum no free lunch}-Theorem ein, welches eine Begrenzung für die Wahrscheinlichkeit beschreibt, dass ein als unitärer Prozess modellierbares und mit Quantendaten optimiertes Quantenbauelement eine falsche Ausgabe für eine zufällige Eingabe herausgibt. Das Theorem gibt uns ein Werkzeug, um das Lernverhalten von quantenneuronalen Netzwerken im Allgemeinen und der DQNNs im Besonderen zu überprüfen.

Darüber hinaus erweitern wir den Anwendungsbereich von DQNNs auf zwei Weisen. Im ersten Fall beziehen wir Informationen zusätzlich zu den Trainingsdaten mit ein: Da Quantenbauelemente immer eine gewisse Struktur haben, sind auch die resultierenden Daten strukturiert. Wir modifizieren den Trainingsalgorithmus der DQNNs so, dass Kenntnisse über die Struktur genutzt werden können und zeigen, dass dies zu einem besseren Trainingsergebnis führen kann.

Sowohl das ursprüngliche DQNN als auch das Graphen-DQNN wird mit Datenpaaren trainiert, um eine zugrunde liegende Relation zu charakterisieren. Als zweite Erweiterung wollen wir jedoch die Eigenschaften einer Menge einzelner Quantenzustände untersuchen, um sie mit Quantenzuständen ähnlicher Eigenschaften zu erweitern. Daher konstruieren wir ein Modell, bei dem zwei DQNNs, Generator und Diskriminator genannt, kompetitiv trainiert werden.

Zusammenfassend stellen wir fest, dass DQNNs nicht nur effizient trainiert, sondern auch, ähnlich wie ihre klassischen Gegenstücke, an unterschiedliche Anwendungen angepasst werden können. 
\begin{flushleft}
	\emph{Schlagwörter:} Quantencomputer, Neuronales Netz, Maschinelles Lernen
\end{flushleft}
\restoregeometry

%% file: text/acknowledgements.tex
\phantomsection\addcontentsline{toc}{section}{Acknowledgements}\section*{Acknowledgements}

Without question, I am very thankful that I was given the chance to be a Ph.D.\ student and write this thesis. This would have been inconceivable without the people named in the following.

Foremost, I would like to thank my supervisor Tobias J.\ Osborne for introducing me to the exciting field of quantum information, guiding me through the last years, and sharing his knowledge and experiences with me. I thank him for creating a well-organised, open-minded and safe working atmosphere, all his helpful advice, and making me sometimes feel a little bit like Pippi Longstocking, see page~\pageref{quote}.

Further, I would like to thank Avishek Anand for introducing me to the topic of deep neural networks in his lecture, Megha Khosla for various helpful discussions, and all people who cooperated with me on several exciting projects.

Many thanks go to all past and current members of the \emph{quantum information group} in Hanover. I explicitly thank Kais Abdelkhalek, Thomas Cope, Friederike Dziemba, Terry Farrelly, Tobias Geib, Alexander Hahn, Wiebke Möller, Laura Niermann, Viktoria Schmiesing, Reinhard Werner, and Ramona Wolf for making our group a warm-hearted and amusing working place. Moreover, I thank Florian Oppermann for being the funniest and most generous IT administrator colleague one could wish for. I also thank all Master and Bachelor students I supervised for knowledge exchange and various enjoyable meetings in the last years. 

I am very thankful for all the opportunities to have thrilling travel experiences and meet inspiring people. Particularly I would like to thank Felipe Montealegre-Mora for the most cheerful conferences. Moreover, I thank Hanover's university groups \emph{Students for Future} and \emph{Scientists for Future} for giving me the chance to learn and discuss science beyond quantum information.

Sometimes life can be hard, hence it is important to have good friends. I want to thank Bianka and Robert Heymann, Luisa Oyen, and all members of \emph{Zirkus Johnass}, my second family. Further, I thank my friends at the acrobatic course of our university, and at the \emph{Gr\"unt\"one Ensemble} for all the fun time we spend together. I thank my cello teacher Jacob Jordan for one certainly joyful hour every week. I also would like to thank Franziska Weeren for various long and understanding phone calls, as well as my therapist and her dog Erwin.

I would like to thank Thomas Cope, Tobias Geib, Dominik Lack, Gabriel M\"uller, Laura Niermann, Viktoria Schmiesing, Marvin Schwiering, Alexander Stottmeister, and Christian Struckmann for finding several typos in my drafts and, hence, helped me a lot to bring this thesis in its final form.

My biggest thanks go to my beloved family, my mum Renata Beer, my dad Christian Beer and my sister Anika Beer for their limitless support and love.

Most importantly, I want to thank my amazing partner, lovely husband and best friend Dominik Lack for being by my side for the last eleven years. I would like to thank him for his kind words and hugs whenever I panicked, for his endless love, and for all the snacks he brought me during the home office times. The last years, and this thesis would not have been possible without him.

%% file: text/publications.tex
\phantomsection\addcontentsline{toc}{section}{Publications} \chapter*{Publications}

\begingroup
\makeatletter
\renewcommand*\bib@heading{\quad
}
\makeatother
\renewcommand{\chapter}[2]{}
\subsection*{Publications whose material appears in this thesis:}

\subsection*{Further publications:}

\endgroup

%% file: text/introduction.tex
\chapter{Introduction}\hypertarget{Intro}{}
\pagenumbering{arabic}
\label{chapter:intro}

Results of \emph{machine learning} (ML) \cite{Murphy2012, Jordan2015, Nielsen2015, Goodfellow2016}, the well-known subfield of \emph{artificial intelligence} where knowledge is gained from experiences rather than from instructions, have carried over into our everyday life in the last decade: our web search engines rearrange and optimise results based on learned user characteristics \cite{Boyan1996, Serrano2016, Serrano2017}, we are used to traffic forecasting and always up-to-date commute-estimating apps \cite{Jiang2019, Ibarz2017}, spam and phishing emails can be automatically classified \cite{Awad2011, Hassanpour2018, Bagui2019, Dada2019} and several apps even offer smart replies \cite{Kannan2016, Henderson2017, Weng2019}, on social media platforms the tools of ML allow recommendation of friends, posts or tags \cite{Nguyen2017,Qu2018}, identification of illegal, unwanted or fake data \cite{Agrawal2018a,Li2019b,Minin2018,Purba2020} and actually allow the platforms to extract informations on the user's personality, interests or mental health \cite{Liu2016a,Gkotsis2017,Xue2018,Sawhney2018,Kim2020}.

Also, generally, the application of ML techniques in neuroscience, medical diagnosis, and healthcare is widespread \cite{Gao2018, Wang2019, Piccialli2021, Ravi2016}. Here especially ML algorithms, which learn from data samples, are common. Such methods can be, for instance, used for side effect prediction of drugs \cite{Zitnik2018} or for analysing or classifying tomography-computed, X-ray, and magnetic resonance images \cite{Roth2015,Yan2016,Anthimopoulos2016,Cao2016,Khan2020,Muhammad2021,Zhang2015,Kleesiek2016}.

Since ML can be applied in nearly every area where enough data is accumulated, it is without question that also a multitude of industries make use of these mechanisms \cite{Ge2017, Akinosho2020, Le2020}. However, ML approaches can even be a powerful tool in reducing greenhouse gas emissions, helping society adapt to a changing climate and model extreme weather events \cite{OGorman2018, Ardabili2019, Rolnick2019}.

Because ML is, as illustrated above, used on so many different problems in various areas, plenty of different methods exist. However, the use of \emph{artificial neural networks} (NNs) \cite{Nielsen2015,Goodfellow2016,Aggarwal2018} is very common. These networks consist of a layered architecture built up of fundamental units, originally inspired by the neurons of a human's neural network. These units are linked via weighted connections, which are updated during a training process. After that, the network ideally is able to master specific tasks, such as classifying pictures \cite{Nielsen2015}. 

The development of such neural networks and machine learning, in general, would be unimaginable without the invention of the modern computer \cite{Ceruzzi2003}. Living in the age of laptops, smartphones and smartwatches and constantly observing arrivals of new technologies can give the expectation of endless growing computational potential. The so-called \emph{Moore's law} \cite{Schaller1997}, describes the observation that the number of transistors on a silicon chip doubles every 18 to 24 months. However, the inevitable end of this law was predicted, and indeed a turning point was reached in 2009 when reducing the transistor dimensionality could not improve the device performance any more \cite{Prati2017}. The demand for new device architectures and information processing methods arose and is even more motivated trough the exponentially increasing amount of data created every day \cite{Hilbert2011}.

One promising candidate can be found in the field of \emph{quantum information} \cite{Nielsen2000}, the study of information processing tasks that can be carried out using quantum mechanical systems: \emph{quantum computers} containing hundreds of quantum bits became experimentally realisable in the last years \cite{Preskill2018, Brooks2019, Arute2019} and give the opportunities to exploit the laws of quantum mechanics to avoid the limits of classical computing. 

These quantum bits, called \emph{qubits}, are physically implemented as two-state devices. Since these have different characteristics than their classical counterparts, it is possible to find quantum algorithms which outperform classical computers for specific tasks \cite{Deutsch1992, Shor1994, Grover1996}. Today's quantum computers do not yet comprise of enough qubits to run the most useful quantum algorithms. However, the arrival and public access \cite{IBMQuantum2021} of these devices is a huge motivation for research in many scientific areas. 

One of them is \emph{quantum machine learning} (QML) \cite{Biamonte2017,Dunjko2018, Cerezo2020}, whose invention was motivated by the great success of classical ML and the big hope invested in quantum computers. Generally we can divide the field into three categories: classical ML with quantum data \cite{Lovett2013,Tiersch2015,Carleo2017}, quantum speed-ups for classical ML\cite{Aimeur2013,Paparo2014,Schuld2014,Wiebe2016}, and quantum algorithms used on quantum data \cite{Sasaki2002,Gambs2008,Sentis2012,Dunjko2016,Alvarez2017,Monras2017,Amin2018,Sentis2019, Du2018, Beer2020}. 

In the first category, quantum data is fed into a classical machine learning algorithm, for instance, to construct representations of many-body systems \cite{Carleo2017} or estimating physical parameters in quantum metrology \cite{Lovett2013}. On the contrary, quantum speed-ups for classical ML are motivated with the hope that quantum information processors, which produce patterns which are classically challenging to create, can probably also recognise patterns, which are difficult to distinguish classically \cite{Biamonte2017}. Here, for example, subroutines of an otherwise classical algorithm are quantised \cite{Aimeur2013} using classical data encoded as quantum states.

In this work, we focus on the last category, which raises the most questions. Here both, the algorithm and the training data are based on quantum mechanics, and in analogy to the classical case, \emph{quantum neural networks} (QNNs) \cite{
	Andrecut2002, 
	Oliveira2008, 
	Panella2011, 
	Silva2016, 
	Cao2017, 
	Wan2017, 
	Alvarez2017, 
	Farhi2018, 
	Killoran2019, 
	Steinbrecher2019, 
	Torrontegui2019, 
	Sentis2019, 
	Tacchino2020, 
	Beer2020, 
	Skolik2020, 
	Zhang2020, 
	Schuld2020, 
	Sharma2020, 
	Zhang2021 
} are a widely spread tool for solving quantum tasks for which no generic quantum algorithm exists. Furthermore, exploiting such QNNs with ML allows to characterise quantum states and operations with fewer samples, compared to quantum state tomography \cite{Vogel1989}, or quantum process tomography \cite{Poyatos1997}, where the number of needed samples increases exponentially with the number of particles \cite{Mohseni2008} and the characterisation of even today's still minimal quantum devices is impossible. Here QNNs can be explicitly useful since they allow to use quantum devices themselves to cope with large amounts of produced quantum data. 

However, the search for the most optimal quantum versions of a neuron, network structure or training algorithm is still ongoing, and the advent of the above mentioned early quantum computers motivates scientists to find executable implementations of QNNs. This work addresses these questions and presents so-called \emph{dissipative quantum neural networks} (DQNNs) \cite{Beer2020}. Their training algorithm is designed for fully quantum learning tasks and allows efficient optimisation with memory requirements scaling only with the width, not the length of the QNN. Moreover, we demonstrate that it can be successfully implemented on today's quantum computers \cite{Beer2021}. 

Due to the rapid progress in quantum learning theory, it is also important to understand the ultimate limits for training methods. Therefore we present the \emph{quantum no free lunch} (QNFL) theorem \cite{Poland2020}. It describes a bound on the probability that a quantum device, which can be modelled as a unitary process and is trained with quantum examples, gives an incorrect output for a random input. This theorem gives us a tool to review the prior demonstrated DQNNs.

The DQNNs discussed before are exclusively trained with training data pairs in form of input and desired output states. We extend this ansatz in two directions. Quantum data is always structured due to the structure of the device producing it. In the first extension we present a variation of the DQNN training algorithm using the graph structure of quantum data \cite{Beer2021a} and demonstrate that we can improve the learning efficiency and the generalisation behaviour by including this additional information.

Whereas both the original DQNN and the DQNN, including graph structure, focus on characterising an unknown quantum operation while learning from the data pairs, we further undertake the task of extending a set of quantum states to states which have similar properties. Hence, in the second extension we follow a generative adversarial approach, where two DQNNs, a generator and a discriminator model, are trained in a competitive manner \cite{Beer2021b}. Learning characteristics of the training data, the generative model is able to produce quantum states with similar properties. The resulting extended quantum data set can be, for example, useful for training other QNN architectures or experiments.  

\subsection*{Outline}
Since the topic is interdisciplinary, we begin with two introductory chapters, whereof one or both can be skipped based on the experience and knowledge of the reader. In \textbf{\cref{chapter:ML}} we introduce classical artificial neurons, neural networks and their training algorithms. This helps to follow the discussion of their quantum analogues. Further, we give an overview of standard methods and applications. 

In contrast, \textbf{\cref{chapter:QI}} introduces the field of quantum information. Here we not only present the characteristics of qubits but also how these can be used in quantum algorithms to outperform classical algorithms. Moreover, we introduce quantum circuits which provides the basis for the implementation of QNNs on today's quantum devices. It is followed by a discussion of quantum computers and an introduction to quantum neural networks.  

The heart of this thesis is \textbf{\cref{chapter:DQNN}}, where the DQNNs are introduced. Here we present an analogue of the classical neuron and explain the architecture of these kinds of QNNs. We define loss functions, which are optimised during the training algorithm. Moreover, we not only show the results of a classical simulation but also of the implementation on actual quantum computers. At the end, we discuss the training behaviour of the DQNN compared to another QNN architecture \cite{Farhi2014, Farhi2016, Hadfield2019}. The following part, \textbf{\cref{chapter:NFL}}, contains an introduction to the classical NFL theorem and the derivative and application of the QNFL theorem.

In \textbf{\cref{chapter:graphs}}, we give a short overview on how graph-structured data is used in classical ML. Afterwards, we explain and demonstrate how to include the graph structure of quantum training data into the optimisation algorithm of the DQNN. Therefore, we formulate new loss functions and rules to update the network during the training and present their usage in classical simulations and quantum device implementation.

We refer to the generative adversarial NNs in \textbf{\cref{chapter:QGAN}}. After explaining their basic concepts in classical ML, we transfer the model to QNNs and construct generative adversarial DQNNs.  Further, we show that these are able to extend data sets to quantum states which have similar properties, given only a few samples. Here we include classical simulations as well as simulations of quantum devices. Finally, \textbf{\cref{chapter:concl}} concludes our results and gives an outlook of potential further research directions.


%% file: text/ML.tex
\chapter{Classical neural networks }\hypertarget{ML}{}
\label{chapter:ML}

For all vertebrate and also most invertebrate animals, the brain is the nervous system's centre. This complex organ and also the whole nervous system is built of billions of fundamental units, referred to as \emph{neurons} \cite{Saladin2011}. The connection of one of these building blocks to others through so-called synapses allows interactions. Whereas the connections within smaller groups of neurons can be recorded, studying the communication between a larger population of these units is very tough \cite{Ahrens2013}.

Therefore in the middle of the 20th century, the first computational models for neural networks were proposed \cite{McClulloch1943,Kleene1956,Householder1958}. Based on these ideas, artificial neural networks, also abbreviated by \emph{neural networks} (NNs)\cite{Farley1954, Rochester1956,Rosenblatt1961} arose and could be performed on the at this time available electronic computers. Soon, NNs were applied to describe biological neural networks. Further, the usage for AI got more attention.

Nowadays, NNs built of many layers of neurons are highly used tools for machine learning (ML) \cite{Murphy2012,Nielsen2015,Goodfellow2016,Aggarwal2018} and applied in endless different fields of research and sectors of industry \cite{Samek2021}. We will discuss some of them throughout the chapter when presenting different NN methods.

In the following we will focus on NN build of simple consecutive layers trained with training data pairs but also refer to other techniques. We start our discussion by explaining the building blocks and different architectures of NNs in \cref{sec:ML_neurons} and \cref{sec:ML_neuralnetworks}, in order to explain the training process later on. It follows an introduction to training data and its usage in \cref{sec:ML_data}. We will further discuss different optimisation techniques used for training the NN architectures in \cref{sec:ML_optimisation}. Since this work's limit allows only giving a brief overview into the huge world of NNs, we refer to \cite{Murphy2012,Nielsen2015,Goodfellow2016} for a comprehensive discussion.

\section{Artificial neurons}
\label{sec:ML_neurons}

The \emph{perceptron}, developed by Frank Rosenblatt in 1958  \cite{Rosenblatt1958}, was the first artificial neural network with complex adaptive behaviour \cite{Rosenblatt1961}. Until today versions of this building block, referred to as \emph{artificial neurons}, are used for NN architectures. 

\begin{figure}[H]
	\centering
	\begin{tikzpicture}[scale=1]
		\node[] (x1) at (0,2) {$x_1$};
		\node[] (x2) at (0,0.75) {$x_2$};
		\node[] (xd) at (0,-0.75) {$\vdots$};
		\node[] (xn) at (0,-2) {$x_n$};
		\node[perceptron0] (p) at (4.5,0) {$\kappa\Big(\sum_i w_i x_i + b\Big)$};
		\node[] (out) at (8,0) {$y$};
		\draw[-stealth,shorten <=7pt, shorten >=7pt] (x1) --node[above=0.05ex] {$w_1$} (p);
		\draw[-stealth,shorten <=7pt, shorten >=7pt] (x2) --node[above=0.05ex] {$w_2$} (p);
		\draw[-stealth,shorten <=7pt, shorten >=7pt] (xn) --node[above=0.05ex] {$w_n$} (p);
		\draw[-stealth,shorten <=7pt, shorten >=7pt] (p) -- (out);
	\end{tikzpicture}
	\caption{\textbf{Artificial neuron.} The building block of NNs takes $n$ inputs and outputs an activation $y$ using the activation function $\kappa$.}
	\label{fig:ML_neuron}
\end{figure}
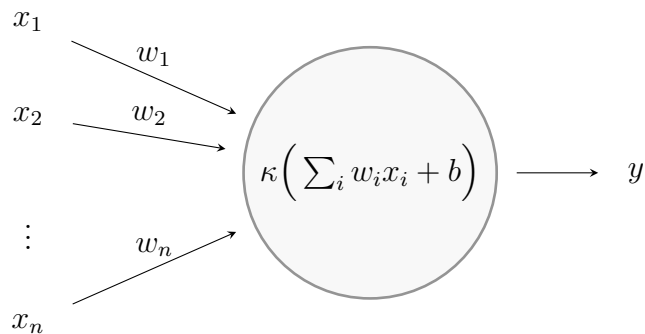

Such a neuron takes $n$ inputs $\{x_1,\hdots,x_n\}$ and has a single binary output $y$, also called \emph{activation} as depicted in \cref{fig:ML_neuron}. Every input has an assigned weight $w_i\in\mathbbm{R}$. Additionally, the neuron is equipped with a bias $b\in\mathbbm{R}$. The neuron's output is computed through 
\begin{equation}
	\mathcolorbox{y=\kappa(z)=\kappa\Big(\sum_i w_i x_i + b\Big),}
	\label{eqn:ML_neuron}
\end{equation}
where $\kappa(z)$ denotes a so-called \emph{activation function}. 

Rosenblatt originally allowed only binary inputs $x_i\in\{0,1\}$ and used the \emph{step function}, see \cref{fig:ML_binarystep}, as the activation function $\kappa$ for the definition of the perceptron. Hence, the output is binary. To intuitively understand the method of a neuron, we can imagine the perceptron's task as deciding between two choices, $0$ and $1$. The inputs $x_i$ can be seen as arguments with different importances $w_i$, where an argument with $w_i x_i<0$ is pro-choice $0$ and $w_i x_i>0$ is pro-choice $1$, respectively. The step function and the bias $b$ describe a threshold. Depending on which side of the threshold the weighted sum $\sum_i w_i x_i$ of the arguments is, the perceptron \enquote{decides} for the output $0$ or $1$.

\FloatBarrier \subsection*{Activation functions}

In the context of NNs, where neurons are layered, we can say that the step function, used in the original definition of the perceptron, \emph{activates} the neuron if the input is above a certain threshold. Otherwise, the neuron is \emph{deactivated}, which means the input will not propagate further through the network, as will be described in \cref{sec:ML_neuralnetworks}. It turns out that using the step function as the activation of a neuron causes problems: sometimes, a slight change in the weights or bias of any single perceptron in the network can flip the perceptron's binary output and therefore also changes completely the output of the NN. This behaviour can damage the training process \cite{Nielsen2015}, and it is aimed that small changes in the input cause only minor changes in the output of NNs. This can be realised by choosing activation functions different from the step function.

\begin{figure}
	\centering
	\begin{subfigure}[b]{0.495\linewidth}
		\centering 
		\begin{equation*}
			\kappa_{\text{step}}(z)=
			\left\{
			\begin{aligned}
				0  & \text{ if } z \geq 0 \\
				1 & \text{ if } z < 0
			\end{aligned}
			\right.
		\end{equation*} 
		\begin{tikzpicture}[scale=.6]
			\begin{axis}%
				[
				grid=major,grid style={color0M},    
				xmin=-5.2,
				xmax=5.2,
				axis x line=middle,
				ymin=-.2,
				ymax=1.2,
				axis y line=middle,
				samples=100,
				]
				\addplot+[const plot, no marks, line width=2, color1] coordinates {(-6,0) (0,0) (0,1) (6,1)} node[] {};
			\end{axis}
		\end{tikzpicture}
		\subcaption{Binary Step}
		\label{fig:ML_binarystep}
	\end{subfigure}
	\begin{subfigure}[b]{0.495\linewidth}
		\centering 
		\vspace{1cm}
		\begin{equation*}
			\kappa_{\text{lin}}(z)=\alpha z
		\end{equation*} 
		\begin{tikzpicture}[scale=.6]
			\begin{axis}%
				[
				grid=major,grid style={color0M},    
				xmin=-5.2,
				xmax=5.2,
				axis x line=middle,
				ymin=-1.2,
				ymax=1.2,
				axis y line=middle,
				samples=100,
				legend style={at={(1,0.5)}}     
				]
				\addplot[color1,line width=2,mark=none]   (x,x);
			\end{axis}
		\end{tikzpicture}
		\subcaption{Linear function}
	\end{subfigure}
	
	\begin{subfigure}[b]{0.495\linewidth}
		\centering 
		\begin{equation*}
			\kappa_{\text{sigmoid}}(z)=\frac{1}{1+e^{-z}}\text{.}
		\end{equation*} 
		\begin{tikzpicture}[scale=.6,declare function={sigma(\x)=1/(1+exp(-\x));
				sigmap(\x)=sigma(\x)*(1-sigma(\x));}]
			\begin{axis}%
				[
				grid=major,grid style={color0M},    
				xmin=-5.2,
				xmax=5.2,
				axis x line=middle,
				ymin=-.2,
				ymax=1.2,
				axis y line=middle,
				samples=100,
				]
				\addplot[color1,line width=2,mark=none]   (x,{sigma(x)});
			\end{axis}
		\end{tikzpicture}
		\subcaption{Sigmoid/logistic function}
	\end{subfigure}
	\begin{subfigure}[b]{0.495\linewidth}
		\centering 
		\vspace{1cm}
		\begin{equation*}
			\kappa_{\text{TanH}}(z) =\frac{2}{1+e^{-2z}}-1
		\end{equation*}
		\begin{tikzpicture}[scale=.6]
			\begin{axis}%
				[
				grid=major,grid style={color0M},    
				xmin=-5.2,
				xmax=5.2,
				axis x line=middle,
				ymin=-1.2,
				ymax=1.2,
				axis y line=middle,
				samples=100,
				legend style={at={(1,0.5)}}     
				]
				\addplot[color1,line width=2,mark=none]   (x,{tanh(\x)});
			\end{axis}
		\end{tikzpicture}
		\subcaption{Hyperbolic tangens}
	\end{subfigure}
	
	\begin{subfigure}[b]{0.495\linewidth}
		\centering 
		\vspace{1cm}
		\begin{equation*}
			\kappa_{\text{ReLU}}(z)=\text{max}(0,z)\text{.}
		\end{equation*} 
		\begin{tikzpicture}[scale=.6]
			\begin{axis}%
				[
				grid=major,grid style={color0M},    
				xmin=-2.2,
				xmax=2.2,
				axis x line=middle,
				ymin=-.2,
				ymax=2.2,
				axis y line=middle,
				samples=100,
				]
				\addplot+[no marks, color1,line width=2,mark=none] coordinates {(-2,0) (0,0) (2,2)} node[] {};
			\end{axis}
		\end{tikzpicture}
		\subcaption{Rectified Linear Unit}
	\end{subfigure}\begin{subfigure}[b]{0.495\linewidth}
		\centering 
		\vspace{1cm}
		\begin{equation*}
			\kappa_{\text{LReLU}}(z)=\text{max}(\alpha z,z)\text{.}
		\end{equation*} 
		\begin{tikzpicture}[scale=.6]
			\begin{axis}%
				[
				grid=major,grid style={color0M},    
				xmin=-2.2,
				xmax=2.2,
				axis x line=middle,
				ymin=-.5,
				ymax=2.2,
				axis y line=middle,
				samples=100,
				]
				\addplot+[no marks, color1,line width=2,mark=none] coordinates {(-2,-.2) (0,0) (2,2)} node[] {};
			\end{axis}
		\end{tikzpicture}
		\subcaption{Leaky Rectified Linear Unit}
		\label{fig:ML_leakyrelu}
	\end{subfigure}
	\vspace{1cm}
	\caption{\textbf{Activation functions.} This figure presents the graphical and mathematical representation of a selections of activation functions.}
	\label{fig:ML_activation}
\end{figure}
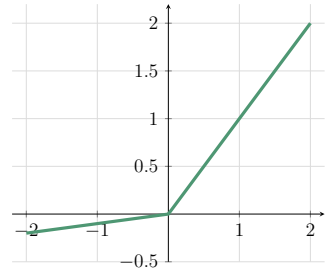

We can generally say that choosing the activation functions wisely is crucial for good training results since the activation functions decide if an input of a neuron is relevant. To better understand the activation process, we discuss the advantages and disadvantages of some standard activation functions in the following. An overview of the named functions can be found in \cref{fig:ML_activation}.

For simplicity, we restrict the discussion to activation functions with one input. This input is often chosen to be the weighted sum $z=\sum_i w_i x_i + b$ of the preceding layer's outputs $x_i$. There exist similar functions with many inputs, defined directly on the outputs $x_i$. One example is the \emph{softmax} function \cite{Goodfellow2016} defined for $n$ outputs $z_i$ via
\begin{equation*}
	\kappa_{\text{softmax}}(z_1,\dots,z_n)_i=
	\frac{\exp^{z_i}}{\sum_{i=j}^n \exp^{z_j}}.
\end{equation*} 

The most straightforward smooth activation would be a \emph{linear function}, for example, the identity. Nevertheless, non-linear activation functions are usually chosen for the activation of neurons, since linear activation functions are problematic. First of all, the derivative of a linear function is a constant. As we will see later when explaining the back-propagation algorithm in \cref{sec:ML_optimisation}, the derivative is used to update the neurons. If the derivative is a constant, it does not include any information about how the network operates on the input data. On the other hand, compositions of linear functions are linear functions again. Thus the whole network, built of many of these neurons, technically collapses to a single-layer network.

On the contrary, non-linear functions create more complex mappings. A common choice is the differentiable and monotonic \emph{Sigmoid function} \cite{Minai1993,Mira1995}, also named \emph{logistic function}. Whereas it transforms the output smoothly into a number $\in[0,1]$, the mean disadvantage of this function is that its gradient is vanishing for very low or high input values. This stops the network's learning process, as will be clear in \cref{sec:ML_optimisation}. Nevertheless neurons using the Sigmoid function for activation, also commonly called \emph{Sigmoid neurons}, find applications, for example in classification tasks \cite{Daqi2005,Tivive2006}. 

The \emph{hyperbolic tangens} (TanH) is similar to the Sigmoid function, but often preferred over it, because it is symmetric around the origin, which makes its output values centred around the zero value. One can express the hyperbolic tangens using the Sigmoid function, namely\begin{align*}
	\kappa_{\text{TanH}}(z) =\frac{2}{1+e^{-2z}}-1 =	2\kappa_{\text{sigmoid}}(2z)-1 \text{.}
\end{align*}
Hence the problems with vanishing gradients are the same.

The \emph{rectified linear unit} (ReLU) deactivates the neuron if the output is smaller than zero. Otherwise, a linear function, for example, the identity, acts. This activation function is computationally very inexpensive. Since the gradient becomes zero for non-positive input values, training algorithms based on the gradients fail in these cases. 

The \emph{leaky rectified linear unit} solves this problem, with slightly descending outputs for inputs smaller than zero, see \cref{fig:ML_leakyrelu}. A further variation of this activation function is the \emph{parametrised} ReLU, where the parameter $\alpha$ is optimised during network training. All versions of ReLU have the disadvantage that the function's output is not always in $[0,1]$, which is often desirable. 


\section{Network architecture}
\label{sec:ML_neuralnetworks}

After the first enthusiasm about perceptrons, it got rapidly clear that these one-layer NNs were quite limited in computational power \cite{Minsky1969}. It was discovered that stacking these early artificial neurons in layers increases the computational power. Whereas with only one of the by Rosenblatt introduced perceptrons, only the learning of linearly separable classes can be performed, i.e.\ a linear hyperplane can divide two classes, with \emph{multi-layer perceptrons} also non-linear classification problems can be solved \cite{HechtNielsen1988,Basheer2000}.

In this section, we describe how to build such multi-layer NNs out of neurons. Since we will only mention a few different NN architectures in the following, but many more exist, we point to \cite{Leijnen2020} for a more complete overview.

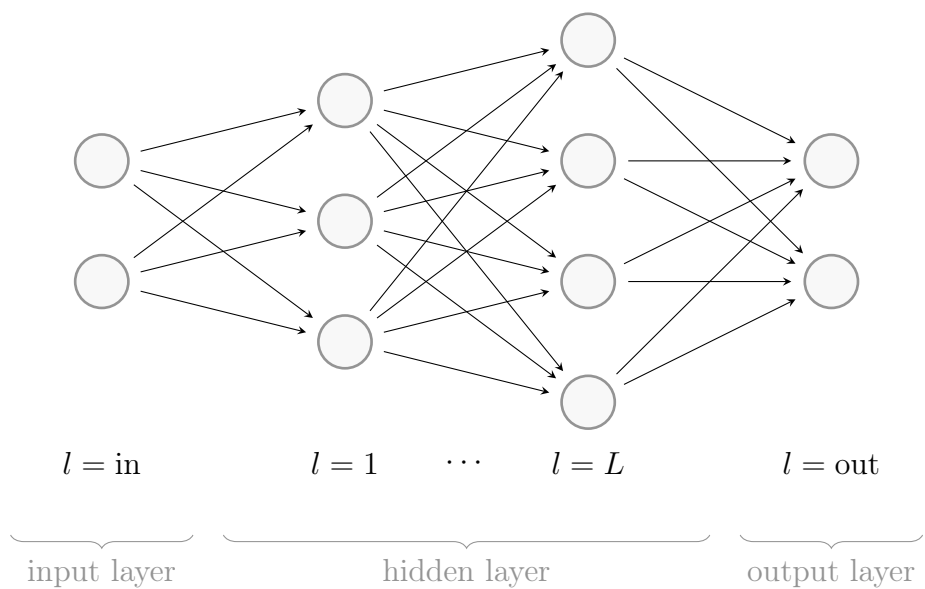
\begin{figure}[h]
	\begin{center}
		\begin{tikzpicture}[scale=1.6]
			\foreach \x in {-.5,.5} {
				\draw[-stealth,shorten <=15pt, shorten >=15pt] (0,\x) -- (2,-1);
				\draw[-stealth,shorten <=15pt, shorten >=15pt] (0,\x) -- (2,0);
				\draw[-stealth,shorten <=15pt, shorten >=15pt] (0,\x) -- (2,1);
			}
			\foreach \x in {-1.5,-0.5, ..., 1.5} {
				\draw[-stealth,shorten <=15pt, shorten >=15pt] (2,-1) -- (4,\x);
				\draw[-stealth,shorten <=15pt, shorten >=15pt] (2,0) -- (4,\x);
				\draw[-stealth,shorten <=15pt, shorten >=15pt] (2,1) -- (4,\x);
			}
			\foreach \x in {-1.5,-0.5, ..., 1.5} {
				\draw[-stealth,shorten <=15pt, shorten >=15pt] (4,\x) -- (6,-0.5);
				\draw[-stealth,shorten <=15pt, shorten >=15pt] (4,\x) -- (6,0.5);
			}
			\foreach \x in {-1,0,1} {
				\node[perceptron0] at  (2,\x) {};
			}
			\foreach \x in {-1.5,-0.5, ..., 1.5} {
				\node[perceptron0] at  (4,\x) {};
			}
			\node[perceptron0] at (0,-0.5){};
			\node[perceptron0] at (0,0.5){};
			\node[perceptron0] at (6,-0.5){};
			\node[perceptron0] at (6,0.5){};
			\node at (0,-2.0){$l=\text{in}$};
			\node at (2,-2.0){$l=1$};
			\node at (3,-2.0){$\cdots$};
			\node at (4,-2.0){$l=L$};
			\node at (6,-2.0){$l=\text{out}$};
			\draw[brace0] 
			(0.75,-2.6) -- node[below=1ex] {input layer} (-0.75,-2.6); 
			\draw[brace0] 
			(5,-2.6) -- node[below=1ex] {hidden layer} (1,-2.6); 
			\draw[brace0] (6.75,-2.6) -- node[below=1ex] {output layer} (5.25,-2.6); 
		\end{tikzpicture}
		\caption{\textbf{Feed-forward NN}. We depict neurons as circles. This figure shows a feed-forward NN with $L$ hidden layers. The connections between the neurons are denoted by arrows.}
		\label{fig:ML_NN}
	\end{center}
\end{figure}

Most often, the neurons in NN are arranged layerwise, see \cref{fig:ML_NN}. We call the first layer of neurons, which get the initial input, the \emph{input layer}. The last layer of neurons is named \emph{output layer}. This layer's output is the final output of the network. The layers in between are called \emph{hidden layers}.

Further, we call networks with many hidden layers \emph{deep} NNs \cite{Goodfellow2016}. While working through a considerable amount of the layers, the original information gets more and more abstract, and in that way, the complex data processing is divided into a series of simple nested assignments.

\FloatBarrier \subsection*{Feed-forward neural networks}

The simplest NN architecture can be found in \emph{feed-forward neural networks} \cite{Bebis1994}. Here the neurons get the output of previous layers neurons as an input, and no loops are built-in. Furthermore, such NNs are often built of \emph{fully connected layers}, i.e.\ layers where all the inputs from one layer are connected to every neuron of the next layer. See \cref{fig:ML_NN} for a depiction.

Despite this simple structure, these networks are widely used, for example, for modelling the spread of COVID-19 \cite{Car2020}, forecasting of wind power \cite{Bhaskar2012}, or studying the drying kinetics of pistachio nuts \cite{Omid2009}. Since many applications of feed-forward NNs are based on pattern recognition\cite{Ramchoun2017,Jin2018,Pelka2019}, we will describe how handwritten digits can be classified using a supervised training ansatz and a feed-forward neural network in \cref{sec:ML_data}.

\FloatBarrier \subsection*{Recurrent neural networks}
\begin{figure}[H]
	\centering
	\begin{tikzpicture}[scale=1.6]
		\foreach \x in {0,1,2} {
			\draw[-stealth,shorten <=17pt, shorten >=17pt] (\x,0) -- (\x,1);
			\draw[-stealth,shorten <=17pt, shorten >=17pt] (\x,1) -- (\x,2);
		}
		\draw[-stealth,shorten <=17pt, shorten >=17pt,color1] (0,1) -- (1,1);
		\draw[-stealth,shorten <=17pt, shorten >=17pt,color1] (1,1) -- (2,1);
		\node[perceptron0, minimum height=0.9cm] at (0,0){$x_0$};
		\node[perceptron0, minimum height=0.9cm] at (1,0){$x_1$};
		\node[perceptron0, minimum height=0.9cm] at (2,0){$x_2$};
		\node[perceptron1, minimum height=0.9cm] at (0,1){};
		\node[perceptron1, minimum height=0.9cm] at (1,1){};
		\node[perceptron1, minimum height=0.9cm] at (2,1){};
		\node[perceptron0, minimum height=0.9cm] at (0,2){$y_0$};
		\node[perceptron0, minimum height=0.9cm] at (1,2){$y_1$};
		\node[perceptron0, minimum height=0.9cm] at (2,2){$y_2$};	
		\draw[-stealth] (-0.5,-0.5) --node[below]{$t$} (2.5,-0.5);
		\draw[-stealth] (-0.5,-0.5) --node[left]{layers} (-0.5,2.5);
	\end{tikzpicture}
	\caption{\textbf{Recurrent neural network.} This figure represents an RNN consisting of a one-neuron input, hidden (colored) and output layer in tree steps $t\in\{0,1,2\}$.}
	\label{fig:ML_RNN}
\end{figure}
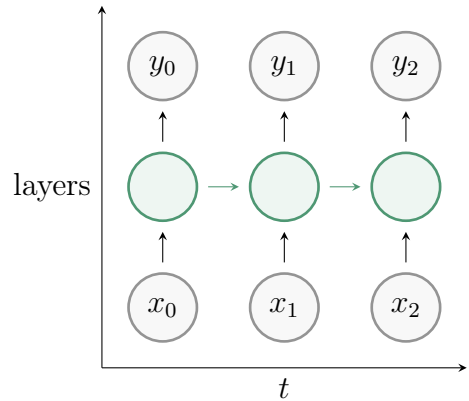

\emph{Recurrent neural networks} (RNNs) \cite{Elman1990, Mikolov2010, Sundermeyer2012, Goodfellow2016, Aggarwal2018} are constructed in a more complicated way than the feed-forward NNs we discussed above. Such RNNs can be used for working with data sequences, which can be, for example of temporal order as in speech recognition \cite{Graves2006,Graves2012,Graves2012a}, video analysis \cite{Baccouche2011,EbrahimiKahou2015,Donahue2015} or language processing tasks\cite{Bahdanau2014,Sutskever2014}. Instead of, for instance, the whole text is being fed into the algorithm as one single input, the RNN allows several inputs $x_t$ during the time the algorithm is running, for example, the single words, and gives also outputs $y_t$ during the running process. These outputs are not only based on the according input but also on the hidden layers of the preceding time step $t-1$. \cref{fig:ML_RNN} shows the simplest version of an RNN. 

\FloatBarrier \subsection*{Convolutional neural networks}
Yet another NN structure can be found in \emph{convolutional neural networks} (CNN) \cite{Lawrence1997,Kalchbrenner2014,Goodfellow2016, Kim2017,Aggarwal2018}. Besides fully connected layers, as we used in the feed-forward NNs, also so-called \emph{convolutional layers} are used to shape networks of this class. These layers use the convolution of the layer's input, often inserted in matrix form,  with another matrix, often called \emph{kernel}. 

A common example is the input of a grey-scale image in form of a matrix. Sharpening the image can be executed by executing a convolution between the image matrix and a kernel, which is in this case a sharpening matrix, see \cref{fig:ML_Convolution}.
\begin{figure}[H]
	\centering
	\begin{tikzpicture}[mmat/.style={matrix of math nodes,column sep=-\pgflinewidth/2,
			row sep=-\pgflinewidth/2,cells={nodes={draw,inner sep=2pt,thin}},draw=#1,thick,inner sep=0pt},
		mmat/.default=black,
		node distance=0.3em]
		\matrix[matrix of nodes,row sep=-\pgflinewidth,style={nodes={rectangle,draw=color0M,minimum width=1.75em}}](mat1){
			3& 3& 4& 3& 3&3\\ 
			3&4& 5& 6& 4&4\\ 
			3&4&18& 20& 15&4\\ 
			4&7& 25& 43& 14&4\\ 
			3&4&18& 16& 11&4\\ 
			3&4&4&6&9&4\\
		};
		\node[fit=(mat1-3-3)(mat1-5-5),inner sep=0pt,draw,color2,thick](f1){};  
		\node[right=of mat1] (mul) {$*$};  
		\matrix[matrix of nodes,row sep=-\pgflinewidth,style={nodes={rectangle,draw=color2,fill=color2L,minimum width=1.5em}},right=of mul](mat2){  
			-1 & -1 & -1 \\ 
			-1 & 5 & -1  \\
			-1 & -1 & -1  \\ };
		\node[above=of mat2] (s) {sharpening kernel}; 		
		\node[right=of mat2] (eq) {$=$};  
		\matrix[matrix of nodes,row sep=-\pgflinewidth,style={nodes={rectangle,draw=color0M,minimum width=2.2em}},right=of eq](mat3){ 
			-23&-37&-42&-38\\
			-49&-24&-30&-24\\
			-44&-5&|[draw=color2,thick,fill=color2L,alias=4]|78&-47\\
			-48&-19&-50&-45\\
		};
		\begin{scope}[on background layer]
			\fill[color2L] (f1.north west) rectangle (f1.south east);
		\end{scope}
	\end{tikzpicture}
	\caption{\textbf{Matrix convolution.} In this example for a convolution of a matrix with a sharpening kernel the computation of one matrix element is highlighted for demonstration. The number $78$ is the sum of the element-wise multiplications of the highlighted matrix elements.}
	\label{fig:ML_Convolution}
\end{figure}

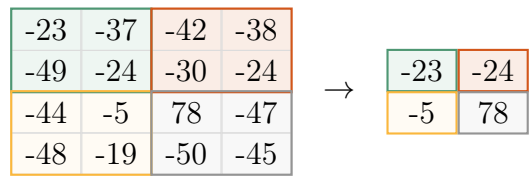
\begin{figure}[H]
	\centering
	\begin{tikzpicture}[mmat/.style={matrix of math nodes,column sep=-\pgflinewidth/2,
			row sep=-\pgflinewidth/2,cells={nodes={draw,inner sep=2pt,thin}},draw=#1,thick,inner sep=0pt},
		mmat/.default=black,
		node distance=0.3em]
		\matrix[matrix of nodes,row sep=-\pgflinewidth,style={nodes={rectangle,draw=color0M,minimum width=2.2em}}](mat1){ 
			-23&-37&-42&-38\\
			-49&-24&-30&-24\\
			-44&-5&78&-47\\
			-48&-19&-50&-45\\
		};
		\node[fit=(mat1-1-1)(mat1-2-2),inner sep=0pt,draw,color1,thick](f1){}; 
		\node[fit=(mat1-3-1)(mat1-4-2),inner sep=0pt,draw,color2,thick](f2){}; 
		\node[fit=(mat1-1-3)(mat1-2-4),inner sep=0pt,draw,color3,thick](f3){}; 
		\node[fit=(mat1-3-3)(mat1-4-4),inner sep=0pt,draw,color0,thick](f4){}; 
		\node[right=of mat1] (eq) {$\rightarrow$};  
		\matrix[matrix of nodes,row sep=-\pgflinewidth,style={nodes={rectangle,draw=color0M,minimum width=2.2em}},right=of eq](mat3){ 
			|[draw=color1,thick,fill=color1L,alias=4]|-23&|[draw=color3,thick,fill=color3L,alias=4]|-24\\
			|[draw=color2,thick,fill=color2L,alias=4]|-5&|[draw=color0,thick,fill=color0L,alias=4]|78\\
		};
		\begin{scope}[on background layer]
			\fill[color1L] (f1.north west) rectangle (f1.south east);
			\fill[color2L] (f2.north west) rectangle (f2.south east);
			\fill[color3L] (f3.north west) rectangle (f3.south east);
			\fill[color0L] (f4.north west) rectangle (f4.south east);
	\end{scope}\end{tikzpicture}
	\caption{\textbf{Max pooling.} An example for pooling is max pooling, where the filter selects the biggest element of every array.}
	\label{fig:ML_pooling}
\end{figure}

In many applications, additional layers which reduce the number of parameters of their input are used. These layers are referred to as \emph{pooling layers}. An example for such a filter is \emph{max pooling}, where the filter selects the biggest value of an array, exemplary depicted in \cref{fig:ML_pooling}. Furthermore, \emph{average pooling} is common, where every array is replaced by its average value.

A typical CNN starts with a convolutional layer, followed by additional convolutional layers or pooling layers and ends with a fully connected output layer. The architecture of CNNs was inspired by the neurobiological model of the visual cortex \cite{Hubel1962}. 

Since CNNs are good tools for analysing image data \cite{Lecun1998}, nowadays CNNs have a big impact on many domains, for example, health informatics \cite{Ravi2016}. These kinds of NNs are used, for example, on tomography-computed images to categorise them by body-parts \cite{Roth2015,Yan2016} or to classify diseases \cite{Anthimopoulos2016}. Other applications are based on X-ray images, for instance to classify tuberculosis \cite{Cao2016} or diagnosis of COVID-19 \cite{Khan2020,Muhammad2021}. Furthermore, magnetic resonance images of the brain are processed \cite{Zhang2015,Kleesiek2016}.

\begin{figure}[h!]
	\centering
	\begin{tikzpicture}[scale=1.85]
		\node[] (pz) at (-1,0) {};
		\node[perceptron1] (a) at (0,0) {};
		\node[perceptron1] (b) at (1,-.5) {};
		\node[perceptron1] (c) at (1,.5) {};
		\node[] (pg) at (2,0) {};
		\draw[-stealth,shorten <=2pt, shorten >=2pt] (a) -- (b);
		\draw[-stealth,shorten <=2pt, shorten >=2pt] (a) -- (c);
		\node[color1] (G) at (.5,0) {$G$};
		\draw[-stealth,shorten <=4pt, shorten >=4pt,color0] (pz) -- (-.25,0);
		\draw[-stealth,shorten <=4pt, shorten >=4pt,color0] (1.3,0) -- (pg);
		
		\node[perceptron0,ellipse] (ellipse) at (-.25,-1.5) {training data};
		\draw[brace0](2.45,.75)--  (2.45,-1.75) ;
		\node[] (pd) at (2,-1.5) {};
		\draw[-stealth,color0,shorten <=10pt] (ellipse) -- (pd);
		
		\begin{scope}[shift={(4,-.55)}]
			\node[] (pd) at (2,0) {};
			\node[perceptron2] (d) at (0,-.5) {};
			\node[perceptron2] (e) at (0,.5) {};
			\node[perceptron2] (f) at (1,0) {};
			\draw[-stealth,shorten <=4pt, shorten >=2pt] (d) -- (f);
			\draw[-stealth,shorten <=4pt, shorten >=2pt] (e) -- (f);
			\node[color2] (D) at (.5,0) {$D$};
			\draw[-stealth,shorten <=4pt, shorten >=4pt,color0] (-1,0) -- (-.2,0);
			\draw[-stealth,shorten <=4pt, shorten >=4pt,color0] (1.2,0) -- (pd);
			
		\end{scope}
	\end{tikzpicture}
	\caption{\textbf{GAN.} The generative NN $G$ produces data, whereas the discriminative NN $D$ has the training goal to distinct between the by $G$ produced data and the training data.}
	\label{fig:ML_adver}
\end{figure}
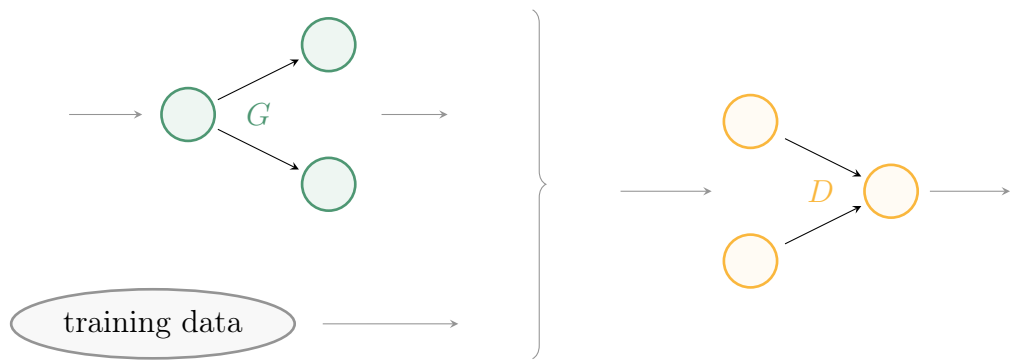

\FloatBarrier \subsection*{Adversarial neural networks}
The task of handwritten digit recognition was mentioned above and will be studied in the following two sections. Such a problem is a typical classification task. A different goal would be to train a NN to be able to \emph{produce} such digit images. We call such a NN model \emph{generative}. An adversarial process, where two NNs are trained, see \cref{fig:ML_adver}, can be used to train such a \emph{generative adversarial network} (GAN) \cite{Goodfellow2014}: the generative model $G$ captures the data distribution and produces data, whereas the other model, referred to as \emph{discriminative model} $D$ estimates the probability $D(\vec{x})$ that a sample $\vec{x}$ came from the training data rather than from $G$. Both of the NNs have different training goals and are well trained at the end. Therefore, we are left with a perfectly well-trained generator. We will explain the training process in detail in \cref{sec:QGAN_classicalGAN}.

\section{Training data and loss functions}
\label{sec:ML_data}

So far, we only described the information flow of input data through the NN. The neurons, and hence the NN, are parametrised through weights and biases, see \cref{eqn:ML_neuron}, and sometimes parameters of the activation function, for example, like in the parametrised ReLu, defined in \cref{fig:ML_leakyrelu}. What is missing now is an algorithm to optimise the parameters in a way that the NN satisfies our standards.

To explain the comparison on which the parameter's update is based, we first have to define a training goal. As mentioned in the introduction, we focus on \emph{supervised learning} tasks in this chapter. These kinds of learning tasks are based on \emph{training data}, most often structured in pairs, each containing an input and the desired output value.

At this point, an example comes in handy: one of the most, if not \emph{the} most famous data set is the MNIST data set introduced in \cite{Lecun1998} and available at \cite{MNIST}. This set contains thousands of greyscale $28\times28=784$-pixel images of handwritten digits and the suiting label of the set $\{0,1,\dots,9\}$.

\begin{figure}[H]
	\centering
	\begin{tikzpicture}
		\node at (0,0){$x_i\equalhat\hspace{1cm},\,\,{d_i}=(0,0,1,0,0,0,0,0,0,0,0)^T$};
		\node at (-2.5,0){\includegraphics{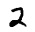}};
	\end{tikzpicture}
	\caption{\textbf{MNIST data set.} In this figure an exemplary input, a 784-pixel image, and desired output, a vector in $\mathbbm{R}^{10}$ are depicted.}\label{fig:ML_MNIST}
\end{figure}

Based on this data set, the task, a typical classification problem, is clear: we want to train the NN so that we can feed it an unseen image saved as a vector ${x}$ of length $784$ and want to get the correct digit of the set $\{0,1,\dots,9\}$. Due to the characteristics of the training algorithm, we convert the digits into vectors of dimension 10. See \cref{fig:ML_MNIST} for an example.

To test the training process it makes sense to divide the provided training data set of length $N$ into a \emph{supervised training set} $\mathcal{S}_\text{SV}$ containing $S$ pairs and in a \emph{unsupervised validation set} $\mathcal{S}_\text{USV}$ build of the remaining data. The latter stays unseen by the network during the training and is only used for validation. We can denote the two sets as
\begin{align*}
	\mathcal{S}_\text{SV}&=\big\{\{{x}_i,{d_i}\}\big\}_{i=1}^{S}\\
	\mathcal{S}_\text{USV}&=\big\{\{{x}_i,{d_i}\}\big\}_{i=S+1}^{N}.
\end{align*}
The original MNIST data contains, for example, $S=60000$ training pairs and $N-S=10000$ testing pairs \cite{MNIST}.

\FloatBarrier \subsection*{Training and validation loss}

Given a NN suiting the MNIST example, i.e.\ one with a 784-neuron input layer and a 10-neuron output layer, we can load an image saved as ${x_i}$ to the network and feed-forward the information as described in \cref{sec:ML_neuralnetworks}. We suppose further that we get a vector of dimension $10$ as an output, consisting of the by the NN estimated probabilities that ${x_i}$ has a specific label. We denote the process with a function $\mathcal{E}_{{w},{b}}:\mathbbm{R}^{784}\rightarrow \mathbbm{R}^{10}$, which depends on the weights and biases, saved in vectors ${w}$ and ${b}$. The output of the network can then be denoted as $\mathcal{E}_{{w},{b}}({x_i})$.

Independently from the supervised training task it is indispensable to compare the network's output $\mathcal{E}_{{w},{b}}({x_i})\in{R}^{10}$ with the desired output ${d_i}\in{R}^{10}$. This is done by a \emph{loss function}, also referred to as \emph{cost function} or \emph{objective function}. Not only should this function compare the two pieces of information, but it should also converge to a global extreme point, which is achieved if the training goal is accomplished. Further, it is optimal if the gradient of the loss is continuous since the below-presented algorithm is based on it. 

A common loss function is the \emph{mean squared error} (MSE)
\cite{Lehmann2006, Wang2009}, also often simply called the \emph{quadratic cost}. Given the network's output $\mathcal{E}_{{w},{b}}({x_i})$ and the desired output ${y_i}$ we can use the MSE to form the \emph{training loss}
\begin{equation}
	\mathcolorbox{\mathcal{L}_\text{SV,MSE}({w},{b})=\frac{1}{S}\sum_{i=1}^S||{d_i}-\mathcal{E}_{{w},{b}}({x_i})||^2.}
	\label{eqn:ML_trainingloss}
\end{equation}
If this loss functions gets small during training we are sure that the algorithm found weights and biases which suite the aim. For validation, we can define the \emph{validation loss}
\begin{equation}
	\mathcolorbox{\mathcal{L}_\text{USV,MSE}({w},{b})=\frac{1}{N-S}\sum_{i=S+1}^N||{d_i}-\mathcal{E}_{{w},{b}}({x_i})||^2.}
	\label{eqn:ML_validationloss}
\end{equation}

Note that the validation is a not neglectable part of the training since, in some cases, the training leads to parameters that let the NN perform very well on the training data set but cannot generalise the same performance to unseen data. We refer to this issue as \emph{overfitting} and can detect it with the validation data set. Overfitting can be avoided with \emph{early stopping}\cite{Goodfellow2016}, where the implementation includes stopping the training to the time where the validation loss $\mathcal{L}_\text{USV}$ begins to rise again, although the training loss error $\mathcal{L}_\text{SV}$ still gets smaller.

\FloatBarrier \subsection*{Choice of the loss function}
The choice of the loss depends on the learning problem and method. For example, the MSE is very sensitive towards outliers, due to the quadratic behaviour. In cases where this causes problems, for example where the input is not normally distributed, the \emph{mean absolute error} (MAE), namely
\begin{equation*}
	\mathcal{L}_\text{SV,MAE}({w},{b})=\frac{1}{S}\sum_{i=1}^S|{d_i}-\mathcal{E}_{{w},{b}}({x_i})|,
\end{equation*}
can be an alternative. Note that the validation loss can be defined for MAE in analogy to \cref{eqn:ML_validationloss}. This also holds for the loss functions presented in the following.

Another common loss function for regression problems is the \emph{smooth mean absolute error}, also called \emph{Huber loss}\cite{Huber1964}. This loss function is defined piecewise and depends on an additional parameter $\zeta$, namely  
\begin{equation*}
	\mathcal{L}_\text{SV,Huber}({w},{b})=\frac{1}{S}\sum_{i=1}^S\begin{cases} \frac{1}{2}({d_i}-\mathcal{E}_{{w},{b}}({x_i}))^2 &\mbox{for } {d_i}-\mathcal{E}_{{w},{b}}({x_i})\le \zeta \\
		\zeta (|{d_i}-\mathcal{E}_{{w},{b}}({x_i})|-\frac{1}{2}\zeta) & \mbox{otherwise. } \end{cases}
\end{equation*}

Loss functions containing the entropy were proved to be useful for classification tasks \cite{Rubinstein1999, Rubinstein2001,Rubinstein2004,Tewari2007}. The most simple case is binary classification, where we have training data pairs, containing an input $x_i$ and a label $d(x_i)=0$ or $d(x_i)=1$. The network's output for such an input is the probability $\mathcal{E}_{{w},{b}}({x_i})=p_{x_i}(0)$ that the input is labelled with $d(x_i)=0$. Since $1-p_{x_i}(0)=p_{x_i}(1)$ we can use the \emph{binary cross entropy} (BCE) as loss function, i.e.\
\begin{equation*}
	\mathcal{L}_\text{SV,BCE}(w,b)=-\frac{1}{S}\sum_{i=1}^S d(x_i)\log(\mathcal{E}_{{w},{b}}({x_i}))+(1-d(x_i))\log(1-\mathcal{E}_{{w},{b}}({x_i})).
\end{equation*}
From this definition, it gets clear that minimising this loss function leads to a correctly labelling NN.

When working with more then two, namely $C$, different classes, the class labels are unit vectors, similar to the example in \cref{fig:ML_MNIST}. The NN's output has to be a $C$ dimensional vector $\mathcal{E}_{{w},{b}}({x_i})=p_{x_i}$ with $p_{x_i,j}\in \mathbbm{R}^+_0$. Here the general \emph{cross entropy} (CE) is used to formulate the loss function
\begin{equation*}
	\mathcal{L}_\text{SV,CE}(w,b)=-\frac{1}{S}\sum_{i=1}^S\sum_{j=1}^C d(x_i)_j\log\left(\mathcal{E}_{{w},{b}}({x_i})_j\right).
\end{equation*}

As an example we can assume $C=3$ classes and the label of a specific training data $i$ input is $d(x_i)=\{0,1,0\}$, i.e.\ it belongs to the second class. The network predicts the probabilities $p_{x_i}=\{0.2,0.5,0.3\}$, hence 50 precent for the correct label. The according summand in $\mathcal{L}_\text{SV,BCE}$ is then $0\times 0.2+1\times 0.5+0 \times 0.3$.

Since classification tasks are a vast field in ML, many other loss functions were studied. Alternatives to the CE are, for example, the \emph{pairwise loss} \cite{Hadsell2006}, which runs over pairs of training data inputs, or the \emph{triplet loss} \cite{Schroff2015} working with the assumption that a training data input is closer to all inputs with the same label, then to those which are labelled differently. More definitions of loss functions can be found, for example, at \cite{Goodfellow2016}.

\FloatBarrier \subsection*{Unsupervised learning approaches}

The so far discussed approaches are supervised \cite{Moller1990,Jordan1992,Caruana2006}. These model the relationships or patterns between the input features and the target prediction output. At the end of the process, the NN can imitate the relation and predict output values for unseen data. For these kinds of models, having enough and proper training data is key.

Beyond that, also \emph{unsupervised} algorithms \cite{Le2013,Radford2015,Srivastava2015} are used. These methods do not rely on labelled input and output data and are especially useful in cases where it is unclear what kind of patterns to look for in the data. An example of unsupervised learning is \emph{$k$-means clustering} \cite{Wagstaff2001,Krishna1999,Jain2010}. The algorithm aims to partition data into $k$ clusters given a set of vectors and minimises the within-cluster sum of squares. Note that $k$ is fixed beforehand. One method used for finding the number of clusters is, for example, the \emph{elbow method} \cite{Bholowalia2014}.

There are also hybrid versions of unsupervised and supervised learning. These are often used for classification problems, where only a small subset of the data is labelled because obtaining labels is expensive or impossible. In these algorithms, the classification is based not only on the labelled part of the data but also on all data bits. Applications for \emph{semi-supervised learning}\cite{Chapelle2009,Kingma2014} are for example in speech analysis \cite{Liu2013}, image search \cite{Fergus2009} or genomics \cite{Shi2011}.

An examples for a semi-supervised ansatz will be presented in \cref{sec:graphs_classical}, where \emph{machine learning with graphs} \cite{Zhu2003,Liu2019a,Hamilton2020} is discussed. A graph describes the connection between different vertices, and only some of the vertices are labelled. These labels can be, for example, of the form of input and output data pairs as we met them in the above explained supervised ansatz. In addition to the supervised data, the training algorithms then also use the information implemented via a graph structure.

A third large area of machine learning alongside supervised and unsupervised learning is \emph{reinforcement learning} \cite{Sutton2018}. This method is modelled as a \emph{Markov decision process}, where we have a set of environment and agent states, a set of actions and the probability of transition from one state to another under a specific action. Further, some rules describe what the agent observes, and a (positive or negative) reward is given after transitioning from one state to another with a specific action. The training goal is to maximise the sum of rewards.  

\section{Optimisation}
\label{sec:ML_optimisation}

We have already argued that layering neurons is very beneficial in contrast to training single neurons. However, these multi-layer networks were not helpful first since no suiting training algorithm could be found. This changed with the invention of the \emph{back-propagation algorithm} \cite{Werbos1974, Rumelhart1986,Werbos1990}, which will be discussed at the end of this section.

The all-over optimisation procedure using such an algorithm can be defined as follows: we feed information through the network and get output data. This output data is compared with some desired output, as discussed in \cref{sec:ML_data}. Based on this comparison, the parameters are updated. This process is repeated until the desired accuracy is reached. We will describe the update of the parameters in the following, assuming to train a simple feed-forward NN. 

\FloatBarrier \subsection*{Gradient descent}
\label{subsec:ML_gradientdescent}
The training algorithm improves the network by changing the before training randomly initialised parameters ${w}_0$ and ${b}_0$. By definition this is done by optimising the training loss discussed in \cref{sec:ML_data}, i.e.\ finding a global extreme point of $\mathcal{L}_\text{SV}( w, b )$. 

The most common method for this purpose is \emph{gradient descent}, which was proposed for non-linear optimisation problems by \cite{Curry1944} based on earlier contributions of Augustin-Louis Cauchy and Jacques Hadamard. With this technique, a local minimum can be found by changing the parameters $w_t$ and $b_t$ into the direction of the gradient of the loss function concerning the parameters, namely
\[
\mathcolorbox{\begin{aligned}
		{w}_{t+1}& ={w}_{t}-\eta \nabla_{{w_t}}\mathcal{L}_\text{SV}( w_t, b_t )\\
		{b}_{t+1}&={b}_{t}-\eta \nabla_{{b_t}}\mathcal{L}_\text{SV}( w_t, b_t ),\nonumber
\end{aligned}}
\]
where $t$ denotes the training \emph{step} or \emph{epoch} and $\eta$ the \emph{learning rate}. Although there is mathematically no guarantee for success, numerical experiments have shown that the method reliably finds global minima of deep NNs \cite{Du2019}.

\FloatBarrier \subsection*{Alternatives to gradient descent}
\label{subsec:ML_alternativesGradient}
Finally, we want to name some optimisation methods different from the above-described gradient descent method. A more detailed overview of such methods and their variations can be found at \cite{Goodfellow2016}, the comparison of different proposals in \cite{Schaul2013, Soydaner2020}. We can summarise the gradient descent method via the rule
\begin{equation*}
	{\vartheta}_{t+1} ={\vartheta}_{t}-\eta \nabla_{{\vartheta_t}}\mathcal{L}_\text{SV}(\vartheta_t),
\end{equation*}
for updating a parameter $\vartheta$ and name some adaptions of this rule in the following. 

To optimise the computing time of each training step, \emph{stochastic gradient descent} (SGD) can be utilised. Whereas the in \cref{subsec:ML_gradientdescent} described method uses the loss function evaluated on all $S$ training pairs, here for each training epoch only one random training pair is chosen, the loss function evaluated and the gradients computed \cite{Bottou1998}. A good compromise is \emph{mini-batch stochastic gradient descent}, where a randomly chosen set of $M<S$ training pairs is selected and used instead of just one \cite{Li2014}.

A faster converging technique is the \emph{momentum method}\cite{Polyak1964, Sutskever2013}, where we update the parameter
is done via ${\vartheta}_{t+1} ={\vartheta}_{t}+M(t)$ while preserving the momentum of the last update via an additional parameter $\mu$ by using
\begin{equation*}
	M(t+1) =\mu M(t)- \eta \nabla_{{\vartheta_t}}\mathcal{L}_\text{SV}(\vartheta_t).
\end{equation*}
For very steep parts of the parameter change, the momentum gets bigger and finding the extrema is accelerated. On the other hand, with this method, the change of the parameter gets small if we get near the turning point, and overshooting it is less likely.

So far, we treated the learning rate $\eta$ as a beforehand fixed and unchanged parameter during the training. On the contrary, some optimisation methods use \emph{adaptive learning rates} changing individually for each parameter during the training \cite{Goodfellow2016}. One of the first proposed adaptive learning technique was the \emph{detla-bar-delta algorithm} \cite{Jacobs1988}, which is based on a simple idea: if the partial derivative of the loss with respect to the parameter remains of the same sign, the learning rate increases, otherwise it decreases.

A very common optimisation algorithm using adaptive learning rates is \emph{Adam}\cite{Kingma2014a}, a synonym for \emph{adaptive moments}. It can be seen as an improvement of two previously invented methods, \emph{AdaGrad}\cite{Duchi2011}, and  \emph{RMSProp}\cite{Hinton2012}: Using AdaGrad, the learning rates scale inversely proportional to the square root of the sum of all squared values of the earlier epoch's gradients. RMSProp is a slight improvement of this technique as it uses the squared gradients of earlier epochs primarily. The method Adam is based on RMSProp, but also includes techniques of the momentum method.

\FloatBarrier \subsection*{Back-propagation algorithm}

When it comes to calculating the gradients of the loss needed for the parameter update, it is often referred to the already above-mentioned \emph{back-propagation algorithm} \cite{Werbos1974, Rumelhart1986,Werbos1990}. This algorithm can be applied for loss functions that can be written as averages over the loss for a specific training pair since the derivatives of the training loss are computed through the averaging of the derivatives of the training loss of single training examples. Further, the loss has to be a function of the NN's output $\mathcal{E}_{{w},{b}}({x_i})$. Both is the case for the in \cref{eqn:ML_trainingloss} defined training loss. A very detailed and graphic description can be found in \cite{Nielsen2015}. In the following, we will present a summary. 

Since the parameters denoted in the vectors $w$ and $b$ change during the training epochs, their value depends on $t$. Due to that also the training loss $\mathcal{L}_\text{SV}$ and the activations $y^l$ depend on $t$. Note that for convenience, we drop the parameter $t$ in the description of the gradient evaluation for a specific epoch, in the following.

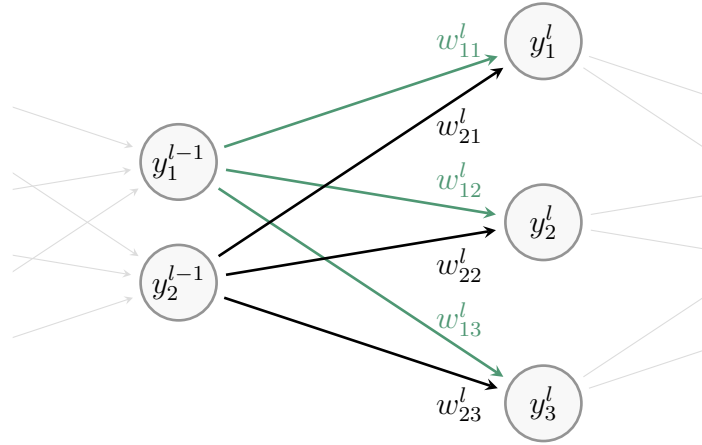
\begin{figure}
	\centering
	\begin{tikzpicture}[scale=1.6]
		\draw[-stealth,shorten <=60pt, shorten >=18pt,color0M] (-3,1.5) -- (0,.5);
		\draw[-stealth,shorten <=60pt, shorten >=18pt,color0M] (-3,1.5) -- (0,-.5);
		\draw[-stealth,shorten <=60pt, shorten >=18pt,color0M] (-3,0) --  (0,.5);
		\draw[-stealth,shorten <=60pt, shorten >=18pt,color0M] (-3,0) -- (0,-.5);
		\draw[-stealth,shorten <=60pt, shorten >=18pt,color0M] (-3,-1.5) --  (0,.5);
		\draw[-stealth,shorten <=60pt, shorten >=18pt,color0M] (-3,-1.5) -- (0,-.5);
		
		\draw[-stealth,shorten <=18pt, shorten >=18pt,line1] (0,.5) -- node[color1,pos=0.77,above=0ex] {$w^l_{11}$}  (3,1.5);
		\draw[-stealth,shorten <=18pt, shorten >=18pt,line1] (0,.5) -- node[color1, pos=0.77,above=0ex] {$w^l_{12}$}  (3,0);
		\draw[-stealth,shorten <=18pt, shorten >=18pt,line1] (0,.5) -- node[color1,pos=0.77,above=0ex] {$w^l_{13}$}  (3,-1.5);
		\draw[-stealth,shorten <=18pt, shorten >=18pt,lineD] (0,-.5) -- node[pos=0.77,below=0ex] {$w^l_{21}$} (3,1.5);
		\draw[-stealth,shorten <=18pt, shorten >=18pt,lineD] (0,-.5) -- node[pos=0.77,below=0ex] {$w^l_{22}$} (3,0);
		\draw[-stealth,shorten <=18pt, shorten >=18pt,lineD] (0,-.5) -- node[pos=0.77,below=0ex] {$w^l_{23}$} (3,-1.5);
		\foreach \x in {-0.5,0.5} {
			\node[perceptron0, minimum width=1cm] at  (0,\x) {};
		}
		\foreach \x in {-1.5,0,1.5} {
			\node[perceptron0, minimum width=1cm] at  (3,\x) {};
		}
		\node[] at (0,0.5){$y^{l-1}_1$};
		\node[] at (0,-0.5){$y^{l-1}_2$};
		\node[] at (3,1.5){$y^{l}_1$};
		\node[] at (3,0){$y^{l}_2$};
		\node[] at (3,-1.5){$y^{l}_3$};
		
		\draw[-stealth,shorten <=18pt, shorten >=60pt,color0M] (3,1.5) -- (6,.5);
		\draw[-stealth,shorten <=18pt, shorten >=60pt,color0M] (3,1.5) -- (6,-.5);
		\draw[-stealth,shorten <=18pt, shorten >=60pt,color0M] (3,0) --  (6,.5);
		\draw[-stealth,shorten <=18pt, shorten >=60pt,color0M] (3,0) -- (6,-.5);
		\draw[-stealth,shorten <=18pt, shorten >=60pt,color0M] (3,-1.5) --  (6,.5);
		\draw[-stealth,shorten <=18pt, shorten >=60pt,color0M] (3,-1.5) -- (6,-.5);
		
		\node[operator1, minimum height=5cm, minimum width=2cm, draw=white, fill=white] at (-2,0) {};
		\node[operator1, minimum height=5cm, minimum width=2cm, draw=white, fill=white] at (5,0) {};
	\end{tikzpicture}
	\caption{\textbf{Notation for back-propagation algorithm.} Two layers $l-1$ and $l$ of a feed-forward NN containing two and three neurons are connected via weights $\{w^l_{11},\dots ,w^l_{23}\}$.}
	\label{fig:ML_backprop}
\end{figure}

For every layer $l$ we save the weights of the neurons connection in a matrix $w^l$, where $w^l_{j,k}$ is the weight from the $k$th neuron in layer number $l-1$ to the $j$th neuron in layer $l$. See \cref{fig:ML_backprop} for an example. Further, we describe the biases in in vectors ${b^l}$, where $b^l_j$ describes the bias of the $j$th neuron in the $l$th layer. 

Since we work with feed-forward NNs, we can use $w^l$, ${b^l}$ and the activation function $\kappa$ to express the activation of layer $l$ trough the activation of layer $(l-1)$, namely
\begin{equation*}
	{y^l}=\kappa\left(z^l\right)=\kappa\left(w^l{y^{l-1}}+{b^l}\right),
\end{equation*}
where the component $z^l_j$ of $z^l$ is the weighted input to the activation for neuron $j$ in layer $l$. 

Since the loss function describes the difference between the desired output and the networks output we define the \emph{error} of the $j$th neuron in layer $l$ as
\begin{equation*}
	\delta^l_j=\frac{\partial \mathcal{L}_\text{SV}( w, b )}{\partial z^l_j},
\end{equation*}
and describe the errors in layer $l$ as vector $\delta^l$.

In the following we will use ${a}\odot{b}$ to denote the element-wise product, i.e.\ $({a}\odot{b})_i=a_i b_i$, also referred to as Hadamard product. Due to the chain rule we can express the error of the last layer, namely layer $L$, as
\begin{equation*}
	\delta^L=\nabla_y \mathcal{L}_\text{SV}( w, b ) \odot \frac{\partial \kappa(z^L)}{\partial z^L}.
\end{equation*}

As the name of the algorithm already revealed we can now back-propagate this error through the NN and hence compute the error of an arbitrary layer $l$ using the error of the preceding layer $l+1$, in other words
\begin{equation*}
	\mathcolorbox{\delta^l=((w^{l+1})^T\delta^{l+1})\odot \frac{\partial \kappa(z^l)}{\partial z^l}.}
\end{equation*}

The components of the gradients $\nabla_{{b}}\mathcal{L}_\text{SV}( w, b )$ and $\nabla_{{w}}\mathcal{L}_\text{SV}( w, b )$ are thus
\[
\mathcolorbox{\begin{aligned}
		\frac{\partial \mathcal{L}_\text{SV}( w, b )}{\partial b^l_j}&=\delta^l_j\\	\frac{\partial \mathcal{L}_\text{SV}( w, b )}{\partial w^l_{jk}}&=y^{l-1}_k\delta^l_j,
\end{aligned}}
\]
where $y^{l-1}=\kappa(z^{l-1})$ is the activation of layer $l-1$. An implementation in Python of the back-propagation algorithm used for classifying handwritten digits can be found at \cite{Nielsen2015}.

In \cref{chapter:DQNN} we will introduce a quantum version of a feed-forward NN, which can be optimised with an algorithm similar to the classical back-propagation method we described above. Before digging deeper into quantum NNs, we will give an introduction to quantum information in the following chapter.

%% file: text/QI.tex
\chapter{Quantum information}\hypertarget{graph}{}
\label{chapter:QI}

As described in \cref{chapter:intro}, in the past decades, the power of classical computers has grown exponentially. Nevertheless, boundaries for classical computers exist \cite{Prati2017}. For instance, the description of more sizeable quantum systems is still impossible on today's classical supercomputers since the dimension of quantum systems scales exponentially with the number of basic building blocks. 

Consequently, in the eighties, Richard Feynman proposed the concept of quantum computers \cite{Feynman1982} with the aim to simulate many-body quantum systems with the use of a quantum system of the same size instead of a classical computer. Building on this idea, the field of quantum computing evolved in the last decades and is now one of the most rapidly developing research areas. Until today many quantum algorithms were proposed with the aim of solving a particular task more efficiently than classical supercomputers. The first breakthroughs happened in the nineties, among them the Deutsch-Jozsa algorithm \cite{Deutsch1992}, Shor's factoring algorithm \cite{Shor1994} and Grover's search algorithm \cite{Grover1996}. 

Whereas the key developments remained on the more theoretical side first, in the last years processing quantum information on quantum computers has become experimentally possibly \cite{Preskill2018, Brooks2019, Arute2019}. These first quantum computers are called \emph{noisy intermediate-scale quantum} (NISQ) devices, comprise up to a few hundred quantum bits and give the opportunities to test quantum algorithms for their behaviour under high noise levels. 

The arrival of these new devices and also their public access \cite{IBMQuantum2021} lead to many attempts of solving problems on these devices. It should be underlined at this point that computations on quantum computers containing only about 100 qubits can be, in many cases, simulated classically. The emergence of NISQ devices should not be underrated but also not overrated, or as John Preskill appropriately formulated in \cite{Preskill2018}: \enquote{The 100-qubit quantum computer will not change the world right away — we should regard it as a significant step toward the more powerful quantum technologies of the future.}

Altogether, we can conclude that we are still at the beginning of the quantum information era. However, instead of feeling disappointed, this should be seen as an even more significant motivation for study and research in the area of quantum information these days, in the view of many so far undiscovered opportunities.  

To make the research presented from \cref{chapter:DQNN} onwards accessible to people new to quantum information, we will discuss the basic definitions in this chapter. We start by introducing the quantum bit and its characteristics differing from the classical bits in \cref{sec:QI_states}. Based on this, we define quantum gates and circuits in \cref{sec:QI_circuits}. In \cref{sec:QI_algorithms} we not only deliver insight into why quantum algorithms can outperform classical algorithms but also give an overview of the most famous quantum algorithms. \cref{sec:QI_QC} explains how quantum circuits can be implemented and describes the state of the art of quantum computers. Since the topic of this thesis is the study of quantum neural networks, we conclude with a review of these in \cref{sec:QI_QNN}. For a wholesome introduction to quantum information, we point to \cite{Nielsen2000}.

\section{Quantum bits}
\label{sec:QI_states}

In classical information theory, basic binary units of information, called bits, are used for information processing. These bits - physically implemented with a two-state device - can only have one of two values, commonly represented as either $0$ or $1$. The quantum equivalent is named \emph{qubit}. It represents a two-level quantum mechanical system, often denoted in terms of $\ket{0}$ and $\ket{1}$. These levels can, for example, be two different polarisations of a photon (vertical and horizontal polarisation), two different alignments of spin (up and down) or two states of an electron orbiting an atom (ground state and excited state). But, they can also be defined by more complex systems based on very cold superconducting electrical circuits in which several electrons move \cite{Nielsen2000,Preskill2018}.

\subsection*{Superposition}
Whereas the state of a classical bit can be either $0$ or $1$, the qubit is allowed to be in a coherent \emph{superposition} of both levels $\ket{0}$ and $\ket{1}$ simultaneously. For a mathematical description of these qubit \emph{states} we identify the levels with two orthonormal basis vectors 
\begin{equation}
\label{eqn:QI_basis}
\ket{0}=\begin{pmatrix}[0.6]1\\0\end{pmatrix},\,\,\,\ket{1}=\begin{pmatrix}[0.6]0\\1\end{pmatrix}.
\end{equation}\label{subsec:QI_superposition}

A qubit in a so called \emph{pure state} can than be described as a normalised vector  $\ket{\phi}$ in a two-dimensional complex Hilbert space $\mathcal{H}=\mathbb{C}^2$. Using the in \cref{eqn:QI_basis} defined basis states we can express such a state as superposition
\begin{equation}\label{eqn:QI_superposition}
\mathcolorbox{\ket{\psi}=\alpha\ket{0}+\beta\ket{1}\text{,}}
\end{equation}
where $\alpha$ and $\beta$ are complex numbers and $|\alpha|^2+|\beta|^2=1$.

In quantum mechanics the \emph{Dirac notation}, read as \emph{bra} $\bra{*}$ and \emph{ket} $\ket{*}$, is the standard notation for states. Here $\bra{\psi}$ denotes the conjugate transpose of the vector $\ket{\psi}$. The eponymous \emph{bra-ket} $\braket{\psi_1|\psi_2}$ expresses the scalar product of the vectors $\bra{\psi_1}$ and $\ket{\psi_2}$. At this point it is worth to accentuate that this description only includes pure states. Density matrices, discussed later in this section, provide a more general notion of statistical ensembles of pure states.

\begin{figure}[h]
\centering
\begin{tikzpicture}[scale=1.4,line cap=round, line join=round]
\clip(-2.19,-2.49) rectangle (2.66,2.58);
\draw [shift={(0,0)}, line3, fill=color3L] (0,0) -- (56.7:0.65) arc (56.7:90.:0.65) -- cycle;
\draw [shift={(0,0)}, line1, fill=color1L] (0,0) -- (-135.7:0.65) arc (-135.7:-33.2:0.65) -- cycle;
\draw[color0](0,0) circle (2cm);
\draw [rotate around={0.:(0.,0.)},densely dashed, color0] (0,0) ellipse (2cm and 0.85cm);
\draw[dotted] (0,0)-- (0.70,1.1);
\draw [->,lineD] (0,0) -- (0,2);
\draw [->,lineD] (0,0) -- (-0.81,-0.79);
\draw [->,lineD] (0,0) -- (2,0);
\draw [dotted] (0.7,1)-- (0.7,-0.46);
\draw [dotted] (0,0)-- (0.7,-0.46);
\node at (-85:0.45) {$\varphi$};
\node at (73:0.45)  {$\theta$};
\node at  (-1.01,-1) {$\hat{x}$};
\node at  (2.3,0) {$\hat{y}$};
\node at  (0,2.3)  {$\ket{0}$};
\node at (0,-2.3) {$\ket{1}$};
\node at (0.9,1.4) {$\ket{\psi}_{\text{Bloch}}$};
\scriptsize
\draw [fill] (0.7,1.1) circle (1.5pt);
\end{tikzpicture}
\caption{\textbf{Bloch sphere}. Pure qubit states can be represented as points on the surface of a sphere.}
\label{fig:QI_bloch}
\end{figure}
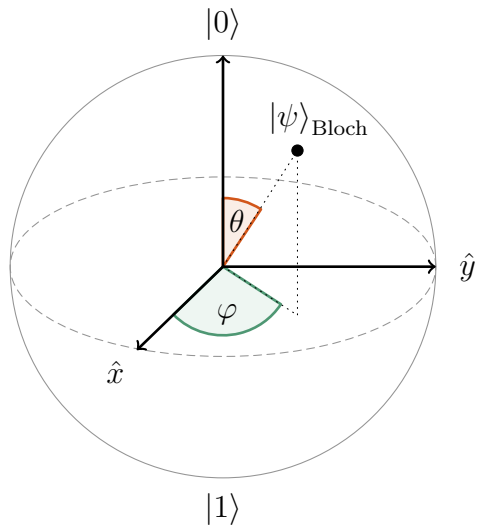

Since the description of a pure qubit state is a vector with norm $1$ we can depict such a state on the surface of a sphere. For this purpose we rewrite \cref{eqn:QI_superposition} as
\begin{equation}
\label{eqn:QI_ket}
\ket{\psi}=e^{i\gamma}\left(\cos\left(\frac{\theta}{2}\right)\ket{0}+e^{i\varphi}\sin\left(\frac{\theta}{2}\right)\ket{1}\right)\text{.}
\end{equation}
The visualisation of quantum states with two numbers $\theta$ and $\phi$, depicted in \cref{fig:QI_bloch}, is named \emph{Bloch sphere}. Note that the condition $|\alpha|^2+|\beta|^2=1$ defines a 3-dimensional sphere. The \emph{Hopf fibration} \cite{Hopf1964}, mapping from the unit 3-sphere to the two-dimensional states space $\mathbbm{C}^2$, is used to describe the sphere with two parameters $\theta$ and $\phi$, namely as
\begin{equation}
\label{eqn:QI_ketbloch}
\ket{\psi}_{\text{Bloch}}=\cos\left(\frac{\theta}{2}\right)\ket{0}+e^{i\varphi}\sin\left(\frac{\theta}{2}\right)\ket{1}\text{.}
\end{equation}
The factor $e^{i\gamma}$ can be ignored, since it has no \emph{observable effects} which is in detail explained in \cite{Nielsen2000}. We will discuss the abandoning of this total phase later in this chapter again.

Like for the classical bits there are two possible outcomes for the measurement of a qubit, also often taken to be $0$ and $1$. After measuring in the basis denoted in \cref{eqn:QI_basis} such a qubit state collapses in the pure state $\ket{0}$ or $\ket{1}$, respectively. It is possible to compute the probabilities of getting a special measurement outcome. With probability
\begin{equation*}
p_0=\braket{\psi|0}\braket{0|\psi} = |\bra{0}\left(\alpha\ket{0}+\beta\ket{1}\right)|^2 = |\alpha \braket{0|0}|^2 = |\alpha|^2
\end{equation*}
the measured value will be $0$. We call $M_0=\ket{0}\bra{0}$ a \emph{measurement operator}. Analogous we get $p_1=|\beta|^2$ with the operator $M_1=\ket{1}\bra{1}$. 

To summarise, a qubit can be in an infinite number of possible states, but after the measurement, we again only receive binary information as with the classic bit and the superposition information seems lost. However, with quantum algorithms, this fundamental property of quantum mechanics can be used effectively and is the reason for quantum speed up, as will be explained in \cref{sec:QI_algorithms}.

Although not relevant for this work, we want to mention at this point the existence of the generalised concept of \emph{qudits}. These describe quantum systems with $d>2$ levels. In the case of $d=3$ the name \emph{qutrit} is established. Albeit most of the studies in quantum information theory operate with qubits, these higher-level bits can be still beneficial in some contexts \cite{Kaszlikowski2003,Kues2017,Suslov2011,Shlyakhov2018,Soltamov2019}.

\subsection*{Multiple qubits}

In the same way, as classical bits unfold their full potential only when used in strings of many bits, a quantum system and, therefore, quantum computing becomes much more attractive when involving more than one qubit.

We describe $n$ qubits with the tensor product of $2^n$ basic states living in the Hilbert space $\mathcal{H}=\left(\mathbb{C}^2\right)^{\otimes n} = \mathbb{C}^{2^n}$, namely
\begin{equation}
\label{eqn:QI_tensorKet}
\ket{q_1}\otimes \ket{q_2} \otimes \hdots \otimes \ket{q_n} = \ket{q_1q_2\hdots q_n}  \text{,}
\end{equation}\label{subsec:QI_multiqubits}
where $q_i$ indicates the state of the $i$th qubit with $q_i\in\{0,1\}$. 

In that way the basic states of a two qubit system can be denoted as $\ket{00}$, $\ket{01}$, $\ket{10}$, and $\ket{11}$. A general pure two-qubit state can be expressed as a superposition
\begin{equation*}
\ket{\psi}=\alpha_{00}\ket{00}+\alpha_{01}\ket{01}+\alpha_{10}\ket{10}+\alpha_{11}\ket{11}
\end{equation*}
in $\mathcal{H}=\mathbb{C}^4$, where $|\alpha_{00}|^2+|\alpha_{01}|^2+|\alpha_{10}|^2+|\alpha_{11}|^2=1$ and $\alpha_{ij}\in \mathbb{C}$. 

In the same manner a pure $n$ qubit state can be described as the superposition
\begin{equation*}
\mathcolorbox{\ket{\psi}= \sum_{q_1,\dots,q_n\in\{0,1\}} \alpha_{q_1,\dots,q_n}\ket{q_1\hdots q_n}}
\end{equation*}
in $\mathcal{H}=\mathbb{C}^{2^n}$ with complex coefficients $\alpha_{q_1,\dots,q_n}$.

At this point, the exponential growth of the Hilbert space with the number of qubits becomes obvious, and with this, the high costs when simulating quantum systems classically fall into place. On the other hand, this exponential relationship indicates the extensive capacity of quantum computing. 

\subsection*{Mixed states}

To this point all mentioned quantum states were pure states, which can be represented as vectors in $\mathbb{C}^{2^n}$. Though in general a system has to be described by a mixtures of these pure states. Such a \emph{mixed state} can be described with a matrix $\rho\in\mathcal{H}=\mathbb{C}^{2^n}\times\mathbb{C}^{2^n}$ with positive eigenvalues, $\tr\left(\rho\right)=1$, a so called \emph{density matrix}. 

If the state $\rho$ is a mixture of $m$ pure states $\ket{\psi_i}$, we can describe as a convex combination of these states, namely as
\begin{equation*}
\mathcolorbox{\rho=\sum_i^m p_i \ket{\psi_i}\bra{\psi_i}}
\end{equation*}
with $\sum_i p_i=1$. We can view a pure state rather as special case of the mixed state and can write it as
\begin{equation*}
\rho=\ket{\psi}\bra{\psi}. 
\end{equation*}
Note that at this point the total phase of $\ket{\psi}$ is lost (compare \cref{eqn:QI_ket} and \cref{eqn:QI_ketbloch}). For the density matrix of a pure state it is $\rho=\rho^2$ and $\tr\left(\rho^2\right)=1$.

As explained when using the bracket notation in \cref{eqn:QI_tensorKet}, we use the tensor product to build the composite of two quantum systems, for example, two qubits. When working with mixed states, for example, a qubit $A$ in the state $\rho_A\in\mathcal{H}_A$ and qubit $B$ in the state $\rho_B\in\mathcal{H}_B$, we can describe the composite of both qubits with the state
\begin{equation*}
\rho_{AB}=\rho_A \otimes \rho_B
\end{equation*} 
in the Hilbert space $\mathcal{H}_A\otimes\mathcal{H}_B$. Not only the composition of systems is frequently used in quantum algorithms, but also the reduction to a subsystem is an important tool: tracing over the space $\mathcal{H}_B$ gives
\begin{equation*}
\rho_A=\tr_B(\rho_{AB}) \in \mathcal{H}_A.
\end{equation*} 
The most simple composition of $n$ systems can be written as
\begin{equation*}
\rho_{1,\hdots,n}=\rho_1\otimes \hdots \otimes \rho_n.
\end{equation*}
When tracing out the $i$th system we get
\begin{equation*}
\text{tr}_i\left(\rho_{1,\hdots,n}\right)=\rho_1\otimes \hdots  \rho_{i-1}\otimes \rho_{i+1}\hdots \otimes \rho_n.
\end{equation*}

We want to close the description of mixed states with two remarks. Since with the density matrix we generalised the description of a quantum state, we have to update the way to calculate probabilities: a mixed state $\rho$ collapses into the state $\ket{\psi}$ after measurement with probability $p=\tr\left(\ket{\psi}\bra{\psi}\rho\right)=\braket{\psi|\rho|\psi}$. More generally we can describe a measurement with an linear operator $M$, called \emph{observable}. The expectation value of the measurement outcome is given by $\tr(\rho M)$ when measuring $\rho$. We point to \cite{Nielsen2000} for a complete discussion of observables.

The second remark concerns the representation of one-qubit states in the Bloch sphere, see \cref{fig:QI_bloch}. Mixed one-qubit states can be represented in this way as well. Whereas the pure states lie on the surface of the Bloch sphere, a general mixed state lies inside. Hence the representation requires the radius $r$ and is given by
\begin{equation*}
\rho=\frac{\mathbb{1}+\vec{r}\vec{\sigma}}{2},
\end{equation*}
where $\vec{r}\in \mathbb{R}^2$ is called the Bloch vector and $\vec{\sigma}=\left(\sigma_x,\sigma_y,\sigma_z\right)^T$ is built from the Pauli matrices, which are denoted in \cref{fig:QI_gates}. 

\subsection*{Entanglement}
Next to superposition, \emph{entanglement} is one of the most popular and fundamental features in quantum mechanics. Both are crucial elements in quantum algorithms. Whereas superposition can be defined on just one particle, entanglement requires two or more qubits. Precisely, a pair or set of qubits are called entangled when each qubit cannot be described independently of the state of the remaining qubits.

For the simplest example of entanglement we discuss the pure two qubit state
\begin{equation}
\label{eqn:QI_bell}
\ket{\Phi^+}=\frac{1}{\sqrt{2}}\left(\ket{00}+\ket{11}\right).
\end{equation}
This state describes the superposition of two cases: after measurement with probability $0.5$, both of the qubits are in te state $\ket{0}$ or with the same probability both of them are in the state $\ket{1}$. When measuring only one qubit, the other qubit will \emph{collapse} into the same state. In this sense the two qubits are entangled.

$\ket{\Psi^+}$ in \label{eqn:QI_bell} is one of the four \emph{Bell states} \cite{Braunstein1992} displayed in \cref{fig:QI_BellB}. These states are maximally entangled superpositions of $\ket{0}$ and $\ket{1}$. For all four states yields: measuring one qubit determines the state of the second qubit. The measurement outcomes are \emph{correlated}. Even if the two qubits are spatially separated, this correlation remains, which is discussed in the \emph{EPR paradox} by Albert Einstein, Boris Podolsky, and Nathan Rosen \cite{Einstein1935}.

In general, we name a quantum state of two or more systems \emph{entangled}, if it is not \emph{separable}. Further, we name a pure state of $n$ subsystems separable if it can be written in the form
\begin{equation*}
\ket{\psi}=	\ket{\psi_1}\otimes\dots\otimes	\ket{\psi_n}.
\end{equation*}
For mixed states the equivalent form is the convex sum
\begin{equation*}
\rho=\sum_i p_i \rho_1^i\otimes\dots\otimes	\rho_n^i.
\end{equation*}
If there exists only a single coefficient with $p_k\neq0$, separable two-particle states are also often called \emph{product states} and can be described in the form $\rho_{AB}=\rho_A\otimes\rho_B$.

\subsection*{State distances}

During quantum algorithms, it is often essential to compare two quantum states. We will see a use case in \cref{subsec:DQNN_trainingloss}, where the update of a quantum neural network algorithm is based on the difference of two states.

Although many different proposals for measuring how close two quantum states exist, the \emph{fidelity} is one of the most referenced options. We define the fidelity of two density matrices as 
\begin{equation}
\mathcolorbox{F\left(\rho,\sigma\right)\equiv\left(\tr\sqrt{\sqrt{\rho}\sigma\sqrt{\rho}}\right)^2,}  \label{eqn:QI_fid}
\end{equation} 
\label{subsec:QI_distance}
where $F\left(\rho,\rho\right)=1$. The fidelity can be expressed as $F\left(\ket{\phi}\bra{\phi},\rho\right)= \braket{\phi|\rho|\phi}$ if one of the states is pure and as $F\left(\ket{\psi},\ket{\phi}\right)=|\braket{\psi|\phi}|^2$ if both of the states are pure.  Note that in some literature the fidelity, in contrast to \cref{eqn:QI_fid}, is defined differently, namely as $\overline{F}\left(\rho,\sigma\right)\equiv \tr\sqrt{\sqrt{\rho}\sigma\sqrt{\rho}}$.

If at least one of the states is pure, the fidelity as a measure of closeness has some advantages: It can be measured using the SWAP test, which we will explain later in this work, see \cref{subsec:DQNN_swap}. Another essential point is that the fidelity measure has an operational meaning: it defines the probability $\braket{\phi|\rho|\phi}$ (or $|\braket{\psi|\phi}|^2$) that $\rho$ (or $\ket{\phi}$) passes the test being the same as $\ket{\psi}$ in measurements. 

However, when comparing two mixed states, the fidelity still has an interpretation: It is the largest fidelity between any two purifications of given states. Armin Uhlmann describes this fact in \cite{Uhlmann1976}: if we assume $\rho$ and $\sigma$ are two states of a quantum system $A$, and $B$ is a second system with dimension greater than or equal to the dimension of $A$, then the fidelity can be described as
\begin{equation*}
F\left(\rho,\sigma\right)=\max |\braket{\psi_0|\phi_0}|
\end{equation*}
where the maximisation runs over all $\ket{\psi_0}$ and $\ket{\phi_0}$ which are purification of $\rho$ and $\sigma$ in $AB$. Although the authors of \cite{Dodd2001} show an operational interpretation of this theorem, the drawback of the fidelity becomes clear at this point: due to the maximisation progress, excessive computational complexity is required to evaluate the fidelity of two mixed states on a quantum computer. 

A good alternative for comparing two mixed states is the \emph{Hilbert-Schmidt} distance \cite{Ozawa2000}
\begin{equation*}
\mathcolorbox{d_{\text{HS}}\left(\rho,\sigma\right) \equiv \tr\left(\left(\rho-\sigma\right)^2\right),}
\end{equation*} 
because it can be evaluated on a quantum computer \cite{GarciaEscartin2013,Cincio2018}. In this work, we use the Hilbert-Schmidt distance as a training loss with mixed states for graph-structured quantum data, see \cref{sec:graph_loss}. It is important to note that this distance reaches its minimum, i.e.\ $d_{\text{HS}}\left(\rho,\rho\right)=0$, if the two compared states coincide. In contrast, the fidelity reaches the maximum, the value $1$, in this case. 


\section{Quantum circuits}
\label{sec:QI_circuits}

To exploit quantum mechanics for quantum computing, so-called \emph{quantum circuits} are needed. These describe the processes of initialising qubit states, applying operations on these qubits and reading out results via measurements. Since we already discussed qubits and measurements, we will start the discussion of quantum circuits by introducing quantum operations, called quantum \emph{gates}. Further, we will explain how these operations and circuits can be depicted. We will conclude this section by discussing exemplary quantum circuits.

\subsection*{One-qubit gates}

Similar to logic gates, which manipulate classical information and are used for classical computing, we define quantum gates. Some of the classical gates have a direct quantum analogue. One example is the $\notgate$ gate transferring the state  $\ket{0}$ in $\ket{1}$ and  $\ket{1}$ in $\ket{0}$. Written out in Dirac notation this also called Pauli $X$-gate is of the form
\begin{align*}
X=\ket{1}\bra{0}+\ket{0}\bra{1},
\end{align*}
or in matrix notation with respect to the basis in \cref{eqn:QI_basis}
\begin{equation*}
\sigma_x=\begin{pmatrix}[0.6]
0&1\\
1&0
\end{pmatrix}.
\end{equation*}
In circuit notation we depict a gate $X$ as
\begin{center}
\begin{tikzpicture}
\draw (-1,0)--(1,0);
\node[operator1] at (0,0) {$X$};
\end{tikzpicture},
\end{center}
where the line denotes the qubit. Applying this gate on the in \cref{eqn:QI_basis} presented basis states gives the expected results, namely
\begin{align*}
X\ket{0}&=\ket{1}\braket{0|0}+\ket{0}\braket{1|0}=\ket{1}\\
X\ket{1}&=\ket{1}\braket{0|1}+\ket{0}\braket{1|1}=\ket{0}.
\end{align*}
Thus the gate $X$ interchanges the roles of the two basis states in an arbitrary pure one-qubit state, i.e.\
\begin{align*}
X\ket{\psi}=X\left(\alpha\ket{0}+\beta\ket{1}\right)=\beta\ket{0}+\alpha\ket{1}=\ket{\psi'}.
\end{align*}

With the $X$-gate as an example, we have shown how one-qubit gates act on pure states. From here, important questions arise: which gates are allowed in quantum computing and how are they constraint? As explained in \cref{sec:QI_states}, every normalised element in $\mathbbm{C}^2$ can describe a pure one-qubit quantum state, namely $\ket{\psi}=\alpha\ket{0}+\beta\ket{1}$ with $|\alpha|^2+|\beta|^2=1$. For the state $U\ket{\psi}=\ket{\psi'}$ resulting from applying the gate $U$ the same constraint has to be fulfilled, specifically $\braket{\psi|\psi}=1=\braket{\psi'|\psi'}$. It follows that the gate has to be \emph{unitary} defined through $U^\dagger U=1$, where $U^\dagger$ denotes the adjoint of the matrix $U$.  

Although every unitary matrix is a legitimate quantum gate, some gates are more frequently used in algorithms than others. In \cref{fig:QI_gates} we list some famous one-qubit gates. The \emph{Hadamard gate} $H$ provides a possibility to build superposition of states. It is also often applied in order to measure in the basis $\{\ket{\pm}=\frac{1}{\sqrt{2}}\left(\ket{0}\pm\ket{1}\right)\}_\pm$. In general it is possible to measure in the basis $\{U\ket{0},U\ket{1}\}$ by applying $U^\dagger$ before measuring in the basis $\{\ket{0},\ket{1}\}$.

\begin{figure}[H]
\centering
\begin{tabular}{cccccc}
\begin{tikzpicture}
\draw (-1,0)--(1,0);
\node[operator1] at (0,0) {$X$};
\end{tikzpicture} & 
\begin{tikzpicture}
\draw (-1,0)--(1,0);
\node[operator1] at (0,0) {$Y$};
\end{tikzpicture} & 
\begin{tikzpicture}
\draw (-1,0)--(1,0);
\node[operator1] at (0,0) {$Z$};
\end{tikzpicture} & 
\begin{tikzpicture}
\draw (-1,0)--(1,0);
\node[operator1] at (0,0) {$H$};
\end{tikzpicture} & 
\begin{tikzpicture}
\draw (-1,0)--(1,0);
\node[operator1] at (0,0) {$T$};
\end{tikzpicture} & 
\begin{tikzpicture}
\draw (-1,0)--(1,0);
\node[operator1] at (0,0) {$S$};
\end{tikzpicture} \\

$\begin{pmatrix}[0.6]
0&1\\
1&0
\end{pmatrix} $ & 
$\begin{pmatrix}[0.6]
0&-i\\
i&0
\end{pmatrix} $ & 
$\begin{pmatrix}[0.6]
1&0\\
0&-1
\end{pmatrix} $ & 
$ \frac{1}{\sqrt{2}}\begin{pmatrix}[0.6]
1&1\\
1&-1
\end{pmatrix} $ & 
$ \begin{pmatrix}[0.6]
1&0\\
0&e^{i\frac{\pi}{4}}
\end{pmatrix} $& 
$ \begin{pmatrix}[0.6]
1&0\\
0&i
\end{pmatrix} $ \\
\end{tabular}
\caption{\textbf{A selection of one-qubit gates.} The gates $X$ (also called $\notgate$), $Y$ and $Z$ are represented by Pauli matrices $\sigma_x$, $\sigma_y$ and $\sigma_z$. Further famous gates are the Hadmard gate $H$, the $T$-gate also known as $\frac{\pi}{8}$-gate and the phase gate $S$.}
\label{fig:QI_gates}
\end{figure}

Note that all elements of $SU(2)$, i.e.\ those unitaries on $\mathbb{C}^2$ with $\det(U)=1$ are given by $SU(2)=\left\{\exp(i n \sigma)\right\}$,
where $n\in \mathbbm{R}^3$ and $\sigma$ is the vector of Pauli matrices $\sigma=\left(\sigma_x, \sigma_y, \sigma_z\right)$, see \cref{fig:QI_gates}. 

\subsection*{Multi-qubit gates}
\begin{figure}[H]
\centering
\begin{subfigure}[t]{0.25\linewidth}
\centering
\begin{tikzpicture}[thick]
\matrix[row sep=0.4cm, column sep=0.8cm] (circuit) { 
\node (in1) {};  
& \node (c1){};
& \node (out1) {};  \\
\node (in2) {};  
& \node (c2){};
& \node (out2) {};  
\\
\node (in3) {};  
& \node (c3){};
& \node (out3) {};  
\\
};
\begin{pgfonlayer}{background}
\draw[] (in1) -- (out1)  
(in2) -- (out2)
(in3) -- (out3);
\draw[] (c1) -- (c2);
\end{pgfonlayer}
\node[operator1,minimum height=2cm] at (0,0) {$U$};
\end{tikzpicture}
\subcaption{3-qubit gate $U$.}
\label{fig:QI_2gatesA}
\end{subfigure}
\begin{subfigure}[t]{0.45\linewidth}
\centering
\begin{tikzpicture}[thick]
\matrix[row sep=0.4cm, column sep=0.8cm] (circuit) { 
\node (in1) {};  
& \node[circlewc] (c1){};
& \node (out1) {};  \\
\node (in2) {};  
& \node[dot] (c2){};
& \node (out2) {};  
\\
};
\begin{pgfonlayer}{background}
\draw[] (in1) -- (out1)  
(in2) -- (out2);
\draw[thick] (c1) -- (c2);
\end{pgfonlayer}
\begin{scope}[xshift=2.8cm]
\node[] at (0,0) (v) {$\widehat{=}\ \begin{pmatrix}[0.6]
1&0&0&0\\0&1&0&0\\0&0&0&1\\0&0&1&0
\end{pmatrix}$};
\end{scope}
\end{tikzpicture}
\subcaption{$\cnot$ gate.}
\label{fig:QI_2gatesB}
\end{subfigure}
\caption{\textbf{Multi-qubit gates.} In general multi-qubit gates are depicted as gates covering diverse lines, symbolising the qubits, see (a). The $\cnot$ gate has a special notation, depicted in (b).}  
\label{fig:QI_2gates}
\end{figure}
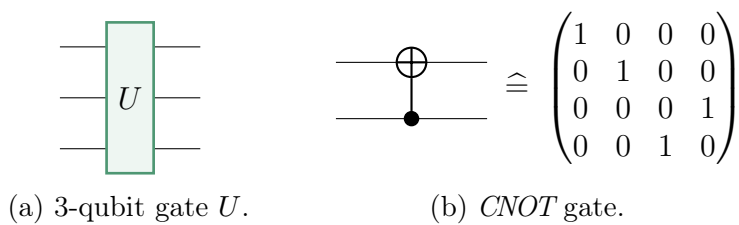

So far, we only discussed gates acting on one-qubit states, but also operations on multiple qubits simultaneously, as depicted in \cref{fig:QI_2gatesA}, are regularly used. An typical example for a multi-qubit gate is the \emph{controlled-not gate}, abbreviated by $\cnot$, see \cref{fig:QI_2gatesB}. This gate affects two qubits, where one is called the controlled qubit, and the other is the target qubit. If the controlled qubit is in the state $\ket{1}$, the basic states of the target qubits get interchanged, i.e.\ we apply the $X$ gate. Otherwise, the identity operates.

Before we describe the circuits comprised of quantum gates, we want to comment on density matrices in this context.  Although the qubits are usually initialised in pure states at the beginning, mixed states occur during quantum circuits. For applying a gate on a mixed state $\rho$ we multiply the gate from the left side and its adjugate from the right side that is $\rho'=X\rho X^\dagger$. Due to unitarity the constraint $\tr\left(\rho\right)=1=\tr\left(\rho'\right)$ stays fulfilled.

\subsection*{Circuits}
\label{subsec:QI_circuits_circ}
Since it is impossible to directly implement an arbitrary number of different gates in a quantum computer, the question naturally arises whether complex operations can be approximated by breaking them down into sequences of elementary gates. Indeed the Solovay-Kitaev theorem \cite{Dawson2005} states that every unitary $U$ can be approximated with an accuracy of $\epsilon$ by a sequence of gates from $\mathcal{G}$ of length $\mathcal{O}\left(\log^c\left(\frac{1}{\epsilon}\right)\right)$, where $\mathcal{G}$ is finite universal gate set and $c$ is a constant based on the choice of $U$. The small error is unavoidable in praxis.

Finite sets of quantum gates are called \emph{universal} if it is possible to build any arbitrary single-qubit gate with them. One of the most common universal sets contains the gates $\{\cnot,H,S,T\}$. The quantum computers used for the numerics in this work fragment the operations into the gates $\{\cnot,RZ,SX,X\}$, where $RZ=e^{-i\frac{\theta}{2}X}$ and $S=\sqrt{X}$. Additionally, the identity operation is usually also directly included in these universal gate sets. The quantum operations built from these sets of gates can be represented in a quantum circuit. Every horizontal line depicts a qubit. Following the lines from left to right depicts the evolution in time.

\begin{figure}[H]
	\centering
	\begin{tikzpicture}[]
		\matrix[row sep=0.3cm, column sep=0.5cm] (circuit) {
			\node(start2){$\rho_\text{A}$};
			& \node[]{}; 
			& \node[]{}; 
			& \node[]{}; 
			& \node[]{}; 
			& \node[meter,scale=0.6] (end2){}; \\
			& \node[]{}; 
			\node(start1){$\rho_\text{B}$};
			& \node[]{}; 
			& \node[]{}; 
			& \node[dcross](end1){}; 
			& \node[]{};  \\
		};
		\begin{pgfonlayer}{background}
			\draw[] (start1) -- (end1)  
			(start2) -- (end2);
			\node[operator1,minimum height=1.5cm] at (.3,0) {$U$};
		\end{pgfonlayer}
	\end{tikzpicture}
	\caption{\textbf{Exemplary quantum circuit.} This circuit depicts building the tensor product of $\rho_\text{A}$ and $\rho_\text{B}$, applying a two-qubit gate $U$, tracing out the qubit $B$ and measuring the resulting state.}
	\label{fig:QI_circuit}
\end{figure}
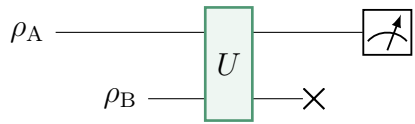

Next to gates also tensor products, tracing out subsystems and measurements can be depicted in quantum circuits. In \cref{fig:QI_circuit}, the circuit describes building the tensor product of two one-qubit states $\rho_\text{A}$ and $\rho_\text{B}$, applying a gate $U$ and reducing the system to the first qubit. The resulting state, namely $\tr_B\left(U\left(\rho_\text{A}\otimes\rho_\text{B}\right)U^\dagger\right)$, is measured.

Further, it follows from the Stinespring dilation theorem \cite{Nielsen2000} that every description of a quantum circuit through tensor products, unitary transformations and reductions to subsystem is equivalent to the description via a \emph{completely positive} (CP) map. We call a bounded linear operator $P$ on a Hilbert space \emph{positive} if it a bounded operator $T$ exists so that $P=T^\dagger T$. Using this, the definition of a CP map can be expressed in the following way:
\begin{defi}[Completely positive map] \label{def:QI_CP}
Let $A$ and $B$ be $C^*$-algebras and $
\mathcal{E}:A \rightarrow B$ a linear map. $\mathcal{E}$ is called \emph{positive} if it maps positive elements (corresponding to the positive operators) to positive elements. We can define a new map 
\begin{equation*}
\text{id}\otimes \mathcal{E}\ : \ \mathcal{C}^{n \times n} \otimes A  \ \rightarrow  \ \mathcal{C}^{n \times n} \otimes B
\end{equation*}
and call $\mathcal{E}$ \emph{completely positive} if $\text{id}\otimes \mathcal{E}$ is positive for all $n$.
\end{defi}
Moreover we can write every CP map in the form $\mathcal{E}(\rho)=\sum_\alpha A_\alpha \rho A_\alpha^\dagger$, where the operators $A_\alpha$ are called \emph{Kraus operators} and satisfy $\sum_\alpha A_\alpha^\dagger A_\alpha=\mathbbm{1}$. This is called \emph{Kraus decomposition} \cite{Kraus1983}.\label{subsec:QI_Kraus}

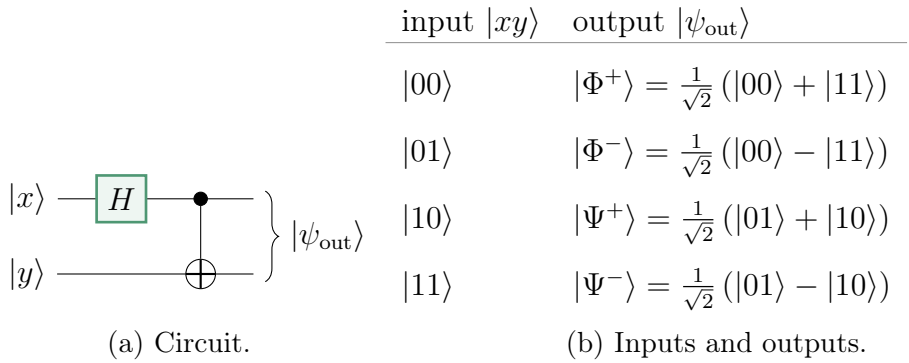
\begin{figure}[H]
\centering
\begin{subfigure}[b]{0.35\linewidth}
\begin{tikzpicture}[]
\matrix[row sep=0.3cm, column sep=0.5cm] (circuit) {
\node(start2){$\ket{x}$};
& \node[operator1]{$H$}; 
& \node[dot](dot){} ; 
& \node[] (end2){}; \\
\node(start1){$\ket{y}$};
& \node[]{}; 
& \node[circlewc](circ){} ; 
& \node[](end1){};; \\
};
\begin{pgfonlayer}{background}
\draw[] (start1) -- (end1)  
(start2) -- (end2)
(dot) -- (circ);
(start2) -- (end2);
\draw[brace0,color=black] (1.7,0.6)-- node [right=1ex] {$\ket{\psi_\text{out}}$}(1.7,-0.6);
\end{pgfonlayer}
\end{tikzpicture}
\caption{Circuit.}
\label{fig:QI_BellA} 
\end{subfigure}
\begin{subfigure}[b]{0.6\linewidth}
\begin{tabular}{l l}
input $\ket{xy}$ & output $\ket{\psi_\text{out}}$ \\
\noalign{
\color{color0}
\hrule 	}%
$\ket{00}$ & $\ket{\Phi^+}=\frac{1}{\sqrt{2}}\left(\ket{00}+\ket{11}\right)$ \\
$\ket{01}$ & $\ket{\Phi^-}=\frac{1}{\sqrt{2}}\left(\ket{00}-\ket{11}\right)$ \\
$\ket{10}$ & $\ket{\Psi^+}=\frac{1}{\sqrt{2}}\left(\ket{01}+\ket{10}\right)$ \\
$\ket{11}$ & $\ket{\Psi^-}=\frac{1}{\sqrt{2}}\left(\ket{01}-\ket{10}\right)$ \\
\end{tabular}
\caption{Inputs and outputs.}
\label{fig:QI_BellB} 
\end{subfigure}
\caption{\textbf{Bell state producing circuit.} The circuit depicted in (a) produces the Bell states according to given input states. The relations between input and output is shown in (b).}
\label{fig:QI_Bell} 
\end{figure}

A more concrete example of a quantum circuit is explained in \cite{Nielsen2000}: the circuit in \cref{fig:QI_BellA} is built of a Hadamard and $\cnot$-gate and outputs the various Bell states, see \cref{fig:QI_BellB}. 

\subsection*{Sampling random unitaries and quantum states}
\label{subsec:QI_random}

We have already argued that all unitaries are legitimate quantum gates. Therefore, these mathematical operations play a central role in quantum information \cite{Nielsen2000}. When it comes to initialising random unitaries or integrating over all possible unitaries, phrases like \enquote{$U$ is uniformly sampled at random with respect to the Haar measure} are often used in the description of quantum algorithms. In the following, we briefly motivate the \emph{Haar measure} \cite{Diestel2014}. For a detailed discussion of the \emph{unitary group} we point to \cite{Duistermaat2012}.

A measure formulates the notion of how objects are distributed. The Haar measure is used to make results of measure theory applicable in group theory. Given a matrix-valued function $f(Y)$ on the unitary group $\mathcal{U}(d)$ we write the integral  $\mathcal{U}(d)$ of $d\times d$ matrices of $f(Y)$ with respect to Haar measure as
\begin{equation*}
I = \int dY\, f(Y).
\end{equation*}
The Haar measure is left- and right-invariance with respect to shifts via multiplication. Hence
\begin{equation*}
\int dY\, f(YU) = \int d(Y'U^\dag)\, f(Y') = \int dY'\, f(Y),
\end{equation*}
where $U\in \mathcal{U}(d)$ is a fixed unitary. The right-invariant can be spelled out, respectively.

Since we can sample unitary operations using the Haar measure, also sampling quantum states uniformly at random is possible: Therefore, we simply generate the unitaries and apply them to a fixed basis state, i.e.\ $d\ket{\psi} \equiv dW\ket{0}$ \cite{Hayden2006}.
In this work, the Haar measure is used for determining an optimal lower bound on the probability that a quantum neural network gives an incorrect output for random input, see \cref{chapter:NFL}.

\section{Quantum algorithms}
\label{sec:QI_algorithms}

The big goal for quantum algorithms is \emph{quantum supremacy}. This aim is reached when a quantum algorithm can solve a specific task that cannot be solved in any feasible time by today's classical computers. On the one hand, this definition is not very strict and has often spark discussions on whether quantum supremacy was reached in a specific case or not \cite{Pednault2019}. On the other hand, proposals of such quantum algorithms are sometimes the motivation for new classical algorithms outperforming the proposed quantum algorithm \cite{Hastings2021}.

However, work in this direction already has also shown that quantum circuits can prepare probability distributions that are not reachable in classical computing \cite{Aaronson2005, Aaronson2011, Fujii2017, Fujii2018, Jozsa2013, Bremner2016, Farhi2016, Boixo2018}. In this section, we expound on the benefits of quantum algorithms compared to classical ones. Therefore, we study a simplified version of Deutsch's algorithm \cite{Nielsen2000}. In the end, we give a short overview of the most famous early quantum algorithms.

\subsection*{Exponential memory capacity}
One advantage of quantum information is the exponential memory capacity: whereas the state space of a classical computer grows with $2^b$, where $b$ is the number of computational bits, a $q$-qubit quantum device is described with a, much larger, $2^b$-dimensional Hilbert space, as was explained in \cref{subsec:QI_multiqubits}. When aiming to solve exponential problems, for example like simulating a quantum system \cite{Feynman1982,Georgescu2014}, classical computers run out of memory capacity very quick, and quantum computers can lead to opportunities. 

\subsection*{Quantum parallelism}
Besides, classical computers compute only classical functions. Assume, for example, a function, which get two bits as an input, i.e.\ the input is $\{0,0\}$, $\{0,1\}$, $\{1,0\}$ or $\{1,1\}$. If we want the output for all four inputs, we need to run the function four times. Due to the superposition principle, described in \cref{subsec:QI_superposition}, one can argue that a quantum computer taking a $q$-qubit quantum state as an input computes the output for $2^q$ inputs in parallel. This principle is called \emph{quantum parallelism} \cite{Deutsch1985,Deutsch1992} and can eventually lead to speed-up over classical computers. 

To be more precise, we assume in the following that the aim is to compute outputs of a classical one-bit function $f(x): \{0,1\}\rightarrow\{0,1\}$ for both inputs, $0$ and $1$ simultaneously. Two qubits are initialised as $\ket{x}=\frac{\ket{0}+\ket{1}}{\sqrt{2}}$ and $\ket{y}=\ket{0}$. Using a two-qubit unitary $U_f$, defined trough $U_f\ket{x,y}=\ket{x,y\oplus f(x)}$, the function $f(x)$ can be worked out on both inputs via
\begin{equation*}
U_f\ket{x,y}=\frac{\ket{0,f(0)}+\ket{1,f(1)}}{\sqrt{2}}
\end{equation*}
Note that $f(x)$ is a binary function and $y\oplus f(x)$ denotes the addition of $y$ and $f(x)$ modulo $2$. 

\begin{figure} [H]
\centering
\begin{tikzpicture}
\matrix[row sep=0.3cm, column sep=1cm] (circuit) {
\node(start2){$\ket{0}$};
& \node[operator0]{$H$}; 
& \node[]{}; 
& \node[]{}; 
& \node[operator0]{H};  
& \node[]{}; 
& \node[] (end2){}; \\
\node(start1){$\ket{1}$};
& \node[operator0]{$H$}; 
& \node[]{};
& \node[]{}; 
& \node[]{}; 
& \node[]{}; 
& \node[](end1){}; \\
};
\begin{pgfonlayer}{background}
\draw[] (start1) -- (end1)  
(start2) -- (end2);
\draw[dashed, color0] (-1.5,1) -- (-1.5,-1) 
(0.7,1) -- (0.7,-1)  
(2.7,1) -- (2.7,-1) ;
\node at (-1.5,-1.4){$ \ket{\psi_1} $};
\node at (0.7,-1.4){$ \ket{\psi_2} $};
\node at (2.7,-1.4){$ \ket{\psi_3} $};
\node[operator1,minimum height=1.5cm] at (-.5,0) {$U_f$};
\end{pgfonlayer}
\end{tikzpicture}	
\caption{\textbf{Deutsch's algorithm.} The quantum two-qubit circuit for Deutsch's algorithm is build of three Hadamard gates and a Unitary operation $U_f$.}
\label{fig:QI_deutsch}
\end{figure}
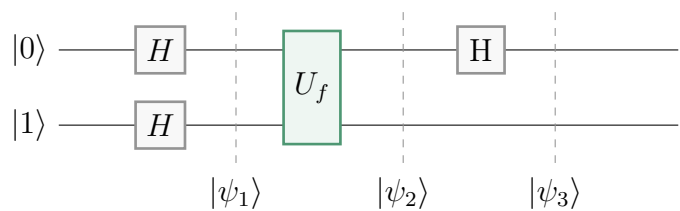

The problem with quantum parallelism is that measurements are required to extract classical data out of the quantum algorithm. The measurement gives only one result, $f(0)$ when $\ket{0}$ is measured and $f(1)$ for $\ket{1}$, respectively. The benefit of parallelism is not available. However, we explain in the following how this problem can be avoided and determining $f(0)\oplus f(1)$ is possible. To this end, we follow \cite{Nielsen2000} to present a simplified version of Deutsch's algorithm \cite{Deutsch1985}, depicted as a quantum circuit in \cref{fig:QI_deutsch}.

\subsection*{Deutsch's algorithm}
Unlike the ansatz above we initialise two qubits in the state $\ket{\psi_1}=\ket{xy}$, where
\begin{align*}
\ket{x}&=H\ket{0}=\frac{\ket{0}+\ket{1}}{\sqrt{2}}\text{,}\\ \ket{y}&=H\ket{1}=\frac{\ket{0}-\ket{1}}{\sqrt{2}}.
\end{align*}
Applying the unitary $U_f$ gives
\begin{equation*}
U_f\ket{\psi_1}=\ket{\psi_2}= 
\begin{cases} 
\pm\left(\frac{\ket{0}+\ket{1}}{\sqrt{2}}\right)\otimes\left(\frac{\ket{0}-\ket{1}}{\sqrt{2}}\right) &\mbox{if } f(0)=f(1)\\
\pm\left(\frac{\ket{0}-\ket{1}}{\sqrt{2}}\right)\otimes\left(\frac{\ket{0}-\ket{1}}{\sqrt{2}}\right) &\mbox{if } f(0)\neq f(1).
\end{cases} 
\end{equation*}
For the derivation it helps to see first that 
\begin{equation*}
U_f\left(\ket{x}\otimes\left(\frac{\ket{0}-\ket{1}}{\sqrt{2}}\right)\right)=(-1)^{f(x)}\ket{x}\otimes\left(\frac{\ket{0}-\ket{1}}{\sqrt{2}}\right).
\end{equation*}
After applying a Hadamard gate on the first qubit of $\ket{\psi_1}$ we end with the resulting state
\begin{equation*}
(H\otimes \mathbbm{1})\ket{\psi_2}=\ket{\psi_3}= 
\begin{cases} 
\pm\ket{0}\otimes\left(\frac{\ket{0}-\ket{1}}{\sqrt{2}}\right) &\mbox{if } f(0)=f(1)\\
\pm\ket{1}\otimes\left(\frac{\ket{0}-\ket{1}}{\sqrt{2}}\right) &\mbox{if } f(0)\neq f(1).
\end{cases} 
\end{equation*}
Since
\begin{equation*}
\ket{\psi_3}= 
\begin{cases} 
f(0)\oplus f(1)=0 &\mbox{if } f(0)=f(1)\\
f(0)\oplus f(1)=1  &\mbox{if } f(0)\neq f(1),
\end{cases} 
\end{equation*}
we can rewrite $\ket{\psi_3}$ as 
\begin{equation*}
\ket{\psi_3}=\pm\ket{f(0)\oplus f(1)}\otimes\left(\frac{\ket{0}-\ket{1}}{\sqrt{2}}\right).
\end{equation*}

Hence when measuring the first qubit of $\ket{\psi_3}$, we can determine $f(0)\oplus f(1)$, which is remarkable since we only used one evaluation of $f(x)$ during the process. Although a classical computer can work with probabilistic methods evaluating $f(0)$ with probability $0.5$ and $f(1)$ with the same probability, a quantum algorithm can work with the interference of the values. In that way, the determination of global information, for example, here $f(0)\oplus f(1)$, is possible.

\subsection*{Famous algorithms}
To conclude this section, we want to give a quick overview of the most famous quantum algorithms.

We already motivated and explained Deutsch's algorithm \cite{Deutsch1985} above, which is able to compute whether a one-bit function $f(x)$ is constant, i.e.\ $ f(0)\oplus f(1)=0 $ or balanced, i.e.\ $ f(0)\oplus f(1)=1$. The generalisation from a one-bit to n-bit function of Deutsch's algorithm is known as \emph{Deutsch-Jozsa algorithm} \cite{Deutsch1992}.

An algorithm that solves a problem of more practical use is \emph{Shor's factoring algorithm} \cite{Shor1994}: using the quantum version of the Fourier transform \cite{Nielsen2000,Coppersmith2002} this algorithm can find the prime factors of a given number. Remarkable is that whereas the best nowadays known classical algorithm runs in sub-exponential time \cite{Buhler1993}, Shor's factoring algorithm runs in polynomial time. 

Next to quantum Fourier transform-based algorithm, quantum search algorithms gained much attention:  the task is to find a particular item given a list of length $N$. Classically this problem requires approximately $N$ operations.  Although not providing a speed-up as sensational as Shor's algorithm, it is noteworthy that \emph{Grover's search algorithm} \cite{Grover1996} solves the task using only approximately $\sqrt{N}$ operations. More detailed reviews of the mentioned algorithms can be found at \cite{Cleve1998}. 

\section{Quantum computers}
\label{sec:QI_QC}

From the view of a theoretician, it is compelling to sink directly into the theory of new algorithms after learning the basics of quantum computing. However, we should bear in mind that all these approaches have the most beneficial effect in physical realisation. Therefore we will give in the following lines a short introduction to the problems but also the successes of the experimental side of quantum computing.  

\subsection*{Qubit implementation}
Since the quantum computer's elementary component is the qubit, implementing the latter is fundamental for building a quantum computer. As explained in \cref{sec:QI_states} such a two-level system can be realised through a physical object that can exist in a superposition of two states. The problem is that, on the one hand, these physical qubits have to be well isolated to preserve their features but, on the other hand, accessible for the computational tasks and measurements \cite{Nielsen2000}. Achieving a good balance of both of these requirements is the leading implementation problem. Moreover, a general rule is that the \emph{decoherence time} has to be longer than the gate operation time. The therm \emph{decoherence} describes here the irreversible interactions of the qubits with the environment, also referred to as \emph{leakage} \cite{Plenio1997,McEwen2021}.

The quality of the qubit implementation can be described by the accuracy with which quantum gates can be performed. The error rate per gate for two-qubit gates with the best hardware for controlling trapped ions \cite{Ballance2016,Bruzewicz2019} or superconducting circuits \cite{Barends2014,Corcoles2015,Ofek2016} is above $0.1\%$. This induces a limit in the number of quantum gates that can be applied within one quantum circuit before the noise overwhelms the signal. Other limiting parameters are the time per gate execution and the measurement error probability. The time per gate execution strongly depends on the specific quantum computer model, whereas the measurement error probability is often about $1\%$ \cite{Preskill2018}.

\subsection*{NISQ devises}
Despite these challenges, in the last years, large-scale quantum computation became possible. Big companies like Microsoft, Google and IBM competed in building quantum computers. In 2019, IBM launched a quantum processor containing $53$ qubits \cite{Pednault2019}. In the same year, Google announced their processor \emph{Sycamore}, also including $53$ qubits based on a superconducting circuit \cite{Arute2019}. 

Today's quantum devices are called \emph{noisy intermediate-scale quantum devices} (NISQ devices) \cite{Preskill2018, Brooks2019, Arute2019}. These early quantum computers are characterised by their high sensitivity to the environment (\emph{noisy}) and limited number of available qubits and applicable gates (\emph{intermediate-scale}). Due to these limitations, today's quantum computers do not achieve fault-tolerant quantum computation \cite{Shor1996,Preskill1998}. However, future generations of these devices may be able to perform tasks which exceed the capabilities of classical supercomputers. 

Noteworthy is moreover that IBM offers the free access to some of their quantum devices \cite{IBMQuantum2021}. Naturally, the availability of NISQ devices leads to new opportunities. One of them is the implementation of quantum learning algorithms.

\section{Quantum neural networks}
\label{sec:QI_QNN}
In the last section, we discussed the exciting and ongoing progress on quantum computers. Beforehand, \cref{chapter:ML} gave an introduction to the widely spreading world of neural networks, including various algorithms, optimisation methods and countless applications. Hence it is not surprising that merging both of these topics in the field of \emph{quantum machine learning} (QML) \cite{Sasaki2002, Schuld2014,Adcock2015,Dunjko2016,Farhi2017,Biamonte2017,Dunjko2018,Ciliberto2018,Du2020} carries great promise for the discovery of new opportunities.

As discussed in \cref{chapter:ML}, the method of artificial neural networks (NNs) is very popular in classical machine learning (ML). Consequently, in the past years, the construction of a quantum analogue and its usage for QML was of great interest. The term \emph{quantum neural network} (QNN) was introduced first by Subhash Kak in 1995 \cite{Kak1995} with the motivation of connecting investigations in the field of neuroscience with characteristics of quantum computation. Whereas the proposals in the nineteenths remain to be of a rather theoretical idea \cite{Kak1995,Menneer1995,Zak1998}, today QNNs are mostly viewed as a subset of practical quantum circuits containing parametrised gates. Many different proposals on how to construct these have been made \cite{
Andrecut2002, 
Oliveira2008, 
Panella2011, 
Silva2016, 
Cao2017, 
Wan2017, 
Alvarez2017, 
Farhi2018, 
Killoran2019, 
Steinbrecher2019, 
Torrontegui2019, 
Sentis2019, 
Tacchino2020, 
Beer2020, 
Skolik2020, 
Zhang2020, 
Schuld2020, 
Sharma2020, 
Zhang2021 
}. Similar to their classical counterparts QNNs are often build of a fundamental building block. These \emph{quantum perceptrons} have been proposed in various ways in the last years \cite{
Lewenstein1994, 
Altaisky2001, 
Siomau2014, 
Schuld2015, 
Silva2016, 
Cao2017, 
Wan2017, 
Mangini2020,
Torrontegui2019, 
Beer2020,  
Zhang2021 
}. Note, that some of the named QNNs are designed for pure quantum tasks, whereas others can be used for classical input. In the latter case, the classical data has to be encoded first. In \cite{Tacchino2021} it is shown, for example, how to encode a binary vector of dimension $d$ using $\log_2 d$ qubits. 

The detailed discussion of all of these proposals would exceed the scope of this section. However, we want to describe two of them here shortly as they will play a role in the rest of the thesis: \cref{chapter:DQNN} considers comprehensively the QNNs proposed in \cite{Beer2020}, called \emph{dissipative quantum neural networks} (DQNNs). In this ansatz, the QNN is built of multiple layers of qubits. The quantum perceptrons connect qubits of two consecutive layers with unitary operations and feed-forward the input state over the layers through the network. On the contrary, other QNN proposals are constructed by just one layer of qubits. The perceptrons are then, for example, defined as a sequence of alternating unitary operators \cite{Farhi2014, Farhi2016, Hadfield2019} which act all on the same qubits. Such an algorithm will be discussed in \cref{sec:DQNN_QAOA}. Note that an overview of the most recent QNN proposals can be found in \cite{Mangini2021}.

\subsection*{Implementation}
\label{subsec:QI_Implementation}
QNNs can be implemented on today's quantum computers as \emph{variational quantum algorithms} (VQA) \cite{Cerezo2020}. Their process is a quantum-classical hybrid: the algorithms themselves are executed on a quantum computer, but the optimisation process is done classically. In particular \emph{parametrised quantum circuits} \cite{Mitarai2018,Benedetti2019,Du2020,Bu2021 } are used for the implementation. These quantum circuits consist of unitary transformations and can be modified using parameters $\vec{\theta}$. Further, a quantum algorithm computes the training loss function. During the training the parameters are updated classically such that the training loss is optimised \cite{Mitarai2018,Ostaszewski2019,Schuld2019,Stokes2020,Beer2021a}, since this task can be efficiently fulfilled in this way. 

A substantial benefit of VQAs is that they can be successfully executed on NISQ devices. As explained in \cref{sec:QI_QC} these devices are highly impaired through noise entering with each quantum gate \cite{Preskill2018}. This limits the number of quantum gates within one quantum circuit before the noise outweighs the algorithm's performance. It follows that only quantum circuits of small depth can be applied. In contrast to famous methods like Shor's factoring algorithm \cite{Shor1994} or Grover's search algorithm \cite{Grover1996} which require too many gates, QNNs can be executed on these early quantum computers. In \cref{sec:DQNN_quantumalg} and \cref{sec:DQNN_QAOA} we will discuss the implementations of two of the QNN proposals mentioned above.

\subsection*{Challenges}
\label{subsec:QI_Challenges}

Although we can train QNNs on today's NISQ devices, the execution remains challenging: both, the limited number of available qubits as well as the high noise levels can lead to problems and limits. Especially when working with quantum circuits of higher depth, the accurate computation of the loss gradients becomes ambitious \cite{Preskill2018,Xue2019,Alam2019,Maciejewski2021}. 

This is not the only challenge we have to accept: \emph{Barren plateaus}, the phenomenon of exponentially vanishing gradients of the training loss function in the number of qubits \cite{McClean2018,Grant2019,Wang2020,Cerezo2020a,Mangini2021}, can lead to significant problems. The update of the parameters $\vec{\theta}$ is based on the gradient of the training function, similar to the classical case which was explained in \cref{sec:ML_optimisation}. When this gradient approaches zero for all components the change of the parameters will stagnate and so the training process. Meeting this case, it is advisable to change the architecture of the QNN and/or the training function.

\subsection*{Opportunities}
As well as their classical analogous, QNNs are able to perform a variety of tasks. The algorithms can be used for tasks including classical data, for example image processing \cite{Tacchino2020,Liu2020}. However, with the advent of quantum computers containing more than just a handful qubits the task of coping with large amounts of quantum data becomes crucial. State tomography \cite{Vogel1989}, or quantum process tomography \cite{Poyatos1997}, where the number of needed samples increases exponentially with the number of particles \cite{Mohseni2008} is out of question. 

Hence, it is exciting that QNNs allow characterisations with fewer samples compared to tomography methods. Specifically, they can be applied for the classification of classical or quantum data  \cite{Sentis2019,Schuld2019a, Schuld2020} or de-noising quantum data \cite{Wan2017, Romero2017, Pepper2019,Bondarenko2020,Achache2020}. Also the learning of graph-structured quantum data \cite{Verdon2018,Sedlak2019,Verdon2019,Cong2019,Arunachalam2017,Beer2021} was studied. This will be discussed in detail in \cref{chapter:graphs}. Moreover the characterisation of quantum devices is possible, often formulated as the assignment of learning an unknown unitary operation \cite{ Verdon2018,Sedlak2019,Beer2020, Beer2021a,Kiani2020,Geller2021}. 

The latter task will be discussed in the next chapter of this work, where the topic of QNNs will be presented in detail using the example of DQNNs. The discussion includes a precise description of the DQNN architecture and algorithm, implementation details and training results. 

%% file: text/DQNN.tex
\chapter{Dissipative quantum neural networks}\hypertarget{DQNN}{}
\label{chapter:DQNN}

The introductions to \emph{machine learning} (ML) and quantum information in \cref{chapter:ML} and \cref{chapter:QI} prepared the reader for the following chapter, probably the heart of this thesis. The here presented quantum analogue of a classical neuron \cite{Beer2020} leads not only to a quantum feed-forward neural network capable of universal quantum computation with remarkable generalisation behaviour and robustness to noisy training data but is also the basis for further ongoing studies. Two of them, the usage of graph-structured data (see \cref{chapter:graphs}) and an ansatz for a quantum generative adversarial network (see \cref{chapter:QGAN}), will be presented later in this work.

As discussed in \cref{chapter:intro}, one can categorise \emph{quantum machine learning} (QML) techniques into classical ML improving quantum tasks, quantum algorithms speeding up classical ML and exploiting quantum computing devices for tasks with quantum data. The \emph{dissipative quantum neural networks} (DQNNs) is integrated into the latter category. It consists of layers of qubits and can be trained with pairs of quantum states. A training data pair consists of an input state and an aimed output according to the training goal. 

The DQNN is built of quantum perceptrons. Such a building block connects two consecutive layers of qubits and can be represented as a completely positive (CP) transition map. The word \emph{dissipative} describes the action of these maps: such a map does not only contain tensoring the state of the current layer to the state of the next layer's qubits and applying unitary operations but also tracing out the qubits from the first of the two layers. In this way, the layer-to-layer transition maps feed-forward input states through the DQNN. The obtained output state can then be compared with the aimed output. This comparison is made through the fidelity of two quantum states and allows conclusions about how the perceptron unitaries must be updated to fulfil the training goal better. 

Both, the simulation on classical computers as well as the implementation on quantum computers provide fast optimisation of the network. We focus on supervised training in this chapter and challenge the algorithm with the job of learning an unknown unitary operation. In addition to that, quantum neural networks (QNNs) of this kind can also be exploited for unsupervised learning task and act as \emph{quantum autoencoders} to de-noise entangled quantum states \cite{Bondarenko2020}. 

One advantage of DQNNs is that the dissipative structure leads to reduced memory requirements, since the number
of required qubits scales with only the width, not the depth of the QNN. This enables us to train deep QNNs, if the quantum device allows qubits to be \enquote{reused}. Furthermore, no disturbing effects of the Barren plateau phenomenon, discussed in \cref{subsec:QI_Challenges}, have been observed in the studies of \cite{Beer2020}. When the quantum neurons are sufficiently local and sparse, the QNNs might totally avoid these feared vanishing gradients of the training loss function \cite{Sharma2020}. Finally, numerical results have shown that DQNNs reach the fundamental information-theoretic limits on quantum learning stated in the form of the \emph{quantum no free lunch} (QNFL) theorem. This bound will be discussed in \cref{chapter:NFL}.

This chapter will introduce the reader to DQNNs in their entirety. The first four sections follow the work of \cite{Beer2020}. We start by explaining the layer-to-layer transition maps and the general network architecture in \cref{sec:DQNN_networkarchitecture}. We focus on the task of learning an unknown unitary, hence in \cref{sec:DQNN_lossfunctions} we formulate the training data for this exemplary task and a suiting loss for training the DQNN. Further, we define a second loss function for testing the behaviour of the QNN after training. The training algorithm itself is explained in \cref{sec:DQNN_trainingalgorithm}. All derivations of the rules used for updating the quantum perceptrons are derived at this point. \cref{sec:DQNN_universality} describes a vivid proof of the above-mentioned universality. The classical simulation of the training leads to promising results, which are presented in \cref{sec:DQNN_classical}. Additionally, the implementation on a NISQ quantum device and training results are discussed in \cref{sec:DQNN_quantumalg}, following \cite{Beer2021a}. We end the chapter with a comparison of the DQNN algorithm with another QNN architecture in \cref{sec:DQNN_QAOA}.

\section{Network architecture}
\label{sec:DQNN_networkarchitecture}

In analogy to a classical neural network (NN), see \cref{sec:ML_neuralnetworks}, the DQNN is build of \emph{quantum perceptrons} acting on qubits arranged in layers. We start with a description of the perceptrons which act as building blocks of the QNN and explain how the QNN is built using these afterwards.

\FloatBarrier \subsection*{Quantum perceptron}

\begin{figure}
\centering
\begin{subfigure}[t]{0.19\linewidth}
\centering
\begin{tikzpicture}[]
\coordinate (n) at (2,0);
\foreach \x in {-1,0,1} {
\draw[line0] (n) -- (0,\x);
\draw[line1] (n) -- (0,\x);
\node[perceptron0] at (0,\x){};}
\node[perceptron0] at (n){};	
\node[white] at ($(n) + (-2,-1.7)$){shift};
\end{tikzpicture}
\subcaption{Network.}
\label{fig:DQNN_percA}
\end{subfigure}
\begin{subfigure}[t]{0.59\linewidth}
\centering
\begin{tikzpicture}[]
\matrix[row sep=0.5cm, column sep=0.5cm] (circuit) {
\node(start4){};
& \node[]{}; 
& \node[]{}; 
& \node[]{}; 
& \node[dcross](end4){}; 
& \node {}; \\
\node(start3){};
& \node[]{}; 
& \node[]{}; 
& \node[]{}; 
& \node[dcross](end3){}; 
& \node {}; \\
\node(start2){};
& \node[]{}; 
& \node[]{}; 
& \node[]{}; 
& \node[dcross](end2){}; 
& \node {}; \\
& \node[]{}; 
\node(start1){$\ket{0}\bra{0}$};
& \node[]{}; 
& \node[]{}; 
& \node[]{}; 
& \node[](end1){$\rho ^\text{out}$}; \\
};
\begin{pgfonlayer}{background}
\draw[] (start1) -- (end1)  
(start2) -- (end2)
(start3) -- (end3)
(start4) -- (end4);
\node[operator1,minimum width=.8cm,minimum height=3cm] at (0,.1) (a) {$U$};
\draw[brace0,color=black] {(-2.9,-0.3) -- node[left=1ex] {$\rho^{\text{in}}$}  (-2.9,1.5)};	
\end{pgfonlayer}
\end{tikzpicture}
\subcaption{Implementation }
\label{fig:DQNN_percB}
\end{subfigure}
\caption{\textbf{Quantum perceptron}. The implementation of the perceptron, here acting on $4$ qubits with $m=3$ and $n=1$, can be depicted in network notation (a). The implementation as quantum circuit includes initialisation of $m$ qubits in the input state and $n$ in the zero state, applying a unitary operation $U$ and tracing out $m$ qubits (b).}
\label{fig:DQNN_perc}
\end{figure}
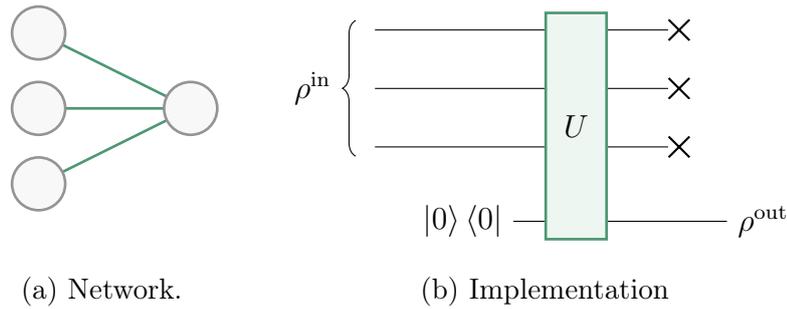

The perceptrons are engineered as arbitrary unitary operators. Such a perceptron unitary acts on $m+n$ qubits and depends on $(2^{m+n})^2-1$ parameters. We define $m$ of the qubits as input qubits and $n$ as output qubits. It is further needed for the algorithm that the input qubits are initialised in a state $\rho^{\text{in}}$ and the output qubits in a product state $\ket{0...0}$, respectively. After applying the perceptron unitary, the $m$ input qubits are traced out and we are left with the $n$-qubit state $\rho^\text{out}$. This output state has in general not the same dimension as the input state. \cref{fig:DQNN_perc} makes clear how an exemplary perceptron can be depicted in network or quantum circuit notation. One single perceptron can be seen as a small DQNN consisting only of two layers of qubits and one unitary operation.

Before describing how to build a large DQNN constructed of many perceptrons, we want to make two remarks: for simplicity, we set $n=1$ here, i.e.\ the perceptrons are $m+1$-qubit unitaries. As we will see in \cref{sec:DQNN_universality} this will not threaten the universality of this model.  Further, although we suppose working with $2$-level qubits, it can be handy to notice that a so defined perceptron can be easily generalised for qudits.

\FloatBarrier \subsection*{Quantum neural network}
\begin{figure}
\centering
\begin{tikzpicture}[scale=1.4]
\begin{scope}[xshift=0.9cm,yshift=1.45cm]
\draw[brace0] 
(-1.25,0) -- node[above=1ex] {$U^1=U_3^1U_2^1U_1^1$}
(1.5,0);   
\end{scope}
\foreach \x in {-.5,.5} {
\draw[line0] (0,\x) -- (2,-1);
\draw[line3] (0,\x) -- (2,-1);
\draw[line0] (0,\x) -- (2,0);
\draw[line2] (0,\x) -- (2,0);
\draw[line0] (0,\x) -- (2,1);
\draw[line1] (0,\x) -- (2,1);
}
\foreach \x in {-1.5,-0.5, ..., 1.5} {
\draw[line0] (2,-1) -- (4,\x);
\draw (2,-1) -- (4,\x);
\draw[line0] (2,0) -- (4,\x);
\draw (2,0) -- (4,\x);
\draw[line0] (2,1) -- (4,\x);
\draw (2,1) -- (4,\x);
}
\foreach \x in {-1.5,-0.5, ..., 1.5} {
\draw[line0] (4,\x) -- (6,-0.5);
\draw (4,\x) -- (6,-0.5);
\draw[line0] (4,\x) -- (6,0.5);
\draw (4,\x) -- (6,0.5);
}
\foreach \x in {-1,0,1} {
\node[perceptron0] at (2,\x) {};
}
\foreach \x in {-1.5,-0.5, ..., 1.5} {
\node[perceptron0] at (4,\x) {};
}
\node[perceptron0] at (0,-0.5) {};
\node[perceptron0] at (0,0.5) {};
\node[perceptron0] at (6,-0.5) {};
\node[perceptron0] at (6,0.5) {};
\node at (0,-2.0){$l=\text{in}$};
\node at (2,-2.0){$l=1$};
\node at (3,-2.0){$\dots$};
\node at (4,-2.0){$l=L$};
\node at (6,-2.0){$l=\text{out}$};
\draw[brace0] 
(0.75,-2.6) -- node[below=1ex] {input layer} (-0.75,-2.6); 
\draw[brace0] 
(5,-2.6) -- node[below=1ex] {hidden layers} (1,-2.6); 
\draw[brace0] (6.75,-2.6) -- node[below=1ex] {output layer} (5.25,-2.6); 
\end{tikzpicture}
\caption{\textbf{Network representation of a DQNN.} This DQNN consists of $L+2$ layers of qubits, and $L+1$ layers of quantum perceptrons represented by unitary operations $U^l_j$.}
\label{fig:DQNN_genQNN}
\end{figure}
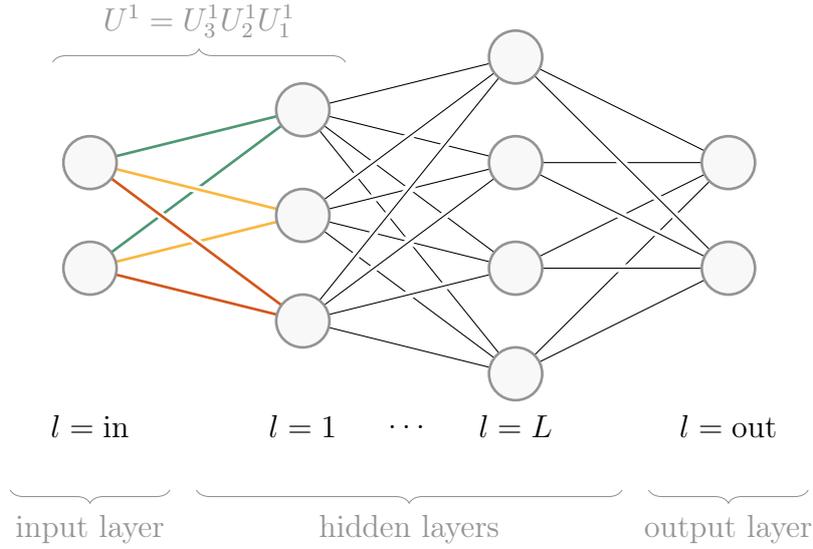

The DQNN is a quantum circuit built out of $L+2$ layers of qubits. We name the first layer \emph{input} layer, the last layer \emph{output} layer and count the $L$ layers in between, namely the \emph{hidden} layers, with the variable $l$, see \cref{fig:DQNN_genQNN}. In each layer, the perceptron unitaries are applied layerwise and from top to bottom. In that way two subsequent layers are fully connected through perceptrons. Remember that to each perceptron an arbitrary unitary is assigned, and therefore the perceptrons do not generally commute. The order of application of the perceptron unitaries is indicated with over- and under-crossings in \cref{fig:DQNN_genQNN}. 

In the introduction to this chapter we already mentioned that the whole network can be expressed as a composition of layer-to-layer transition maps. This notation will be highly important through the whole chapter. According to the definition of a single perceptron, we can phrase the output state of the DQNN as 
\begin{equation}\label{eq:DQNN_rhoOut}
\rho^\text{out}=\mathcal{E}\left(\rho^{\text{in}}\right)= \mathcal{E}^{L+1}\left(\mathcal{E}^{L}\left(\dots \mathcal{E}^{2}\left(\mathcal{E}^{1}\left(\rho^{\text{in}}\right)\right)\dots\right)\right),
\end{equation}
using the CP maps $\mathcal{E}^l$ defined via
\begin{equation}\label{eq:DQNN_E}
\mathcolorbox{\mathcal{E}^{l}(X^{l-1}) \equiv \tr_{l-1}\big(\prod_{j=m_l}^{1} U^l_j (X^{l-1}\otimes \ket{0...0}_l\bra{ 0...0})\prod_{j=1}^{m_l} {U_j^l}^\dag\big),}
\end{equation}
where $U_j^l$ is assigned to the $j$th perceptron acting on the qubit layers $l-1$ and $l$, and $m_l$ is the total number of perceptrons acting on layers $l-1$ and $l$. See \cref{def:QI_CP} for an introduction to CP maps. From this notation, it becomes clear that the information propagates from the input to the output layer and a quantum feed-forward NN is implemented. This fact is the groundwork for the back-propagation algorithm, which we will discuss in \cref{sec:DQNN_trainingalgorithm}.

We want to close the discussion of the network architecture with a short observation: it is possible to see the quantum circuit of the network as a single unitary $\mathcal{U}=U^\text{out}U^LU^{L-1}\dots U^1$, where $U^l=U_{m_l}^l \dots U_1^l$ are the layer unitaries, comprised of a product of quantum perceptrons acting on the qubits in layers $l-1$ and $l$, see \cref{fig:DQNN_genQNN}. However, note that every $U^l_k$ has to be extended by identities for the remaining qubits to get the correct dimensions when implementing the architecture. We leave these off for a cleaner notation here. Taking this into account the formula for the output state can be expressed as 
\begin{equation*}
\mathcal{E}\left(\rho^{\text{in}}\right)\equiv\tr_\text{in,hid}(\mathcal{U}(\rho^\text{in} \otimes \ket{0...0}_\text{hid,out}\bra{0...0})\mathcal{U}^\dagger).
\end{equation*} 

\section{Loss functions}
\label{sec:DQNN_lossfunctions}

After describing the structure of the DQNN and before explaining the algorithm, we need to formulate a training goal. In this section, we do so by formulating two loss functions: one is aimed to be maximised during training, and a second one is used to check the trained DQNN afterwards. 

\FloatBarrier\subsection*{Training and validation data}
We start by specifying a learning task. Therefore, imagine the following setting: we have access to a not-trusted or uncharacterised quantum device or a similar source acting as a unitary $Y$ on $m$-qubit input states. This device can be repeatedly initialised and applied to arbitrary states.  In that way it is possible to prepare training data pairs, also called supervised data, structured as $\{\ket{\phi^{\text{in}}_x}, \ket{\phi^{\text{SV}}_x}\} $ with $\ket{\phi^\text{SV}_x} = Y\ket{\phi^\text{in}_x}$ and $x=1,2,\dots,N$. Notice that the input state generally does not have to be pure, but we district our discussion to this case for simplicity here. In the following we use the notation $\rho^{\text{in}}_x\equiv\ket{\phi^{\text{in}}_x}\bra{\phi^{\text{in}}_x}$ for the input. Notice further that after generating $N$ training data pairs, we will only use $S$ of them for training. $N-S$ pairs will not be used for the training algorithm, but for testing the generalisation behaviour of the trained network as explained later.

At this point we have to mention a crucial point: in our scenario and using the in the afterwards discussed algorithm, it is essential that we can request multiple copies of each of our $S$ training pairs to overcome quantum projection noise in evaluating the derivative of the loss function. Since \emph{cloning} an unknown quantum state is not possible \cite{Nielsen2000}, we can not simply copy the training data pairs. Instead we assume we have a device, where it is possible to press the button $x$ and we get the training data pair $\{\ket{\phi^{\text{in}}_x}, \ket{\phi^{\text{SV}}_x}\}$. It is worth mentioning at this point that this procedure in combination with QML is still more efficient then characterising a $n$-qubit device via tomography \cite{Vogel1989, Poyatos1997}, where the number of needed samples scales exponentially with the number of qubits $n$.

The more neurons we use to build the QNN, the more copies per training round are needed. In detail we need $n^{\text{proj}}\times n^{\text{par}}$ copies, where $n^{\text{par}}$ is the total number of parameters in the network and $n^{\text{proj}}$ the factor coming from repetition of measurements to reduce projection noise. $n^{\text{par}}$ is linear in the number of neurons. It is given by 
\begin{align*}
n^\text{DQNN} & =\sum_{l=1}^{L+1} \sum_{j=1}^{m_l} \text{\#param} (U^l_j)  =\sum_{l=1}^{L+1} \sum_{j=1}^{m_l} (4^{(m_{l-1}+1)}-1)  =\sum_{l=1}^{L+1} m_l\times (4^{(m_{l-1}+1)}-1),
\end{align*} where $\text{\#param} (U^l_j)$ denotes the number of parameters used to describe $U^l_j$ and $m_l$ is the number of perceptrons acting on the qubit layers $l-1$ and layer $l$. The $-1$ term appears because the overall phase of the unitaries is unimportant. Moreover, we assume that the used DQNN has number of input and output qubits suiting the training data, i.e.\ $m=m_{0}=m_{L+1}$.

\FloatBarrier\subsection*{Training loss}
\label{subsec:DQNN_trainingloss}
Since we already discussed the training data set and goal, we can now investigate the loss functions. Given an architecture in the form of a DQNN, the aim is now to exploit the training pairs to learn the action of the given quantum device and perfectly reproduce it through the network. This means the loss function should not only be meaningful to the problem, but reach a global extremum if the training goal is reached. During the training it is to avoid that the the gradient of the loss function vanishes. Furthermore, we require that the function be efficiently computed on a quantum computer.  

The training algorithm, explained in \cref{sec:DQNN_trainingalgorithm}, naturally includes updating the DQNN. This is the centrepiece of the training and the main step to reach our training goal which can be formulated the following way: we desire the network's output $\mathcal{E}\left(\rho^\text{in}_x\right)$ to be as close as possible to the correct output $\ket{\phi^\text{SV}_x}$ for a specific input $\ket{\phi^\text{in}_x}$. We use an essentially unique measure of closeness for (pure) quantum states, to quantify this, and define the \emph{training loss function} as the fidelity $F$ between the QNN output and the desired output, averaged over the training data. We aim to optimise, more precisely to maximise, the loss function during training. For pure supervised states the loss function takes the form  
\begin{equation*}
\mathcolorbox{\mathcal{L}_\text{SV}=\frac{1}{S}\sum_{x=1}^S F(\ket{\phi^{\text{SV}}_x}\bra{\phi^{\text{SV}}_x},\rho_x^{\text{out}}) = \frac{1}{S}\sum_{x=1}^S \braket{\phi^{\text{SV}}_x|\rho_x^{\text{out}}|\phi^{\text{SV}}_x}}
\end{equation*}
and has the domain $[0,1]$. We aim for the value $1$ during training, since the fidelity reaches $1$, if the two compared states are equal.

Choosing the fidelity as a distance measure of a pure state and a mixed state has a few advantages. Not only is the fidelity a good generalisation of the risk function considered in training classical NN, but it can also be measured and has an operational meaning. See \cref{subsec:QI_distance} for a more detailed discussion.

\FloatBarrier\subsection*{Validation loss}
Quantum data is generally very limited. Therefore it is of great interest to analyse how well the network performs on unseen data. After the training we can do this with checking how well the network predicts the $N-S$ unsupervised outcomes. For this task we specify the \emph{validation loss} 
\begin{equation*}
\mathcolorbox{\mathcal{L}_\text{USV}=\frac{1}{N-S}\sum_{x=S+1}^{N} \langle\phi^{\text{USV}}_x\rvert\rho_x^{\text{out}}\lvert\phi^{\text{USV}}_x\rangle.}
\end{equation*}
This loss functions also helps to analyse, if the training the DQNN leads to overfitting, i.e.\ the model matches the training data but does not perform accurately on unseen data, which is generally to avoid. 

\section{Training algorithm}
\label{sec:DQNN_trainingalgorithm}

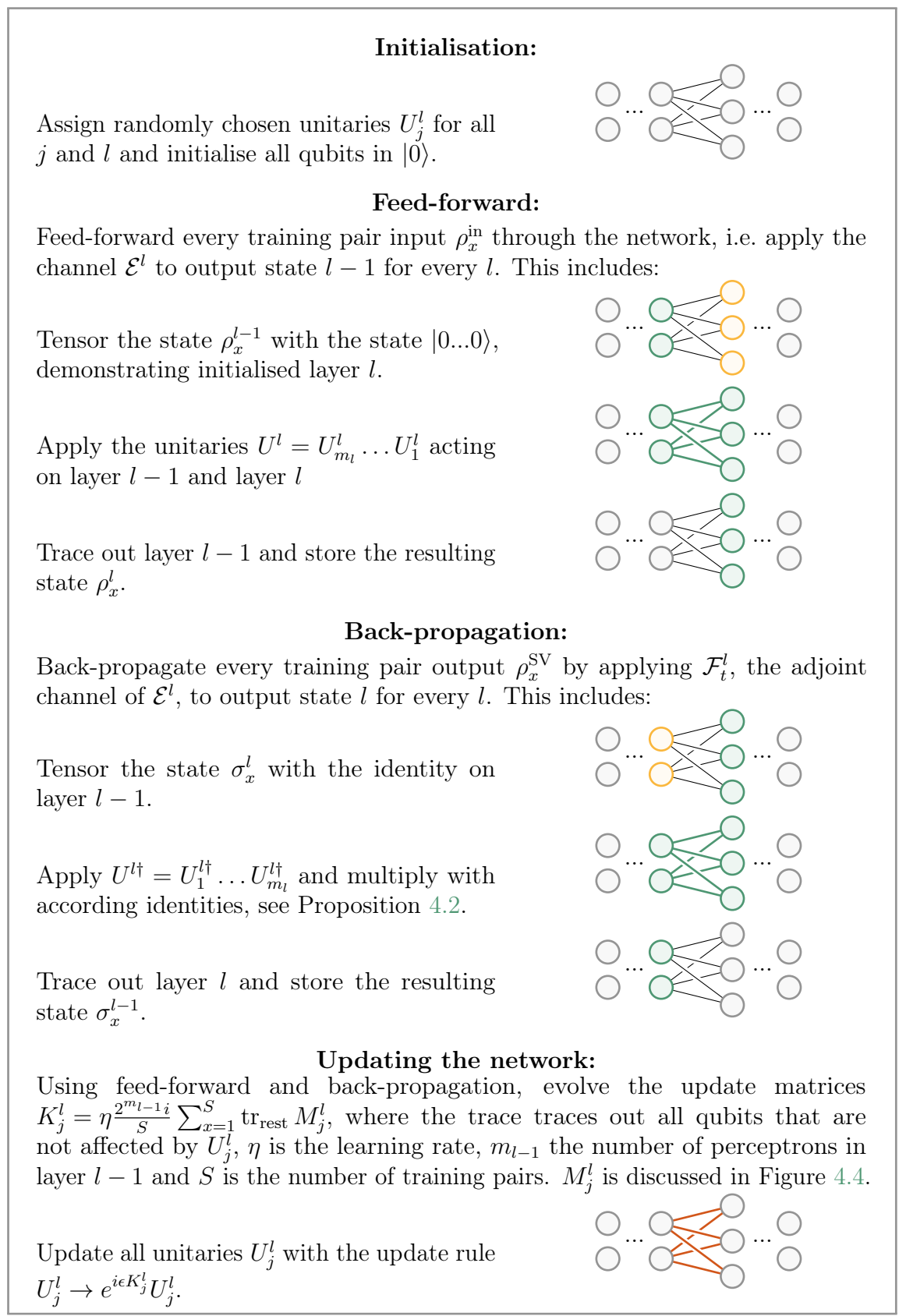
\begin{figure}
\centering
\fcolorbox{color0}{white}{
\begin{minipage}{\textwidth}
\begin{tabularx}{\textwidth}{ X C  }
\heading{Initialisation:}\\
Assign randomly chosen unitaries $U_j^l$ for all $j$ and $l$ and initialise all qubits in $\ket{0}$. & 	\begin{tikzpicture}[scale=0.6]
\foreach \x in {-.5,.5} {
\draw[line0] (0,\x) -- (2,-1);
\draw (0,\x) -- (2,-1);
\draw[line0] (0,\x) -- (2,0);
\draw (0,\x) -- (2,0);
\draw[line0] (0,\x) -- (2,1);
\draw (0,\x) -- (2,1);}
\node[perceptronS0] at (-1.5,-0.5) {};
\node[perceptronS0] at (-1.5,0.5) {};
\node at (-0.75,0) {...};
\node[perceptronS0] at (0,-0.5) {};
\node[perceptronS0] at (0,0.5) {};
\node[perceptronS0] at (2,-1) {};
\node[perceptronS0] at (2,0) {};
\node[perceptronS0] at (2,1) {};
\node at (2.85,0) {...};
\node[perceptronS0] at (3.6,-0.5) {};
\node[perceptronS0] at (3.6,0.5) {};
\end{tikzpicture}\\
\heading{Feed-forward:}\\
\headingtext{Feed-forward every training pair input $\rho_x^\text{in}$ through the network, i.e.\ apply the channel $\mathcal{E}^l$ to output state $l-1$ for every $l$. This includes:}\\
Tensor the state $\rho_x^{l-1}$ with the state $\ket{0...0}$, demonstrating initialised layer $l$.	& \begin{tikzpicture}[scale=0.6]
\foreach \x in {-.5,.5} {
\draw[line0] (0,\x) -- (2,-1);
\draw (0,\x) -- (2,-1);
\draw[line0] (0,\x) -- (2,0);
\draw (0,\x) -- (2,0);
\draw[line0] (0,\x) -- (2,1);
\draw (0,\x) -- (2,1);}
\node[perceptronS0] at (-1.5,-0.5) {};
\node[perceptronS0] at (-1.5,0.5) {};
\node at (-0.75,0) {...};
\node[perceptronS1] at  (0,-0.5) {};
\node[perceptronS1] at  (0,0.5) {};
\node[perceptronS2] at  (2,-1) {};
\node[perceptronS2] at  (2,0) {};
\node[perceptronS2] at  (2,1) {};
\node at (2.85,0) {...};
\node[perceptronS0] at (3.6,-0.5) {};
\node[perceptronS0] at (3.6,0.5) {};
\end{tikzpicture}\\
Apply the unitaries $U^l=U_{m_l}^l\dots U_1^l$ acting on layer $l-1$ and layer $l$&	\begin{tikzpicture}[scale=0.6]
\foreach \x in {-.5,.5} {
\draw[line0] (0,\x) -- (2,-1);
\draw[line1] (0,\x) -- (2,-1);
\draw[line0] (0,\x) -- (2,0);
\draw[line1] (0,\x) -- (2,0);
\draw[line0] (0,\x) -- (2,1);
\draw[line1] (0,\x) -- (2,1);}
\node[perceptronS0] at (-1.5,-0.5) {};
\node[perceptronS0] at (-1.5,0.5) {};
\node at (-0.75,0) {...};
\node[perceptronS1] at  (0,-0.5) {};
\node[perceptronS1] at  (0,0.5) {};
\node[perceptronS1] at  (2,-1) {};
\node[perceptronS1] at  (2,0) {};
\node[perceptronS1] at  (2,1) {};
\node at (2.85,0) {...};
\node[perceptronS0] at (3.6,-0.5) {};
\node[perceptronS0] at (3.6,0.5) {};
\end{tikzpicture}\\
Trace out layer $l-1$ and store the resulting state $\rho_x^l$. &\begin{tikzpicture}[scale=0.6]
\foreach \x in {-.5,.5} {
\draw[line0] (0,\x) -- (2,-1);
\draw (0,\x) -- (2,-1);
\draw[line0] (0,\x) -- (2,0);
\draw (0,\x) -- (2,0);
\draw[line0] (0,\x) -- (2,1);
\draw (0,\x) -- (2,1);}
\node[perceptronS0] at (-1.5,-0.5) {};
\node[perceptronS0] at (-1.5,0.5) {};
\node at (-0.75,0) {...};
\node[perceptronS0] at (0,-0.5) {};
\node[perceptronS0] at (0,0.5) {};
\node[perceptronS1] at  (2,-1) {};
\node[perceptronS1] at  (2,0) {};
\node[perceptronS1] at  (2,1) {};
\node at (2.85,0) {...};
\node[perceptronS0] at (3.6,-0.5) {};
\node[perceptronS0] at (3.6,0.5) {};
\end{tikzpicture}\\
\heading{Back-propagation:}\\
\headingtext{Back-propagate every training pair output $\rho_x^\text{SV}$ by applying $\mathcal{F}^l_t$, the adjoint channel of $\mathcal{E}^l$, to output state $l$ for every $l$. This includes:}\\
Tensor the state $\sigma_x^{l}$ with the identity on layer $l-1$.	& \begin{tikzpicture}[scale=0.6]
\foreach \x in {-.5,.5} {
\draw[line0] (0,\x) -- (2,-1);
\draw (0,\x) -- (2,-1);
\draw[line0] (0,\x) -- (2,0);
\draw (0,\x) -- (2,0);
\draw[line0] (0,\x) -- (2,1);
\draw (0,\x) -- (2,1);}
\node[perceptronS0] at (-1.5,-0.5) {};
\node[perceptronS0] at (-1.5,0.5) {};
\node at (-0.75,0) {...};
\node[perceptronS2] at  (0,-0.5) {};
\node[perceptronS2] at  (0,0.5) {};
\node[perceptronS1] at  (2,-1) {};
\node[perceptronS1] at  (2,0) {};
\node[perceptronS1] at  (2,1) {};
\node at (2.85,0) {...};
\node[perceptronS0] at (3.6,-0.5) {};
\node[perceptronS0] at (3.6,0.5) {};
\end{tikzpicture}\\
Apply $U^{l\dagger}=U_1^{l\dagger}\dots U_{m_l}^{l\dagger} $ and multiply with according identities, see \cref{prop:DQNN_F}. &	\begin{tikzpicture}[scale=0.6]
\foreach \x in {-.5,.5} {
\draw[line0] (0,\x) -- (2,-1);
\draw[line1] (0,\x) -- (2,-1);
\draw[line0] (0,\x) -- (2,0);
\draw[line1] (0,\x) -- (2,0);
\draw[line0] (0,\x) -- (2,1);
\draw[line1] (0,\x) -- (2,1);}
\node[perceptronS0] at (-1.5,-0.5) {};
\node[perceptronS0] at (-1.5,0.5) {};
\node at (-0.75,0) {...};
\node[perceptronS1] at  (0,-0.5) {};
\node[perceptronS1] at  (0,0.5) {};
\node[perceptronS1] at  (2,-1) {};
\node[perceptronS1] at  (2,0) {};
\node[perceptronS1] at  (2,1) {};
\node at (2.85,0) {...};
\node[perceptronS0] at (3.6,-0.5) {};
\node[perceptronS0] at (3.6,0.5) {};
\end{tikzpicture}\\
Trace out layer $l$ and store the resulting state $\sigma_x^{l-1}$. &\begin{tikzpicture}[scale=0.6]
\foreach \x in {-.5,.5} {
\draw[line0] (0,\x) -- (2,-1);
\draw (0,\x) -- (2,-1);
\draw[line0] (0,\x) -- (2,0);
\draw (0,\x) -- (2,0);
\draw[line0] (0,\x) -- (2,1);
\draw (0,\x) -- (2,1);}
\node[perceptronS0] at (-1.5,-0.5) {};
\node[perceptronS0] at (-1.5,0.5) {};
\node at (-0.75,0) {...};
\node[perceptronS1] at  (0,-0.5) {};
\node[perceptronS1] at  (0,0.5) {};
\node[perceptronS0] at (2,-1) {};
\node[perceptronS0] at (2,0) {};
\node[perceptronS0] at (2,1) {};
\node at (2.85,0) {...};
\node[perceptronS0] at (3.6,-0.5) {};
\node[perceptronS0] at (3.6,0.5) {};
\end{tikzpicture}\\
\heading{Updating the network:}\\
\headingtext{Using feed-forward and back-propagation, evolve the update matrices ${K_j^l=\eta\frac{ 2^{m_{l-1}}i}{S}\sum_{x=1}^S \tr_\text{rest}M_j^l,}$
where the trace traces out all qubits that are not affected by $U_j^l$, $\eta$ is the learning rate, $m_{l-1}$ the number of perceptrons in layer $l-1$ and $S$ is the number of training pairs. $M_j^l$ is discussed in \cref{fig:DQNN_commutator}.}\\
Update all unitaries $U_j^l$ with the update rule $U_j^l\rightarrow e^{i\epsilon K_j^l} U_j^l$.&\begin{tikzpicture}[scale=0.6]
\foreach \x in {-.5,.5} {
\draw[line0] (0,\x) -- (2,-1);
\draw[line3] (0,\x) -- (2,-1);
\draw[line0] (0,\x) -- (2,0);
\draw[line3] (0,\x) -- (2,0);
\draw[line0] (0,\x) -- (2,1);
\draw[line3] (0,\x) -- (2,1);}
\node[perceptronS0] at (-1.5,-0.5) {};
\node[perceptronS0] at (-1.5,0.5) {};
\node at (-0.75,0) {...};
\node[perceptronS0] at (0,-0.5) {};
\node[perceptronS0] at (0,0.5) {};
\node[perceptronS0] at (2,-1) {};
\node[perceptronS0] at (2,0) {};
\node[perceptronS0] at (2,1) {};
\node at (2.85,0) {...};
\node[perceptronS0] at (3.6,-0.5) {};
\node[perceptronS0] at (3.6,0.5) {};
\end{tikzpicture}\\
\end{tabularx}
\end{minipage}}
\caption{\textbf{DQNN training algorithm.} The steps feed-forward, back-propagation and updating  are repeated until the training loss reaches its maximum.}
\label{fig:DQNN_algorithm}
\end{figure}

With the loss function and QNN ansatz in hand, we can explain how the training proceeds. In the following, we will lead through the algorithm which can be found as a summarised version in \cref{fig:DQNN_algorithm}. The rest of this section will be somewhat technical. Nonetheless, we encourage the reader not to skip these parts, since the proofs allow a deeper understanding of the algorithm and structure of the DQNN. We begin with the derivation of the update matrix (see \cref{prop:DQNN_K}). It follows a derivation of the adjoint channel $\mathcal{F}^{l}_t$ of $\mathcal{E}^l_t$ (see \cref{prop:DQNN_F}) as well as a revealing discussion on the form of the update matrix, which can be expressed in terms of the layer-to-layer channels. We close the section with calculating of the change in the training loss during training (see \cref{prop:DQNN_DeltaLoss}) in a way that is similar to the classical back-propagation explained in \cref{sec:ML_optimisation}.

\FloatBarrier\subsection*{Overview}
In the first step, the \emph{initialisation}, all perceptron unitaries $U_j^l$ the DQNN is build of are initialised by randomly chosen unitaries and all qubit states in all layers are initialised in the zero state. The copies of training data pairs are prepared.

The \emph{feed-forward} part of the algorithm promotes the input of every training pair through the network, which is described with \cref{eq:DQNN_rhoOut}. For every layer, the state of the previous layer $\rho_x^{l-1}$ is tensored with the layer $l$ initialised in the state $\ket{0...0}$. The unitaries  $U^l=U_{m_l}^l\dots U_1^l$ acting on layer $l-1$ and layer $l$ are applied. Finally, the layer $l-1$ is traced out and the resulting state $\rho_x^l$ is saved and used for the next feed-forward step.

The next piece of the algorithm, the \emph{back-propagation}, is very similar to the previous part. The only differences are that we propagate the \emph{output} part of the training pair, and that the propagation direction is the opposite. This means we start at the end of the network with the state $\rho_x^\text{SV}$ and use the adjoint channel of $\mathcal{E}^l$ instead of $\mathcal{E}^l$ itself for the propagation. The resulting states are saved as $\sigma_x^{l-1}$. We will derive the exact form of the adjoint channel $\mathcal{F}^{l}_t(X^l)$ in \cref{prop:DQNN_F}. 

The information gained in the feed-forward and back-propagation parts of the algorithm, saved as states $\rho_x^l$ and $\sigma_x^{l-1}$, is used for \emph{updating} the network's unitaries in a way that optimises the training loss. Assuming that the unitaries depend on a parameter $t$ and using gradient descent leads to the update rule
\begin{equation*}
	\mathcolorbox{U_j^l(t+\epsilon)=e^{i\epsilon K_j^l(t)} U_j^l(t),}
\end{equation*}
where $\epsilon$ is the training step size. It follows that also the layer-to-layer map $\mathcal{E}^{l}(t)\equiv\mathcal{E}^{l}_t$ and its adjoint $\mathcal{F}^{l}(t)\equiv\mathcal{F}^{l}_t$ depend on the parameter $t$. We will determine the update matrix in \cref{prop:DQNN_K}. Note that the steps feed-forward, back-propagation and updating are repeated until the training loss reaches its maximum. We will discuss training results in \cref{sec:DQNN_classical} (classical simulation) and \cref{sec:DQNN_quantumalg} (NISQ device implementation). 

\FloatBarrier\subsection*{Derivation of the update matrix}
The above mentioned update rule is based on the update matrices. In the following it will be shown of which form these matrices have to be to improve the perceptrons' performance on the training data.
\begin{prop}
\label{prop:DQNN_K}
The update matrix for a QNN trained with pure states $\ket{\phi^\text{SV}_x}$  has to be of the form
\begin{equation*}
\mathcolorbox{K^l_j(t) = \frac{\eta 2^{m_{l-1}}i}{S}\sum_x\tr_\text{rest}\big(M^l_{j}(x,t)\big),}
\end{equation*}
where 
\begin{align*}
M^l_{j}(x,t)=&\left[U_j^l(t)\dots U_1^1(t)\ \left(\rho_x^\text{in}\otimes \ket{0...0} \bra{0...0}\right) {U_1^1}^\dagger(t)\dots{U_j^l}^\dagger(t),\right.\\
&\hspace{15pt}\left.{U_{j+1}^l}^\dagger(t)\dots {U_{m_{L+1}}^{L+1}}^\dagger(t)\left(\mathbbm{1}_\text{in,hid}\otimes\ket{\phi^\text{SV}_x}\bra{\phi^\text{SV}_x}\right)U_{m_{L+1}}^{L+1}(t)\dots U_{j+1}^l(t)\right],
\end{align*} $U_j^l$ is assigned to the $j$th perceptron acting on the qubit layers $l-1$ and $l$ and $\eta$ is the learning rate.
\end{prop}
\begin{proof}
To show the statement we derive $\text{d}\mathcal{L}_\text{SV}(t)/dt$. Therefore the output state of the update step $t+\epsilon$ is needed. Note that the unitaries act on the current qubit layers and we do not write out the identity matrices for a cleaner notation, e.g.\ $U^2_1$ is the short form of $U^2_1\otimes \mathbbm{1}^2_{2,3,\dots m_2}$. 
\begin{align*}
\rho_x^\text{out}(t+\epsilon)=&\tr_\text{in,hid}\Big(e^{i\epsilon K_{m_{L+1}}^{L+1}(t)}U_{m_{L+1}}^{L+1}(t)\dots e^{i\epsilon K_1^1(t)}U_1^1(t)\big(\rho_x^\text{in}\otimes\ket{0...0}_\text{hid,out}\bra{0...0}\big)  \ \\
&{U_1^1}^\dagger(t)e^{-i\epsilon K_{1}^1(t)}\dots {U_{m_{L+1}}^{L+1}}^\dagger(t)e^{-i\epsilon K_{m_{L+1}}^{L+1}(t)}\Big)\\
=&\rho_x^\text{out}(t)+i\epsilon\  \tr_\text{in,hid}\Big(\big[K_{m_{L+1}}^{L+1}(t), U_{m_{L+1}}^{L+1}(t)\dots U_1^1(t)\\
&\ \big(\rho_x^\text{in}\otimes\ket{0...0}_\text{hid,out}\bra{0...0}\big)  \ {U_1^1}^\dagger(t)\dots{U_{m_{L+1}}^{L+1}}^\dagger(t)\big]+\dots\\
&+U_{m_{L+1}}^{L+1}(t)\dots U_2^1(t) \big[K_1^1(t),U_1^1(t)\ \big(\rho_x^\text{in}\otimes\ket{0...0}_\text{hid,out}\bra{0...0}\big)  \ \\
&{U_1^1}^\dagger(t)\big]{U_2^1}^\dagger(t)\dots{U_{m_{L+1}}^{L+1}}^\dagger(t)\Big)+\mathcal{O}\big(\epsilon^2\big).
\end{align*}
The derivative of the loss function, up to the first order in $\epsilon$, can be written as
\begin{align*}
\frac{d\mathcal{L}_\text{SV}(t)}{dt}=&\lim_{\epsilon\rightarrow 0}\frac{\mathcal{L}_\text{SV}(t+\epsilon)-\mathcal{L}_\text{SV}(t)}{\epsilon}\\
=&\lim_{\epsilon\rightarrow 0}\frac{\frac{1}{S}\sum_{x=1}^{S}\bra{\phi_x^{SV}}(\rho_x^{out}(t+\epsilon)-\rho_x^{out}(t))\ket{\phi_x^{SV}}}{\epsilon}\\
=&\frac{1}{S}\sum_{x=1}^{S} \tr\Bigg((\mathbbm{1}_\text{in,hid}\otimes\ket{\phi^\text{SV}_x}\bra{\phi^\text{SV}_x})\bigg(\Big[iK_{m_{L+1}}^{L+1}(t), U_{m_{L+1}}^{L+1}(t)\dots U_1^1(t)\\&\big(\rho_x^\text{in}\otimes\ket{0...0}_\text{hid,out}\bra{0...0}\big){U_1^1}^\dagger(t)\dots{U_{m_{L+1}}^{L+1}}^\dagger(t)\Big]+\dots\\
&+U_{m_{L+1}}^{L+1}(t)\dots U_2^1(t) \Big[iK_1^1(t),U_1^1(t)\big(\rho_x^\text{in}\otimes\ket{0...0}_\text{hid,out}\bra{0...0}\big){U_1^1}^\dagger(t)\Big]\\&{U_2^1}^\dagger(t)\dots{U_{m_{L+1}}^{L+1}}^\dagger(t)\bigg)\Bigg)\\
=&\frac{i}{S}\sum_{x=1}^{S} \tr\bigg(\Big[U_{m_{L+1}}^{L+1}(t)\dots\big(\rho_x^\text{in}\otimes\ket{0...0}_\text{hid,out}\bra{0...0}\big)\dots{U_{m_{L+1}}^{L+1}}^\dagger(t),\\&\mathbbm{1}_\text{in,hid}\otimes\ket{\phi^\text{SV}_x}\bra{\phi^\text{SV}_x}\Big]iK_{m_{L+1}}^{L+1}(t)+\dots\\
&+\Big[U_1^1(t)\big(\rho_x^\text{in}\otimes\ket{0...0}_\text{hid,out}\bra{0...0}\big){U_1^1}^\dagger(t),\\
&{U_2^1}^\dagger(t)\dots{U_{m_{L+1}}^{L+1}}^\dagger(t)\big(\mathbbm{1}_\text{in,hid}\otimes\ket{\phi^\text{SV}_x}\bra{\phi^\text{SV}_x}\big)U_{m_{L+1}}^{L+1}(t)\dots U_2^1(t)\Big]iK_1^1(t)\bigg)\\
=&\frac{i}{S}\sum_{x=1}^{S} \tr\Big(M_{m_{L+1}}^{L+1}(t) K_{m_{L+1}}^{L+1}(t)+\ \dots\ +M_1^1(t)K_1^1(t)\Big),
\end{align*}
where \begin{align*}M_{m_{L+1}}^{L+1}(t)\equiv&\Big[U_{m_{L+1}}^{L+1}(t)\dots\big(\rho_x^\text{in}\otimes\ket{0...0}_\text{hid,out}\bra{0...0}\big)\dots{U_{m_{L+1}}^{L+1}}^\dagger(t),\\&\mathbbm{1}_\text{in,hid}\otimes\ket{\phi^\text{SV}_x}\bra{\phi^\text{SV}_x}\Big],\\ M_1^1(t)\equiv&\Big[U_1^1(t)\big(\rho_x^\text{in}\otimes\ket{0...0}_\text{hid,out}\bra{0...0}\big){U_1^1}^\dagger(t),\\&{U_2^1}^\dagger(t)\dots{U_{m_{L+1}}^{L+1}}^\dagger(t)\big (\mathbbm{1}_\text{in,hid}\otimes\ket{\phi^\text{SV}_x}\bra{\phi^\text{SV}_x}\big)U_{m_{L+1}}^{L+1}(t)\dots U_2^1(t)\Big].	
\end{align*}
With $\ket{0...0}_\text{hid,out}\bra{0...0}$ we denote the initialised qubits of the hidden layers and the output layer. We describe the identity on the space of the input and hidden layer's qubits with $\mathbbm{1}_\text{in,hid}$. With the usage of the Pauli matrices $\sigma\equiv\{\mathbbm{1},\sigma^x,\sigma^y,\sigma^z\}$, defined in \cref{fig:QI_gates}, the parameter matrices $K_l^j$ can be parametrised as
\begin{equation*}
K_j^l(t)=\sum_{\alpha_1,\alpha_2,\dots,\alpha_{m_{l-1}},\beta}K^l_{j,\alpha_1,\dots,\alpha_{m_{l-1}},\beta}(t)\big(\sigma^{\alpha_1}\otimes\ \dots\ \otimes\sigma^{\alpha_{m_{l-1}}}\otimes\sigma^\beta\big).
\end{equation*}
The index $\alpha_i$ counts the qubits in the previous layer, $\beta$ describes the current qubit in layer $l$. In the next step, we maximise $\frac{d\mathcal{L}_\text{SV}}{dt}$ because we aim to maximise the loss function in every step of the training algorithm. Therefore we introduce a Lagrange multiplier $\lambda\in\mathbbm{R}$ to ensure a finite solution. Without this condition the extrema would be $\pm\infty$.
\begin{align*}
\max_{K^l_{j,\alpha_1,\dots,\beta}}&\Big(\frac{d\mathcal{L}_\text{SV}(t)}{dt}-\lambda\sum_{\alpha_i,\beta}{K^l_{j,\alpha_1,\dots,\beta}}(t)^2\Big)\\
=&\max_{K^l_{j,\alpha_1,\dots,\beta}}\Big(\frac{i}{S}\sum_{x=1}^{S} \tr\big(M_{m_{L+1}}^{L+1}(t) K_{m_{L+1}}^{L+1}(t)+\ \dots\ +M_1^1(t)K_1^1(t)\big)\\
&-\lambda\sum_{\alpha_1,\dots,\beta}{K^l_{j,\alpha_1,\dots,\beta}}(t)^2\Big)\\
=&\max_{K^l_{j,\alpha_1,\dots,\beta}}\Big(\frac{i}{S}\sum_{x=1}^{S}\tr_{\alpha_1,\dots,\beta}\big(\tr_\text{rest}\big(M_{m_{L+1}}^{L+1}(t) K_{m_{L+1}}^{L+1}(t)+\ \dots\ +M_1^1(t)K_1^1(t)\big)\big)\\
&-\lambda\sum_{\alpha_1,\dots,\beta}{K^l_{j,\alpha_1,\dots,\beta}}(t)^2\Big).
\end{align*}
The notation $\tr_\text{rest}$ describes tracing out all qubits which are not effected by $K^l_{j,\alpha_1,\dots,\beta}$. As a next step we take the derivative of the gained expression with respect to $K^l_{j,\alpha_1,\dots,\beta}$:
\begin{align*}
\frac{i}{S}\sum_{x=1}^{S}\tr_{\alpha_1,\dots,\beta}\Big(\tr_\text{rest}\big(M_j^l(t)\big)\big(\sigma^{\alpha_1}\otimes\ \dots\ \otimes\sigma^\beta\big)\Big)-2\lambda K^l_{j,\alpha_1,\dots,\beta}(t)=0,
\end{align*}
This is equivalent to
\begin{align*}
K^l_{j,\alpha_1,\dots,\beta}(t)=\frac{i}{2S\lambda}\sum_{x=1}^{S}\tr_{\alpha_1,\dots,\beta}\Big(\tr_\text{rest}\big(M_j^l(t)\big)\big(\sigma^{\alpha_1}\otimes\ \dots\ \otimes\sigma^\beta\big)\Big).
\end{align*}
We finally can express the parameter matrices as
\begin{align*}
K_j^l(t)=&\sum_{\alpha_1,\dots,\beta}K^l_{j,\alpha_1,\dots,\beta}(t)\big(\sigma^{\alpha_1}\otimes\ \dots\ \otimes\sigma^\beta\big)\\
=&\frac{i}{2S\lambda}\sum_{\alpha_1,\dots,\beta}\sum_{x=1}^{S}\tr_{\alpha_1,\dots,\beta}\Big(\tr_\text{rest}\big(M_j^l(t)\big)\big(\sigma^{\alpha_1}\otimes\ \dots\ \otimes\sigma^\beta\big)\Big)\big(\sigma^{\alpha_1}\otimes\ \dots\ \otimes\sigma^\beta\big)\\
=&\frac{\eta 2^{m_{l-1}}i}{S}\sum_{x=1}^{S}\tr_\text{rest}\big(M_j^l(t)\big),
\end{align*}
where $\eta=1/\lambda$ is the learning rate and 
\begin{align*}
M_j^l(t)=&\Big[U_j^l(t)U_{j-1}^l(t)\dots U_1^1(t)\ \big(\rho_x^\text{in}\otimes\ket{0...0}_{\text{hid,out}}\bra{0...0}\big) {U_1^1}^\dagger(t)\dots{U_{j-1}^l}^\dagger(t){U_j^l}^\dagger(t),\\
&{U_{j+1}^l}^\dagger(t)\dots {U_{m_{L+1}}^{L+1}}^\dagger(t)\big(\mathbbm{1}_\text{in,hid}\otimes\ket{\phi^\text{SV}_x}\bra{\phi^\text{SV}_x}\big)U_{m_{L+1}}^{L+1}(t)\dots U_{j+1}^l(t)\Big].
\end{align*}
\end{proof}

\FloatBarrier\subsection*{Derivation of the adjoint channel}
Knowing how the update is done, we can describe the entire training algorithm. Nevertheless, we aim to understand the structure of the update matrices more deeply. To write the commutator $M_j^l(t)$ in terms of layer-to-layer channels, we need the adjoint channel of $\mathcal{E}$ which is derived in the following. 

\begin{prop}
\label{prop:DQNN_F}
\begin{sloppypar}
The adjoint channel $\mathcal{F}_t(X) $ for the CP map ${\mathcal{E}^{l}_t(X^{l-1}) = \tr_{l-1}\big( U^l(t) (X^{l-1}\otimes \ket{0...0}_l\bra{ 0...0}) {U^l}^\dag(t)\big)}$ is of the form
\end{sloppypar}
\begin{equation*}
\mathcolorbox{\mathcal{F}^{l}_t(X^l) = \tr_{l}\big(\big(\mathbbm{1}_{l-1}\otimes|0...0\rangle_l\langle0...0|_l\big){U^l}^\dag(t)\big(\mathbbm{1}_{l-1}\otimes X^l\big)U^l(t)\big).}
\end{equation*}
\end{prop}
\begin{proof}
In order to show which form the adjoint channel $\mathcal{F}^l_t$ has, we  express $\mathcal{E}^{l}_t$ in its Kraus representation \cite{Kraus1983}, which was introduced in \cref{subsec:QI_Kraus}. For any operator $X^{l-1}$ on the $(l-1)$th layer we can phrase
\begin{equation*}
\mathcal{E}^{l}_t(X^{l-1}) = \sum_\alpha A^l_\alpha(t) X^{l-1}{A^l_\alpha}^\dag(t),
\end{equation*}
where the Kraus operators $A_\alpha$ are maps from the $(l-1)$th layer of $m_{l-1}$ qubits to the $l$th layer of $m_l$ qubits.
Naturally the adjoint channel $\mathcal{F}^{l}_t$ can be written as 
\begin{equation}
\label{eqapdx:F}
\mathcal{F}^{l}_t(X^{l}) = \sum_\alpha {A^l_\alpha}^\dag(t) X^{l}A^l_\alpha(t),
\end{equation}
for any operator $X^l$ on the $l$th layer.
Our aim is to express  the Kraus operators $A^l_\alpha$. We choose $\{\ket{\alpha}\}_\alpha$ to be an orthonormal basis in the $(l-1)$th layer. Further we assume $\ket{b},\ket{c}$ are any vectors in the $(l-1)$th layer and $\ket{d},\ket{e}$ any vectors in the $l$th layer. Using \cref{eq:DQNN_E} and $U^l(t) = U^l_{m_l}(t)\dots U^l_1(t)$ we get
\begin{align*}
\big\langle d\big|\, \mathcal{E}^{l}_t\big(\ket{b}\bra{c}\big)\big| e\big\rangle =&\Big\langle d\Big|\,\tr_{l-1}\big(U^l(t)\big(\ket{b}\bra{c}\otimes\ket{0...0}_l\bra{0...0}\big){U^l}^\dagger(t)\big)\Big| e \Big\rangle  \\
=& \sum_\alpha\big\langle \alpha,d\big|\,U^l(t)\big(\ket{b}\bra{c}\otimes\ket{0...0}_l\bra{0...0}\big){U^l}^\dagger(t)\big|\alpha, e \big\rangle \\
=& \sum_\alpha\big\langle \alpha,d\big|\,U^l(t)\big| b, 0...0\big\rangle\big\langle n, 0...0\big|{U^l}^\dagger(t)\big|\alpha, e \big\rangle.
\end{align*}
Defining $A_\alpha^l(t)$ with $\bra{d}A_\alpha^l(t)\ket{b} = \big\langle \alpha,d\big|\,U^l(t)\big| b, 0...0\big\rangle$ and using Eq.~\eqref{eqapdx:F} we reach the expression
\begin{align*}
\bra{b}\mathcal{F}^{l}_t(\ket{d}\bra{e})\ket{c} =& \sum_\alpha \bra{b}{A_\alpha^l}^\dagger(t) \ket{d}\bra{e}A_\alpha^l(t)\ket{c}\\
=& \sum_\alpha\big\langle b,0...0\big|\,{U^l}^\dag(t)\big| \alpha, i\big\rangle\big\langle \alpha, e \big|U^l(t)\big|c,0...0 \big\rangle \\
=& \big\langle b,0...0\big|\,{U^l}^\dag(t)\big(\mathbbm{1}_{l-1}\otimes\ket{d}\bra{e}\big)U^l(t)\big|c,0...0 \big\rangle \\
=& \Big\langle b\Big|\,\tr_{l}\big(\mathbbm{1}_{l-1}\otimes|0...0\rangle_l\langle0...0|_l{U^l}^\dag(t)\big(\mathbbm{1}_{l-1}\otimes\ket{d}\bra{e}\big)U^l(t)\big)\Big|c\Big\rangle.
\end{align*}
Extracting the action of $\mathcal{F}^{l}(t)$ on a general operator $X^l$ leads to
\begin{equation*}
\mathcal{F}^{l}_t(X^l) = \tr_{l}\big(\mathbbm{1}_{l-1}\otimes|0...0\rangle_l\langle0...0|_l{U^l}^\dag(t)\big(\mathbbm{1}_{l-1}\otimes X^l\big)U^l(t)\big).
\end{equation*}
\end{proof}

\FloatBarrier\subsection*{Beneficial form of the update matrix}
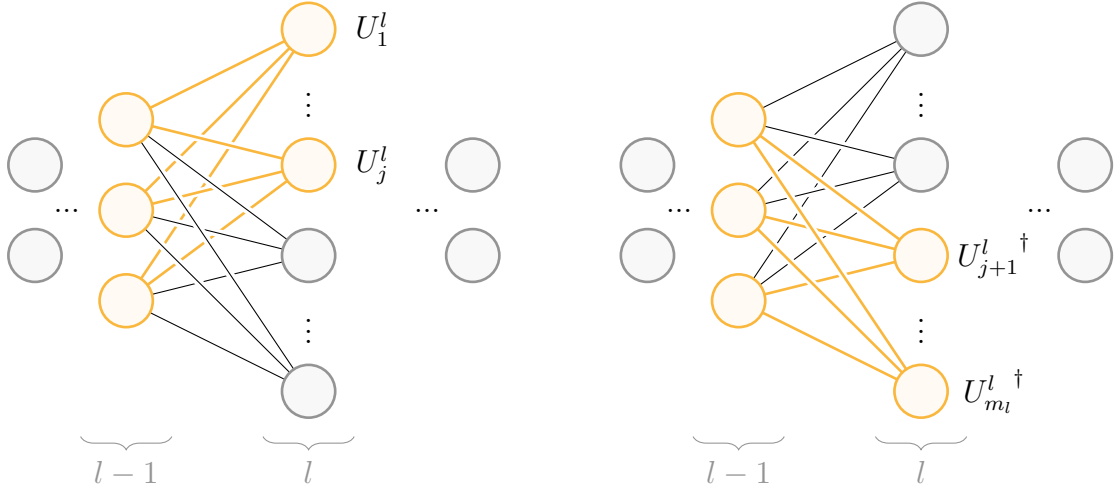
\begin{figure}[h!]
\centering
\begin{subfigure}{0.49\textwidth}\centering
\begin{tikzpicture}[scale=1.2]
\foreach \x in {-1,0,1} {
\draw[line0] (0,\x) -- (2,-2);
\draw (0,\x) -- (2,-2);
\draw[line0] (0,\x) -- (2,-0.5);
\draw (0,\x) -- (2,-0.5);
\draw[line0] (0,\x) -- (2,0.5);
\draw[line2] (0,\x) -- (2,0.5);
\draw[line0] (0,\x) -- (2,2);
\draw[line2] (0,\x) -- (2,2);
}
\node[perceptron0] at (-1,-0.5) {};
\node[perceptron0] at (-1,0.5) {};
\node at (-0.65,0) {...};
\node[perceptron2] at  (0,-1) {};
\node[perceptron2] at  (0,0) {};
\node[perceptron2] at  (0,1) {};
\node[perceptron0] at (2,-2) {};
\node[perceptron0] at (2,-0.5) {};
\node[perceptron2] at (2,0.5) {};
\node[perceptron2] at (2,2) {};
\node at (3.3,0) {...};
\node[perceptron0] at (3.8,-0.5) {};
\node[perceptron0] at (3.8,0.5) {};
\draw[brace0] (0.5,-2.5) --node[below=1ex] {$l-1$} (-0.5,-2.5) ; 
\draw[brace0] (2.5,-2.5) --node[below=1ex] {$l$} (1.5,-2.5) {}; 
\node at (2.7,0.5) {$U_j^l$};
\node at (2.7,2) {$U_1^l$};
\node at (2,1.25) {\vdots};
\node at (2,-1.25) {\vdots}; 
\node[white] at (4.3,0) {shift}; 
\end{tikzpicture}
\subcaption{First element of the commutator.}
\end{subfigure}
\begin{subfigure}{0.49\textwidth}\centering
\begin{tikzpicture}[scale=1.2]
\node[white] at (-1.5,0) {shift}; 
\foreach \x in {-1,0,1} {
\draw[line0] (0,\x) -- (2,-2);
\draw[line2] (0,\x) -- (2,-2);
\draw[line0] (0,\x) -- (2,-0.5);
\draw[line2] (0,\x) -- (2,-0.5);
\draw[line0] (0,\x) -- (2,0.5);
\draw (0,\x) -- (2,0.5);
\draw[line0] (0,\x) -- (2,2);
\draw (0,\x) -- (2,2);
}
\node[perceptron0] at (-1,-0.5) {};
\node[perceptron0] at (-1,0.5) {};
\node at (-0.65,0) {...};
\node[perceptron2] at  (0,-1) {};
\node[perceptron2] at  (0,0) {};
\node[perceptron2] at  (0,1) {};
\node[perceptron2] at  (2,-2) {};
\node[perceptron2] at  (2,-0.5) {};
\node[perceptron0] at (2,0.5) {};
\node[perceptron0] at (2,2) {};
\node at (3.3,0) {...};
\node[perceptron0] at (3.8,-0.5) {};
\node[perceptron0] at (3.8,0.5) {};
\draw[brace0] (0.5,-2.5) --node[below=1ex] {$l-1$} (-0.5,-2.5) ; 
\draw[brace0] (2.5,-2.5) --node[below=1ex] {$l$} (1.5,-2.5) {}; 
\node at (2.8,-0.5) {${U_{j+1}^l}^\dagger$};
\node at (2.8,-2) {${U_{m_l}^l}^\dagger$};
\node at (2,1.25) {\vdots};
\node at (2,-1.25) {\vdots};
\end{tikzpicture}
\subcaption{Second element of the commutator.}
\end{subfigure}
\caption{\textbf{The commutator $M_j^l$}. The actions of the two parts of $M_j^l$ become clear in network notation.}
\label{fig:DQNN_commutator}
\end{figure}

We successfully derived the adjoint channel of $\mathcal{E}$. At this point, we can discuss the structure of 
\begin{align*}
M^l_{j}(x,t)=&\left[U_j^l(t)\dots U_1^1(t)\ \left(\rho_x^\text{in}\otimes \ket{0...0} \bra{0...0}\right) {U_1^1}^\dagger(t)\dots{U_j^l}^\dagger(t),\right.\\
&\hspace{15pt}\left.{U_{j+1}^l}^\dagger(t)\dots {U_{m_{L+1}}^{L+1}}^\dagger(t)\left(\mathbbm{1}_\text{in,hid}\otimes\ket{\phi^\text{SV}_x}\bra{\phi^\text{SV}_x}\right)U_{m_{L+1}}^{L+1}(t)\dots U_{j+1}^l(t)\right],
\end{align*} 
see \cref{prop:DQNN_K}, in more detail. At a second glance, the commutator expression presents an exciting structure: the first part of the commutator is the input state propagated through the network until we reach the unitary $U_j^l$ we wish to update. The second part is obtained by back-propagation of the matching desired supervised output state with stopping right before $U_j^l$. This observation is visualised in \cref{fig:DQNN_commutator}. 

Using the the derived channels we can rewrite $M^l_{j}(x,t)$ as
\[
\mathcolorbox{\begin{aligned}
M^l_{j}(x,t)=&\left[U_j^l(t)\dots U_l^1(t)\ \left(\rho_x^{l-1}(t)\otimes \ket{0...0} \bra{0...0}\right) {U_l^1}^\dagger(t)\dots{U_j^l}^\dagger(t),\right.\\
&\hspace{15pt}\left.{U_{j+1}^l}^\dagger(t)\dots {U_{m_l}^l}^\dagger(t)\left(\mathbbm{1}_\text{in,hid}\otimes\sigma_x^{l}(t)\right)U_{m_l}^l(t)\dots U_{j+1}^l(t)\right],
\end{aligned}}
\]
where the density matrix of the $l$ layer concerning the $x$th training data can be expressed via $\rho^{l-1}_x(t)= \mathcal{E}^{l-1}_t\big(\dots\allowbreak \mathcal{E}^{2}_t\big(\mathcal{E}^{1}_t\big(\rho^{\text{in}}_x\big)\big)\dots\big)$. In an analogous way we write $\sigma^l_x(t)=\mathcal{F}^{l+1}_t\big(\dots \mathcal{F}^{L}_t\big(\mathcal{F}^{L+1}_t\big(\ket{\phi_x^\text{SV}}\bra{\phi_x^\text{SV}}\big)\big)\dots\big)$ using back-propagation. Note here again that both, the feed-forward channels $\mathcal{E}$ as well as the corresponding adjoint channels, depend on the unitary parameter $t$ and change during the training.

At this point, it becomes clear why the layer structure of the network is beneficial: to update a perceptron unitary $U_j^l$, we need to evaluate $K_j^l$. Therefore only the output state of the previous layer, $\rho^{l-1}$, obtained by feed-forward propagation through the network, and the state of the following layer $\sigma^l$, obtained by back-propagation of the desired output up to the current layer is needed. It follows that the parameter matrices can be received with only accessing two qubit layers at any time, and there is no need to access the whole network for updating a single perceptron. This fact allows us to train deep DQNNs.

\FloatBarrier\subsection*{Change of the training loss}
Connecting to \cref{sec:ML_optimisation} where the classical back-propagation algorithm is described, we want to close this technical section with the derivation of a handy formula for the change in the training loss function using the layer-to-layer channels. 

\begin{prop}
\label{prop:DQNN_DeltaLoss}
The change in the loss function can be written as
\begin{equation*}
\mathcolorbox{\frac{\text{d}\mathcal{L}_\text{SV}(t)}{dt}=\frac{i}{S}\sum_{x=1}^N \sum_{l=1}^{L+1}\tr\Big(\sigma_x^{l}(t)\mathcal{D}^{l}_t\big(\rho_x^{l-1}(t)\big)\Big),}
\end{equation*}
where $\mathcal{D}^{l}_t=\partial\mathcal{E}^{l}_t/\partial t$ is the derivative of the channel $\mathcal{E}^{l}_t$.
\end{prop}
\begin{proof}
We already evaluated $\frac{\text{d}\mathcal{L}_\text{SV}(t)}{dt}$ in the proof of \cref{prop:DQNN_K}. As a primary step we translate the gained expressions to the channel formalism and get
\begin{align*}
\frac{\text{d}\mathcal{L}_\text{SV}(t)}{dt}=&\frac{i}{S}\sum_{x=1}^{S}\tr\bigg(\mathbbm{1}_\text{in,hid}\otimes\ket{\phi^\text{SV}_x}\bra{\phi^\text{SV}_x}\\
&\Big(\Big[K_{m_{L+1}}^{L+1}(t), U_{m_{L+1}}^{L+1}(t)\dots U_1^1(t)\big(\rho_x^\text{in}\otimes\ket{0...0}_\text{hid,out}\bra{ 0...0}\big){U_1^1}^\dagger(t)\dots{U_{m_{L+1}}^{L+1}}^\dagger(t)\Big]\\
&+\dots+U_{m_{L+1}}^{L+1}(t)\dots U_2^1(t) \Big[K_1^1(t),U_1^1(t)\big(\rho_x^\text{in}\otimes \ket{0...0}_\text{hid,out}\bra{0...0}\big){U_1^1}^\dagger(t)\Big]\\
&{U_2^1}^\dagger(t)\dots{U_{m_{L+1}}^{L+1}}^\dagger(t)\Big)\bigg)\\
=& \frac{i}{S}\sum_{x=1}^{S}\sum_{l=1}^{L+1}\sum_{j=1}^{m_l} \tr \bigg( \Big( \mathbbm{1}_\text{in,hid}  \otimes \ket{\phi_x^\text{SV}}\bra{\phi_x^\text{SV}}\Big)U_{m_{L+1}}^{L+1}\dots U_{j+1}^{l}\Big[ K_{j}^l,\\
&  U_{j}^l\dots U_{1}^1 \big( \rho_x^\text{in}\otimes  \ket{0...0}_\text{hid,out} \bra{0...0}\big)U_{1}^{1^\dagger}\dots U_{j}^{l^\dagger}\Big] U_{j+1}^{l^\dagger} \dots U_{m_{L+1}}^{L+1^\dagger} \bigg)\\
=&\frac{i}{S}\sum_{x=1}^N \sum_{l=1}^{L+1}\sum_{j=1}^{m_l}\tr\bigg({U_{1}^{l+1}}^\dagger(t)\dots{U_{m_{L+1}}^{L+1}}^\dagger(t)\big(\mathbbm{1}_L\otimes \ket{\phi_x^\text{SV}}\bra{\phi_x^\text{SV}}\big)U_{m_{L+1}}^{L+1}(t)\dots U_{1}^{l+1}(t)\\
&U_{m_j}^l(t)\dots U_{j+1}^l(t)\Big[K_j^l(t),U_j^l(t)\dots U_1^l(t)\big(\rho_x^{l-1}\otimes \ket{0...0}_l \bra{0...0}\big){U_1^l}^\dagger(t)\dots{U_j^l}^\dagger(t)\Big]\\
&{U_{j+1}^l}^\dagger(t)\dots{U_{m_j}^l}^\dagger(t)\bigg).
\end{align*}	
For a shorter notation we define
\begin{align*}
A=&U_{x}^{l+1^\dagger}\dots U_{L+1^\dagger} \big( \mathbbm{1}_{l,...,L} \otimes  \ket{\phi_x^\text{SV}}\bra{\phi_x^\text{SV}}\big) U_{x}^{L+1}\dots U_{x}^{l+1} \\
B=&U_{m_{l}}^{l}\dots U_{j+1}^{l} \Big[ K_{j}^l, U_{j}^l\dots U_{1}^1 \big( \rho_x^\text{in} \otimes \ket{0...0}_\text{1,...,l} \bra{0...0}\big)U_{1}^{1^\dagger}\dots U_{j}^{l^\dagger}\Big] U_{j+1}^{l^\dagger} \dots U_{m_{l}}^{l^\dagger}
\end{align*}
and rewrite the change of the loss function as
\begin{align}
\frac{\text{d}\mathcal{L}_\text{SV}}{dt}
=& \frac{i}{S}\sum_{x=1}^{S}\sum_{l=1}^{L+1}\sum_{j=1}^{m_l}  \tr \Big( \big(\mathbbm{1}_{0,...,l-1}\otimes A\big)  \big(B\otimes \ket{0...0}_{l+1,...,L+1}\bra{0...0}\big) \Big)\nonumber\\
=& \frac{i}{S}\sum_{x=1}^{S}\sum_{l=1}^{L+1}\sum_{j=1}^{m_l}  \tr \Big( A  \big(\tr_{0,...,l-1}(B)\otimes \ket{0...0}_{l+1,...,L+1}\bra{0...0}\big) \Big)\nonumber\\
=& \frac{i}{S}\sum_{x=1}^{S}\sum_{l=1}^{L+1}\sum_{j=1}^{m_l}  \tr \Big( A  \big(\tr_{0,...,l-1}(B)\otimes \mathbbm{1}_{l+1,...,L+1}\big) \big(\mathbbm{1}_l\otimes\ket{0...0}_{l+1,...,L+1}\bra{0...0}\big)\Big) \nonumber\\
=& \frac{i}{S}\sum_{x=1}^{S}\sum_{l=1}^{L+1}\sum_{j=1}^{m_l} \tr \Big(\big(\mathbbm{1}_l\otimes\ket{0...0}_{l+1,...,L+1}\bra{0...0}\big) A  \big(\tr_{0,...,l-1}(B)\otimes \mathbbm{1}_{l+1,...,L+1}\big) \Big) \nonumber\\
=& \frac{i}{S}\sum_{x=1}^{S}\sum_{l=1}^{L+1}\sum_{j=1}^{m_l}  \tr \Big(\tr_{l+1,...,L+1}\big(\big(\mathbbm{1}_l\otimes\ket{0...0}_{l+1,...,L+1}\bra{0...0}\big) A \big) \tr_{0\dots l-1}(B) \Big).\label{eq:DQNN_changeAB}
\end{align}
After propagating the state $\rho_x^\text{in}$ trough the first $l-1$ qubit layers of the network we get the expression
\begin{align*}
\rho_x^{l-1}=&\mathcal{E}^{l-1}\big(\dots \mathcal{E}^1\big( \rho_x^\text{in}\big)\dots\big)\\
=&\tr_{l-2}\Big(\big(U^{l-1}\Big( \mathcal{E}^{l-2}\big(\dots\mathcal{E}^{l-3}\big( \ket{\phi_x^\text{SV}}\bra{\phi_x^\text{SV}}\big)\dots\big)\otimes\mathbbm{1}_{l-1}\Big)  U^{l-1\dagger}\Big)\nonumber \\
=&\dots\nonumber\\
=& \tr_{1,...,l-2} \Big(U^{l-1}\dots U^1 \Big(\mathcal{E}^1\Big(\rho_x^\text{in}\Big) \otimes \ket{0...0}_\text{2,...,l-1} \bra{0...0}\Big) U^{2^\dagger }\dots U^{l-1^\dagger}\Big)\\
=& \tr_{1,...,l-2} \Big(U^{l-1}\dots U^2 \Big(\tr_{0} \Big( U^1 \Big( \rho_x^\text{in} \otimes \ket{0...0}_\text{1} \bra{0...0}\Big)U^{1^\dagger}\Big) \otimes \ket{0...0}_\text{2,...,l-1} \bra{0...0}\Big)\\ & U^{2^\dagger }\dots U^{l-1^\dagger}\Big)\\
=&\tr_{0,...,l-2} \Big(U^{l-1}\dots  U^1 \Big( \rho_x^\text{in}\otimes \ket{0...0}_\text{1,...,l-1} \bra{0...0}\Big)U^{1^\dagger} \dots U^{l-1^\dagger}\Big).\\
\end{align*} 
This can be used to express $\tr_{0,...,l-1}(B)$ via
\begin{align}
\tr_{0,...,l-1}(B)=&\tr_{0,...,l-1} \Big(U_{m_{l}}^{l}\dots U_{j+1}^{l} \Big[ K_{j}^l, U_{j}^l\dots U_{1}^1 \Big( \rho_x^\text{in} \otimes \ket{0...0}_\text{1,...,l} \bra{0...0}\Big)U_{1}^{1^\dagger}\dots U_{j}^{l^\dagger}\Big]\nonumber\\ & U_{j+1}^{l^\dagger} \dots U_{m_{l}}^{l^\dagger}\Big)\nonumber\\
=&\tr_{l-1} \Big(U_{m_{l}}^{l}\dots U_{j+1}^{l} \Big[ K_{j}^l, U_{j}^l\dots U_{1}^l \Big( \tr_{0,...,l-2} \Big(U_{x}^{l-1}\dots  U^1 \nonumber\\ &\Big( \rho_x^\text{in} \otimes \ket{0...0}_\text{1\dots l-1} \bra{0...0}\Big)U^{1^\dagger} \dots U^{l-1^\dagger}\Big)\nonumber\\
&\otimes \ket{0...0}_l \bra{0...0} \Big) U_{1}^{l^\dagger} \dots U_{j}^{l^\dagger}\Big] U_{j+1}^{l^\dagger} \dots U_{m_{l}}^{l^\dagger}\Big)\nonumber\\
=&\tr_{l-1} \Big(U_{m_{l}}^{l}\dots U_{j+1}^{l} \Big[ K_{j}^l, U_{j}^l\dots U_{1}^l \Big( \rho_x^{l-1}\otimes \ket{0...0}_l \bra{0...0} \Big) U_{1}^{l^\dagger} \dots U_{j}^{l^\dagger}\Big] \nonumber\\ & U_{j+1}^{l^\dagger} \dots U_{m_{l}}^{l^\dagger}\Big)\label{eq:DQNN_changeA}.
\end{align}	
In the proof of \cref{prop:DQNN_F} it was shown that the structure of the adjoint channel of \(\mathcal{E}^l\) is 
\begin{equation*}
\mathcal{F}^l(X^l)=\tr_l\big( \big(\mathbbm{1}_{l-1}\otimes \ket{0...0}_l\bra{0...0}\big)U^{l^\dagger}\big(\mathbbm{1}_{l-1}\otimes X^l\big)U^l\big).
\end{equation*}
With this in hand we can back propagate a supervised state $\ket{\phi_x^\text{SV}}\bra{\phi_x^\text{SV}}$ $L-l$ qubit layers through the network and ending up with the state
\begin{align*}
\sigma_x^l	=& \mathcal{F}^{l+1}\big(\dots\mathcal{F}^{L+1}\big(\ket{\phi_x^\text{SV}}\bra{\phi_x^\text{SV}}\big)\dots\big)\\
=&\tr_{l+1}\Big(\big(\mathbbm{1}_l\otimes\ket{0...0}_{l+1}\bra{0...0}\big)U^{l+1^\dagger}\Big(\mathbbm{1}_{l}\otimes \mathcal{F}^{l+2}\big(\dots\mathcal{F}^{L+1}\big( \ket{\phi_x^\text{SV}}\bra{\phi_x^\text{SV}}\big)\dots\big)\Big)  U^{l+1}\Big)\nonumber \\
=&\dots\nonumber\\
=&\tr_{l+1,...,L}\Big(\big(\mathbbm{1}_l\otimes\ket{0...0}_{l+1,...,L}\bra{0...0}\big)\\& U^{l+1^\dagger}\dots U^{L^\dagger}\Big(\mathbbm{1}_{l,...,L-1}\otimes \mathcal{F}^{L+1}\big( \ket{\phi_x^\text{SV}}\bra{\phi_x^\text{SV}}\big)\Big) U^{L}\dots U^{l+1}\Big)\nonumber \\
=&\tr_{l+1,...,L}\Big(\big(\mathbbm{1}_l\otimes\ket{0...0}_{l+1,...,L}\bra{0...0}\big)\\
&U^{l+1^\dagger}\dots U^{L^\dagger}\Big(\mathbbm{1}_{l,...,L-1}\otimes \tr_{L+1}\Big(\big(\mathbbm{1}_{L}\otimes\ket{0...0}_{L+1}\bra{0...0}\big) \\
& U^{L+1^\dagger} \big( \mathbbm{1}_{L} \otimes \ket{\phi_x^\text{SV}}\bra{\phi_x^\text{SV}}\big)  U^{L+1}\Big)\Big)  U^{L}\dots U^{l+1}\Big)\\
=&\tr_{l+1,...,L+1}\Big(\big(\mathbbm{1}_l\otimes\ket{0...0}_{l+1,...,L}\bra{0...0} \otimes \mathbbm{1}_{L+1}\big)\\
&\big(U^{l+1^\dagger}\dots U^{L^\dagger}\otimes \mathbbm{1}_{L+1}\big)\big(\mathbbm{1}_{l,...,L}\otimes\ket{0...0}_{L+1}\bra{0...0}\big) \\
& \big(\mathbbm{1}_{l,...,L-1}\otimes  U^{L+1^\dagger}\big) \big( \mathbbm{1}_{l,...,L}  \otimes \ket{\phi_x^\text{SV}}\bra{\phi_x^\text{SV}}\big) \big(\mathbbm{1}_{l,...,L-1}\otimes  U^{L+1}\big) \big( U^{L}\dots U^{l+1}\otimes \mathbbm{1}_{L+1}\big)\Big)\\
=&\tr_{l+1,...,L+1}\Big(\big(\mathbbm{1}_l\otimes\ket{0...0}_{l+1,...,L}\bra{0...0} \otimes \mathbbm{1}_{L+1}\big)\big(\mathbbm{1}_{l,...,L}\otimes\ket{0...0}_{L+1}\bra{0...0}\big)\\
& \big(U_{x}^{l+1^\dagger}\dots U^{L^\dagger}\otimes \mathbbm{1}_{L+1}\big)\big(\mathbbm{1}_{l,...,L-1}\otimes  U^{L+1^\dagger}\big) \big( \mathbbm{1}_{l,...,L}  \otimes \ket{\phi_x^\text{SV}}\bra{\phi_x^\text{SV}}\big) \big(\mathbbm{1}_{l,...,L-1}\otimes  U^{L+1}\big)\\
& \big( U^{L}\dots U^{l+1}\otimes \mathbbm{1}_{L+1}\big)\Big)\\
=&\tr_{l+1,...,L+1}\big(\big(\mathbbm{1}_l\otimes\ket{0...0}_{l+1,...,L+1}\bra{0...0}\big) U^{l+1^\dagger}\dots U^{L+1^\dagger} \big( \mathbbm{1}_{l\dots L}  \otimes \ket{\phi_x^\text{SV}}\bra{\phi_x^\text{SV}}\big) \\& U^{L+1}\dots U^{l+1} \big)\\
=&\tr_{l+1,...,L+1}\big(\big(\mathbbm{1}_l\otimes\ket{0...0}_{l+1,...,L+1}\bra{0...0}\big) A \big).
\end{align*}
With this,  \cref{eq:DQNN_changeAB} and \cref{eq:DQNN_changeA} we acquire the expression
\begin{align*}
\frac{\text{d}\mathcal{L}_\text{SV}(t)}{dt}=&\frac{i}{S}\sum_{x=1}^N \sum_{l=1}^{L+1}\tr\big(\sigma_x^l\sum_{j=1}^{m_j}U_{m_j}^l(t)\dots U_{j+1}^l(t)\big[K_j^l(t),U_j^l(t)\dots U_1^l(t)\\
&\big(\rho_x^{l-1}\otimes \ket{0...0}_l \bra{0...0}\big){U_1^l}^\dagger(t)\dots{U_j^l}^\dagger(t)\big]{U_{j+1}^l}^\dagger(t)\dots{U_{m_j}^l}^\dagger(t)\big).\\          
\end{align*}	
It is possible to reformulate the latter equation as 
\begin{align*}
\frac{\text{d}\mathcal{L}_\text{SV}(t)}{dt}=&\frac{i}{S}\sum_{x=1}^N \sum_{l=1}^{L+1}\tr\Big(\sigma_x^{l}(t)\mathcal{D}^{l}_t\big(\rho_x^{l-1}(t)\big)\Big),
\end{align*}
where  $\mathcal{F}^{l}_t$ denotes the adjoint channel of the channel $\mathcal{E}^{l}_t$ and the quantum state $\sigma_x^l(t)$ is defined as $\sigma_x^l(t)=\mathcal{F}_t^{l+1}\big(\dots\mathcal{F}_t^{\text{out}}\big(\ket{\phi^\text{SV}_x}\bra{\phi^\text{SV}_x}\big)\dots\big)$.  Further $\mathcal{D}^{l}_t=\partial\mathcal{E}^{l}_t/\partial t$ is the derivative of the channel $\mathcal{E}^{l}_t$, defined as
\begin{align*}
\mathcal{D}^{l}_t\big(X^{l-1}\big)=&\sum_{j=1}^{m_j}\tr_{l-1}\big(U_{m_j}^l\dots U_{j+1}^l\big[K_j^l,U_j^l\dots U_1^l\big(\rho_x^{l-1}\otimes\ket{0...0}_l\bra{0...0}\big){U_1^l}^\dagger\dots{U_j^l}^\dagger\big]\\&{U_{j+1}^l}^\dagger\dots{U_{m_j}^l}^\dagger\big).
\end{align*}
\end{proof}

Due to the length of this section, we take a moment to conclude the results. In the preceding pages, we explained the training algorithm of the DQNN. We not only derived the desired necessary update rule for optimising training loss, defined \cref{sec:DQNN_lossfunctions}, but also showed that for the update of the $l$th layer of the network, only the quantum states of two neighbouring layers are needed. Finally, we derived an expression for the change of the mentioned loss during training in equivalence to the classical back-propagation. Before we proceed with presenting the DQNN architecture with convincing numerical results, we discuss the previously mentioned universality of the DQNN.

\section{Universality}
\label{sec:DQNN_universality}

Classical NNs composed of classical perceptrons can represent any function \cite{Wolf2018}. In the following, we discuss that the in \cref{sec:DQNN_networkarchitecture} presented quantum perceptron holds the same feature for QNNs. 

\begin{figure}[H]
\centering
\begin{tikzpicture}[]
\matrix[row sep=0.3cm, column sep=0.5cm] (circuit) {
\node(start1){$\rho^\text{in}$};
& \node[]{}; 
& \node[]{}; 
& \node (end1){$\rho ^\text{out}$}; \\
};
\begin{pgfonlayer}{background}
\draw[] (start1) -- (end1);
\node[operator2] at (0,0) {$\mathcal{E}$};
\end{pgfonlayer}
\draw[stealth-stealth] (2.75,0) -- (3.23,0);
\begin{scope}[xshift=7.5cm]
\matrix[row sep=0.3cm, column sep=0.5cm] (circuit) {
\node(start2){$\rho^\text{in}$};
& \node[]{}; 
& \node[]{}; 
& \node[]{}; 
& \node[]{}; 
& \node (end2){$\rho ^\text{out}$}; \\
& \node[]{}; 
\node(start1){$\omega_\text{ancilla}$};
& \node[]{}; 
& \node[]{}; 
& \node[dcross](end1){}; 
& \node[]{};  \\
};
\begin{pgfonlayer}{background}
\draw[] (start1) -- (end1)  
(start2) -- (end2);
\node[operator1,minimum height=1.5cm] at (.5,0) {$U$};
\end{pgfonlayer}
\end{scope}
\end{tikzpicture}
\caption{\textbf{Stinespring theorem in quatum circuits.} A CP map $\mathcal{E}$ can be represented using tensoring an ancilla state, applying a unitary transformation and tracing out a subsystem.}
\label{fig:DQNN_Stinespring}
\end{figure}
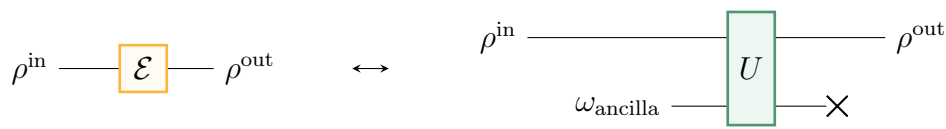

As already motivated in \cref{subsec:QI_circuits_circ}, from the Stinespring dilation theorem \cite{Nielsen2000} directly follows that every description of a quantum circuit through tensor products, unitary transformations and reductions to subsystems is equivalent to the description as a CP map. \cref{fig:DQNN_Stinespring} describes this relation using a quantum circuit. The only task is to find the suiting unitary. This is exactly what the algorithm presented in \cref{sec:DQNN_trainingalgorithm} does. Therefore it is clear that we can implement any CP map with the DQNN algorithm.

However, it is more remarkable that a QNN, comprised of the DQNN quantum perceptrons acting on 4-level qudits (equivalent to pairs of 2-level qubits) that commute within each layer, is capable of carrying out universal quantum computation. In the following, we use such a construction to illustrate that it is possible to construct a DQNN simulating an arbitrary quantum circuit consisting out of these perceptrons. Without loss of generality only two qubit gates on neighbouring qubits are used in the circuit. 

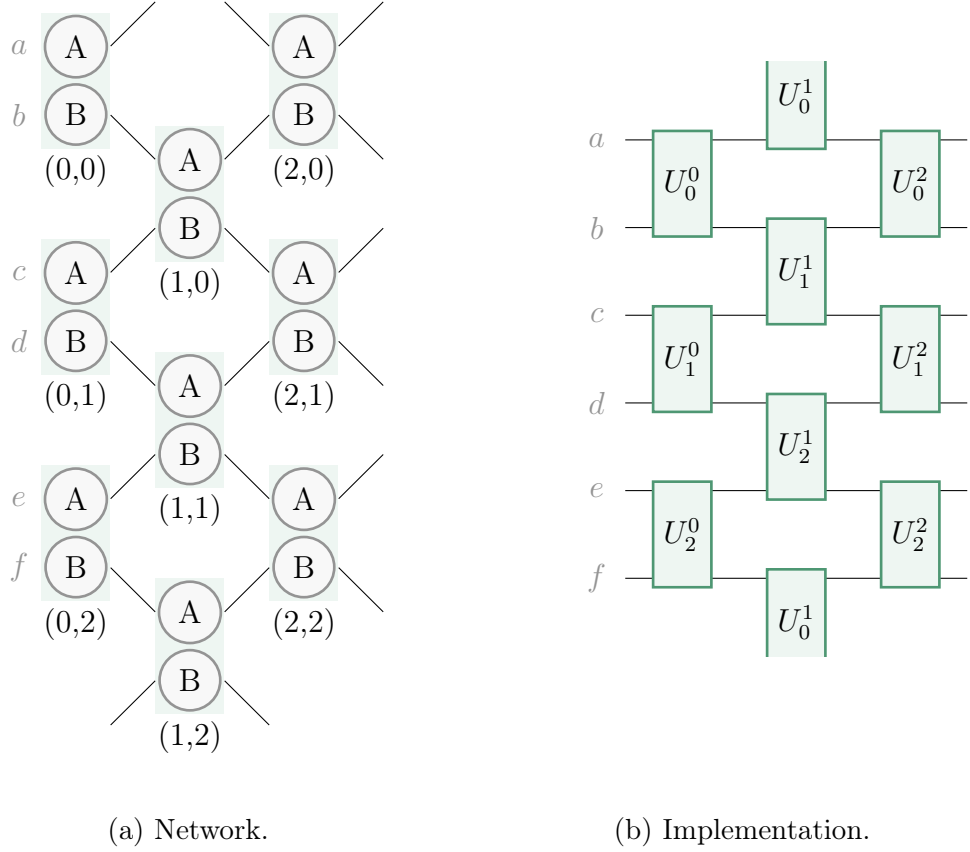
\begin{figure}[H]
\centering
\begin{subfigure}[t]{0.49\linewidth}
\centering
\begin{tikzpicture}[scale=1.5]
\node[fill=color1L,minimum width=0.9cm,minimum height=1.8cm]  at (0,3) (a1) {};
\node[fill=color1L,minimum width=0.9cm,minimum height=1.8cm]  at (0,1) (a2) {};
\node[fill=color1L,minimum width=0.9cm,minimum height=1.8cm]  at (0,-1) (a3) {};
\node[draw=white,fill=white,minimum width=0.9cm,minimum height=1cm] at (0,-3) (a4) {};
\node[draw=white,fill=white,minimum width=0.9cm,minimum height=1cm] at (-1,-3) (x) {};
\node[draw=white,fill=white,minimum width=0.9cm,minimum height=1cm]  at (1,4) (b1) {};
\node[fill=color1L,minimum width=0.9cm,minimum height=1.8cm]  at (1,2) (b2) {};
\node[fill=color1L,minimum width=0.9cm,minimum height=1.8cm]  at (1,0) (b3) {};
\node[fill=color1L,minimum width=0.9cm,minimum height=1.8cm]  at (1,-2) (b4) {};
\node[fill=color1L,minimum width=0.9cm,minimum height=1.8cm]  at (2,3) (c1) {};
\node[fill=color1L,minimum width=0.9cm,minimum height=1.8cm]  at (2,1) (c2) {};
\node[fill=color1L,minimum width=0.9cm,minimum height=1.8cm]  at (2,-1) (c3) {};
\node[draw=white,fill=white,minimum width=0.9cm,minimum height=1cm] at (2,-3) (c4) {};
\node[draw=white,fill=white,minimum width=0.9cm,minimum height=1cm]  at (3,4) (d1) {};
\node[draw=white,fill=white,minimum width=0.9cm,minimum height=1cm]  at (3,2) (d2) {};
\node[draw=white,fill=white,minimum width=0.9cm,minimum height=1cm] at (3,0) (d3) {};
\node[draw=white,fill=white,minimum width=0.9cm,minimum height=1cm] at (3,-2) (d4) {};
\node[color0] at ([xshift=-.5cm,yshift=.3cm]a1){$a$};
\node[color0] at ([xshift=-.5cm,yshift=-.3cm]a1){$b$};
\node[color0] at ([xshift=-.5cm,yshift=.3cm]a2){$c$};
\node[color0] at ([xshift=-.5cm,yshift=-.3cm]a2){$d$};
\node[color0] at ([xshift=-.5cm,yshift=.3cm]a3){$e$};
\node[color0] at ([xshift=-.5cm,yshift=-.3cm]a3){$f$};
\node[perceptron0] at ([yshift=.3cm]a1){A};
\node[perceptron0]  at ([yshift=-.3cm]a1){B};
\node at ([yshift=-.8cm]a1){(0,0)};
\node[perceptron0]  at ([yshift=.3cm]a2){A};
\node[perceptron0]  at ([yshift=-.3cm]a2){B};
\node at ([yshift=-.8cm]a2){(0,1)};
\node[perceptron0]  at ([yshift=.3cm]a3){A};
\node[perceptron0]  at ([yshift=-.3cm]a3){B};
\node at ([yshift=-.8cm]a3){(0,2)};
\node[perceptron0]  at ([yshift=.3cm]b2){A};
\node[perceptron0]  at ([yshift=-.3cm]b2){B};
\node at ([yshift=-.8cm]b2){(1,0)};
\node[perceptron0]  at ([yshift=.3cm]b3){A};
\node[perceptron0]  at ([yshift=-.3cm]b3){B};
\node at ([yshift=-.8cm]b3){(1,1)};
\node[perceptron0]  at ([yshift=.3cm]b4){A};
\node[perceptron0]  at ([yshift=-.3cm]b4){B};
\node at ([yshift=-.8cm]b4){(1,2)};
\node[perceptron0]  at ([yshift=.3cm]c1){A};
\node[perceptron0]  at ([yshift=-.3cm]c1){B};
\node at ([yshift=-.8cm]c1){(2,0)};
\node[perceptron0]  at ([yshift=.3cm]c2){A};
\node[perceptron0]  at ([yshift=-.3cm]c2){B};
\node at ([yshift=-.8cm]c2){(2,1)};
\node[perceptron0]  at ([yshift=.3cm]c3){A};
\node[perceptron0]  at ([yshift=-.3cm]c3){B};
\node at ([yshift=-.8cm]c3){(2,2)};
\draw[] (a1) -- (b1);
\draw[] (a1) -- (b2);
\draw[] (a2) -- (b2);
\draw[] (a2) -- (b3);
\draw[] (a3) -- (b3);
\draw[] (a3) -- (b4);
\draw[] (a4) -- (b4);
\draw[] (b1) -- (c1);
\draw[] (b2) -- (c1);
\draw[] (b2) -- (c2);
\draw[] (b3) -- (c2);
\draw[] (b3) -- (c3);
\draw[] (b4) -- (c3);
\draw[] (b4) -- (c4);
\draw[] (c1) -- (d1);
\draw[] (c1) -- (d2);
\draw[] (c2) -- (d2);
\draw[] (c2) -- (d3);
\draw[] (c3) -- (d3);
\draw[] (c3) -- (d4);
\end{tikzpicture}
\subcaption{Network.}
\label{fig:DQNN_universalityQNN}
\end{subfigure}
\begin{subfigure}[t]{0.49\linewidth}
\centering
\begin{tikzpicture}[scale=1.5,yscale=1.55,decoration={markings,mark=at position 0.5 with {\arrow{>}}} ]
\draw (-.5,.25)--(2.5,.25);
\draw (-.5,-.25)--(2.5,-.25);
\draw (-.5,-.75)--(2.5,-.75);
\draw (-.5,-1.25)--(2.5,-1.25);
\draw (-.5,-1.75)--(2.5,-1.75);			
\draw (-.5,-2.25)--(2.5,-2.25);
\node[operator1,minimum height=1.4cm]  at (0,0) (b) {$U_0^0$};			
\node[operator1,minimum height=1.4cm]  at (0,-1) (a) {$U_1^0$};
\node[operator1,minimum height=1.4cm]  at (0,-2) (a) {$U_2^0$};
\node[operator1,minimum height=1.4cm]   at (1,.5) (d) {$U_0^1$};
\node[operator1,minimum height=1.4cm]   at (1,-.5) (c) {$U_1^1$};
\node[operator1,minimum height=1.4cm]   at (1,-1.5) (c) {$U_2^1$};
\node[operator1,minimum height=1.4cm]   at (1,-2.5) (d) {$U_0^1$};
\node[operator1,minimum height=1.4cm]   at (2,0) (b) {$U_0^2$};			
\node[operator1,minimum height=1.4cm]  at (2,-1) (a) {$U_1^2$};
\node[operator1,minimum height=1.4cm]  at (2,-2) (a) {$U_2^2$};
\node[fill=white,minimum width=1cm,minimum height=1.4cm]  at (1,1) (g) {};
\node[fill=white,minimum width=1cm,minimum height=1.4cm]  at (1,-3) (g) {};
\node[fill=white,minimum width=1cm,minimum height=1.4cm]  at (1,-3.2) (g) {};
\node[white] at (-1.25,.25) (b) {$x$};
\node[color0] at (-.75,.25) (b) {$a$};
\node[color0] at (-.75,-.25) (a) {$b$};
\node[color0] at (-.75,-.75) (b) {$c$};
\node[color0] at (-.75,-1.25) (a) {$d$};
\node[color0] at (-.75,-1.75) (b) {$e$};
\node[color0] at (-.75,-2.25) (b) {$f$};
\end{tikzpicture}
\subcaption{Implementation.}
\label{fig:DQNN_universalityCirc}
\end{subfigure}
\caption{\textbf{DQNN universality proof.} A DQNN constructed out of pairs of perceptrons using unitary operations $U_j^l$ and $\swap$ gates (a) is equivalent to a quantum circuit of two-qubit gates $U_j^l$ acting on alternating pairs of qubits (b).}
\label{fig:DQNN_universality}
\end{figure}

To construct a QNN that is equivalent to that circuit, we first number the neurons of the QNN, the 4-level qudits, by two indices. Neuron $(l,j)$ is the $j$th neuron in $l$th layer. We assume there are $m_l$ neurons in the $l$th layer. Further we suppose that every neuron $(l,j)$ is connected to neurons $(l-1,j)$ and $(l+1,j-(-1)^l\mod m_l)$ and other connections do not exist. The neurons are marked as rectangles in \cref{fig:DQNN_universalityQNN}.

We further construct the neurons in a way that every neuron corresponds to two qubits. Each neurons qubits are labelled $A$ and $B$ and initialised in the state $\ket{00}$.

We consider that every operation of the network has the form
\begin{equation*}\label{eq:DQNN_qcdecomp}
\rho^l=\tr_{l-1}(U^{l-1}(\rho^{l-1}\otimes[\bigotimes_{j=0}^{m_l-1}\ket{00}_{(l,j)}\bra{00}]){U^{l-1}}^\dagger).
\end{equation*}
$U^l=\prod_{j=m_l-1}^{0}U^l_j$ consists of the perceptrons acting on the qubit layers $l$ and $l+1$. Note that, different to the preceding chapters, we let the indices $l$ and $j$ start at $0$. This results in a much simpler expression of $U_j^l$ when using the modulo notation. The operation acting on the neuron $(l,j)$ are defined as
\begin{align*}
U^l_j =& V_j^l\ \text{SWAP}[(l,j,A),(l+1,j-\frac{1-(-1)^l}{2}\mod m_l,B)]\\& \text{SWAP}[(l,j,B),(l+1,j+\frac{1+(-1)^l}{2}\mod m_l,A)],
\end{align*}
see \cref{fig:DQNN_universalityQNN}. Notice that the SWAP operators acting on two qubits: on one qubit that corresponds to the neuron $(l,j)$ and one of the $l+1$th layer. $V_j^l$ is a unitary that acts on the qubits of the neuron $(l,j)$. All the SWAP operations commute for a fixed $l$, because they act on different pairs of qubits. 

An in that way constructed DQNN is equivalent to the quantum circuit of two-qubit gates $V_j^l$ that act on alternating pairs of qubits like depicted in \cref{fig:DQNN_universalityCirc}. Qubit $A$ in neuron $(0,0)$ on the left hand side diagram corresponds to the qubit labelled by $a$ on the right hand side.  Similarly, the $(0,0)$ $B$ qubit corresponds to the qubit labelled by $b$ and so on.

It is known that two-qubit gates are universal \cite{Nielsen2000}. The SWAP operation is a two-qubit operation, thus the pictured quantum circuit is universal. It is clear that the construction depicted in \cref{fig:DQNN_universality} is by far not the most efficient ones but shows the universality of the in a very neat way.

\section{Classical simulation}
\label{sec:DQNN_classical}

Different aspects of the DQNN training algorithm have been discussed in this chapter so far. However, no training results were presented heretofore. The algorithm described in \cref{sec:DQNN_trainingalgorithm} can be indeed fully simulated on a classical computer. Note that since the Hilbert space dimension scales exponentially with the number of qubits, the simulation is even on supercomputers restricted to a few qubits. The simulation is still useful to study the behaviour of the algorithm.

Hence in this section, we describe a classical simulation of the DQNN algorithm using QuTip \cite{Qutip}, a quantum toolbox in Python. The code can be found at \cite{GithubKerstin}. We start with a short description of the code. A welcome side effect is getting a good overview of the parameters and data needed during the algorithm. We conclude the section with numerical results.

\FloatBarrier\subsection*{Algorithm}
\begin{algorithm}
\begin{algorithmic}[1]
\State Set parameters step size $\epsilon$, learning rate $\eta$ and number of epochs $r_T$
\State Set $s=0$
\State Assign the QNN unitaries $U_j^l(0)$ randomly
\State Provide $S$ training data pairs $\{\ket{\phi^{\text{in}}_x}, \ket{\phi^{\text{SV}}_x}\} $
\State Provide $N-S$ validation data pairs $\{\ket{\phi^{\text{in}}_x}, \ket{\phi^{\text{T}}_x}\} $
\For {$r_T$ training epochs}
\State Save the values of training loss $\mathcal{L}_\text{SV}$ and validation loss $\mathcal{L}_\text{USV}$
\For {all l and x}
\State Feed-forward the input state $\rho^{l-1}_x(t)= \mathcal{E}^{l-1}_t\big(...\mathcal{E}^{1}_t\big(\rho^{\text{in}}_x\big)...\big)$
\State Back-propagate the output states ${\sigma^l_x(t)=\mathcal{F}^{l+1}_t\big(... \mathcal{F}^{L+1}_t\big(\ket{\phi_x^\text{SV}}\bra{\phi_x^\text{SV}}\big)...\big)}$
\EndFor
\For {all l and j}
\State Calculate the update matrix $K^l_j(t) = \frac{\eta 2^{m_{l-1}}i}{S}\sum_x\tr_\text{rest}\big(M^l_{j}(x,t)\big)$
\State Update the unitaries $U_j^l(t+\epsilon)=e^{i\epsilon K_j^l(t)}U_j^l(t)$
\EndFor
\EndFor
\State Feed-forward the input state $\rho^{l-1}_x(t)= \mathcal{E}^{l-1}_t\big(\dots\mathcal{E}^{1}_t\big(\rho^{\text{in}}_x\big)\dots\big)$
\State Save the values of training loss $\mathcal{L}_\text{SV}$ and validation loss $\mathcal{L}_\text{USV}$
\end{algorithmic}
\caption{Classical simulation of the DQNN algorithm.}
\label{fig:DQNN_classcialAlgorithm}
\end{algorithm}

\begin{figure}
\centering
\begin{tikzpicture}
\begin{axis}[
xmin=0,   xmax=10,
ymin=0,   ymax=1,
width=.8\linewidth, 
height=.5\linewidth,
grid=major,grid style={color0M},
xlabel= Training epochs $r_T$, 
xticklabels={0,0,100,200,300,400,500,600,700,800,900,1000},
ylabel=$\mathcal{L}(t)$,legend pos=south east,legend cell align={left},legend style={draw=none,legend image code/.code={\filldraw[##1] (-.5ex,-.5ex) rectangle (0.5ex,0.5ex);}}]
\addplot[line width=2pt, color=color2] table [x=step times epsilon, y=SsvTestingUsv, col sep=comma] {numerics/randomUnitary_100pairs20sv_2-3-2network_delta0_lda1_ep0i01_plot.csv};
\addlegendentry[mark size=10 pt,scale=1]{Validation loss $\mathcal{L}_\text{USV}$} 
\addplot[line width=2pt, color=color1] table [x=step times epsilon, y=SsvTraining, col sep=comma] {numerics/randomUnitary_100pairs20sv_2-3-2network_delta0_lda1_ep0i01_plot.csv};
\addlegendentry[scale=1]{Training loss $\mathcal{L}_\text{SV}$} 
\end{axis}
\end{tikzpicture}
\caption{\textbf{Training a DQNN.} The validation and training loss converge to the value $1$ during the training of a \protect\twothreetwo QNN in $r_T=1000$ steps using $10$ training pairs and $90$ testing pairs (based on a unitary $Y \in \mathcal{U}(4)$), $\eta=1$ and $\epsilon=0.01$.}
\label{fig:DQNN_training}
\end{figure}
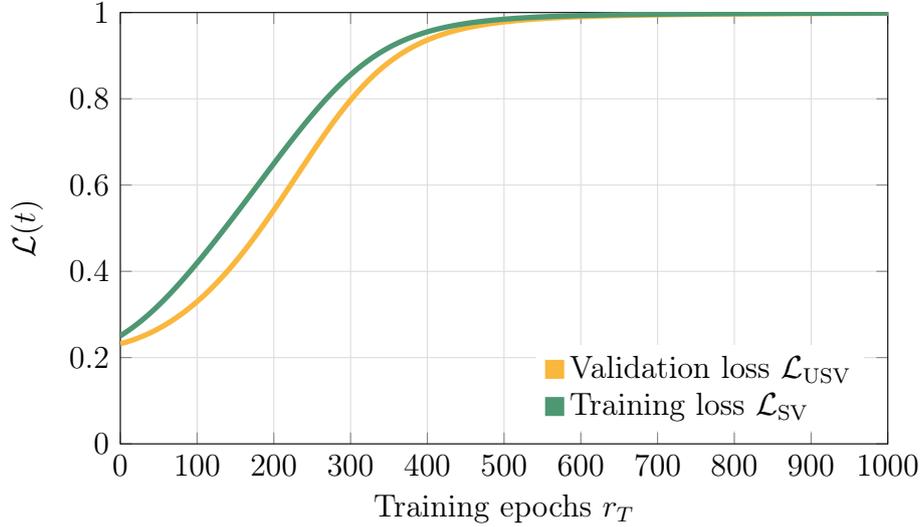

A description of the code can be found in \cref{fig:DQNN_classcialAlgorithm}. We explained the parts of the algorithm procedure in detail in \cref{sec:DQNN_trainingalgorithm}. Hence we will focus on technical details here only. 

The behaviour of the algorithm can be controlled with the parameters step size $\epsilon$, learning rate $\eta$ and number of epochs $r_T$. The latter was not discussed in this work until now and determines how often the training algorithm using all data pairs and updating all QNN unitaries is repeated. 

Beyond the parameters, the data pairs can be seen as inputs of the algorithm as well. However, the code in \cite{GithubKerstin} includes preparing the training and validation data using a randomly chosen unitary $Y$, which is aimed to simulate with the DQNN after sucessful training. 

The output of the algorithm is the trained network, saved in network unitaries and lists of the values of the loss functions, which will be exploited in the following to discuss the training success.

In \cref{fig:DQNN_training} the training and validation loss during the whole training algorithm of an exemplary QNN is plotted. The training loss increases faster during the process, but also the validation loss reaches $0.9$ after only $r_T=363$ epochs. The training and the validation loss end with a value of nearly $1$. Since the losses are based on the fidelity of the desired states and the output states of the DQNN, we can conclude that it simulates satisfyingly the unknown unitary $Y$ on the training input states as well as on the unseen validation input states. 

\subsection*{Generalisation analysis}
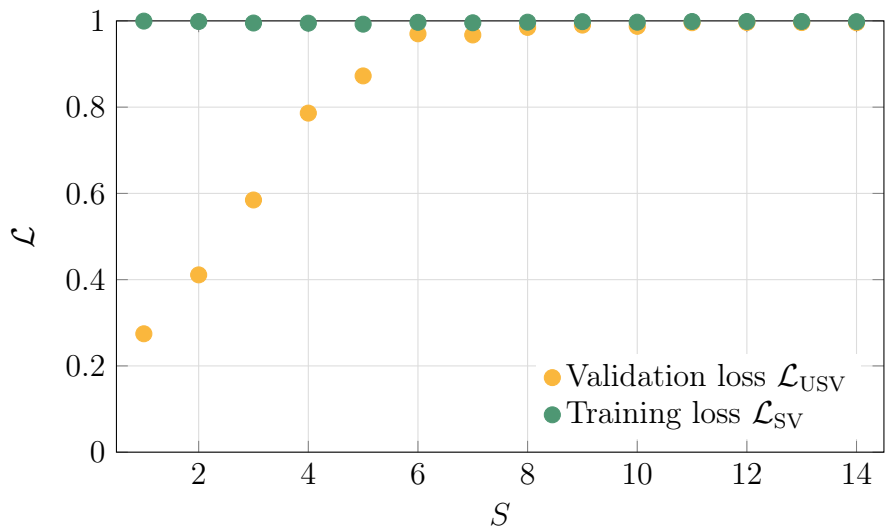
\begin{figure} 
	\centering
	\begin{tikzpicture}[scale=1]
		\begin{axis}[
			xmin=0.5,   xmax=14.5,
			ymin=0,   ymax=1,
			width=.8\linewidth, 
			height=.5\linewidth,
			grid=major,
			grid style={color0M},
			xlabel= $S$, 
			ylabel=$\mathcal{L}$,legend pos=south east,legend cell align={left},legend style={draw=none}]
			\addplot[color=color2, only marks, mark size=3 pt,mark phase=0] table [x=numberSupervisedPairsList, y=SsvTestingUsvMeanList, col sep=comma] {numerics/randomUnitary_100pairs_2-3-2network_delta0_lda1_ep0i01_rounds1000_shots10_plotmean.csv};
			\addlegendentry[mark size=3 pt,scale=1]{Validation loss $\mathcal{L}_\text{USV}$} 
			\addplot[color=color1, only marks, mark size=3 pt,mark phase=0] table [x=numberSupervisedPairsList, y=SsvTrainingMeanList, col sep=comma] {numerics/randomUnitary_100pairs_2-3-2network_delta0_lda1_ep0i01_rounds1000_shots10_plotmean.csv};
			\addlegendentry[scale=1]{Training loss $\mathcal{L}_\text{SV}$} 
		\end{axis}
	\end{tikzpicture}
	\caption{\textbf{Generalisation analysis of a DQNN.} This plot shows the training and validation loss after training a \protect\twothreetwo QNN in $r_T=1000$ steps with $\eta=1$ and $\epsilon=0.01$ using $100$ data pairs (based on a unitary $Y$), where $S$ pairs were used for training and $100-S$ pairs for testing. We vary the number of $S$. All values are averaged over $10$ individual training attempts.}
	\label{fig:DQNN_generalisation}
\end{figure}

In the plot mentioned above $S=10$ of $100$ data pairs build the training set. The remaining pairs are saved for validation. To study how well the QNN is able to generalise, $S$ can be varied. Therefore we only compare the end value of the loss function, namely the values obtained in the last line of \cref{fig:DQNN_classcialAlgorithm}. We averaged these values over $10$ individual training attempts for all values of $S \in [1,14]$ to get a more valid result. \cref{fig:DQNN_generalisation} presents this scenario. As expected, the validation loss increase with the number of training pairs. It is remarkable that using only six of $100$ data pairs for training leads to a validation loss over $0.95$.

\subsection*{Noise analysis}
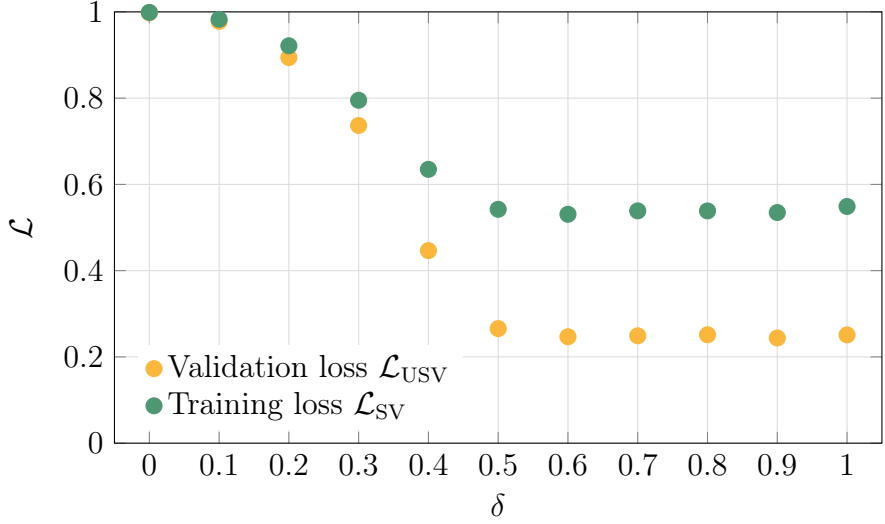
\begin{figure} 
	\centering
	\begin{tikzpicture}[scale=1]
		\begin{axis}[
			xmin=-0.05,   xmax=1.05,
			ymin=0,   ymax=1,
			width=.8\linewidth, 
			height=.5\linewidth,
			grid=major,
			grid style={color0M},
			xlabel= $\delta$, 
			ylabel=$\mathcal{L}$,legend pos=south west,legend cell align={left},legend style={draw=none}]
			\addplot[color=color2, only marks, mark size=3 pt,mark phase=0] table [x=deltaList, y=SsvTestingUsvMeanList, col sep=comma] {numerics/randomUnitary_100pairs_20sv_2-3-2network_deltaVar_lda1_ep0i01_rounds1000_shots10_plotmean.csv};
			\addlegendentry[mark size=3 pt,scale=1]{Validation loss $\mathcal{L}_\text{USV}$} 
			\addplot[color=color1, only marks, mark size=3 pt,mark phase=0] table [x=deltaList, y=SsvTrainingMeanList, col sep=comma] {numerics/randomUnitary_100pairs_20sv_2-3-2network_deltaVar_lda1_ep0i01_rounds1000_shots10_plotmean.csv};
			\addlegendentry[scale=1]{Training loss $\mathcal{L}_\text{SV}$} 
		\end{axis}
	\end{tikzpicture}
	\caption{\textbf{Testing the noise robustness of a DQNN.} The training and validation loss after training a \protect\twothreetwo QNN changes with the noise parameter $\delta$. The training was done in $r_T=1000$ steps with $\eta=1$ and $\epsilon=0.01$ using $20$ training and $80$ testing pairs (based on a unitary $Y$). The loss values are averaged over $10$ individual training attempts.}
	\label{fig:DQNN_noise}
\end{figure}

We have already seen a very good generalisation behaviour. Since the approach is to implement the DQNN on currently available quantum devices, we test the algorithm for robustness to noise. Therefore we replace the supervised states with $\ket{\phi_{x,\delta}^\text{SV}}$, using a new parameter $\delta$ and a for every training pair randomly choosen state $\ket{\psi^\text{random}}$ of the same dimension as $\ket{\phi_x^\text{SV}}$, i.e.\ we build
\begin{equation*}
\ket{\phi_{x,\delta}^\text{SV}}= \frac{(1-\delta)\ket{\phi_x^\text{SV}} + \delta\ket{\psi^\text{random}}}{||(1-\delta)\ket{\phi_x^\text{SV}} + \delta\ket{\psi^\text{random}}||}.
\end{equation*}
In the same way as when observing the generalisation behaviour in \cref{fig:DQNN_generalisation} we compare the last values of the loss functions and average over $10$ completely independent trainings. In \cref{fig:DQNN_noise} we plot the validation loss for different values of $\delta$.

We can observe that both, the training and the validation loss, take higher values then $0.7$ for $\delta \le 0.3$. If the noise exceeds this value of delta, the training loss rests at a value around $0.55$ and the validation loss around $0.25$.

Discussing only a few examples, we could already observe the great learning behaviour of the DQNN with outstanding generalisation possibility and robustness to noise. More numerical experiments, including $1$- and $3$-qubit training states and deeper networks, can be found in the appendix in \cref{apdx:DQNN}.

\section{NISQ device implementation}
\label{sec:DQNN_quantumalg}

In the preceding paragraphs, we have already discussed the training results of the DQNN algorithm using a simulation running on a classical computer. However, in this section, we explain the implementation of the same algorithm on a quantum computer, more precisely on a NISQ device. A short introduction to NISQ devices was given in \cref{sec:QI_QC}. Since the implementation presented in \cref{sec:DQNN_classical} differs from the implementation on these early quantum devices we denote the latter with \DQNNNISQ.

\FloatBarrier\subsection*{Implementation of \DQNNNISQ}
We already explained the quantum perceptron and the network architecture in detail in \cref{sec:DQNN_networkarchitecture}. Since the implementation on the NISQ device is based on this, we want to shortly remind the reader of the basic idea at this point. The quantum perceptron is defined as a unitary $U^l_j$ acting on $m_{l-1}+1$ qubits: $m_{l-1}$ qubits, placed in layer ${l-1}$, are the input $\rho^{l-1}$ of the perceptron, and the last qubit belongs to layer $l$ and is initialised in the zero state. We get the perceptron's output state after applying the unitary $U^l_j$ to the $m_{l-1}+1$ qubits and tracing out the $m_{l-1}$ input qubits. The perceptron layers of the DQNN can be summarised in unitaries $U^l=U_{m_{l}}^l \dots U_1^l$. We will use this layer notation to describe the implementation of the DQNN on a NISQ device. As before, we assume $m=m_{0}=m_{L+1}$, hence we train the DQNN to imitate a unitary operation. 

As a first and essential step to build a \DQNNNISQ, the DQNN's perceptron has to be implemented. In the classical simulation discussed in \cref{sec:DQNN_classical} the perceptrons were defined by unitary matrices, whose entries would be updated during the training algorithm. As explained in \cref{subsec:QI_Implementation}, QNNs can be implemented on a quantum computer via \emph{parametrised quantum circuits} \cite{Mitarai2018,Benedetti2019,Du2020,Bu2021} consisting of parametrised quantum gates. For this, two aspects need to be considered: the realisation should be universal and the number of gates and parameters small. Here, we present the work of \cite{Beer2021a}, where a balance of these objectives leads to good training results on NISQ devise as plotted later in this section. A detailed discussion of over-parametrisation can be found at \cite{Larocca2021}.

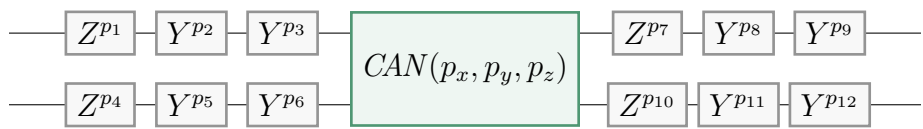
\begin{figure} [H]
\centering
\begin{tikzpicture}[xscale=1.2, yscale=0.95]
\draw (0,1)-- (10,1);
\draw (0,0)-- (10,0);
\node[operator0,minimum height=0.5cm] at (1,1){$Z^{p_1}$};
\node[operator0,minimum height=0.5cm] at (2,1){$Y^{p_2}$};
\node[operator0,minimum height=0.5cm] at (3,1){$Y^{p_3}$};
\node[operator0,minimum height=0.5cm] at (1,0){$Z^{p_4}$};
\node[operator0,minimum height=0.5cm] at (2,0){$Y^{p_5}$};
\node[operator0,minimum height=0.5cm] at (3,0){$Y^{p_6}$};
\node[operator1,minimum height=1.5cm] at (5,0.5){$\can(p_x,p_y,p_z)$};
\node[operator0,minimum height=0.5cm] at (7,1){$Z^{p_7}$};
\node[operator0,minimum height=0.5cm] at (8,1){$Y^{p_8}$};
\node[operator0,minimum height=0.5cm] at (9,1){$Y^{p_9}$};
\node[operator0,minimum height=0.5cm] at (7,0){$Z^{p_{10}}$};
\node[operator0,minimum height=0.5cm] at (8,0){$Y^{p_{11}}$};
\node[operator0,minimum height=0.5cm] at (9,0){$Y^{p_{12}}$};
\end{tikzpicture}
\caption{\textbf{Implementation of a two-qubit unitary.} An arbitrary two-qubit unitary operation can be implemented using one two-qubit gate and twelve single qubit gates.}
\label{fig:DQNN_CAN12}
\end{figure}

To express the perceptron unitaries we apply a result of studies on the implementation of two-qubit gates \cite{Zhang2003,Zhang2004,Blaauboer2008,Watts2013,Crooks2019, Peterson2020}: using a two-qubit canonical gate and twelve single qubit gates, see \cref{fig:DQNN_CAN12}, every arbitrary two-qubit unitary can be simulated.
Hereby the two-qubit canonical gate $\can (p_x, p_y, p_z)$  is defined as the composition of three two-qubit gates
\begin{align*}\label{can_definition}
\can (p_x, p_y, p_z) =& e^{-i\frac{\pi}{2}p_x X\otimes X} e^{-i\frac{\pi}{2}p_y Y\otimes Y} e^{-i\frac{\pi}{2}p_z Z\otimes Z} \\
=& \rxx (p_x \pi) \ryy (p_y \pi) \rzz (p_z \pi).
\end{align*}
using the Pauli matrices  $X = \big(\begin{smallmatrix}
0 & 1\\
1 & 0
\end{smallmatrix}\big),$ 
$Y = \big(\begin{smallmatrix}
0 & -i\\
i & 0
\end{smallmatrix}\big),$ and  
$ Z = \big(\begin{smallmatrix}
1 & 0\\
0 & -1
\end{smallmatrix}\big)$ and the parameters ${p_x,p_y,p_z}\in\mathbbm{R}$\cite{Crooks2019}. It is worth mentioning that the gates $\rxx (p)$, $\ryy (p)$ and $\rzz (p)$ are standard in quantum computing libraries. The single qubit gates $Y^p$ and $Z^p$ are parametrised Pauli operators and, up to a phase, equivalent to rotations around the $y$ and $z$ axis, namely
\begin{align*}
Y^p&\approx R_Y(\pi p)=e^{-i\frac{\pi}{2}p Y}\\
Z^p&\approx R_Z(\pi p)=e^{-i\frac{\pi}{2}p Z}.
\end{align*}
The sequences of single gates $Y^p$ and $Z^p$ with parameters $p_1,\dots,p_{3}\in \mathbbm{R}$, see \cref{fig:DQNN_CAN12}, can be rephrased in form of the gate
\begin{equation*}\label{u_definition}
u(p_1, p_2, p_3) =  \begin{pmatrix}
\cos (p_1 /2) & -e^{ip_3}\sin (p_1 /2)\\
e^{i p_2} \sin (p_1 /2) & e^{i(p_3+p_2)} \cos (p_1 /2)
\end{pmatrix},
\end{equation*}
which is a standard gate in most of the quantum computing libraries as well. We can proceed in the same way for the  parameters $p_4,\dots,p_{12}\in \mathbbm{R}$.

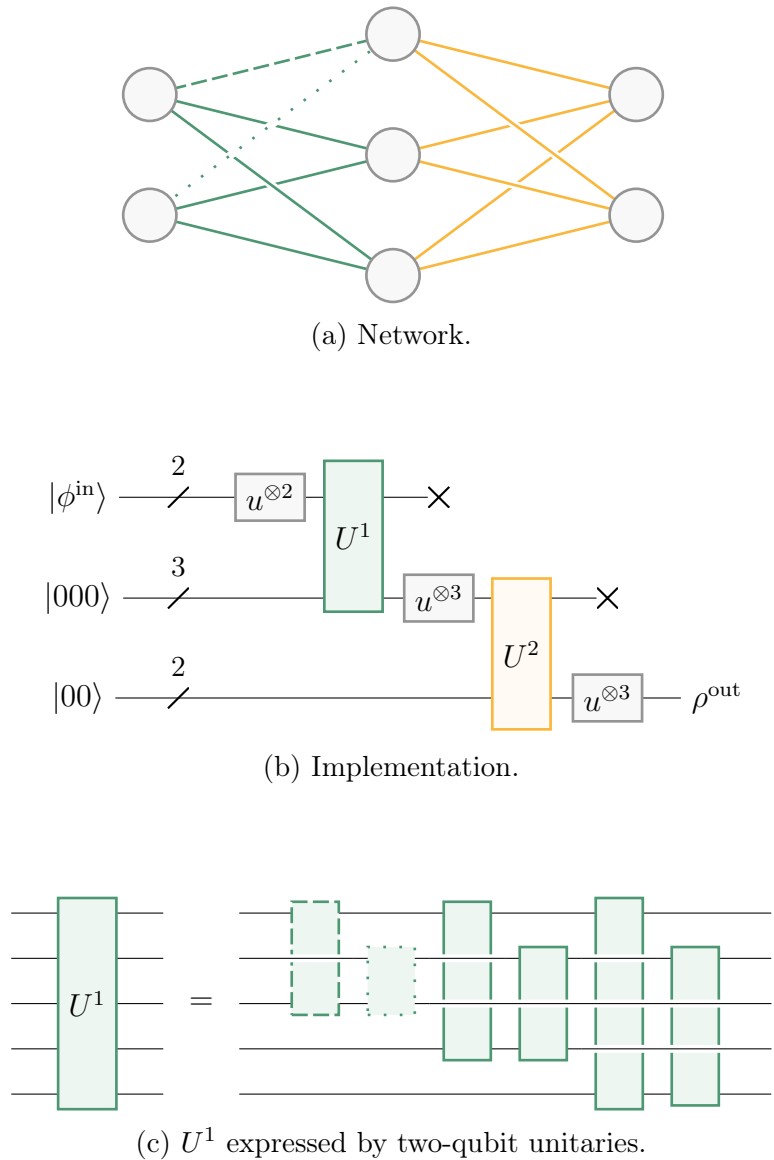
\begin{figure}
\centering
\begin{subfigure}[t]{1\linewidth}
\centering
\begin{tikzpicture}[scale=1.6]
\foreach \x in {-.5,.5} {
\draw[line0] (0,\x) -- (2,-1);
\draw[line1] (0,\x) -- (2,-1);
\draw[line0] (0,\x) -- (2,0);
\draw[line1] (0,\x) -- (2,0);
\draw[line0] (0,\x) -- (2,1);
}
\draw[line0] (0,-.5) -- (2,1);
\draw[line1,dash pattern=on 1pt off 5pt] (0,-.5) -- (2,1);
\draw[line1, dash pattern=on 6pt off 2pt] (0,.5) -- (2,1);
\foreach \x in {-1,0,1} {
\draw[line0] (2,\x) -- (4,-0.5);
\draw[line2] (2,\x) -- (4,-0.5);
\draw[line0] (2,\x) -- (4,0.5);
\draw[line2] (2,\x) -- (4,0.5);
}
\node[perceptron0] at (0,-0.5) {};
\node[perceptron0] at (0,0.5) {};
\node[perceptron0] at (2,-1) {};
\node[perceptron0] at (2,0) {};
\node[perceptron0] at (2,1) {};
\node[perceptron0] at (4,-0.5) {};
\node[perceptron0] at (4,0.5) {};
\end{tikzpicture}
\subcaption{Network. }
\label{fig:DQNN_qnncircuitA}
\end{subfigure}
\vspace*{10mm}

\begin{subfigure}[t]{1\linewidth}
\centering
\begin{tikzpicture}[scale=1.3]
\matrix[row sep=0.3cm, column sep=0.5cm] (circuit) {
\node(start3){$\ket{\phi^\text{in}}$};  
& \node[halfcross,label={\small 2}] (c13){};
& \node[operator0] (c23){$u^{\otimes 2}$};
& \node[]{}; 
& \node[dcross](end3){}; 
& \node[]{}; 
& \node[]{}; 
& \node[]{}; \\
\node(start2){$\ket{000}$};
& \node[halfcross,label={\small 3}] (c12){};
& \node[]{}; 
& \node[]{}; 
& \node[operator0] (c32){$u^{\otimes 3}$};
& \node[]{}; 
& \node[dcross](end2){}; 
& \node[]{};  \\
\node(start1){$\ket{00}$};
& \node[halfcross,label={\small 2}] (c11){};
& \node[]{}; 
& \node[]{}; 
& \node[]{}; 
& \node[]{}; 
& \node[operator0] (c41){$u^{\otimes 3}$};
& \node (end1){$\rho ^\text{out}$}; \\
};
\begin{pgfonlayer}{background}
\draw[] (start1) -- (end1)  
(start2) -- (end2)
(start3) -- (end3);
\node[operator1, minimum height=2cm] at (-.4,0.5) {$U^1$};
\node[operator2,minimum height=2cm] at (1.3,-.7) {$U^2$};
\end{pgfonlayer}
\end{tikzpicture}
\subcaption{Implementation. }
\label{fig:DQNN_qnncircuitB}
\end{subfigure}
\vspace*{10mm}

\begin{subfigure}[t]{1\linewidth}
\centering
\begin{tikzpicture}[yscale=.6]
\draw (0,4)-- (2,4);
\draw (0,3)-- (2,3);
\draw (0,2)-- (2,2);
\draw (0,1)-- (2,1);
\draw (0,0)-- (2,0);
\node[operator1,minimum height=2.8cm] at (1,2){$U^1$};
\node[] at (2.5,2){$=$};
\begin{scope}[xshift=3cm]
\draw[white] (0,5)-- (7,5);
\draw (0,4)-- (7,4);
\draw (0,3)-- (7,3);
\draw (0,2)-- (7,2);
\draw (0,1)-- (7,1);
\draw (0,0)-- (7,0);
\node[operator1,minimum height=1.5cm,line width=1pt, dash pattern=on 6pt off 2pt] at (1,3){};
\node[operator1,minimum height=0.9cm, line width=1pt, dash pattern=on 1pt off 5pt] at (2,2.5){};
\node[operator1,minimum height=2.1cm] at (3,2.5){};
\node[operator1,minimum height=1.5cm] at (4,2){};
\node[operator1,minimum height=2.8cm] at (5,2){};
\node[operator1,minimum height=2.1cm] at (6,1.5){};
\draw[line0] (0,3)-- (1.5,3);
\draw (0,3)-- (1.5,3);
\draw[line0] (2.5,3)-- (3.5,3);
\draw (2.5,3)-- (3.5,3);
\draw[line0] (4.5,3)-- (5.5,3);
\draw (4.5,3)-- (5.5,3);
\draw[line0] (2.5,2)-- (7,2);
\draw (2.5,2)-- (7,2);
\draw[line0] (4.5,1)-- (7,1);
\draw (4.5,1)-- (7,1);
\end{scope}
\end{tikzpicture}
\subcaption{$U^1$ expressed by two-qubit unitaries. }
\label{fig:DQNN_qnncircuitC}
\end{subfigure}
\vspace*{10mm}
\caption{\textbf{Implementation of a \DQNNNISQ.} A \DQNNNISQ consisting of two layers of quantum perceptrons and seven qubits (a) can be implemented as quantum circuit using $u$-gates and unitary operations representing the layers of the network (b). The first layer including three perceptrons can be decomposed in six two-qubit gates (c). The perceptron $U^1_1$ is marked exemplarily in dashed and dotted lines (see a and c).}
\label{fig:DQNN_qnncircuit}
\end{figure}

So far, we have only discussed how to implement two-qubit unitaries. The quantum perceptron unitaries $U_j^l$ are in general $m_{l-1}+1$-qubit unitaries. These qubit unitaries connect $m_{l-1}$ qubits from layer $l-1$ with one qubit of layer $l$. This motivates to replace one of these $m_{l-1}+1$-qubit unitaries with $m_{l-1}$ two-qubit unitaries acting on one of the qubits in layer $l-1$ and the qubit in layer $l$. Numerical results presented later in this section justify this idea.

After clarifying the implementation of the quantum perceptron, we can build the network circuit layerwise. Every \DQNNNISQ consists of $M=\sum ^{L+1}_{l=0} m_l$ qubits. The first $m$ qubits are initialised in the input state $\ket{\phi^\text{in}}$. The remaining qubits are initialised in the zero state $\ket{0}$.  For every layer, we first apply the single-qubit gates $u(p_1, p_2, p_3)$ to all qubits of the layer, followed by the unitary $U^l$, expressed via $m_{l-1}$ of the $\can$-gates. After layer $l$ is complete the $m_{l-1}$ input qubits are neglected and the $m_{l}$ input qubits act as input qubits for the new layer $l+1$. Different to the classical simulation the feed-forward process is not at the end when the last layer is reached: an extra layer of single-qubit gates follows, see \cref{fig:DQNN_qnncircuitB}. After that, the output of the circuit is a $m_{L+1}$-qubit quantum state $\rho^\text{out}$. 

See \cref{fig:DQNN_qnncircuit} for an example: the implementation of the DQNN depicted in \cref{fig:DQNN_qnncircuitA} is visualised in \cref{fig:DQNN_qnncircuitB}. The first layer unitary $U^1=U_{3}^1 U_{2}^1 U_1^1$ is written out in two-qubit unitaries (implemented via $\can$-gates) in \cref{fig:DQNN_qnncircuitC}. The two two-qubit unitaries needed for expressing the perceptron unitary $U_1^1$ are marked with dashed and dotted lines in \cref{fig:DQNN_qnncircuitA} and \cref{fig:DQNN_qnncircuitC}.

To summarise the description of the implementation we shortly discuss the number of parameters. The $u$-gates, placed before every unitary layer and additionally at the end to the output state are described by $3\sum ^{L+1}_{l=1} m_{l-1} + 3m_{L+1}$ parameters. The $\can$-gates are parametrised via $3\sum ^{L+1}_{l=1} m_{l-1} m_{l}$ real numbers. It results that the quantum circuit is described by  
\begin{equation*}
n^\text{\DQNNNISQ}=3\sum ^{L+1}_{l=1} m_{l-1} (1+m_{l})+ 3m_{L+1}
\end{equation*} parameters.

\FloatBarrier\subsection*{Implementation of the training algorithm}

As the reader may notice, we so far only described the processing of the input state through the DQNN and becoming the output state. This is only one component, the feed-forward part, of the training algorithm. In the following lines, the rest of the algorithm will be explained.

Analogously to the preceding discussions we use the same learning task, i.e.\ learn an unknown unitary $Y \in \mathcal{U}(2^{m})$ from a given training set $\{\ket{\phi^\text{in}_x},\ket{\phi^\text{SV}_x} \}^{S}_{x=1}$ where $\ket{\phi^\text{SV}_x} = Y \ket{\phi^\text{in}_x}$. Further we use the fidelity as a training loss function, namely
\begin{equation*}
\mathcal{L}_\text{SV} ({\omega}_t)= \frac{1}{S}\sum_{x=1}^S \braket{\phi^{\text{SV}}_x|\rho_x^{\text{out}}({\omega}_t)|\phi^{\text{SV}}_x},
\end{equation*}
where $\rho_x^{\text{out}}$ denotes the network's output and the vector ${\omega}_t=(\omega_1(t), ...,\omega_{n}(t))^T$ with $n=n^\text{\DQNNNISQ}$ comprises the parameters describing the quantum circuit. 

In the classical simulation, the training takes place by updating the entries of unitary matrices following an update rule, including an updated matrix. This update matrix was derived using the derivative of the loss function and requires the knowledge of the states of each layer's qubits. Since this is not possible on the NISQ device, the implementation of the algorithm uses parametrised quantum gates. The parameters change during the training in order to maximise the loss function. Gradient descent is a good tool to find out in which direction the change should be done. Note that the gradient descent method can be technically exchanged by any other optimising algorithm.

More precicely the $n^\text{\DQNNNISQ}$ parameters of the quantum circuit are initialised as ${\omega}_0$. After every training epoche all parameters are updated by ${\omega}_{t+1} = {\omega}_{t} + {d \omega}_{t}$, where ${d\omega}_{t} = \eta {\nabla} \mathcal{L}_\text{SV} \left({\omega}_t\right)$ using the learning rate $\eta$ and the gradient is of the form
\begin{equation*}
\mathcolorbox{\nabla _k \mathcal{L}_\text{SV} \left({\omega}_t\right) = \frac{\mathcal{L}_\text{SV}\left({\omega}_t + \epsilon{e}_k\right) - \mathcal{L}_\text{SV}\left({\omega}_t - \epsilon{e}_k\right)}{2\epsilon} + \mathcal{O}\left(\epsilon^2\right).}
\end{equation*}
The vectors ${e}_{k}$ are defined as $e_{k}^j=\delta_{k,j}$, $k,j=1,...,n^\text{\DQNNNISQ}$. The parameters step size $\epsilon > 0$ and learning rate $\eta$ are used analogously to \cref{sec:DQNN_classical}. Note at this point that for calculating the gradient, the training loss function, and therefore the whole quantum circuit, has to be evaluated for the parameters ${\omega}_t + \epsilon{e}_k$ and ${\omega}_t - \epsilon{e}_k$ for every $k$.

The overall procedure of the algorithm resembles with \cref{sec:DQNN_networkarchitecture}, where the algorithm was presented in detail. The training and validation data, i.e.\ the set of training pairs and the set of testing pairs, are initialised. The reader is invited to think again of an uncharacterised quantum device acting as a unitary $Y$ on $m$-qubit input states. Further, the step size $\epsilon$ and learning rate $\eta$ are determined conformable to the circumstances, and the parameters ${\omega}_0$ initialised randomly. As a second step the quantum circuit in \cref{fig:DQNN_qnncircuit} is executed for all the states in the training set: $m$ qubits are initialised in a state $\ket{\phi^\text{SV}_x}$ and another $m$ qubits in $\ket{\phi^\text{in}_x}$. Note again that it is $m=m_{0}=m_{L+1}$ for learning a unitary. All the other qubits are initialised in the zero state $\ket{0}$. The network circuit, depicted in \cref{fig:DQNN_qnncircuitB}, is evaluated.

\begin{figure}
	\centering
	\begin{tikzpicture}[]
		\matrix[row sep=0.1cm, column sep=0.5cm] (circuit) {
			\node(start6){};
			& \node[circlewc](circle1){};
			& \node[]{}; 
			& \node[]{}; 
			& \node[]{}; 
			& \node[]{}; 
			& \node[meter,scale=0.5](end6){}; \\
			\node(start5){};
			& \node[]{};
			& \node[circlewc](circle2){}; 
			& \node[]{}; 
			& \node[]{}; 
			& \node[]{}; 
			& \node[meter,scale=0.5](end5){}; \\
			\node{};
			& \node[]{};
			& \node[]{}; 
			& \node[]{}; 
			& \node[](){}; 
			& \node[]{}; 
			& \node[](){}; \\
			\node(start4){};
			& \node[]{};
			& \node[]{}; 
			& \node[]{}; 
			& \node[circlewc](circle3){}; 
			& \node[]{}; 
			& \node[meter,scale=0.5](end4){}; \\
			\node(start3){};
			& \node[dot](dot1){};
			& \node[]{}; 
			& \node[]{}; 
			& \node[]{}; 
			& \node[operator0]{$H$};  
			& \node[meter,scale=0.5](end3){}; \\
			\node(start2){};
			& \node[]{};
			& \node[dot](dot2){}; 
			& \node[]{}; 
			& \node[]{}; 
			& \node[operator0]{$H$}; 
			& \node[meter,scale=0.5](end2){}; \\
			\node{};
			& \node[]{};
			& \node[]{}; 
			& \node[]{}; 
			& \node[](){}; 
			& \node[]{}; 
			& \node[](){}; \\
			\node(start1){};
			& \node[]{};
			& \node[]{}; 
			& \node[]{}; 
			& \node[dot](dot3){}; 
			& \node[operator0]{$H$}; 
			& \node[meter,scale=0.5](end1){}; \\
		};
		\begin{pgfonlayer}{background}
			\draw[] (start1) -- (end1)  
			(start2) -- (end2)
			(start3) -- (end3)
			(start4) -- (end4)
			(start5) -- (end5)
			(start6) -- (end6);
			\draw[] (dot1) -- (circle1)
			(dot2) -- (circle2)
			(dot3) -- (circle3);
		\end{pgfonlayer}
		\node[operator0, minimum height=4.5cm, minimum width=1cm, draw=white, fill=white] at (-.25,0) {};
		\node at (-.22,-.22) {$\dots$};
		\node at (-2.5,-1.4) {$\vdots$};
		\node at (-2.5,1.1) {$\vdots$};
		\draw[brace0,color=black] {(-2.9,-2.2) -- node[left=1ex] {$\rho^\text{out}$}(-2.9,-0.1)};
		\draw[brace0,color=black] {(-2.9,0.4) -- node[left=1ex] {$\ket{\phi^{\text{SV}}_x}$}(-2.9,2.2)};
		\draw[brace0,color=black] {(3.2,2.4)}  -- node[right=1ex] {$\rightarrow{p}$}(3.2,-2.4);
	\end{tikzpicture}
	\caption{\textbf{Implementation of the of destructive swap test}. $m$ $\cnot$ gates, $m$ Hadamard gates and $2m$ measurements are used to calculate the fidelity $\braket{\phi^{\text{SV}}_x|\rho^\text{out}|\phi^{\text{SV}}_x}=p\cdot c$. }
	\label{fig:DQNN_swap}
\end{figure}
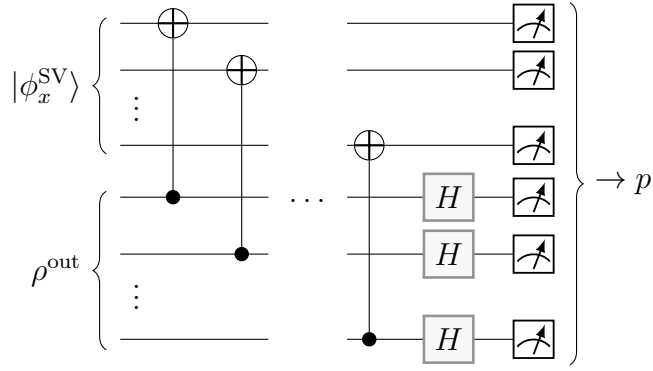\label{subsec:DQNN_swap}

At this point $m$ qubits are in the state $\rho^\text{out}$ and another $m$ qubits in the state $\ket{\phi^\text{SV}_x}$. We conveniently label the qubits in this order  $q_1, \dots q_{2m}$. The next step is to calculate the training loss to update the parameters ${\omega}_t$. For this purpose for all training data pairs, the circuit is executed and the so-called destructive $\swap$-test \cite{Buhrman2001,GarciaEscartin2013,Cincio2018} is used to execute the fidelity: $m$ $\cnot$ gates and $m$ Hadamard gates are applied as depicted in \cref{fig:DQNN_swap}. All qubits are measured in the computational basis and the results are alternating saved in a list, i.e.\ $q_1, q_m,q_2, q_{m+1}, \dots  q_{m-1} q_{2m}$. If we exemplary assume $m=1$, i.e.\ when comparing two one-qubit states, the list can only be of the form $\{0,0\}$, $\{0,1\}$, $\{1,0\}$ or $\{1,1\}$. Due to the quantum projection noise, we execute these measurements $M$ times and save the ratio of the occurrence of each list as components of a vector ${p}$.  It is reasoned in \cite{Cincio2018} that the fidelity of the states $\ket{\phi^{\text{SV}}_x}$ and $\rho_x^{\text{out}}$ can be obtained by
\begin{equation*}
F(\ket{\phi^{\text{SV}}_x}\bra{\phi^{\text{SV}}_x},\rho_x^{\text{out}})={p}\cdot {c},
\end{equation*}
where ${c}=(1,1,1,-1)^{\otimes m}$.

Using the  $\swap$-test not only the parameters can be updated via ${\omega}_{t+1} = {\omega}_{t} + {d \omega}$. Also the validation loss 
\begin{equation*}
\mathcal{L}_\text{USV}({\omega}_t)=\frac{1}{N-S}\sum_{x=S+1}^{N} \langle\phi^{\text{USV}}_x\rvert\rho_x^{\text{out}}({\omega}_t)\lvert\phi^{\text{USV}}_x\rangle
\end{equation*}
is calculated in the same way.

\begin{figure}[H]
\centering
\begin{tikzpicture}[]
\matrix[row sep=0.3cm, column sep=0.5cm] (circuit) {
\node(start3){$\ket{0}^{\otimes m}$};  
& \node[halfcross,label={\small m}] (c13){};
& \node[operator0, fill=white, minimum width=1.2cm] (c23){$\ket{\phi^\text{SV}}$};
& \node[]{}; 
& \node[]{}; 
& \node[]{}; 
& \node[]{}; 
& \node[circlewc](circle){}; 
& \node[]{}; 
& \node[meter](end3){};  \\
\node(start2){$\ket{0}^{\otimes m}$};
& \node[halfcross,label={\small m}] (c12){};
& \node[operator0, fill=white, minimum width=1.2cm] (c22){$\ket{\phi^\text{in}}$};
& \node[]{}; 
& \node[]{}; 
& \node[]{}; 
& \node[halfcross,label={\small n}]{}; 
& \node[](end2){}; 
& \node[]{};
& \node[]{};  \\
\node(start1){$\ket{0}^{\otimes n}$};
& \node[halfcross,label={\small n}] (c11){};
& \node[]{}; 
& \node[]{}; 
& \node[]{}; 
& \node[]{}; 
& \node[halfcross,label={\small m}]{}; 
& \node[dot](dot){}; 
& \node[operator0](c41){$H$};
& \node[meter](end1){}; \\
};
\begin{pgfonlayer}{background}
\draw[] (start1) -- (end1)  
(start2) -- (end2)
(start3) -- (end3);
\node[operator0, minimum width=1cm,draw=white, fill=white] at (2.4,0) {};
\draw[] (dot) -- (circle);
\node[operator1, minimum height=2cm] at (-.1,-0.75) {DQNN};

\end{pgfonlayer}
\end{tikzpicture}
\caption{\textbf{Implementation of training a \DQNNNISQ}. The training process includes initialising the qubits, performing the \DQNNNISQ quantum circuit and executing the $\swap$-test. The number of used qubits is $2m+n$, where $m=m_{0}=m_{L+1}$ and $n = \sum_{l=1}^{L} m_l$. }
\label{fig:DQNN_fullCircuit}
\end{figure}
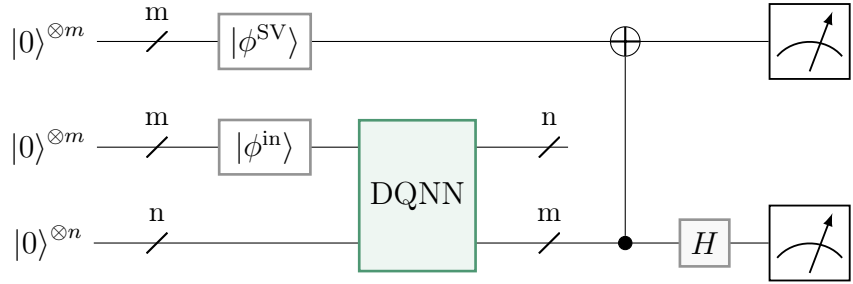

\begin{sloppypar}
In \cref{fig:DQNN_fullCircuit} the complete quantum circuits, including initialising, the \DQNNNISQ algorithm and the $\swap$-test, is depicted. The box with the title \enquote{DQNN} represents the in the above paragraph described DQNN part of the quantum circuit, see \cref{fig:DQNN_qnncircuit}. For a better overview and due to its analogy to the calculation of the training loss, the validation data and the validation process are not depicted.
\end{sloppypar} 

\FloatBarrier\subsection*{Device execution}
Before plotting some results, we shortly have to mention a third loss function, next to the training and validation loss. The \emph{identity loss} is worked out using $Y=\mathbbm{1}$ and parameters which make the network, assuming there would be no noise, act as the identity. Nevertheless, indeed, all the gates are still applied and add noise to the circuit. Hence this loss gives a good insight for the best possible training loss an ideally trained network could generate.

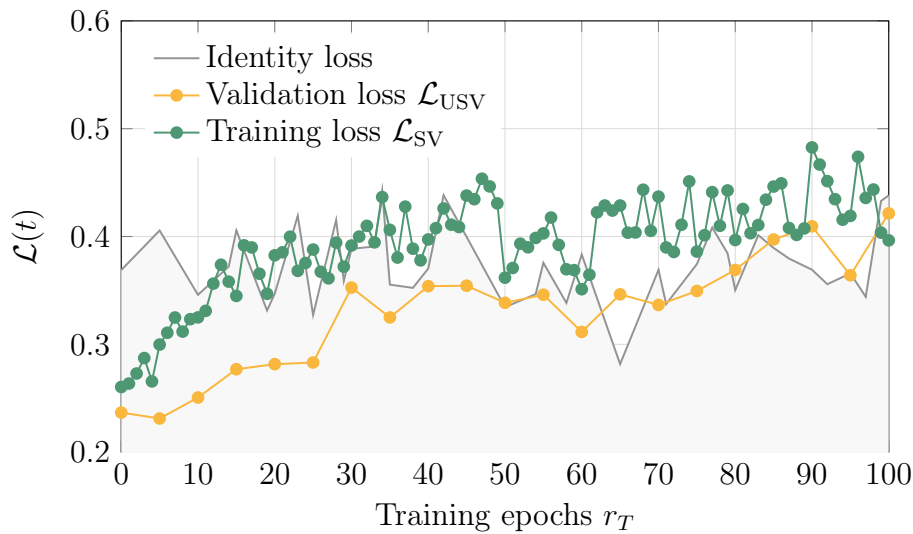
\begin{figure} 
\centering
\begin{tikzpicture}
\begin{axis}[
xmin=-0, xmax=100,
ymin=0.2, ymax=0.6,
width=.8\linewidth, 
height=.5\linewidth,
grid=major,
grid style={color0M},
xlabel= Training epochs $r_T$, 
ylabel=$\mathcal{L}(t)$,legend pos=north west,legend cell align={left},legend style={draw=none}]
\path [fill=color0L, line width=1pt]
(axis cs:0,0.368896484375)
--(axis cs:0,0)
--(axis cs:5,0)
--(axis cs:10,0)
--(axis cs:14,0)
--(axis cs:15,0)
--(axis cs:19,0)
--(axis cs:20,0)
--(axis cs:23,0)
--(axis cs:25,0)
--(axis cs:28,0)
--(axis cs:29,0)
--(axis cs:30,0)
--(axis cs:33,0)
--(axis cs:34,0)
--(axis cs:35,0)
--(axis cs:38,0)
--(axis cs:40,0)
--(axis cs:42,0)
--(axis cs:45,0)
--(axis cs:50,0)
--(axis cs:54,0)
--(axis cs:55,0)
--(axis cs:58,0)
--(axis cs:60,0)
--(axis cs:65,0)
--(axis cs:70,0)
--(axis cs:71,0)
--(axis cs:75,0)
--(axis cs:77,0)
--(axis cs:79,0)
--(axis cs:80,0)
--(axis cs:83,0)
--(axis cs:85,0)
--(axis cs:87,0)
--(axis cs:90,0)
--(axis cs:92,0)
--(axis cs:95,0)
--(axis cs:97,0)
--(axis cs:99,0)
--(axis cs:100,0)
--(axis cs:100,0.43817138671875)
--(axis cs:100,0.43817138671875)
--(axis cs:99,0.4329833984375)
--(axis cs:97,0.3441162109375)
--(axis cs:95,0.3658447265625)
--(axis cs:92,0.35577392578125)
--(axis cs:90,0.36920166015625)
--(axis cs:87,0.37957763671875)
--(axis cs:85,0.389404296875)
--(axis cs:83,0.4012451171875)
--(axis cs:80,0.35028076171875)
--(axis cs:79,0.3843994140625)
--(axis cs:77,0.4083251953125)
--(axis cs:75,0.3743896484375)
--(axis cs:71,0.33740234375)
--(axis cs:70,0.36907958984375)
--(axis cs:65,0.28167724609375)
--(axis cs:60,0.38330078125)
--(axis cs:58,0.33837890625)
--(axis cs:55,0.37567138671875)
--(axis cs:54,0.3465576171875)
--(axis cs:50,0.33447265625)
--(axis cs:45,0.399658203125)
--(axis cs:42,0.43829345703125)
--(axis cs:40,0.37030029296875)
--(axis cs:38,0.35235595703125)
--(axis cs:35,0.35540771484375)
--(axis cs:34,0.4423828125)
--(axis cs:33,0.39080810546875)
--(axis cs:30,0.3885498046875)
--(axis cs:29,0.35894775390625)
--(axis cs:28,0.415283203125)
--(axis cs:25,0.3272705078125)
--(axis cs:23,0.41925048828125)
--(axis cs:20,0.346923828125)
--(axis cs:19,0.3314208984375)
--(axis cs:15,0.40557861328125)
--(axis cs:14,0.37213134765625)
--(axis cs:10,0.34588623046875)
--(axis cs:5,0.40557861328125)
--(axis cs:0,0.368896484375)
--cycle;

\addplot [line width=0.75pt, color0]
table {%
0 0.368896484375
5 0.40557861328125
10 0.34588623046875
14 0.37213134765625
15 0.40557861328125
19 0.3314208984375
20 0.346923828125
23 0.41925048828125
25 0.3272705078125
28 0.415283203125
29 0.35894775390625
30 0.3885498046875
33 0.39080810546875
34 0.4423828125
35 0.35540771484375
38 0.35235595703125
40 0.37030029296875
42 0.43829345703125
45 0.399658203125
50 0.33447265625
54 0.3465576171875
55 0.37567138671875
58 0.33837890625
60 0.38330078125
65 0.28167724609375
70 0.36907958984375
71 0.33740234375
75 0.3743896484375
77 0.4083251953125
79 0.3843994140625
80 0.35028076171875
83 0.4012451171875
85 0.389404296875
87 0.37957763671875
90 0.36920166015625
92 0.35577392578125
95 0.3658447265625
97 0.3441162109375
99 0.4329833984375
100 0.43817138671875
};
\addlegendentry{Identity loss}
\addplot [line width=0.75pt, color2, mark=*, mark size=2, mark options={solid}]
table {%
0 0.2366943359375
5 0.23126220703125
10 0.25067138671875
15 0.27691650390625
20 0.2816162109375
25 0.28314208984375
30 0.3526611328125
35 0.324951171875
40 0.35394287109375
45 0.3543701171875
50 0.33856201171875
55 0.34600830078125
60 0.3115234375
65 0.34637451171875
70 0.3365478515625
75 0.3494873046875
80 0.3690185546875
85 0.39727783203125
90 0.40948486328125
95 0.36407470703125
100 0.4215087890625
};
\addlegendentry{Validation loss  $\mathcal{L}_\text{USV}$}
\addplot [line width=0.75pt, color1, mark=*, mark size=2, mark options={solid}]
table {%
0 0.26043701171875
1 0.263671875
2 0.27294921875
3 0.28729248046875
4 0.265625
5 0.2998046875
6 0.31072998046875
7 0.32489013671875
8 0.31182861328125
9 0.3233642578125
10 0.32501220703125
11 0.3310546875
12 0.3563232421875
13 0.37384033203125
14 0.358154296875
15 0.34490966796875
16 0.391845703125
17 0.389892578125
18 0.365478515625
19 0.34686279296875
20 0.382568359375
21 0.3853759765625
22 0.39984130859375
23 0.3681640625
24 0.37542724609375
25 0.387939453125
26 0.3673095703125
27 0.3612060546875
28 0.39434814453125
29 0.37200927734375
30 0.39166259765625
31 0.39996337890625
32 0.409912109375
33 0.39459228515625
34 0.43658447265625
35 0.40618896484375
36 0.3804931640625
37 0.427734375
38 0.38873291015625
39 0.37786865234375
40 0.39727783203125
41 0.40777587890625
42 0.426025390625
43 0.41094970703125
44 0.4088134765625
45 0.43798828125
46 0.43463134765625
47 0.45361328125
48 0.44647216796875
49 0.43072509765625
50 0.36181640625
51 0.37078857421875
52 0.39337158203125
53 0.3900146484375
54 0.398681640625
55 0.40289306640625
56 0.4176025390625
57 0.39227294921875
58 0.36962890625
59 0.36883544921875
60 0.35113525390625
61 0.36456298828125
62 0.42242431640625
63 0.42877197265625
64 0.424072265625
65 0.42877197265625
66 0.403564453125
67 0.40374755859375
68 0.44342041015625
69 0.40521240234375
70 0.43701171875
71 0.389892578125
72 0.38568115234375
73 0.41094970703125
74 0.45111083984375
75 0.38604736328125
76 0.4012451171875
77 0.441162109375
78 0.409912109375
79 0.4427490234375
80 0.396728515625
81 0.425537109375
82 0.403076171875
83 0.41064453125
84 0.43414306640625
85 0.44635009765625
86 0.4493408203125
87 0.407958984375
88 0.4014892578125
89 0.40753173828125
90 0.482666015625
91 0.46673583984375
92 0.45135498046875
93 0.43463134765625
94 0.41558837890625
95 0.41912841796875
96 0.47393798828125
97 0.435791015625
98 0.443603515625
99 0.403564453125
100 0.39642333984375
};
\addlegendentry{Training loss $\mathcal{L}_\text{SV}$} 
\end{axis}
\end{tikzpicture}
\caption{\textbf{Training a \DQNNNISQ.}  A \protect\twotwo \DQNNNISQ network has been trained on \textit{ibmq\_casablanca} with  $\epsilon = 0.5$ and $\eta = 1.0$ in $100$ epochs. The training loss is computed using $4$ training pairs (based on a unitary $Y \in \mathcal{U}(4)$). After every fifth epoch, the identity loss using $4$ output training states and the validation loss is additionally measured using $4$ validation data pairs, are analysed as well.}
\label{fig:DQNN_realdevice}
\end{figure}

In \cref{fig:DQNN_realdevice} the training of an exemplary \DQNNNISQ learning an unknown unitary $Y\in \mathcal{U}(4)$ is depicted. Based on the quantum no free lunch theorem, see \cref{chapter:NFL}, four training pairs are used to achieve the depicted training success. Another four data pairs were used to calculate the validation loss. 

Before the training algorithm was executed, the parameters were initialised in the range $\left[0,2\pi\right)$. The algorithm was carried out on the $7$-qubit quantum device \textit{ibmq\_casablanca} hosted by IBM \cite{IBMQuantum2021}. For the implementation and execution of the algorithm, we used the open-source SDK Qiskit \cite{Qiskit}, which allows the decomposition of the above-explained parametrised quantum circuits to a circuit able to execute by the IBM quantum devices. It is noteworthy at this point to underline again that the training took place in a hybrid manner, meaning that the loss functions were evaluated indeed by the quantum implementation however the update of the parameters happened classically. 

It looks like the training and validation loss are correlated to the variation of the identity loss. This strengthens the assumption that the identity loss arises from the noise of the quantum device and is not just based on statistical errors. We can understand from the plot that the training loss exceeds the identity loss after a few training epochs and come to the result that the network is able to generalise the information provided through the training data despite the high noise levels. Because of the small number of parameters, effects of barren plateaus are not encountered.

\section{Comparison to quantum approximate optimisation algorithm}
\label{sec:DQNN_QAOA}

We want to close this chapter with a comparison of the above comprehensively discussed \DQNNNISQ with another QNN architecture, the \emph{quantum approximate optimisation algorithm} (QAOA)\cite{Farhi2014,Farhi2016,Hadfield2019}. For this, we follow the work of \cite{Beer2021a} where not only both QNNs were challenged to learn an unknown unitary, but also the noise tolerance of the two approaches are compared, when implemented on an IBM NISQ device \cite{IBMQuantum2021}. The QAOA leads not only to solutions of combinatorial problems \cite{Farhi2015,Wecker2016,Lin2016,Farhi2017,Otterbach2017,Wang2018,Streif2019,Lechner2020,Maciejewski2021} and is universal for quantum computation \cite{Lloyd2018}, but was also successfully used to learn unknown unitaries \cite{Kiani2020}. Since the latter was the leading learning example in this chapter, a comparison of the \DQNNNISQ and QAOA suggests itself. We begin with a short introduction to the QAOA. A comparison with numerical results ensues.

\FloatBarrier\subsection*{Quantum approximate optimisation algorithm}
The QAOA can be interpreted in various ways. In the following, we discuss the QAOA according to \cite{Farhi2014}. There, this algorithm is described as a QNN and this suits the comparison to the \DQNNNISQ the best. Despite this similar implementation, these two approaches differ in crucial aspect: whereas the \DQNNNISQ works in a dissipative manner and a perceptron acts on layers of different qubits, in the QAOA setting, a perceptron is defined as a sequence of operations, and all perceptrons act on the identical qubits.

\begin{figure} 
\centering
\begin{subfigure}[t]{1\linewidth}
\centering
\begin{tikzpicture}[scale=1.5,decoration={markings,mark=at position 0.8 with {\arrow{>}}}]
\foreach \x in {0,1} {
\draw[line0] (0,\x) -- (2,0);
\draw[line1] (0,\x) -- (2,0);
\draw[line0] (0,\x) -- (2,1);
\draw[line1] (0,\x) -- (2,1);
}
\foreach \x in {0,1} {
\draw[line0] (2,\x) -- (4,0);
\draw[] (2,\x) -- (4,0);
\draw[line0] (2,\x) -- (4,1);
\draw[] (2,\x) -- (4,1);
}
\foreach \x in {0,1} {
\draw[line0] (4,\x) -- (6,0);
\draw[line2] (4,\x) -- (6,0);
\draw[line0] (4,\x) -- (6,1);
\draw[line2] (4,\x) -- (6,1);
}
\node[perceptron0] at (0,0) {$q_2$};
\node[perceptron0] at (0,1) {$q_1$};
\node[perceptron0] at (2,0) {$q_2$};
\node[perceptron0] at (2,1) {$q_1$};
\node[perceptron0] at (4,0) {$q_2$};
\node[perceptron0] at (4,1) {$q_1$};
\node[perceptron0] at (6,0) {$q_2$};
\node[perceptron0] at (6,1) {$q_1$};
\node[operator0, minimum height=1.5cm, fill=white, draw=white] at (3,0.5) {};
\node at (3,0.5) {...  };
\end{tikzpicture}
\subcaption{Network. }
\label{fig:QAOA_A}
\end{subfigure}

\begin{subfigure}[t]{1\linewidth}
\centering
\begin{tikzpicture}[xscale=1.7]
\draw (0,0) -- (6,0);
\draw (0,1) -- (6,1);
\node[operator0, minimum height=2cm, fill=white, draw=white] at (3,0.5) {};
\node at (3,0) {...  };
\node at (3,1) {...  };
\node[operator1, minimum height=1.5cm, minimum width=1.5cm] at (.75,0.5) {$e^{-i A p_1}$};
\node[operator1,minimum height=1.5cm, minimum width=1.5cm] at (2,0.5) {$e^{-i B k_1}$};
\node[operator2, minimum height=1.5cm, minimum width=1.5cm] at (4,0.5) {$e^{-i A p_\tau}$};
\node[operator2,minimum height=1.5cm, minimum width=1.5cm] at (5.25,0.5) {$e^{-i B k_\tau}$};
\draw[brace0,color=black] {(-0.1,-.1) -- node[left=1ex] {$\ket{\phi^{\text{in}}}$}  (-0.1,1.1)};
\draw[brace0,color=black] {(6.1,1.1) -- node[right=1ex] {$\rho^\text{out}$}  (6.1,-.1)};
\end{tikzpicture}
\subcaption{Implementation. }
\label{fig:QAOA_B}
\end{subfigure}
\caption{\textbf{Implementation of the QAOA.} A two-qubit QAOA can be represented as a $(\tau+1)$-layer QNN, where the same qubits are used in every layer (a). The implementation as a quantum circuit consistis of a sequence of alternating unitary operations (b).}
\label{fig:QAOA}
\end{figure}
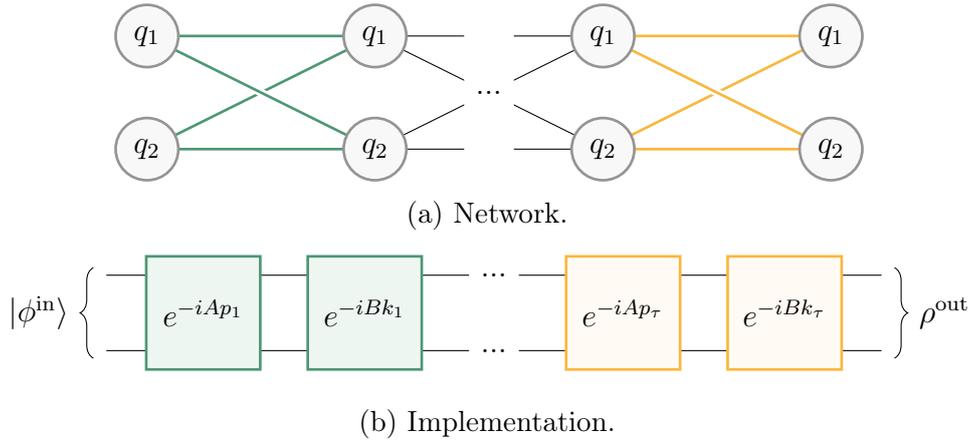

The QAOA is sometimes also called \emph{quantum alternating operator ansatz}, due to the fact that it is working with a sequence of alternating unitary operations $e^{-iAp_l}, e^{-iBk_l} \in \mathcal{U}(d)$, where $p_l,k_l \in \mathbbm{R}$. The matrices $A$ and $B$ are hermitian and initialised using the Gaussian unitary ensemble \cite{Farhi2014}. $\tau$ of these operator pairs can be phrased as
\begin{equation*}
\mathcal{U} = e^{-iBk_\tau} e^{-iAp_\tau} \cdots e^{-iBk_1} e^{-iAp_1}.
\end{equation*}
The output state of the algorithm can be written as $\rho^{out} = \mathcal{U} \ket{\phi^\text{in}}\bra{\phi^\text{in}}\mathcal{U}^\dagger$.

Such an operator sequence can be interpreted as a $\left(\tau+1\right)$-layer QNN, where the number of neurons per layer $m$ is equal in every layer. A layer $l$ is defined as an operator pair $e^{-iAp_l}$, $e^{-iBk_l}$ acting on all $m$ neurons of layer $l-1$. The changed neurons are the inputs for the layer $l+1$. This procedure is depicted for two neurons $q_1$ and $q_2$ in \cref{fig:QAOA_A}.

The implementation of QAOA as a quantum circuit is very simple. The $m$ qubits are initialise in the input state $\ket{\phi^\text{in}}$ with dimension $d = 2^m$. The number of sequences $\tau$ is chosen to be $d^2/2$. In that way the QAOA leads to an optimal solution \cite{Kiani2020} and the number of parameters is $n^\text{QAOA} = d^2=4^m$. See \cref{fig:QAOA_B} for the illustration of the implementation of an exemplary QAOA circuit.

%
\begin{figure}[H]
\centering
\begin{tikzpicture}[]
\matrix[row sep=0.3cm, column sep=0.5cm] (circuit) {
\node(start3){$\ket{0}^{\otimes m}$};  
& \node[halfcross,label={\small m}] (c13){};
& \node[operator0, fill=white, minimum width=1.2cm] (c23){$\ket{\phi^\text{SV}}$};
& \node[]{}; 
& \node[]{}; 
& \node[]{}; 
& \node[]{}; 
& \node[circlewc](circle){}; 
& \node[]{}; 
& \node[meter](end3){};  \\
\node(start1){$\ket{0}^{\otimes m}$};
& \node[halfcross,label={\small m}] (c11){};
& \node[operator0, fill=white, minimum width=1.2cm] (c22){$\ket{\phi^\text{in}}$};
& \node[]{}; 
& \node[]{}; 
& \node[]{}; 
& \node[]{}; 
& \node[dot](dot){}; 
& \node[operator0](c41){$H$};
& \node[meter](end1){}; \\
};
\begin{pgfonlayer}{background}
\draw[] (start1) -- (end1)  
(start3) -- (end3);
\draw[] (dot) -- (circle);
\node[operator2, minimum height=1.5cm] at (0.5,-0.75) {QAOA};
\end{pgfonlayer}
\end{tikzpicture}
\caption{\textbf{Implementation of training the QAOA}. After the qubits are initialised the QAOA quantum circuit can be executed. The $\swap$-test allows calculating the values of the loss functions. The number of used qubits is $2m$. }
\label{fig:QAOA_fullCircuit}
\end{figure}
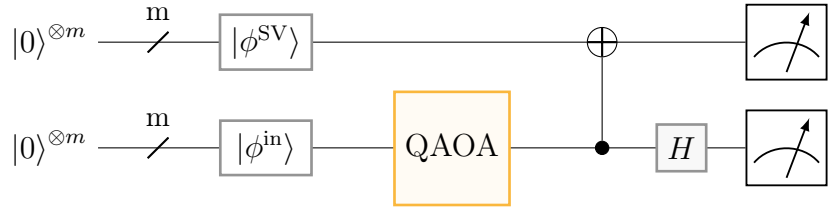
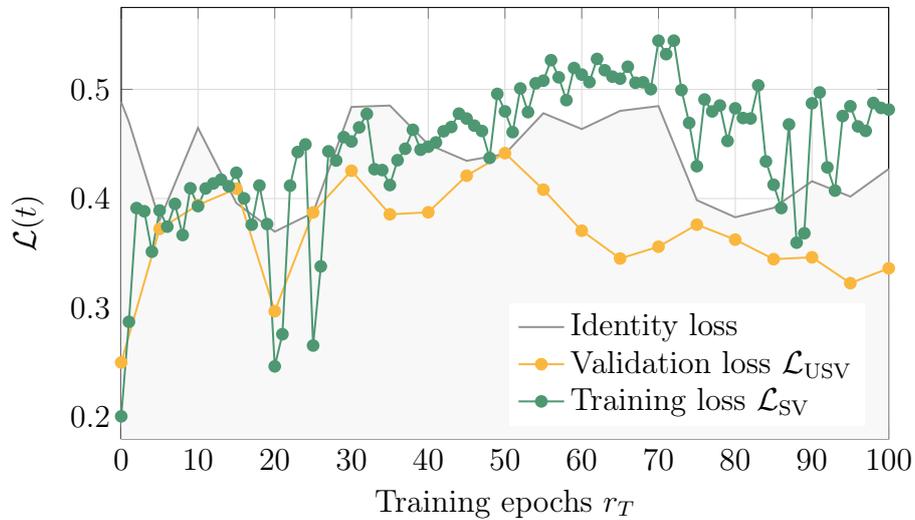
\begin{figure}
\centering
\begin{tikzpicture}
\begin{axis}[
xmin=-0, xmax=100,
ymin=0.18, ymax=0.575,
width=.8\linewidth, 
height=.5\linewidth,
grid=major,
grid style={color0M},
xlabel=Training epochs $r_T$, 
ylabel=$\mathcal{L}(t)$,legend pos=south east,legend cell align={left},legend style={draw=none}]
\path [fill=color0L, line width=1pt]
(axis cs:0,0.48834228515625)
--(axis cs:0,0)
--(axis cs:1,0)
--(axis cs:2,0)
--(axis cs:3,0)
--(axis cs:4,0)
--(axis cs:5,0)
--(axis cs:6,0)
--(axis cs:7,0)
--(axis cs:8,0)
--(axis cs:9,0)
--(axis cs:10,0)
--(axis cs:11,0)
--(axis cs:12,0)
--(axis cs:13,0)
--(axis cs:14,0)
--(axis cs:15,0)
--(axis cs:16,0)
--(axis cs:17,0)
--(axis cs:18,0)
--(axis cs:19,0)
--(axis cs:20,0)
--(axis cs:21,0)
--(axis cs:22,0)
--(axis cs:23,0)
--(axis cs:24,0)
--(axis cs:25,0)
--(axis cs:26,0)
--(axis cs:27,0)
--(axis cs:28,0)
--(axis cs:29,0)
--(axis cs:30,0)
--(axis cs:31,0)
--(axis cs:32,0)
--(axis cs:33,0)
--(axis cs:34,0)
--(axis cs:35,0)
--(axis cs:36,0)
--(axis cs:37,0)
--(axis cs:38,0)
--(axis cs:39,0)
--(axis cs:40,0)
--(axis cs:41,0)
--(axis cs:42,0)
--(axis cs:43,0)
--(axis cs:44,0)
--(axis cs:45,0)
--(axis cs:46,0)
--(axis cs:47,0)
--(axis cs:48,0)
--(axis cs:49,0)
--(axis cs:50,0)
--(axis cs:51,0)
--(axis cs:52,0)
--(axis cs:53,0)
--(axis cs:54,0)
--(axis cs:55,0)
--(axis cs:56,0)
--(axis cs:57,0)
--(axis cs:58,0)
--(axis cs:59,0)
--(axis cs:60,0)
--(axis cs:61,0)
--(axis cs:62,0)
--(axis cs:63,0)
--(axis cs:64,0)
--(axis cs:65,0)
--(axis cs:66,0)
--(axis cs:67,0)
--(axis cs:68,0)
--(axis cs:69,0)
--(axis cs:70,0)
--(axis cs:71,0)
--(axis cs:72,0)
--(axis cs:73,0)
--(axis cs:74,0)
--(axis cs:75,0)
--(axis cs:76,0)
--(axis cs:77,0)
--(axis cs:78,0)
--(axis cs:79,0)
--(axis cs:80,0)
--(axis cs:81,0)
--(axis cs:82,0)
--(axis cs:83,0)
--(axis cs:84,0)
--(axis cs:85,0)
--(axis cs:86,0)
--(axis cs:87,0)
--(axis cs:88,0)
--(axis cs:89,0)
--(axis cs:90,0)
--(axis cs:91,0)
--(axis cs:92,0)
--(axis cs:93,0)
--(axis cs:94,0)
--(axis cs:95,0)
--(axis cs:96,0)
--(axis cs:97,0)
--(axis cs:98,0)
--(axis cs:99,0)
--(axis cs:100,0)
--(axis cs:100,0.4271240234375)
--(axis cs:100,0.4271240234375)
--(axis cs:95,0.40179443359375)
--(axis cs:90,0.41583251953125)
--(axis cs:85,0.3914794921875)
--(axis cs:80,0.38287353515625)
--(axis cs:75,0.39849853515625)
--(axis cs:70,0.48468017578125)
--(axis cs:65,0.48028564453125)
--(axis cs:60,0.463623046875)
--(axis cs:55,0.4781494140625)
--(axis cs:50,0.4407958984375)
--(axis cs:45,0.4345703125)
--(axis cs:40,0.4495849609375)
--(axis cs:35,0.48516845703125)
--(axis cs:30,0.48388671875)
--(axis cs:25,0.386962890625)
--(axis cs:20,0.36968994140625)
--(axis cs:15,0.3958740234375)
--(axis cs:10,0.46466064453125)
--(axis cs:5,0.38006591796875)
--(axis cs:0,0.48834228515625)
--cycle;

\addplot [line width=0.75pt, color0]
table {%
0 0.48834228515625
1 0.47027587890625
5 0.38006591796875
10 0.46466064453125
15 0.3958740234375
20 0.36968994140625
25 0.386962890625
30 0.48388671875
35 0.48516845703125
40 0.4495849609375
45 0.4345703125
50 0.4407958984375
55 0.4781494140625
60 0.463623046875
65 0.48028564453125
70 0.48468017578125
75 0.39849853515625
80 0.38287353515625
85 0.3914794921875
90 0.41583251953125
95 0.40179443359375
100 0.4271240234375
};
\addlegendentry{Identity loss}
\addplot [line width=0.75pt, color2, mark=*, mark size=2, mark options={solid}]
table {%
0 0.2498779296875
5 0.37225341796875
10 0.39404296875
15 0.4090576171875
20 0.2967529296875
25 0.38726806640625
30 0.42547607421875
35 0.38555908203125
40 0.38739013671875
45 0.4208984375
50 0.44158935546875
55 0.40814208984375
60 0.37054443359375
65 0.3450927734375
70 0.35589599609375
75 0.3760986328125
80 0.36236572265625
85 0.344482421875
90 0.3461914062500000000
95 0.322509765625
100 0.3361206054687500000
};
\addlegendentry{Validation loss  $\mathcal{L}_\text{USV}$}
\addplot [line width=0.75pt, color1, mark=*, mark size=2, mark options={solid}]
table {%
0 0.2005615234375
1 0.287109375
2 0.39129638671875
3 0.38836669921875
4 0.351318359375
5 0.38909912109375
6 0.374267578125
7 0.3951416015625
8 0.3665771484375
9 0.409423828125
10 0.39306640625
11 0.40924072265625
12 0.413818359375
13 0.4173583984375
14 0.4112548828125
15 0.42364501953125
16 0.400146484375
17 0.37603759765625
18 0.41192626953125
19 0.37664794921875
20 0.246337890625
21 0.2757568359375
22 0.41180419921875
23 0.442626953125
24 0.4495849609375
25 0.265380859375
26 0.337890625
27 0.4432373046875
28 0.434814453125
29 0.45623779296875
30 0.4522705078125
31 0.46514892578125
32 0.4775390625
33 0.4268798828125
34 0.42608642578125
35 0.412353515625
36 0.43511962890625
37 0.44561767578125
38 0.46295166015625
39 0.4447021484375
40 0.44744873046875
41 0.451416015625
42 0.46160888671875
43 0.46551513671875
44 0.477783203125
45 0.4732666015625
46 0.466796875
47 0.46185302734375
48 0.43719482421875
49 0.495849609375
50 0.4798583984375
51 0.4608154296875
52 0.5008544921875
53 0.4791259765625
54 0.50555419921875
55 0.5079345703125
56 0.5267333984375
57 0.51104736328125
58 0.4901123046875
59 0.51953125
60 0.5135498046875
61 0.50665283203125
62 0.52789306640625
63 0.51751708984375
64 0.51165771484375
65 0.509765625
66 0.520751953125
67 0.50604248046875
68 0.5064697265625
69 0.5001220703125
70 0.5445556640625
71 0.5322265625
72 0.54461669921875
73 0.4993896484375
74 0.46923828125
75 0.42962646484375
76 0.49072265625
77 0.47979736328125
78 0.4852294921875
79 0.4527587890625
80 0.48260498046875
81 0.4737548828125
82 0.47332763671875
83 0.503662109375
84 0.43402099609375
85 0.41265869140625
86 0.39129638671875
87 0.4678955078125
88 0.359619140625
89 0.36822509765625
90 0.48724365234375
91 0.4971923828125
92 0.42852783203125
93 0.40728759765625
94 0.4755859375
95 0.48455810546875
96 0.466064453125
97 0.46197509765625
98 0.487548828125
99 0.48297119140625
100 0.48150634765625
};
\addlegendentry{Training loss $\mathcal{L}_\text{SV}$} 
\end{axis}
\end{tikzpicture}
\caption{\textbf{Training a QAOA.}  The plot shows the training of an $m=2$ QAOA on \textit{ibmq\_casablanca} with  $\epsilon = 0.15$ and $\eta = 1.0$ in $100$ epochs. The training loss is computed using $4$ training pairs. After every fifth epoch, the identity loss using $4$ output training states and the validation loss is additionally measured using $4$ validation data pairs, are analysed as well.}
\label{fig:QAOA_realdevice}
\end{figure}

\begin{sloppypar}
The QAOA circuit can be trained equivalently to the training of the \DQNNNISQ circuit including the following steps: The training and validation data set consisting of pairs of states $\{\ket{\phi^\text{in}_x},\ket{\phi^\text{SV}_x} \}^{N}_{x=1}$ are prepared, where $\ket{\phi^\text{SV}_x} = Y \ket{\phi^\text{in}_x}$. The circuit parameters are initialised randomly. The input states are carried through the quantum circuit and the resulting output states and the supervised states are used to calculate the validation with the  $\swap$-test. Gradient descent is performed to update the parameters. Despite the equivalence to the \DQNNNISQ training, the training algorithm of a QAOA is depicted as a circuit in \cref{fig:QAOA_fullCircuit}.
\end{sloppypar}

In \cref{fig:QAOA_realdevice} the training of an $m=2$ QAOA is depicted while learning an unknown unitary $Y\in \mathcal{U}(4)$. Four training pairs are used to achieve the depicted training success. Another four data pairs were used to calculate the validation loss. The identity loss is defined as explained in \cref{sec:DQNN_quantumalg}. In analogy to \cref{fig:DQNN_realdevice} the QAOA was carried out on the $7$-qubit quantum device \textit{ibmq\_casablanca} by IBM \cite{IBMQuantum2021}.

In comparison to \cref{fig:DQNN_realdevice}, where the same device was used to train the \DQNNNISQ, the validation loss does not exceed the identity loss in the process of training the QAOA. However, it seems to be correlated to the identity loss. Since comparing two results of single training sessions is not very fair, we cite in the following the results on generalisation and gate noise analysis of \cite{Beer2021a}. 

\FloatBarrier \subsection*{Comparison of \DQNNNISQ and QAOA}
In the following, we compare both networks, the \DQNNNISQ and the QAOA, with  $m=2$ input and output qubits. The \DQNNNISQ algorithm comes in the form of four qubits, two of them in both of the layers. The QAOA includes only two qubits, and we choose $\tau=8$. To calculate the loss functions, two more qubits for training each of the QNNs are needed. 

For the training $S=4$ pairs of training data, $N-S=4$ pairs of validation data are used. Additional four state pairs are used to evaluate the identity loss.  

A difference is, that the $n^\text{\DQNNNISQ}=24$ parameters of the \DQNNNISQ are initialised in the range $\left[0,2\pi\right)$, where the $n^\text{QAOA}=16$ parameters needed for the QAOA are initialised in the range $\left[-1,1\right]$. Further, different parameters $\eta$ and $\epsilon$ has been found as optimal (see the captions of \cref{fig:QAOA_generalisation} and \cref{fig:QAOA_errors} for examples). 

As already discussed in \cref{sec:DQNN_lossfunctions} we are interested in the networks generalisation capabilities and use the validation loss $\mathcal{L}_\text{USV}$ for studying it. Both network's quantum circuits were executed on a simulator of \textit{ibmq\_casablanca} imitating the real-time noise of this NISQ device. After repeating the same training with the same circumstances and the same number of supervised pairs $S$ but different initial parameters, we get an average of the testing loss for each $S$. The results of this numerical experiment can be found in \cref{fig:QAOA_generalisation}.  

First of all, we can say that both, the \DQNNNISQ and the QAOA, can generalise the information given through the training data pairs. As expected, the validation loss gets larger with the increasing number of training pairs. It also becomes clear that the \DQNNNISQ reaches higher values which have to be observed together with the higher identity loss. Since the training results seem to be strongly affected by the noise of the executing quantum device, a noise analysis of both of the QNN attempts will be discussed in the succeeding paragraph.

Two primary types of noise are the readout noise, occurring during the measurements, and the gate noise \cite{Nachman2020}. Since both networks are trained, read out and updated with the same methods but consist of different gates we focus only on the gate noise in the following. 

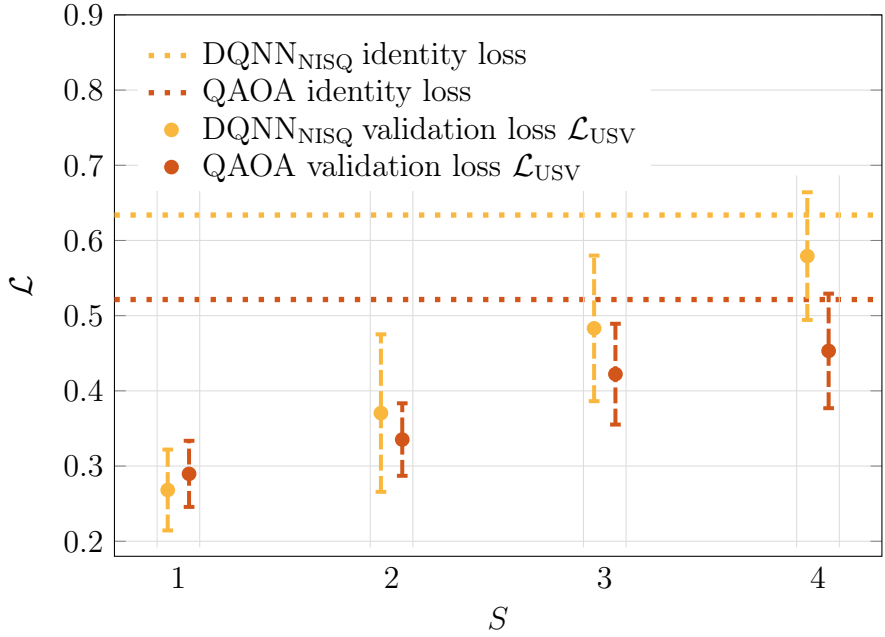
\begin{figure}
\centering
\begin{tikzpicture}
\begin{axis}[
xmin=0.7, xmax=4.3,
xtick={1,2,3,4},
x grid style={white},
xtick style={color=white},
ymin=0.18, ymax=0.9,
width=.8\linewidth, 
height=.6\linewidth,
grid=major,
grid style={color0M},
xlabel=$S$, 
ylabel=$\mathcal{L}$,legend pos=north west,legend cell align={left},legend style={draw=none}]
\addplot [line width=2pt, color2,loosely dotted]
table {%
0.7 0.633802290209349
4.3 0.633802290209349
};
\addlegendentry{\DQNNNISQ identity loss} 
\addplot [line width=2pt, color3, loosely dotted]
table {%
0.7 0.521471089131208
4.3 0.521471089131208
};
\addlegendentry{QAOA identity loss} 
\path [draw=color2, line width=1.5pt, dash pattern=on 6pt off 2pt]
(axis cs:0.95,0.214371036562412)
--(axis cs:0.95,0.321928096738369);

\path [draw=color2, line width=1.5pt, dash pattern=on 6pt off 2pt]
(axis cs:1.95,0.265711397990693)
--(axis cs:1.95,0.475150113239776);

\path [draw=color2, line width=1.5pt, dash pattern=on 6pt off 2pt]
(axis cs:2.95,0.386311747777952)
--(axis cs:2.95,0.579900715600954);

\path [draw=color2, line width=1.5pt, dash pattern=on 6pt off 2pt]
(axis cs:3.95,0.494361005837437)
--(axis cs:3.95,0.664096940939907);

\path [draw=color3, line width=1.5pt, dash pattern=on 6pt off 2pt]
(axis cs:1.05,0.245657753823828)
--(axis cs:1.05,0.333608603598047);

\path [draw=color3, line width=1.5pt, dash pattern=on 6pt off 2pt]
(axis cs:2.05,0.287018815193189)
--(axis cs:2.05,0.383458479728686);

\path [draw=color3, line width=1.5pt, dash pattern=on 6pt off 2pt]
(axis cs:3.05,0.355178206324434)
--(axis cs:3.05,0.489235550999785);

\pgfplotsinvokeforeach {0.9,1.1,1.9,2.1,2.9,3.1,3.9,4.1}{
\path [draw=color0M] (axis cs:#1,0.191884741343538) --(axis cs:#1,0.686583236158782);}

\path [draw=color3, line width=1.5pt, dash pattern=on 6pt off 2pt]
(axis cs:4.05,0.376951544422389)
--(axis cs:4.05,0.529295403819798);

\addplot [line width=1.5pt, color2, mark=*, mark size=2, mark options={solid}, only marks]
table {%
0.95 0.268149566650391
1.95 0.370430755615234
2.95 0.483106231689453
3.95 0.579228973388672
};
\addlegendentry{\DQNNNISQ validation loss $\mathcal{L}_\text{USV}$} 
\addplot [line width=1.5pt, color3, mark=*, mark size=2, mark options={solid}, only marks]
table {%
1.05 0.289633178710938
2.05 0.335238647460938
3.05 0.422206878662109
4.05 0.453123474121094
};
\addlegendentry{QAOA validation loss $\mathcal{L}_\text{USV}$} 

\addplot [line width=1.5pt, color2, mark=-, mark size=2, mark options={solid}, only marks]
table {%
0.95 0.214371036562412
1.95 0.265711397990693
2.95 0.386311747777952
3.95 0.494361005837437
};

\addplot [line width=1.5pt, color2, mark=-, mark size=2, mark options={solid}, only marks]
table {%
0.95 0.321928096738369
1.95 0.475150113239776
2.95 0.579900715600954
3.95 0.664096940939907
};

\addplot [line width=1.5pt, color3, mark=-, mark size=2, mark options={solid}, only marks]
table {%
1.05 0.245657753823828
2.05 0.287018815193189
3.05 0.355178206324434
4.05 0.376951544422389
};
\addplot [line width=1.5pt, color3, mark=-, mark size=2, mark options={solid}, only marks]
table {%
1.05 0.333608603598047
2.05  0.383458479728686
3.05  0.489235550999785
4.05  0.529295403819798
};
\end{axis}
\end{tikzpicture}
\caption{\textbf{Generalisation analysis.} This figure compares the generalisation behaviours of a \protect\twotwo \DQNNNISQ ($\epsilon = 0.5$,  $\eta = 1.0$) and a $m=2$ QAOA ($\epsilon = 0.15$,  $\eta = 0.1$), each trained with $S$ training pairs for $r_T=1000$ epochs on a \textit{ibmq\_casablanca} simulator.}
\label{fig:QAOA_generalisation}
\end{figure}
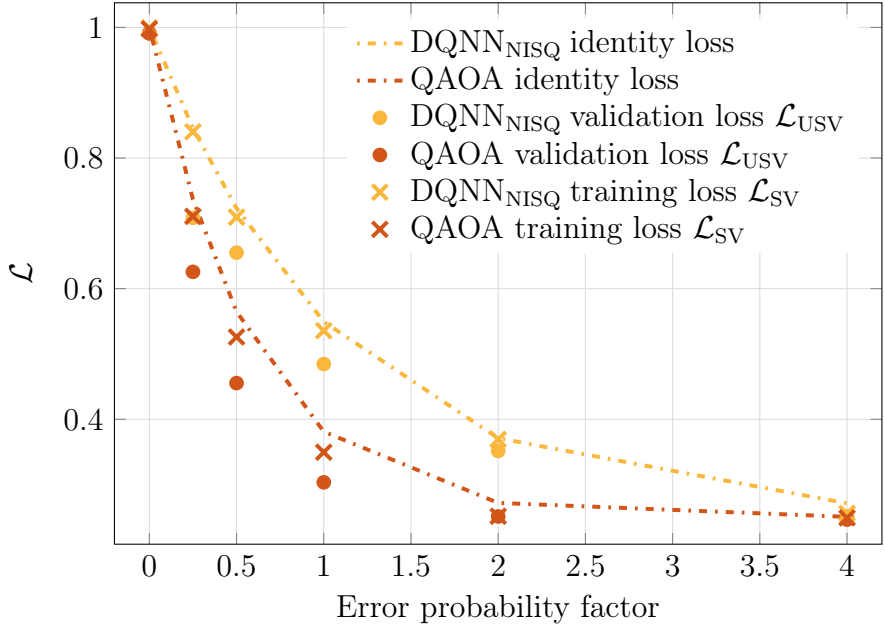
\begin{figure}
\centering
\begin{tikzpicture}
\begin{axis}[
xmin=-0.2, xmax=4.2,
ymin=0.208798980712891, ymax=1.03767623901367,
width=.8\linewidth, 
height=.6\linewidth,
grid=major,
grid style={color0M},
xlabel={Error probability factor},
ylabel=$\mathcal{L}$,legend pos=north east,legend cell align={left},legend style={draw=none}]
\addplot [line width=1.5pt, color2, dash pattern=on 1pt off 3pt on 3pt off 3pt]
table {%
0 1
0.25 0.845605927480682
0.5 0.723037063256346
1 0.549311368717472
2 0.371134262544757
4 0.271346265007468
};
\addlegendentry{\DQNNNISQ identity loss} 
\addplot [line width=1.5pt, color3, dash pattern=on 1pt off 3pt on 3pt off 3pt]
table {%
0 1
0.25 0.736971044275058
0.5 0.564618045909574
1 0.381029807946013
2 0.272007748162076
4 0.250687708075817
};
\addlegendentry{QAOA identity loss} 
\addplot [line width=1.5pt, color2, mark=*, mark size=2, mark options={solid}, only marks]
table {%
0 0.991644287109375
0.25 0.708514404296875
0.5 0.655133056640625
1 0.48455810546875
2 0.351556396484375
4 0.2514892578125
};
\addlegendentry{\DQNNNISQ validation loss $\mathcal{L}_\text{USV}$} 
\addplot [line width=1.5pt, color3, mark=*, mark size=2, mark options={solid}, only marks]
table {%
0 0.990715026855469
0.25 0.625755310058594
0.5 0.455421447753906
1 0.303504943847656
2 0.251235961914062
4 0.246475219726562
};
\addlegendentry{QAOA validation loss $\mathcal{L}_\text{USV}$} 
\addplot [line width=1.5pt, color2, mark=x, mark size=4, mark options={solid}, only marks]
table {%
0 0.999339599609375
0.25 0.84021728515625
0.5 0.709481201171875
1 0.535692138671875
2 0.369239501953125
4 0.255997314453125
};
\addlegendentry{\DQNNNISQ training loss $\mathcal{L}_\text{SV}$} 
\addplot [line width=1.5pt, color3, mark=x, mark size=4,mark options={solid}, only marks]
table {%
0 0.997624206542969
0.25 0.710673522949219
0.5 0.525996398925781
1 0.349626159667969
2 0.251597595214844
4 0.249176025390625
};
\addlegendentry{QAOA training loss $\mathcal{L}_\text{SV}$} 
\end{axis}
\end{tikzpicture}
\caption{\textbf{Gate noise analysis.} Here, the robustness to gate noise of a \protect\twotwo \DQNNNISQ ($\epsilon = 0.25$,  $\eta = 0.5$) and a $m=2$ QAOA ($\epsilon = 0.05$,  $\eta = 0.05$) is compared. Each are trained with $S$ training pairs for $r_T=1000$ epochs on a \textit{ibmq\_16\_melbourne} simulator.}
\label{fig:QAOA_errors}
\end{figure}

Using a depolarising quantum error channel \cite{Nielsen2000} we are able to test the influence of gate noise on the QNNs. In the numerical experiments, the channel was parametrised by the depolarisation probabilities
$\lambda^\text{g}=k \lambda^\text{g}_0$ for the basis gates $g=\text{CNOT, SX, RZ}$ and scaling factor $k$. To simulate the gate errors of a special NISQ devise, the parameter $\lambda^\text{g}_0$ has to be chosen appropriately, for example $\lambda^\text{CNOT}_0=3.14\times10^{-2}$, $\lambda^\text{SX}_0=1.18\times10^{-3}$ and $\lambda^\text{RZ}_0=0$ for the approximation of \textit{ibmq\_16\_melbourne} \cite{Magesan2012}.

\begin{sloppypar}
In \cref{fig:QAOA_errors} the values of identity, validation and training losses of the \DQNNNISQ and QAOA for different error probability factors $k$ are shown. To value the results, the reader must consider that the noise factor $k=1$ corresponds to the gate noise of currently available NISQ devices. The results are gained using the of the qubit coupling map IBM device \textit{ibmq\_16\_melbourne}.
\end{sloppypar}

Both attempts give excellent results, i.e.\ all losses equal one, for the noise-free case $k=0$. When increasing the gate noise, the \DQNNNISQ leads to a higher identity loss compared to the QAOA. This concedes in a higher training and validation loss.

To conclude, we can say that both QNNs succeed in learning an unknown unitary operation. Albeit it also becomes clear that the \DQNNNISQ algorithm attends this learning task more reliably than the QAOA and is less susceptible to gate noise. The results indicate that the \DQNNNISQ is more suitable for learning an unknown unitary operation on a NISQ device of the current stage compared to the QAOA. Nevertheless, it needs to be said that the noise still forbids reaching high values in the loss functions in both algorithms. 

In the preceding sections, we studied the behaviour of the DQNN algorithm characterising a unitary operation, using numerical results gained with classical simulation or the execution on a NISQ device, where we could see that the DQNNs can be successfully trained and perform well on unseen data. In contrast, the following chapter discusses the theoretical boundaries of such a learning process and allows observing the performance of the DQNN from a new perspective.

%% file: text/NFL.tex
\chapter{No free lunch theorem}\hypertarget{NFL}{}
\label{chapter:NFL}


Due to the continuous rapid progress in the field of quantum learning theory, it is also significant to understand the ultimate limits for quantum learning devices and methods. The preceding chapter introduced a \emph{quantum neural network} (QNN) structure that can successfully learn an unknown unitary $Y$ even with limited data. Before we continue in \cref{chapter:graphs} and \cref{chapter:QGAN} with discussing some further opportunities these QNNs offer, we want to take a step back and look at the boundaries such a learning process has imposed by the \emph{quantum no free lunch} (QNFL) theorem\cite{Poland2020}. 

Therefore we assume the studied device can be modelled as a unitary process and is trained with quantum examples. We will describe in the following how to find an optimal lower bound on the probability that such a QNN gives an incorrect output for a random input. This bound equips us with a helpful metric to check how well particular \emph{quantum machine learning} (QML) architectures and algorithms perform. 

The here discussed result can be classed within the research on fundamental information-theoretic limits on quantum learning  \cite{Sasaki2001, Sasaki2002, Gammelmark2009, Sentis2012, Arunachalam2017, Monras2017} and is related to the work on the optimal quantum learning of unitary operations considering storage and later retrieval of unknown quantum processes. The version of the QNFL theorem we discuss here was stated in \cite{Poland2020} and assumes the usages of training data consisting of quantum states for the goal of learning an unknown unitary process. This work was later generalised by \cite{Sharma2020a} to the case where these quantum states can be entangled to a reference system. 

Since the here presented bound is based on the quantum version of the \emph{no free lunch} (NFL) theorem \cite{Wolpert1997}, a result of classical learning theory, we begin with a brief explanation of the latter in \cref{sec:NFL_classical}. In \cref{sec:NFL_quantum}, we derive the QNFL theorem following \cite{Poland2020}. During the derivation, some Haar measure integral identities for the unitary group are used. We present the proofs of these collectively after the derivation for a neat overview. The derived bound is then used to check the in \cref{chapter:DQNN} presented \emph{dissipative quantum neural network} (DQNN) algorithm. We end the chapter with a short but essential note on orthonormal training pairs.

\section{Classical no free lunch theorem}
\label{sec:NFL_classical}

The classical NFL theorem states that an optimisation algorithm that performs better for one class of problems must perform worse for another class. We simply get no class "for free". Despite different mathematical formulations of this theorem existing, we follow the notation of \cite{Wolf2018} hereafter.

After introducing the principle of supervised ML in \cref{chapter:ML}, the following setting should be very familiar to the reader: let $X$ and $Y$ be two finite sets. We name $X$ the \emph{input} and $Y$ the \emph{output set}. For an unknown function $f:X\rightarrow Y$ we call the function $h:X\rightarrow Y$ a \emph{hypothesis}. This hypothesis is based on a subset $\mathcal{S}\subset X\times Y$ containing training pairs $\mathcal{S} = \{(x_x,f(x_x))\,|\, x_x\in X, f(x_x)\in Y, j = 1,2, \ldots S\}$, where $f(x_x)$ is the desired output given input $x_x$. It should yield $h_S(x_x) = f(x_x)$ for all $j = 1,2, \ldots S\}$. Note that we assume $S < |X|$, otherwise there would be nothing to predict. 

The big question leading to the NFL theorem is: how well does a given hypothesis perform? To determine this we define the \emph{risk} $R_f(h)$ as the probability that $h_S$ gives the wrong answer, namely
\begin{equation*}
\label{eq:classicalRisk}
R_f(h) \equiv \mathbbm{P}[h_S(x)\not=f(x)].
\end{equation*}
When averaging over all possible training sets with $S$ elements ($\mathbb{E}_\mathcal{S}$ ) and over all possible functions from $X$ to $Y$ ($\mathbb{E}_f$), the NFL theorem takes the shape
\begin{equation}
\label{eq:classNFL}
\mathcolorbox{\mathbb{E}_f\left[\mathbb{E}_\mathcal{S}\left[R_f(h_S)\right]\right]\ge\left(\frac{|X|-S}{|X|}\right)\left(\frac{|Y|-1}{|Y|}\right)= \left(1-\frac{S}{|X|}\right)\left(1-\frac{1}{|Y|}\right),}
\end{equation}
where $h_S$ denotes an optimal hypothesis given the training data $\mathcal{S}$. It follows naturally that if a learning algorithm performs better at predicting $f$ for one part of the problem, there are other parts where the algorithm will perform worse. 

Inspecting \cref{eq:classNFL} in detail, we can understand the classical NFL theorem in the following way: if $h_S$ is perfectly trained for $S$ data pairs, getting the wrong answer can be explained in two steps. First of all, $x\not\in \mathcal{S}$, otherwise the optimal hypotheses would give us the correct answer. This circumstance is reflected in the first factor. The second factor describes the probability that the hypothesis guesses any point $y \in Y$ but the right one. 

\FloatBarrier\subsection*{No free lunch theorem for invertible functions}

When comparing the classical and the, in the next section introduced, QNFL theorem, we should bear in mind a crucial point: a classical function $f$ does not have to be invertible. If we have $f$ determined on some subset of the inputs, we have no information about the action of $f$ on the complement of this subset. In contrast, unitary quantum operations are always invertible. Thus if we already have the operation determined on one subspace, we know that the operation takes the complementary subspace to the complementary subspace of the output. 

Therefore in \cite{Poland2020} it is argued toward comparing the QNFL theorem with a \emph{classical NFL theorem for invertible functions}. This bound is derived as  
\begin{equation*}
\mathcolorbox{\mathbb{E}_f\left[\mathbb{E}_\mathcal{S}\left[R_f(h_S)\right]\right]\ge \left(\frac{|X|-S}{|X|}\right)\left(\frac{|X|-S-1}{|X|}\right)= 1 - \frac{S+1}{|X|}.}
\end{equation*}

The above-gained intuition of \cref{eq:classNFL} helps us to understand this new bound quickly: if our perfectly trained hypothesis is invertible, it is in particular injective assuming $|X|=|Y|$. It follows that we have fewer chances of being wrong when guessing the correct point in $Y$ because $S$ of these training points are already omitted: we get the classical NFL theorem for invertible functions if we substitute the cardinality of Y by $|X|-S$ in the numerator of the second factor, where $S$ is the number of data pairs used for training the hypothesis.

\section{Quantum no free lunch theorem}
\label{sec:NFL_quantum}
After gaining a good intuition of the classical NFL theorem, we will derive the quantum analogue in the coming paragraphs following the work of \cite{Poland2020}. 

For this purpose we replace the sets $X$ and $Y$ mentioned in the last section with an \emph{input Hilbert space} $\mathcal{H}_{\text{in}}$ and \emph{output Hilbert space} $\mathcal{H}_{\text{out}}$. To replace the role of the expressions $|X|$ and $|Y|$  we use  $d=\dim(\mathcal{H}_{\text{in}})$ and $d'=\dim(\mathcal{H}_{\text{out}})$. 

Moreover, the aim is not to define an unexplored function $f$ but to determine an unknown unitary operation $Y$. As we are familiar with from \cref{chapter:DQNN} the training data is of the form $\{\ket{\phi^{\text{in}}_x}, \ket{\phi^{\text{SV}}_x}\} $, $x=1,2,\dots, N$, where $\ket{\phi^{\text{in}}_x}\in\mathcal{H}_{\text{in}}$ and  $\ket{\phi^\text{SV}_x} = Y\ket{\phi^\text{in}_x}\in\mathcal{H}_{\text{out}}$ and we formulate the hypothesis as a unitary $U$. We assume that the unitary $U$ can exactly reproduce the action of the unknown unitary $Y$  on the training pairs of the set $\mathcal{S}$, namely 
\begin{equation*}
U\ket{\phi^{\text{in}}_x} = \ket{\phi^{\text{SV}}_x} = Y\ket{\phi^{\text{in}}_x}, \quad x=1,2,\ldots,N.
\end{equation*}
The most interesting part of generalising the classical NFL theorem is to find a good \emph{quantum risk} analogously to the classical risk in \cref{eq:classicalRisk}. Here we choose the square of the trace norm distance $\|A\|_1 \equiv \tfrac12\tr|A|$  between the outputs of $Y$ and $U$ applied to the same input, averaged over all pure states. A discussion of the risk in the quantum setting can be found in \cite{Monras2017}, details concerning the trace norm in \cite{Nielsen2000}. Hence we formulate the risk as
\begin{align}
\label{eq:NFL_quantumRisk}
R_{Y}(U) &\equiv \int d\ket{\psi} \,\| Y\ket{\psi}\bra{\psi} Y^\dag - U\ket{\psi}\bra{\psi}U^\dag \|_1^2 \nonumber\\
=& 1-\int d\ket{\psi} \, |\bra{\psi}Y^\dag U\ket{\psi}|^2\nonumber\\
=& 1-\frac{1}{d(d+1)}\left(d+|\tr(Y^\dag U)|^2\right)\\
=& \frac{d}{d+1} - \frac{1}{d(d+1)} |\tr(Y^\dag U)|^2\nonumber ,
\end{align}
where the integral runs over pure states induced by a Haar-measure-distributed unitary $W$ applied to an arbitrary state $d\ket{\psi} \equiv dW\ket{0}$ \cite{Duistermaat2012,Hayden2006}. A short introduction to the Haar-measure can be found \cref{subsec:QI_random}. We explicitly evaluate the integral $	S_4 \equiv \int dY \, Y^\dag\otimes Y^\dag\otimes Y\otimes Y $ and thus the identity  used above $\int d\ket{\psi} \, |\bra{\psi}Y^\dag U\ket{\psi}|^2=-\frac{1}{d(d+1)}\left(d+|\tr(Y^\dag U)|^2\right)$ in \cref{prop:NFL_haarMeasureS4}. In the same way as in the classical case the risk defines the probability that the hypothesis fails to reproduce the action of $Y$ or in other words the probability of incorrectly learning a unitary process.

The next step is to average the risk $R_{Y}(U)$ over all possible training sets $\mathcal{S}$ and unitaries $Y$. The first average is trivial. When averaging over all unitaries $Y$, we need to have in mind that $U$ is the best guess for our unknown unitary operation $Y$ given the training set $S$ and therefore $U$ depends on $Y$ in a not obvious way. Consequently we formulate
\begin{equation*}
\int dY\, R_{Y}(U) = \frac{d}{d+1} - \frac{1}{d(d+1)} \int dY |\tr(Y^\dag U)|^2.
\end{equation*}
To evaluate the integral, we use the fact that $U$ and $Y$ act on the training set $\mathcal{S}$ in the same way. Due to linearity we know further that the unitaries $Y$ and $U$ are equal on the subspace $\mathcal{H}_{S} \equiv \text{span}(\mathcal{S})$. 

Let $\mathcal{H}^{\perp}_{S}$ be the subspace complementary to $\mathcal{H}_S$. Although we have no insight into the action of $Y$ on this space, we can use the properties $Y$ has as a unitary. To understand these constraints, we take a closer look at the unitary $Y^\dag U$. Due to the properties of the training set and the direct sum decomposition $\mathcal{H}_{\text{in}} = \mathcal{H}_S\oplus \mathcal{H}_{S}^{\perp}$ we can write $Y^\dag U$ in the following block decomposition:
\begin{equation*}
Y^{\dagger} U =	\left( \begin{array}{c|c}\mathbbm{1}_n 	& A\\
\hline 	B& W\end{array} \right) 
=\left( \begin{array}{c|c}\mathbbm{1}_n & \mathbf{0}\\
\hline \mathbf{0} & W \end{array} \right) 
= \mathbbm{1}_n \oplus W.
\end{equation*}
$\mathbbm{1}_n$ denotes the $S$-dimensional identity on the subspace $\mathcal{H}_S$, and $A$, $B$, and $W$ stand for $S\times (d-S)$, $(d-S)\times S$, and $(d-S)\times (d-S)$ block matrices. For the simple reason that $Y^\dag U$ is a unitary, the norm of each row and column vector has to be equal to $1$. It follows that $A=B=0$ and $W$ is unitary. 
Because of this form, the trace of $Y^\dag U$ can be written as a sum of traces over  $\mathcal{H}_S$ and $\mathcal{H}^{\perp}_{S}$, namely
\begin{align*}
|\tr(Y^{\dagger} U)|^2 =& |\tr_{\mathcal{H}_S}(Y^{\dagger} U) + \tr_{\mathcal{H}_S^{\perp}}(Y^{\dagger} U)|^2  \\
=& |n + \tr_{\mathcal{H}_S^{\perp}}(Y^{\dagger} U)|^2  \\
=& S^2 + 2 S \text{Re}(\tr_{\mathcal{H}_S^{\perp}}(Y^{\dagger} U)) + |\tr_{\mathcal{H}_S^{\perp}}(Y^{\dagger} U)|^2  \\
=& S^2 + 2 S \text{Re}(\tr(W)) + |\tr(W)|^2. 
\end{align*}  
We have no further constraints on the unitary $W$ and need to guess $W$ randomly with respect to Haar measure on the unitary group  $\mathcal{U}(d-S)$. It follows that the average over $Y$ is cut down to the average of $W$ over the unitary group $\mathcal{U}(d-S)$. As a result we get
\begin{equation}
\label{eq:NFLintegral}
\int dY |\tr(Y^\dag U)|^2 =  \\ \int dW \left(n^2 + 2 S \text{Re}(\tr(W)) + |\tr(W)|^2\right).
\end{equation}
The second integrand on the right hand side is linear in $W$ and vanishes. To demonstrate this one can simply substitute $W'=\exp^{-i\theta}dW$ and get
\begin{align*}
I=&\int dW \tr (W)\\
=&\int dW'\exp^{-i\theta} \exp^{-i\theta}  \tr (W')\\
=&\int dW'\exp^{-2i\theta}   \tr (W')\\
=&\exp^{-2i\theta}  I.
\end{align*}
It follows $I=0$. The last summand in \cref{eq:NFLintegral} equals $\int dY \, |\tr(Y)|^2 \equiv \tr(S_2) = \frac{1}{d}\tr(\swap) = 1$. The identity $S_2=\frac{1}{d}\swap$ is shown in \cref{prop:NFL_haarMeasureS2}and the integral results in
\begin{equation*}
\int dY |\tr(Y^\dag U)|^2 =  S^2 +1.
\end{equation*}
Overall the QNFL theorem is of the form
\begin{equation*}
\mathcolorbox{\mathbb{E}_Y[\mathbb{E}_\mathcal{S}[R_Y(U)]] \ge 1-\frac{1}{d(d+1)}(S^2+d+1).}
\end{equation*}

\FloatBarrier\subsection*{Haar measure itegral identities for the unitary group}
\label{sec:haarMeasure}

Above, we described the derivation of a quantum analogue to the classical NFL theorem. In this context we evaluated integrals running over pure states induced by a Haar-measure-distributed unitary $W$ applied to an arbitrary state $d\ket{\psi} \equiv dW\ket{0}$ \cite{Duistermaat2012,Hayden2006}. More precisely, we need two results for the derivation, in the following formulated and proven as \cref{prop:NFL_haarMeasureS2} and \cref{prop:NFL_haarMeasureS4}. 

As already introduced in \cref{subsec:QI_random}, we write the integral over the unitary group $\mathcal{U}(d)$ of $d\times d$ matrices of $f(Y)$ with respect to Haar measure as
\begin{equation*}
I = \int dY\, f(Y), 
\end{equation*}
where  $f(Y)$ is a matrix-valued function on $\mathcal{U}(d)$. Recall that
The Haar measure is left- and right-invariant with respect to shifts via multiplication. With this in mind we show that $S_2$ can be expressed using the $\swap$ operation. Note that we use tensor network diagrams for the derivations. An introduction to the topic can be find at \cite{Bridgeman2017}.
\begin{prop}
\label{prop:NFL_haarMeasureS2}
\begin{equation*}
S_2 \equiv \int dY \, Y^\dag \otimes Y = \frac1d \swap
\end{equation*}
\end{prop}
\begin{proof}
To proof this identity we note that $S_2$ fulfils
\begin{equation*}
S_2 = (\mathbb{1}\otimes e^{i\epsilon X})S_2(e^{-i\epsilon X}\otimes \mathbb{1})
\end{equation*}
for any hermitian operator $X$ and $\epsilon>0$ infinitesimally small. Expanding the right hand side to the first order in $\epsilon$ leads to
\begin{equation*}
0 = i\epsilon(\mathbb{1}\otimes X)S_2 - i\epsilon S_2(X\otimes \mathbb{1}),
\end{equation*}
and further to 
\begin{equation*}
S_2(X\otimes \mathbb{1}) = (\mathbb{1}\otimes X)S_2.
\end{equation*}
The expression is valid for any hermitian operator. Therefore we can assume in the following that $X$ is each of a Hilbert-Schmidt orthonormal hermitian operator basis $\lambda^{\alpha}$, $\alpha = 0,1, \ldots, d^2-1$, with $\tr(\lambda^{\alpha}\lambda^{\beta}) = \delta^{\alpha\beta}$. If we replace $X$ with $\lambda^\alpha$,  multiply the expression on the right by $\lambda^\alpha\otimes \mathbb{1}$ and sum over $\alpha$ we end up with the equation
\begin{equation}
\label{eq:s2identity}
\sum_{\alpha}S_2(\lambda^\alpha\lambda^\alpha\otimes \mathbb{1}) = \sum_{\alpha}(\mathbb{1}\otimes \lambda^\alpha)S_2(\lambda^\alpha\otimes \mathbb{1}).
\end{equation}
We can express the identity $\swap = \sum_{\alpha} \lambda^\alpha\otimes \lambda^\alpha$ in such a diagram as
\begin{center}
\begin{tikzpicture}[scale=0.8]  
\draw[line1]  (0,0) to[out=0,in=240](1,.5)to[out=70,in=180](2,1);
\draw[line0] (0,1)to[out=0,in=90](1,.5)to[out=270,in=180](2,0);
\draw[line1]  (0,1)to[out=0,in=110](1,.5)to[out=290,in=180](2,0);
\node[networkcircle1, label=center:$\lambda^\alpha$] at (4.6, 0)   (b) {};
\node[networkcircle1, label=center:$\lambda^\alpha$] at (4.6, 1)   (c) {};
\draw[line1]  (3.6,0)--(b) -- (5.5,0);
\draw[line1]  (3.6,1)--(c) -- (5.5,1);			
\node[] at (2.8, .35)   (a) {$\displaystyle =\sum_\alpha$};
\end{tikzpicture}
\end{center}
When wiring together the outputs of this diagram we end up with 
\begin{center}
\begin{tikzpicture}[scale=0.8]  
\node[networkcircle1, label=center:$\lambda^\alpha$] at (1.5, 0)   (b) {};
\node[networkcircle1, label=center:$\lambda^\alpha$] at (3, 0)   (c) {};
\draw[line1]  (.5,0)--(b) -- (c) --(4,0);
\begin{scope}[xshift=5.4cm,yshift=0cm,scale=.75]

\draw[line1]  (1,1)--(3,1);
\draw[line2] (3,0) arc(-90:90:.5);
\draw[line2] (1,0)-- (3,0);
\draw[line2] (1,0) arc(90:270:.5);
\draw[line1]  (1,-1)--(3,-1);
\node[networkcircle1, label=center:$\lambda^\alpha$] at (2, 1)   (e) {};
\node[networkcircle1, label=center:$\lambda^\alpha$] at (2, -1)   (f) {};
\end{scope}
\begin{scope}[xshift=9.3cm,yshift=-.35cm,scale=1.25]
\draw[line2] (0,0) arc(90:270:.25) to[out=0,in=240](1,0)to[out=70,in=180](2,.5) arc(-90:90:.25);
\draw[line1]  (0,0) to[out=0,in=240](1,.5)to[out=70,in=180](2,1);
\draw[line0] (0,1)to[out=0,in=110](1,.5)to[out=290,in=180](2,0);
\draw[line1]  (0,1)to[out=0,in=110](1,.5)to[out=290,in=180](2,0);
\end{scope}
\node[anchor=west] at (-.5, -.2)   (a) {$\displaystyle\sum_\alpha$\hspace{3cm}$\displaystyle=\sum_\alpha$\hspace{2.2cm}$=$\hspace{3cm}$=d  \,\mathbb{1}$};	
\end{tikzpicture}
\end{center}
\cref{eq:s2identity} can be represented as
\begin{center}
\begin{tikzpicture}[scale=0.8]  
\begin{scope}[xshift=.5cm,yshift=-.5cm]
\draw[line1]  (0,1)--(4,1);
\draw[line1]  (0,0)--(4,0);
\node[networkcircle1, label=center:$\lambda^\alpha$] at (2, 1)   (b) {};
\node[networkcircle1, label=center:$\lambda^\alpha$] at (3, 1)   (c) {};
\node[networkellipse1,label=center:$S_2$] at (1, .5)   (c) {};
\end{scope}
\begin{scope}[xshift=6.25cm,yshift=-.5cm]
\draw[line1]  (0,1)--(4,1);
\draw[line1]  (0,0)--(4,0);
\node[networkcircle1, label=center:$\lambda^\alpha$] at (1, 0)   (b) {};
\node[networkellipse1,label=center:$S_2$] at (2, .5)   (c) {};
\node[networkcircle1, label=center:$\lambda^\alpha$] at (3, 1)   (c) {};
\end{scope}
\node[anchor=west] at (-.76, -.2)   (a) {$\displaystyle\sum_\alpha$\hspace{3.65cm}$\displaystyle=\sum_\alpha$};	
\end{tikzpicture}
\end{center}
Using the above described identities leads to
\begin{center}
\begin{tikzpicture}[scale=0.8]  
\begin{scope}[xshift=.5cm,yshift=-.5cm]
\draw[line1]  (0,1)--(2,1);
\draw[line1]  (0,0)--(2,0);
\node[networkellipse1,label=center:$S_2$] at (1, .5)   (c) {};
\end{scope}
\begin{scope}[xshift=4cm,yshift=-.75cm,scale=.5]
\draw[line1]  (-1,3)--(3,3);
\draw[line1]  (1,0)--(6,0);
\node[networkellipse1] at (2, 1.5)   (c) {};
\node at (2, 2.6)   (c) {$S_2$};
\draw[line0] (1,2)--(4,2);
\draw[line2] (1,2)--(4,2);
\draw[line2] (4,2) to[out=0,in=240](5,2.5)to[out=70,in=180](6,3);
\draw[line0] (1,1)--(3,1);
\draw[line2] (1,1)--(3,1);
\draw[line0] (3,1) arc(-90:90:1);
\draw[line2] (3,1) arc(-90:90:1);
\draw[line2] (1,1) arc(90:270:.5);
\draw[line2] (-1,0) to[out=0,in=240](0,1)to[out=70,in=180](1,2);
\end{scope}
\node[anchor=west] at (-.3, 0)   (a) {$d$\hspace{2.15cm}$=$};	
\end{tikzpicture}
\end{center}
Now we can use the integral representation of $S_2$, namely
\begin{center}
\begin{tikzpicture}[scale=0.8]  
\begin{scope}[xshift=.5cm,yshift=-.5cm]
\draw[line1]  (0,1)--(2,1);
\draw[line1]  (0,0)--(2,0);
\node[networkellipse1,label=center:$S_2$] at (1, .5)   (c) {};
\end{scope}
\begin{scope}[xshift=4.6cm,yshift=-.5cm]
\draw[line1]  (0,1)--(2,1);
\draw[line1]  (0,0)--(2,0);
\node[networkcircle1, label=center:$Y^\dagger$] at (1, 1)   (b) {};
\node[networkcircle1, label=center:$Y$] at (1, 0)   (b) {};
\end{scope}
\node[anchor=west] at (2.5, 0)   (a) {$\displaystyle=\int dY$};	
\end{tikzpicture}
\end{center}
and connect the outputs. In tensor network diagrams this can be depicted as
\begin{center}
\begin{tikzpicture}[scale=0.8]  
\begin{scope}[scale=.75]
\draw[line1]  (1,1)--(3,1);
\draw[line1]  (1,-1)--(3,-1);
\node[networkellipse1] at (2, 0)   (c) {};
\node at (2, 0.5)   (c) {$S_2$};
\draw[line0] (3,0) arc(-90:90:.5);
\draw[line0] (1,0)-- (3,0);
\draw[line0] (1,0) arc(90:270:.5);
\draw[line2] (3,0) arc(-90:90:.5);
\draw[line2] (1,0)-- (3,0);
\draw[line2] (1,0) arc(90:270:.5);
\end{scope}
\begin{scope}[xshift=4.25cm,scale=.75]
\draw[line1]  (1,1)--(3,1);
\draw[line2] (3,0) arc(-90:90:.5);
\draw[line2] (1,0)-- (3,0);
\draw[line2] (1,0) arc(90:270:.5);
\draw[line1]  (1,-1)--(3,-1);
\node[networkcircle1, label=center:$Y^\dagger$] at (2, 1)   (e) {};
\node[networkcircle1, label=center:$Y$] at (2, -1)   (f) {};
\end{scope}
\begin{scope}[xshift=7.75cm]
\draw[line1]  (0,0)--(2,0);
\end{scope}
\node[anchor=west] at (2.55, 0)   (a) {$\displaystyle=\int dY$\hspace{2.15cm}$=$};
\end{tikzpicture}
\end{center}

Finally we end up with 
\begin{center}
\begin{tikzpicture}[scale=0.8]  
\begin{scope}[xshift=-2cm]
\draw[line1]  (0,1)--(2,1);
\draw[line1]  (0,0)--(2,0);
\node[networkellipse1,label=center:$S_2$] at (1, .5)   (c) {};
\end{scope}
\begin{scope}[xshift=1.6cm]
\draw[line1]  (0,0) to[out=0,in=240](1,.5)to[out=70,in=180](2,1);
\draw[line0] (0,1)to[out=0,in=90](1,.5)to[out=270,in=180](2,0);
\draw[line1]  (0,1)to[out=0,in=110](1,.5)to[out=290,in=180](2,0);
\end{scope}
\node[anchor=west] at (0.2, 0.5)   (a) {$\displaystyle=\frac{1}{d}$};
\end{tikzpicture}
\end{center}
which is another way to denote $\int dY \, Y^\dag \otimes Y = \frac1d \swap$.
\end{proof}

Using network diagram also the second needed identity used in the proof of the QNFL theorem can be proven.
\begin{prop}
\begin{equation*}
\begin{tikzpicture}[scale=0.4] 
\node[anchor=west] at (0, 3)   (d) {$S_4 \equiv \int dY \, Y^\dag\otimes Y^\dag\otimes Y\otimes Y$};
\node[anchor=west] at (0,0)   (d) {$\hspace{0.5cm}\displaystyle=\frac{1}{d^2-1}\hspace{1.2cm}-\frac{1}{d(d^2-1)}\hspace{1.2cm}+\frac{1}{d^2-1}\hspace{1.2cm}-\frac{1}{d(d^2-1)}$};
\begin{scope}[xshift=6.5cm,yshift=-1.5cm]
\draw[line0] (0,1)--(2,3);
\draw[line1]  (0,1)--(2,3);
\draw[line0] (0,0)--(2,2);
\draw[line1]  (0,0)--(2,2);
\draw[line0] (0,3)--(2,1);
\draw[line1]  (0,3)--(2,1);
\draw[line0] (0,2)--(2,0);
\draw[line1]  (0,2)--(2,0);
\end{scope}
\begin{scope}[xshift=15cm,yshift=-1.5cm]
\draw[line0] (0,2)--(2,0);
\draw[line1]  (0,2)--(2,0);
\draw[line0] (0,1)--(2,2);
\draw[line1]  (0,1)--(2,2);
\draw[line0] (0,0)--(2,3);
\draw[line1]  (0,0)--(2,3);
\draw[line0] (0,3)--(2,1);
\draw[line1]  (0,3)--(2,1);
\end{scope}
\begin{scope}[xshift=22cm,yshift=-1.5cm]
\node at (-2, 1.5)   (d) {$$};
\draw[line0] (0,1)to[bend right=40](2,2);
\draw[line1]  (0,1)to[bend right=40](2,2);
\draw[line0] (0,2)to[bend left=40](2,1);
\draw[line1]  (0,2)to[bend left=40](2,1);
\draw[line0] (0,0)--(2,3);
\draw[line1]  (0,0)--(2,3);
\draw[line0] (0,3)--(2,0);
\draw[line1]  (0,3)--(2,0);
\end{scope}
\begin{scope}[xshift=30.5cm,yshift=-1.5cm]
\node at (-2.5, 1.5)   (d) {$\displaystyle$};
\draw[line0] (0,1)--(2,3);
\draw[line1]  (0,1)--(2,3);
\draw[line0] (0,0)--(2,2);
\draw[line1]  (0,0)--(2,2);
\draw[line0] (0,2)to[bend right=40](2,1);
\draw[line1]  (0,2)to[bend right=40](2,1);
\draw[line0] (0,3)--(2,0);
\draw[line1]  (0,3)--(2,0);
\end{scope}
\end{tikzpicture}
\end{equation*}
\label{prop:NFL_haarMeasureS4}
\end{prop}

\begin{proof}
The tensor-network diagram of the stated identity is
\begin{center}
\begin{tikzpicture}[scale=0.8]  
\begin{scope}[xshift=.5cm,yshift=-.5cm]
\draw[line1]  (0,3)--(2,3);
\draw[line1]  (0,2)--(2,2);
\draw[line1]  (0,1)--(2,1);
\draw[line1]  (0,0)--(2,0);
\node[networkellipseX,label=center:$S_4$] at (1, 1.5)   (c) {};
\end{scope}
\begin{scope}[xshift=4.5cm,yshift=-.5cm]
\draw[line1]  (0,3)--(2,3);
\draw[line1]  (0,2)--(2,2);
\draw[line1]  (0,1)--(2,1);
\draw[line1]  (0,0)--(2,0);
\node[networkcircle1, label=center:$Y^\dagger$] at (1, 3)   (b) {};
\node[networkcircle1, label=center:$Y^\dagger$] at (1, 2)   (b) {};
\node[networkcircle1, label=center:$Y$] at (1, 1)   (b) {};
\node[networkcircle1, label=center:$Y$] at (1, 0)   (b) {};
\end{scope}
\node[anchor=west] at (2.5, 1)   (a) {$\displaystyle=\int dY$};
\end{tikzpicture}
\end{center}
In analogy to the proof of \cref{prop:NFL_haarMeasureS2} we get the equation 
\begin{equation*}
S_4 (X\otimes\mathbb{1}\otimes\mathbb{1}\otimes\mathbb{1}) + S_4 (\mathbb{1}\otimes X\otimes\mathbb{1}\otimes\mathbb{1}) = (\mathbb{1}\otimes\mathbb{1}\otimes X\otimes\mathbb{1}) S_4 + (\mathbb{1}\otimes\mathbb{1}\otimes\mathbb{1}\otimes X) S_4
\end{equation*}
by mapping  $Y$ to $e^{i\epsilon X} Y$ with infinitesimal small $\epsilon$. Further we follow the same proof strategy and set $X = \lambda_\alpha$, multiply on the right by $\lambda_\alpha \otimes\mathbb{1}\otimes\mathbb{1}\otimes\mathbb{1}$, and sum over $\alpha$. We end up with
\begin{center}
\begin{tikzpicture}[scale=0.8]  
\begin{scope}[xshift=.5cm,yshift=-.5cm]
\draw[line1]  (0,3)--(3.8,3);
\draw[line1]  (0,2)--(3.8,2);
\draw[line1]  (0,1)--(3.8,1);
\draw[line1]  (0,0)--(3.8,0);
\node[networkellipseX,label=center:$S_4$] at (.8, 1.5)   (c) {};
\node[networkcircle1, label=center:$\lambda^\alpha$] at (2, 3)   (b) {};
\node[networkcircle1, label=center:$\lambda^\alpha$] at (3, 3)   (b) {};
\end{scope}
\begin{scope}[xshift=5.5cm,yshift=-.5cm]
\draw[line1]  (0,3)--(2.8,3);
\draw[line1]  (0,2)--(2.8,2);
\draw[line1]  (0,1)--(2.8,1);
\draw[line1]  (0,0)--(2.8,0);
\node[networkellipseX,label=center:$S_4$] at (.8, 1.5)   (c) {};
\node[networkcircle1, label=center:$\lambda^\alpha$] at (2, 3)   (b) {};
\node[networkcircle1, label=center:$\lambda^\alpha$] at (2, 2)   (b) {};
\end{scope}
\begin{scope}[xshift=10cm]
\begin{scope}[xshift=0cm,yshift=-.5cm]
\draw[line1]  (.2,3)--(3.8,3);
\draw[line1]  (.2,2)--(3.8,2);
\draw[line1]  (.2,1)--(3.8,1);
\draw[line1]  (.2,0)--(3.8,0);
\node[networkellipseX,label=center:$S_4$] at (2, 1.5)   (c) {};
\node[networkcircle1, label=center:$\lambda^\alpha$] at (.8, 1)   (b) {};
\node[networkcircle1, label=center:$\lambda^\alpha$] at (3, 3)   (b) {};
\end{scope}
\begin{scope}[xshift=5cm,yshift=-.5cm]
\draw[line1]  (.2,3)--(3.8,3);
\draw[line1]  (.2,2)--(3.8,2);
\draw[line1]  (.2,1)--(3.8,1);
\draw[line1]  (.2,0)--(3.8,0);
\node[networkellipseX,label=center:$S_4$] at (2, 1.5)   (c) {};
\node[networkcircle1, label=center:$\lambda^\alpha$] at (1, 0)   (b) {};
\node[networkcircle1, label=center:$\lambda^\alpha$] at (3, 3)   (b) {};
\end{scope}
\end{scope}	
\node[anchor=west] at (-0.5, 0.85)   (a) {$\displaystyle\sum_\alpha\hspace{3.35cm}+\hspace{2.7cm}=\sum_\alpha\hspace{3.4cm}+$};
\end{tikzpicture}
\end{center}
Defining an operator $M$ as
\begin{center}
\begin{tikzpicture}[scale=0.8]  
\begin{scope}[xshift=-2cm]
\draw[line1]  (0,1)--(2,1);
\draw[line1]  (0,0)--(2,0);
\node[networkellipse1,label=center:$M$] at (1, .5)   (c) {};
\end{scope}
\begin{scope}[xshift=1cm]
\draw[line1]  (0,1)--(1,1);
\draw[line1]  (0,0)--(1,0);
\end{scope}
\begin{scope}[xshift=3.5cm]
\draw[line1]  (0,0) to[out=0,in=240](1,.5)to[out=70,in=180](2,1);
\draw[line0] (0,1)to[out=0,in=90](1,.5)to[out=270,in=180](2,0);
\draw[line1]  (0,1)to[out=0,in=110](1,.5)to[out=290,in=180](2,0);
\end{scope}
\node[anchor=west] at (0.1, 0.5)   (a) {$=$\hspace{1.4cm}$\displaystyle+\frac{1}{d}$};
\end{tikzpicture}
\end{center}
and using the identities derived in the proof of \cref{prop:NFL_haarMeasureS2} leads to
\begin{center}
\begin{tikzpicture}[scale=0.8]  

\begin{scope}[xshift=-5.5cm,yshift=-.5cm]
\draw[line1]  (0,3)--(4,3);
\draw[line1]  (0,2)--(4,2);
\draw[line1]  (0,1)--(4,1);
\draw[line1]  (0,0)--(4,0);
\node[networkellipseX,label=center:$S_4$] at (.8, 1.5)   (c) {};
\node[networkellipse1,label=center:$M$] at (2.5, 2.5)   (b) {};
\end{scope}
\begin{scope}[xshift=0cm,yshift=-.5cm]
\draw[line1]  (0,3)--(3,3);
\draw[line1]  (3.75,3)--(4,3);
\draw[line1]  (0,2)--(4,2);
\draw[line1]  (0,1)--(.25,1);
\draw[line1]  (1,1)--(4,1);
\draw[line1]  (0,0)--(4,0);
\node[networkellipseX,label=center:$S_4$] at (2, 1.5)   (c) {};
\draw[line0] (3,3)to[out=0,in=90](3.3,2.8)to[out=270,in=30](3,2.5)--(1,1.5)to[out=210,in=90](0.7,1.2)to[out=270,in=180](1,1);
\draw[line2] (3,3)to[out=0,in=90](3.3,2.8)to[out=270,in=30](3,2.5)--(1,1.5)to[out=210,in=90](0.7,1.2)to[out=270,in=180](1,1);
\draw[line0] (.25,1)to[out=0,in=180](1.5,.5)to[out=0,in=270](3.5,2)to[out=90,in=180](3.75,3);
\draw[line2] (.25,1)to[out=0,in=180](1.5,.5)to[out=0,in=270](3.5,2)to[out=90,in=180](3.75,3);
\end{scope}
\begin{scope}[xshift=5cm,yshift=-.5cm]
\draw[line1]  (0,3)--(3,3);
\draw[line1]  (3.75,3)--(4,3);
\draw[line1]  (0,2)--(4,2);
\draw[line1]  (0,1)--(4,1);
\draw[line1]  (0,0)--(.25,0);
\draw[line1]  (1,0)--(4,0);
\node[networkellipseX] at (2, 1.5)   (c) {};
\node at (2, 2.5)   (c) {$S_4$};
\draw[line0] (.25,0)to[out=0,in=180](1.5,.5)to[out=0,in=270](3.5,2)to[out=90,in=180](3.75,3);
\draw[line2] (.25,0)to[out=0,in=180](1.5,.5)to[out=0,in=270](3.5,2)to[out=90,in=180](3.75,3);
\draw[line0] (3,3)to[out=0,in=90](3.2,2.8)to[out=270,in=50](3,2.5)--(1,0.5)to[out=230,in=90](0.8,0.2)to[out=270,in=180](1,0);
\draw[line2] (3,3)to[out=0,in=90](3.2,2.8)to[out=270,in=50](3,2.5)--(1,0.5)to[out=230,in=90](0.8,0.2)to[out=270,in=180](1,0);
\end{scope}
\node[anchor=west] at (-6.5, 1)   (a) {$d\hspace{4cm}=\hspace{3.75cm}+$};
\end{tikzpicture}
\end{center}
For the first summand we can use the identity
\begin{center}
\begin{tikzpicture}[scale=0.8]  

\begin{scope}[xshift=1cm,yshift=-1.5cm]
\draw[line1]  (0,3)--(2,3);
\draw[line1]  (0,2)--(4,2);
\draw[line1]  (1,1)--(4,1);
\draw[line1]  (0,0)--(4,0);
\draw[line0] (0,1)to[out=0,in=180](1,.5)to[out=0,in=180](4,3);
\draw[line2] (0,1)to[out=0,in=180](1,.5)to[out=0,in=180](4,3);
\draw[line0] (2,3)to[out=0,in=0](1,1.5)to[out=180,in=180](.4,1)--(1,1);
\draw[line2] (2,3)to[out=0,in=0](1,1.5)to[out=180,in=180](.4,1)--(1,1);
\node[networkcircle1, label=center:$Y^\dagger$] (b)  at (1, 3)  { };
\node[networkcircle1, label=center:$Y^\dagger$] at (1, 2)   (b) {};
\node[networkcircle1, label=center:$Y$] at (1, 1)   (b) {};
\node[networkcircle1, label=center:$Y$] at (1, 0)   (b) {};
\end{scope}
\begin{scope}[xshift=6cm,yshift=-1.5cm]
\draw[line2] (0,2)--(4,2);
\draw[line2] (0,0)--(4,0);
\node[networkellipseM,draw=color2] at (3, 1)   (c) {};
\node at (3, 1.5)   (c) {$S_2$};
\draw[line0](0,3)to[out=0,in=110](1,2)to[out=290,in=180](2,1)--(4,1);
\draw[line1]  (0,3)to[out=0,in=110](1,2)to[out=290,in=180](2,1)--(4,1);
\draw[line0] (0,1) to[out=0,in=240](1,2)to[out=70,in=180](2,3)--(4,3);
\draw[line1]  (0,1) to[out=0,in=240](1,2)to[out=70,in=180](2,3)--(4,3);
\end{scope}
\begin{scope}[xshift=11.5cm,yshift=-1.5cm]
\draw[line0] (0,1)--(2,3);
\draw[line1]  (0,1)--(2,3);
\draw[line0] (0,0)--(2,2);
\draw[line2]  (0,0)--(2,2);
\draw[line0] (0,3)--(2,1);
\draw[line1]  (0,3)--(2,1);
\draw[line0] (0,2)--(2,0);
\draw[line2]  (0,2)--(2,0);
\end{scope}
\node[anchor=west] at (-0.75, 0)   (a) {$\displaystyle \int dY\hspace{3.6cm}=\hspace{3.65cm}=\frac{1}{d}$};
\end{tikzpicture}
\end{center}
The second summand can be rewritten similar, and we end up with
\begin{center}
\begin{tikzpicture}[scale=0.8]  

\begin{scope}[xshift=-5.5cm,yshift=-.5cm]
\draw[line1]  (0,3)--(4,3);
\draw[line1]  (0,2)--(4,2);
\draw[line1]  (0,1)--(4,1);
\draw[line1]  (0,0)--(4,0);
\node[networkellipseX,label=center:$S_4$] at (.8, 1.5)   (c) {};
\node[networkellipse1,label=center:$M$] at (2.5, 2.5)   (b) {};
\end{scope}
\begin{scope}[xshift=0cm,yshift=-.5cm]
\draw[line0] (0,1)--(2,3);
\draw[line1]  (0,1)--(2,3);
\draw[line0] (0,0)--(2,2);
\draw[line1]  (0,0)--(2,2);
\draw[line0] (0,3)--(2,1);
\draw[line1]  (0,3)--(2,1);
\draw[line0] (0,2)--(2,0);
\draw[line1]  (0,2)--(2,0);
\end{scope}
\begin{scope}[xshift=3.4cm,yshift=-.5cm]
\draw[line0] (0,1)to[bend right=40](2,2);
\draw[line1]  (0,1)to[bend right=40](2,2);
\draw[line0] (0,2)to[bend left=40](2,1);
\draw[line1]  (0,2)to[bend left=40](2,1);
\draw[line0] (0,0)--(2,3);
\draw[line1]  (0,0)--(2,3);
\draw[line0] (0,3)--(2,0);
\draw[line1]  (0,3)--(2,0);
\end{scope}
\node[anchor=west] at (-6.5,1)   (a) {$\displaystyle d \hspace{3.8cm}=\frac{1}{d}\hspace{2cm}+\frac{1}{d}$};
\end{tikzpicture}
\end{center}
By assuming the inverse is as well a linear combination of the identity and the SWAP operation, and using $\text{SWAP}\times\text{SWAP}=\mathbb{1}$ we can show that the inverse of $M$ is given by
\begin{center}
\begin{tikzpicture}[scale=0.8]  
\begin{scope}[xshift=-2.2cm]
\draw[line1]  (0,1)--(2,1);
\draw[line1]  (0,0)--(2,0);
\node[networkellipse1,minimum width=1cm, label=center:$M^{-1}$] at (1, .5)   (c) {};
\end{scope}
\begin{scope}[xshift=2.25cm]
\draw[line1]  (0,1)--(1,1);
\draw[line1]  (0,0)--(1,0);
\end{scope}
\begin{scope}[xshift=5.5cm]
\draw[line1]  (0,0) to[out=0,in=240](1,.5)to[out=70,in=180](2,1);
\draw[line0] (0,1)to[out=0,in=90](1,.5)to[out=270,in=180](2,0);
\draw[line1]  (0,1)to[out=0,in=110](1,.5)to[out=290,in=180](2,0);
\end{scope}
\node[anchor=west] at (-.1,0.5)   (a) {$\displaystyle=\frac{d^2}{d^2-1}\hspace{0.9cm}-\frac{d}{d^2-1}$};
\end{tikzpicture}
\end{center}
It remains to multiply both sides by $M^{-1}\otimes \mathbb{1}\otimes \mathbb{1}$ and the result is the aimed identity
\begin{center}
\begin{tikzpicture}[scale=0.6] 
\begin{scope}[xshift=.5cm,yshift=-.5cm]
\draw[line1]  (0,3)--(2,3);
\draw[line1]  (0,2)--(2,2);
\draw[line1]  (0,1)--(2,1);
\draw[line1]  (0,0)--(2,0);
\node[networkellipseX,label=center:$S_4$] at (1, 1.5)   (c) {};
\end{scope}
\begin{scope}[xshift=5.7cm,yshift=-.5cm]
\draw[line0] (0,1)--(2,3);
\draw[line1]  (0,1)--(2,3);
\draw[line0] (0,0)--(2,2);
\draw[line1]  (0,0)--(2,2);
\draw[line0] (0,3)--(2,1);
\draw[line1]  (0,3)--(2,1);
\draw[line0] (0,2)--(2,0);
\draw[line1]  (0,2)--(2,0);
\end{scope}
\begin{scope}[xshift=11.3cm,yshift=-.5cm]
\draw[line0] (0,2)--(2,0);
\draw[line1]  (0,2)--(2,0);
\draw[line0] (0,1)--(2,2);
\draw[line1]  (0,1)--(2,2);
\draw[line0] (0,0)--(2,3);
\draw[line1]  (0,0)--(2,3);
\draw[line0] (0,3)--(2,1);
\draw[line1]  (0,3)--(2,1);
\end{scope}
\begin{scope}[xshift=16cm,yshift=-.5cm]
\draw[line0] (0,1)to[bend right=40](2,2);
\draw[line1]  (0,1)to[bend right=40](2,2);
\draw[line0] (0,2)to[bend left=40](2,1);
\draw[line1]  (0,2)to[bend left=40](2,1);
\draw[line0] (0,0)--(2,3);
\draw[line1]  (0,0)--(2,3);
\draw[line0] (0,3)--(2,0);
\draw[line1]  (0,3)--(2,0);
\end{scope}
\begin{scope}[xshift=22cm,yshift=-.5cm]
\draw[line0] (0,1)--(2,3);
\draw[line1]  (0,1)--(2,3);
\draw[line0] (0,0)--(2,2);
\draw[line1]  (0,0)--(2,2);
\draw[line0] (0,2)to[bend right=40](2,1);
\draw[line1]  (0,2)to[bend right=40](2,1);
\draw[line0] (0,3)--(2,0);
\draw[line1]  (0,3)--(2,0);
\end{scope}
\node[anchor=west] at (1.7,0.9)   (d) {$\hspace{0.5cm}\displaystyle=\frac{1}{d^2-1}\hspace{1.2cm}-\frac{1}{d(d^2-1)}\hspace{1.2cm}+\frac{1}{d^2-1}\hspace{1.2cm}-\frac{1}{d(d^2-1)}$};
\end{tikzpicture}
\end{center}
\end{proof}

For the sake of completeness we shortly note how this representation of  $S_4$ helps to derive \cref{eq:NFL_quantumRisk} using $X\equiv Y^\dag U$, namely
\begin{align*}
\int d\ket{\psi} \, |\bra{\psi}X\ket{\psi}|^2=&\int dY \bra{0}Y^\dagger X^\dag Y \ket{0}\bra{0}Y^\dagger X Y \ket{0}\displaystyle\\
=&\int dY	\begin{tikzpicture}[scale=1,baseline={([yshift=-.5ex]current bounding box.center)}] 
\begin{scope}[xshift=5.5cm,yshift=-1.5cm]
\draw[line1]  (0,3)--(2,3);
\draw[line1]  (0,2)--(2,2);
\draw[line1]  (1,1)--(2,1);
\draw[line1]  (1,0)--(2,0);
\draw[line2] (2,2)--(3,2);
\draw[line2] (3,1.5) arc(-90:90:.25);
\draw[line2] (0,1.5)--(3,1.5);
\draw[line2] (0,1.5) arc(90:270:.75);
\draw[line2] (0,0)--(1,0);
\draw[line2] (2,3)--(3,3);
\draw[line2] (3,2.5) arc(-90:90:.25);
\draw[line2] (0,2.5)--(3,2.5);
\draw[white, line width=3pt] (0,2.5) arc(90:270:.75);
\draw[line2] (0,2.5) arc(90:270:.75);
\draw[line2] (0,1)--(1,1);
\node[networkcircle1, label=center:$Y^\dagger$] at (1, 3)   (b) {};
\node[networkcircle1, label=center:$Y^\dagger$] at (1, 2)   (b) {};
\node[networkcircle1, label=center:$Y$] at (1, 1)   (b) {};
\node[networkcircle1, label=center:$Y$] at (1, 0)   (b) {};
\node[networkcircle1, label=center:$\bra{0}$] at (0, 3)   (b) {};
\node[networkcircle1, label=center:$\bra{0}$] at (0, 2)   (b) {};
\node[networkcircle1, label=center:$X^\dag$] at (2, 3)   (b) {};
\node[networkcircle1, label=center:$X$] at (2, 2)   (b) {};
\node[networkcircle1, label=center:$\ket{0}$] at (2, 1)   (b) {};
\node[networkcircle1, label=center:$\ket{0}$] at (2, 0)   (b) {};
\end{scope}
\end{tikzpicture}\\
=&\frac{|\tr(X)|}{d^2+1}-\frac{d}{d(d+1)}+\frac{d}{d^2+1}-\frac{|\tr(X)|}{d(d+1)}\\
=&-\frac{1}{d(d+1)}\left(d+|\tr(X)|^2\right).
\end{align*} 

\FloatBarrier\subsection*{Numerical results}

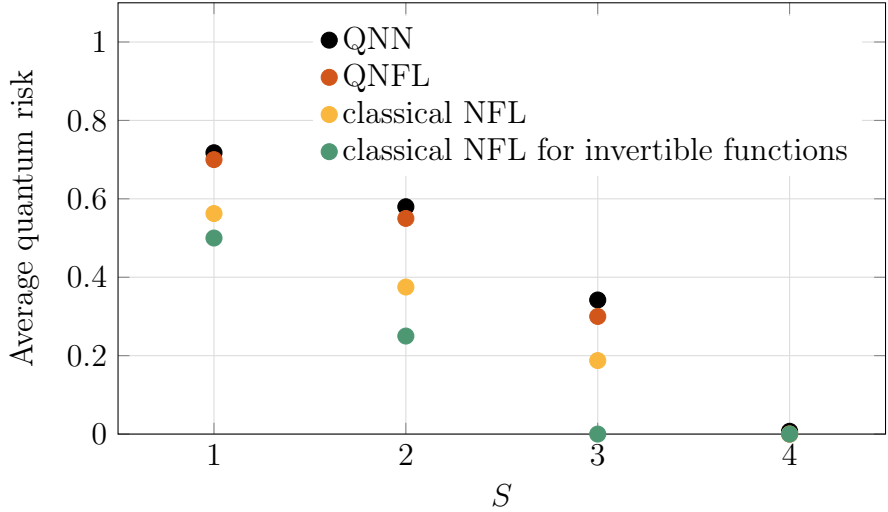
\begin{figure}
\centering
\begin{tikzpicture}[scale=1]			
\begin{axis}[
xmin=0.5,	xmax=4.5,
xtick={1,2,3,4},
ymin=0,   ymax=1.1,
width=.8\linewidth, 
height=.5\linewidth,
grid=major,
grid style={color0M},
xlabel=$S$,
ylabel={Average quantum risk},
legend pos=north east,legend cell align={left},legend style={draw=none}]
\addplot[color=black, only marks, mark size=3 pt,] table[x=n, y=numeric,  col sep=comma] {numerics/2_2_network_raw.txt};
\addlegendentry{ QNN} 
\addplot[color=color3, only marks, mark size=3 pt,] table[x=n, y=nflQ,  col sep=comma] {numerics/2_2_network_nflQ.txt};
\addlegendentry{ QNFL} 
\addplot[color=color2, only marks, mark size=3 pt,] table[x=n, y=nflC,  col sep=comma] {numerics/2_2_network_nflC.txt};
\addlegendentry{ classical NFL}
\addplot[color=color1, only marks, mark size=3 pt,] table[x=n, y=nflCinv,  col sep=comma] {numerics/2_2_network_nflCinv.txt};
\addlegendentry{ classical NFL for invertible functions}
\end{axis}
\end{tikzpicture}
\caption{\textbf{NFL theorem bounds.} The figure shows the average quantum risk based on the classical NFL theorem, the classical NFL theorem for invertible functions and the QNFL theorem for $|X|=4$ or respectively for a quantum data set of four state pairs. Further numerical results of the risk of a under the same conditions trained \protect\twotwo DQNN are plotted. }
\label{fig:NFL_risk}
\end{figure}

At this point, three bounds on the training success of a learning algorithm are stated: the NFL theorem, the NFL theorem for invertible functions and the QNFL theorem. In the following, we not only want to compare these three results but also numerical results based on the DQNN algorithm introduced in \cref{chapter:DQNN}.

In \cref{fig:NFL_risk}, where the bounds are plotted, it becomes clear that the behaviour of NFL bound for invertible functions is very similar to the standard classical NFL theorem. The different slope shows the additionally available information when assuming the invertibility of the unknown function. 

Further, one can see that the QNFL bound gives a stronger lower bound than its classical analogue. A paragraph in \cite{Poland2020} describes the case of a single qubit as a good example to gain intuition: one could be wrong and assume that we can completely determine the action of $Y$ when using just a single training pair $\{\ket{\phi^{\text{in}}_x},Y\ket{\phi^\text{in}_x}\}$. Hence, the complement of the subspace $\mathcal{K}$ spanned by $\ket{\phi^{\text{in}}_x}$ has to be mapped to the complementary subspace determined by $Y\ket{\phi^{\text{in}}_x}$. However, his does not apply for the reason that there is nevertheless the freedom of a phase. One can get a feeling of the big impact of this freedom when noticing that we evaluate the risk averaged over the Hilbert space. In the here described case this is the average over all superposition inputs $\alpha\ket{\phi^{\text{in}}_x} + \beta\ket{\phi^{\text{in}}_x}$. Therefore the freedom of a phase affects the action of $Y$ on nearly all inputs. In the classical case the analogue would be a function $f$ on $\{0,1\}$. Having the information of a single training pair, the action of $f$ is determined for half of the inputs. Hence for half of the cases, we are $100\%$ sure. For the other input we can guess with being correct $50\%$ of the time. In total, our hypothesis predicts the right output in $75\%$ of the cases. 

We so far only compared the QNFL theorem to the classical counterpart. As mentioned, the QNFL theorem is a bound on the training success of the in \cref{chapter:DQNN} described training algorithm. Consequently, \cref{fig:NFL_risk} also depicts the risk for an exemplary DQNN performed with the algorithm presented in \cref{sec:DQNN_classical}. Therefore we choose a unitary $Y$ uniformly at random, build $4$ data pairs and train the QNN in $r_T=1000$ steps using $S=1,\dots,4$ supervised training pairs. After the training, $C=10$ randomly chosen input states $\ket{\phi^{\text{R}}_x}$ are used to evaluate the quantum risk by empirical average, namely 
\begin{equation*}
R_Y(U)=\frac{1}{C} \sum_{x=1}^{C}\| Y\ket{\phi^{\text{R}}_x}\bra{\phi^{\text{R}}_x}Y^\dag - U\ket{\phi^{\text{R}}_x}\bra{\phi^{\text{R}}_x}U^\dag \|_1^2,
\end{equation*}
where the unitary $U$ describes the trained DQNN. The values describing the behaviour of the DQNN in \cref{fig:NFL_risk} are gained by averaging the risks $R_Y(U)$ for $10$ different unitaries $Y$.

We can note that the numerical results of the DQNN are close to achieving the QNFL bound. Since empirical averages are included in the process of evaluating the quantum risk and the DQNN is not trained to the maximum value of the training loss function, the slight discrepancy seen in \cref{fig:NFL_risk} was to be expected.

\FloatBarrier\subsection*{Comment on orthonormal training pairs}
We want to conclude the chapter with a short but essential note on using orthonormal training pairs, i.e.\ $(\ket{\phi^{\text{in}}_x}, Y\ket{\phi^{\text{in}}_x}) \text{ with } \braket{\phi_x|\phi_k}=\delta_{xk}$, for learning an unknown unitary operstion since this case can lead to problems. This circumstance can be explained with the above chose quantum risk function, hence a remark at this point is suitable.

We want to start this discussion with the known case of not orthonormal $\ket{\phi^{\text{in}}_x}$. The aim is to train the network until it is $U=Y$, but indeed it is always possible that $U$ is only equal to $Y$ up to a global phase, namely $U=Y \exp^{i\theta}$. Applying the in \cref{eq:NFL_quantumRisk} defined risk function makes this clear, since $$\sum_x \| Y\ket{\phi^{\text{in}}_x}\bra{\phi^{\text{in}}_x}Y^\dag - U\ket{\phi^{\text{in}}_x}\bra{\phi^{\text{in}}_x}U^\dag \|_1^2=0.$$

However, in the case of orthonormal training pairs, also local phases are possible, i.e.\ cases exist where the phase $\theta_x$ depends on the training pair, namely $U\ket{\phi^{\text{in}}_x}=\exp^{i\theta_x}Y \ket{\phi^{\text{in}}_x}$. This by far worse case when learning a unitary operation is only possible, if the training pairs are orthogonal. To make this clear we can study the following case: the operation we want to learn is the identity, $Y=\mathbb{1}$ and the training pairs are basis states $\{\ket{\phi^{\text{in}}_x}\}_x=\{\ket{0...0},\ket{0...1},...,\ket{1...1}\}$. If global phases come across the risk function cannot tell the difference between the identity and the diagonal matrix with local phases written as $U=\text{Diag}(\exp^{i\theta_1},...,\exp^{i\theta_d})$.
We can reason from this short discussion that orthonormal training pairs should be avoided in training algorithms as described above.

To summarise, the QNFL theorem gives us a tool to review the in \cref{chapter:DQNN} presented DQNN algorithm. This algorithm uses training data for characterising a unitary operation. Here, the information for updating the DQNN parameters was based on training data pairs in the form of input states and desired output states. In the following chapter, we extend this supervised ansatz to the usage of the training data's possible underlying graph structure.

%% file: text/graphs.tex
\chapter{Training with graph-structured quantum data}
\label{chapter:graphs}

Graph-structured data is omnipresent throughout the social and natural sciences. In nearly every scientific area graphs, sets of in some way connected or related objects, are indispensable. These correlations, called edges, between the objects, named vertices, can be for example from spatial, temporal or causal nature. Whether it comes to representing computational devices, studying social media platforms \cite{Pitas2016}, sampling road networks \cite{Jepsen2019} or learning interactions between proteins \cite{Zitnik2018}, graphs provide a theoretical framework to describe these and many more complex systems in a helpful way \cite{Zhou2020}. 

Consequently, approaches including graph-structured data are widely spread in the area of classical \emph{machine learning} (ML). At one hand unsupervised learning algorithms \cite{Perozzi2014,Qiu2018,Liu2019,Khosla2019} exist. Such methods exhibit good learning behaviour with well-encoded graph structure and aim tasks like finding missing graph connections or labels. Further, there are semi-supervised methods \cite{Zhu2003, Belkin2006}. These use loss functions with one part making use of the graph structure of the data and another supervised loss term using supervised training pairs. Further, also methods encoding the graph structure directly in the representations, called graph convolutional neural networks \cite{Kipf2017,Velickovic2018,Hamilton2017,Xu2019}, exist. A summary of some common classical approaches is presented in \cref{sec:graphs_classical}.

In the field of \emph{quantum machine learning} (QML), research on how to work with graph-structured classical and quantum data was done as well. This includes quantum algorithms for classical graph-structured data like quantum walks \cite{Dernbach2019}, a quantum analogue of the random walk. Further, the usage of graph-structured data in quantum convolutional neural networks \cite{Arunachalam2017,Cong2019} has received much attention. 

Nevertheless, when it comes to using graph structure for \emph{quantum neural networks} (QNNs), previous work has mainly focused on building the graph structure into the QNN \cite{Cong2019,Verdon2019}. However, we present in this chapter an approach on how to make an arbitrary, not to the graph structure of the data aligned, QNN learn the graph structure of noisy and unreliable quantum data sources during training \cite{Beer2021}. Based on the \emph{dissipative quantum neural network} (DQNN) architecture presented in \cref{chapter:DQNN} we not only introduce a new loss function and suiting update rules for DQNNs, but also present examples that prove that the learning performance can be improved by learning and considering the graph structure in contrast to a simple supervised learning ansatz. We will see that exploiting this additional information can be highly beneficial, especially when data for supervised learning is rare but information on the structure of the problem is available.  

To introduce the reader to the topic, we start this chapter with the basic graph theory definitions in \cref{sec:graph_basic} and give a survey of problems and methods in the field of classical ML on graphs \cref{sec:graphs_classical}. It follows a general discussion of quantum sources with graph structure, see \cref{sec:graph_data}, and the presentation of appropriate information-theoretic loss functions in \cref{sec:graph_loss} for their characterisation according to the approach of \cite{Beer2021}. Further, we describe how to adjust the training algorithm described in \cref{sec:DQNN_trainingalgorithm} to utilise the new loss functions, see \cref{sec:graph_quantumAlg}. We conclude this chapter with some numerical results: the classical simulation in \cref{sec:graph_classicalSim}, as well as the training on actual quantum computers \cref{sec:graph_quantumAlg}, show the benefits of using graph information. 

\section{Basic definitions}
\label{sec:graph_basic}

This section will present the most fundamental definitions of graph theory. For a complete introduction to this field, we point to \cite{West2001, Trudeau2013, Bollobas2013}.

\begin{figure}[H]
\centering
\begin{tikzpicture}[scale=.95]
\node[vertex0] at (0,1)  (a1)  {$v_1$}; 
\node[vertex2] at (0,-1)  (a2) {$v_2$};
\node at ([shift={(180:1.8)}]a2) (a2l) {$\xi_2$};
\draw[color0,dashed] (a2) -- (a2l.east);
\node[vertex0] at (3,0) (a3) {$v_3$};
\node[vertex2] at (6,1) (a4) {$v_4$};
\node at ([shift={(0:1.8)}]a4) (a4l) {$\xi_4$};
\draw[color0,dashed] (a4) -- (a4l.west);
\node[vertex0] at (6,-1) (a5) {$v_5$};
\node[vertex0] at (9,-1) (a6) {$v_6$};
\draw[line3](a1) -- node[above=1ex] {$e_{13}$} (a3);		
\draw[line3](a1) -- node[left=1ex] {$e_{12}$} (a2);
\draw[line3](a3) -- node[below=1ex] {$e_{23}$} (a2);
\draw[line3](a3) -- node[above=1ex] {$e_{34}$} (a4);
\draw[line3](a3) -- node[below=1ex] {$e_{35}$}  (a5) -- node[below=1ex] {$e_{56}$} (a6);
\end{tikzpicture}
\caption{\textbf{Exemplary graph}. This connected graph $G_1$ consists of eight vertices $V(G_1)=\{v_1,\hdots,v_8\}$, which are represented as heptagons and connected through the edges $E(G_1)=\{e_{13},\dots e_{56}\}$. The vertices $v_2$ and $v_4$ are labelled with labels $\xi_2$ and $\xi_4$.}
\label{fig:graph_example}
\end{figure}
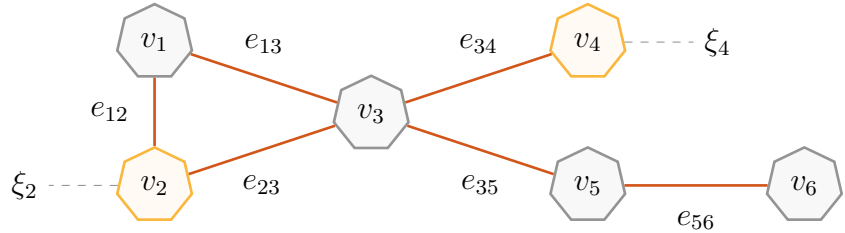

A \emph{graph} $G\{V,E\}$ is defined through the \emph{vertex set}, also called \emph{node set}, $V(G)=\{v_1,\hdots,v_N\}$, and the \emph{edge set} $E(G)=\{e_{wx}\}_{w,x}$. An edge $e_{wx}$ connects vertices $v_w$ and $v_x$ if they are neighbours, referred to as $v_w\sim v_x$. These connections denote usually the relationship of the vertices, describes for example spacially, causal or temporal closeness of the vertices.  The number of neighbours of a vertex $v_x$ is called the \emph{degree} of $v_x$.

A \emph{path} of $k$ vertices is a sequence of $k$ distinct vertices such that consecutive vertices are adjacent. We further denote a graph as \emph{complete} if there is an edge between every pair of vertices. This definition should be not confused with a \emph{connected} graph, where simple every pair of vertices can be connected with a path.

To be useful for ML graph-structure has to be encoded into continuous low-dimensional representations. A helpful tool is the \emph{adjacency matrix}. For a graph with $N$ vertices this $N\times N$ matrix is defined by
\begin{equation*}
A = \begin{cases}
A_{wx} = 1 & \,\text{if there is an edge } e_{wx} \\
A_{wx} = 0 & \, \text{if there is no edge}\\
A_{xx} = 0 .
\end{cases}
\end{equation*}
The graph depicted in \cref{fig:graph_example} can be described with an adjacency matrix
\begin{equation*}
A(G_1)=	\begin{pmatrix}[0.6]
0 & 1 & 1& 0 & 0& 0\\
1 & 0 & 1& 0 & 0& 0\\
1 & 1 & 0& 1 & 1& 0\\
0 & 0 & 1& 0 & 0& 0\\
0 & 0 & 1& 0 & 0& 1\\
0 & 0 & 0& 0 & 1& 0
\end{pmatrix}.
\end{equation*}

Note that in this work, we only discuss so-called \emph{simple graphs}, where at most one edge between a pair of vertices is allowed. Further, we only study \emph{undirected graphs}, where the existence of an edge $e_{wx}$ is equivalent to the presence of an edge $e_{xw}$. Moreover, we assume that the relevance or meaning of all edges is equal. In contrast, there is the concept of \emph{weighted graphs}, where the adjacency matrix' entries can be arbitrary values rather than just $\in\{0,1\}$. 

Every vertex can optionally be assigned to a \emph{label}. Such a label can stand for a category, a number, a user profile in a social network, or some quantum data, to name a few examples.  In the classical case, labels often are formulated as binary \emph{label vectors} $\xi(v_x)$.

\section{Classical machine learning with graph-structured data}
\label{sec:graphs_classical}

Exploiting the graph structure of complex data sets has significant potential for new scientific breakthroughs. However, the challenge is to activate this potential by choosing the right techniques. Not only network analysis methods \cite{Newman2018} but also ML approaches \cite{Hamilton2020} were utilised for this. 

ML techniques are often categorised into supervised tasks with the aim of getting the desired output given an input, or unsupervised tasks engaging with the assignment of learning patterns. Also semi-supervised tasks exist, which can be seen as a combination of both approaches. In the field of ML on graphs, these labels are used as well, just as mentioned in the introduction of this chapter. However, due to the characteristics of graphs, further (sub-)categories turned to be reasonable and valuable. We will present a selection of them shortly in the following paragraphs. This will supply the reader with a short but comprehensive overview of the field of ML on graphs. For a more detailed and comparing review we point to \cite{Hamilton2020,Khosla2021,Wu2021}. Note further that there exist methods, preparing graphs for such algorithms, for example by removing unnecessary neighbors \cite{Rathee2021}.

\subsection*{Node classification}
A widespread problem category is \emph{node classification}\cite{Bhagat2011,Rong2019,Li2019a,Khosla2021}, where the aim is to use the information of the graph to find missing vertex labels. Imagine building a graph where the vertices refer to publications and the edges symbolise citations two papers have in common. If we then label some of the vertices with topics treated in the according to publication, it is possible to forecast missing labels by exploiting the graph structure \cite{Kipf2016}. Since usually the training set is built of the full graph (including the labelled and unlabelled vertices) and the existing labels, vertex classification is often classified as a semi-supervised task.

\subsection*{Link prediction}
Not only absent labels but also missing edges can be the focus of a problem. A well-known example for \emph{link prediction}\cite{Zhang2017,Teru2020,Khosla2021} can be found in polypharmacy: the usage of drug combinations is common for beating complex or co-existing diseases but also offers a higher risk of side effects. Data on these effects is usually scarce. However, \cite{Zitnik2018} presents an approach named \emph{Decagon} for modelling them. It uses a graph with protein-protein, drug-protein and drug-drug (polypharmacy) interactions, including many different edge types that stand for distinct side effects. The algorithm is evidentially able to find missing edges and therefore predict side effects of drug combinations. In the same way as vertex classification, relation prediction is often categorised as semi-supervised.

\subsection*{Community detection}
Whereas in the last two categories, missing graph data information was gained through the learning algorithm using some supervised data in the vertex labels, \emph{community detection}\cite{Liu2020a} is an unsupervised process. The algorithms only input is the graph $G=(V,E)$ itself. In many contexts, graphs naturally are clustered: vertices included in such a cluster are much more likely to be connected with other included vertices than with those from outside the cluster. An example is the connection of user profiles on social media platforms: these will be likely clustered, for instance, by the users home location. Algorithms finding such clusters in graphs can be used, for example, to discover fraudulent anomalies in financial transaction networks given the histories of transactions of people using the networks \cite{Pandit2007}. 

\subsection*{Graph classification}
On the contrary to the three above mentioned categories \emph{graph classification}\cite{Errica2019} is not based on one single graph but is instead feed with different graphs in order to categorise them. For example, we assume a given graph-based representation of the structure of some molecules. The algorithm can be trained such that it reads the graph and classifies the molecule in some way, for instance, by its toxicity \cite{Gilmer2017}. 

\subsection*{Graph-convolution networks}
The last method we want to mention in this short overview is directly encoding the graph structure in the representations. These so-called \emph{graph-convolution networks} (GCNs) \cite{Kipf2017,Velickovic2018,Hamilton2017,Xu2019} are one of the most prominent graph deep learning models and especially useful when other graph-based ML methods face problems. The techniques can also be used on data types that are not structured initially as a graph, such as image or text data, and graph-structured data with very complex patterns. A concrete example is building a representation of video input as a space-time region graph and recognising actions in these videos, for example like "opening a book" \cite{Wang2018a}. A review of GCNs can be found in \cite{Zhang2019}.

\section{Quantum graph-structured data}
\label{sec:graph_data}

In the last section, we gained an insight into the success of classical graph-structured data processed in classical ML algorithms. Nevertheless, not only these achievements are the motivation to study the use of quantum graph-structured data. The properties of the data itself prompt this kind of ansatz: quantum data, produced by structured devices, will always be structured since spatial and causal arrangements lead to different correlations between the states. In the following, we give some examples of how graph-structured data can emerge. We close this section by developing a notation describing the graph structure of the quantum data set in a way that is useful for the training of QNNs.  

\FloatBarrier\subsection*{Motivational examples}

One can imagine a set of $N$ quantum information processors represented by the vertices of a graph. We assume we know in which way these processors are arranged. For every processor, some of the other processors are closer than the rest. We can relate particularly close processors with edges between the concerning vertices encoding the correlations between the processors. We further assume we not only know the structure of the processors, but also have access to a training data set of $S<N$ outputs and inputs $\{(\rho^\text{in}_x, \rho^\text{out}_x)\,|\, x = 1,2, \ldots, S\}$ defined as follows: we know that processing the state $\rho^\text{in}_x$ with the processor $x$ can lead to the state $\rho^\text{out}_x$. In other words, we own a training pair consisting of an input and output state for some of the processors, but not all of them. The aim would be now to learn the input and output relations for all the $N$ processors. This scenario can be seen as the foundation of various physically relevant situations. The quantum NISQ device clusters \cite{Preskill2018} are one example where the quantum data is in some way spacially graph-structured. 

An other example is structure due to time steps. Imagine simulating a quantum system with Hamiltonian $H \in \mathcal{B}(\mathcal{H})$ for some period of time in steps $t\in \{0,\epsilon, 2\epsilon, \ldots (N-1)\epsilon\}$. Based on an initial state $\ket{\rho^\text{in}}\in \mathcal{H}$ the states $\ket{\psi_t}\equiv e^{it H/\hbar}\ket{\rho^\text{in}}$ evolve. Such a structure can also be interpreted as a line graph with $N$ vertices. 

\FloatBarrier\subsection*{Notation}
After motivating graph-structured quantum data, we introduce some convenient notation to describe such in the following. For this, we tie in with \cref{sec:DQNN_trainingalgorithm} and assume having access to one or more quantum devices producing uncharacterised quantum states. We further suppose that the quantum information device can be described by a completely positive map $\mathcal{E}$ mapping from an input state $\rho_x$ to an output state  $\sigma_x = \mathcal{E}(\rho_x)$. The device can be described with $\{(p_x, \rho_x)\}_{x\in V}$, where $\rho_x$ occurs with probability $p_x$.

Additionally to the setting in \cref{chapter:DQNN} we presume the quantum states $\rho_x$ are linked to the vertices of a graph $G=(V,E)$, namely
\begin{equation*}
\rho:V\rightarrow \mathcal{D}(\mathcal{H}).
\end{equation*}
where $V$ denote the vertices of the graph and $\mathcal{D}(\mathcal{H})$ the density matrices on $\mathcal{H}$. More specifically, we assume that the graph structure is associated with the device's outputs.

We further exabit the graph-structure with an edge set $E$. In this sense two states $\rho_w$ and $\rho_x$ which are \emph{information-theoretic close}, defined by some metric $d(\rho_w, \rho_x) \sim \epsilon$, are connected with an corresponding edge $(w,x)\in E$. We discuss the choice of the metric in \cref{sec:graph_loss}. These edges can be described in an adjacency matrix $A$. 

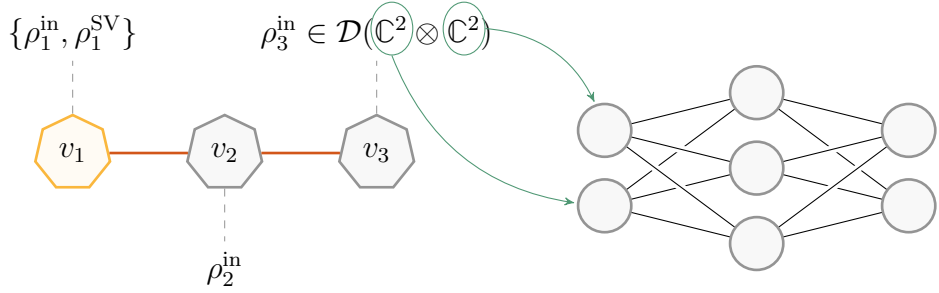
\begin{figure}[h!]
\centering
\begin{tikzpicture}[node distance=2cm,>=stealth',bend angle=30,auto]
\begin{scope}
\node [vertex2, pin={[pin distance=7mm,pin edge={color0,dashed}]90:$\{\rho^\text{in}_1,\rho^\text{SV}_1\}$}] (v1)      {$v_1$};
\node [vertex0, pin={[pin distance=7mm,pin edge={color0,dashed}]270:$\rho^\text{in}_2$}] (v2) [right of=v1]   {$v_2$}
edge[line3] node[swap] {} (v1);
\node [vertex0, pin={[pin distance=7mm,pin edge={color0,dashed}]90:$\rho^\text{in}_3\in \mathcal{D}(\mathbb{C}^2\otimes \mathbb{C}^2)$}] (v3)  [right of=v2]  {$v_3$}	edge[line3] node[swap] {} (v2);
\node (connector1)  [ellipse, draw=color1, right of=v2, yshift=1.65cm, xshift=0.2cm, minimum width=5.5mm, minimum height=7mm]  {};
\node (connector2)  [ellipse, draw=color1, right of=v2, yshift=1.65cm, xshift=1.15cm, minimum width=5.5mm, minimum height=7mm]  {};
\end{scope}	
\begin{scope}[xshift=7cm, yshift=-0.2cm]
\foreach \x in {-.5,.5} {
\draw[line0] (0,\x) -- (2,-1);
\draw (0,\x) -- (2,-1);
\draw[line0] (0,\x) -- (2,0);
\draw (0,\x) -- (2,0);
\draw[line0] (0,\x) -- (2,1);
\draw (0,\x) -- (2,1);
}
\foreach \x in {-0.5,0.5} {
\draw[line0] (2,-1) -- (4,\x);
\draw (2,-1) -- (4,\x);
\draw[line0] (2,0) -- (4,\x);
\draw (2,0) -- (4,\x);
\draw[line0] (2,1) -- (4,\x);
\draw (2,1) -- (4,\x);
}
\foreach \x in {-1,0,1} {
\node[perceptron0] at (2,\x) {};
}
\foreach \x in {-0.5,0.5} {
\node[perceptron0] at (4,\x) {};
}
\node[perceptron0] at (0,0.5) (p1) {}	edge [pre, color1, bend right] (connector2.east);
\node[perceptron0] at (0,-0.5) (p2) {} edge [pre, color1, bend left] (connector1.south);
\end{scope}
\end{tikzpicture}
\caption{\textbf{Graph and QNN.} A line graph with $N=3$ vertices whereof $S=1$ is supervised is depicted on the left-side of the figure. The labels of the vertices are either input states (unsupervised vertex) or pairs of input and output states (supervised vertex). The dimension of the input state should match the input layer of the QNN, the dimension of the supervised states the output layer, respectively. }\label{fig:graph_QNN}
\end{figure}

Additionally to this $N\times N$ matrix we have a training data set on hand. Therefore this setting can be categorised as semi-supervised. The training data set is structured as
\begin{equation}
\label{eqn:graph_trainingData}
\Big\{ \big(\rho_1,\ket{\phi^\text{SV}_1}\bra{\phi^\text{SV}_1}\big),\ldots,\big(\rho_S,\ket{\phi^\text{SV}_S}\bra{\phi^\text{SV}_S}\big),\rho_{S+1},\ldots,\rho_N\Big\},
\end{equation}
where the first $S$ entries are the quantum state pairs describing the supervised data (as in the supervised setting discussed in \cref{chapter:DQNN}) and the remaining $N-S$ entries describe unsupervised vertices without information about the output. Throughout this chapter, we will call the vertices labelled with the input state and the desired output state as \emph{supervised} and the remaining vertices as unsupervised. An example containing one supervised vertex and two unsupervised vertices is depicted in \cref{fig:graph_QNN}.

We aim to find the output states of the unsupervised vertices by exploiting the supervised vertices and the graph structure. Hence we can categorise the problem as a semi-supervised node classification question.

For the validation process of our examples presented in \cref{sec:graph_classicalSim} we assume moreover to have access to the \emph{overall} data set, i.e.\
\begin{align*}
\Big\{& \big(\rho_1,\ket{\phi^\text{SV}_1}\bra{\phi^\text{SV}_1}\big),\ldots,\big(\rho_S,\ket{\phi^\text{SV}_S}\bra{\phi^\text{SV}_S}\big),\\
&\big(\rho_{S+1},\ket{\phi^\text{USV}_{S+1}}\bra{\phi^\text{USV}_{S+1}}\big),\ldots,\big(\rho_N,\ket{\phi^\text{USV}_N}\bra{\phi^\text{USV}_N}\big)\Big\}.
\end{align*}
Using this additional data, we can test if the new graph loss function, presented in the next section, increases the generalisation behaviour of the DQNN training algorithm. In the use case, this additional data is not necessary, and the data in the form of \cref{eqn:graph_trainingData} is sufficient. Note that we assume that the training and validation data consist of pure states in the following, but mixed states are possible in general as well. 

\section{Loss functions}
\label{sec:graph_loss}

As in \cref{chapter:DQNN} we use pairs of quantum states as training data to learn unknown quantum information processes. In addition to that, we include here information about the problem's graph structure. The task is now to find a way that the in \cref{chapter:DQNN} presented DQNN can learn and generalise from this graph data. Hence in the following, we will guide through different loss functions used in the proposal of \cite{Beer2021}. To remind the reader, we describe the learning architecture we train with the function $\mathcal{E}$, i.e.\ $\rho^{\text{out}}=\mathcal{E}\left(\rho^{\text{in}}\right)$.

\FloatBarrier\subsection*{Training loss}
We already discussed how to train a DQNN with pure supervised states using the fidelity in \cref{sec:DQNN_trainingalgorithm}, namely with the \emph{supervised loss}
\begin{equation*}
\mathcal{L}_\text{SV} \equiv \frac{1}{S}\sum\limits_{x=1}^S\bra{\phi^\text{SV}_x}\mathcal{E}\big(\rho^\text{in}_x\big)\ket{\phi^\text{SV}_x}.
\end{equation*}

In addition, we define here a new loss in order to exploit the graph structure of the output states of the device. These output states are, in general, mixed. Even though the fidelity is defined for mixed states, the computation complexity required to evaluate it is immense. Therefore, we choose a different measure here, namely the \emph{Hilbert-Schmidt} distance
\begin{equation*}
d_{\text{HS}}(\rho,\sigma) \equiv \text{tr}((\rho-\sigma)^2).
\end{equation*} 
Keep in mind that the Hilbert-Schmidt distance reaches its minimum if the two compared states resemble. In contrast, the fidelity reaches the maximum, the value $1$, in this case. See \cref{subsec:QI_distance} for a more detailed discussion of the fidelity and the Hilbert-Schmidt distance. 

We use the adjacency matrix $A$ of the graph $G$ to include the graph structure information during the learning process and define the \emph{graph-based loss} as
\begin{equation*}
\mathcolorbox{
\mathcal{L}_{G} \equiv \sum_{w,x\in V} [A]_{wx} d_{\text{HS}}(\mathcal{E}(\rho_w^\text{in}),\mathcal{E}(\rho_x^\text{in}),}
\end{equation*}
where and $[A]_{wx}$ denotes the matrix element of $A$ relating to vertices $v$ and $w$. 

The full \emph{training loss function} is now specified as the combination of supervised and graph-based loss, with the graph part controlled by a Lagrange multiplier $\gamma$:
\begin{equation*}
\mathcolorbox{
\mathcal{L}_\text{SV+G}=\mathcal{L}_\text{SV} + \gamma \mathcal{L}_{G}.}
\end{equation*}   
Defining the training loss in that way forces the network to map supervised input states to the desired output states and also considers the graph structure of the data. 

In the same way as represented in \cref{sec:DQNN_trainingalgorithm} where we described the training of a neural network without using graph-structure, the training task is to maximise the training loss function, here $\mathcal{L}_\text{SV+G}$ with $\gamma \leq 0$. The maximum depends on the Lagrange multiplier, and by tuning it, we can decide the importance of the graph structure while training. 

\FloatBarrier\subsection*{Validation loss}
We aim the quantum neural network to lead to the right output state for a given input state in the same way as in \cref{chapter:DQNN} and therefore use the equivalent \emph{validation} loss, namely
\begin{equation*}
\mathcal{L}_\text{USV}=\frac{1}{N-S}\sum_{x=S+1}^{N} \langle\phi^{\text{USV}}_x\rvert\mathcal{E}\big(\rho^\text{in}_x\big)\lvert\phi^{\text{USV}}_x\rangle,
\end{equation*}
to check the trained quantum neural networks behaviour using the unsupervised data pairs. We will use this function to study numerical examples in \cref{sec:graph_classicalSim}.

\section{Training algorithm}	
\label{sec:graph_algorithm}

For training the DQNN, including the graph information, we use an algorithm of the same structure as described in \cref{sec:DQNN_trainingalgorithm}. The only difference is the new training loss function $\mathcal{L}_\text{SV+G}=\mathcal{L}_\text{SV} + \gamma \mathcal{L}_{G}$ leads to new update rules which we formulate in Hermitian matrices $K^l_{j\text{,SV+G}}(s)$ for $j$th qubit in $l$th layer. Following the discussions in \cref{chapter:DQNN} we recognise that all calculations are linear in the loss function, and therefore we can make the ansatz
\begin{equation*}
K^l_{j\text{,SV+G}}(s) = K^l_{j\text{,SV}}(s) + \gamma \cdot K^l_{j\text{,G}}(s),
\end{equation*}  
where $K^l_{j\text{,SV}}(s)$ denotes the update matrix derived in \cref{sec:DQNN_trainingalgorithm}. It remains to derive the unsupervised update matrix $ K^l_{j\text{,G}}(s)$ to clarify the training algorithm when using graph-structured quantum data for training a DQNN. 

\begin{prop}
\label{prop:graph_K}
The update matrix for training a QNN  with a graph structure between the output states  $\{\rho^\text{out}_w,\rho^\text{out}_x\}$ encoded via the adjacency matrix $[A]_{wx}$ (and without any supervised states) is
\begin{equation*}
\mathcolorbox{K^l_{j\text{,G}}(s)=2^{m_{l-1}+1}i\eta\sum\limits_{v\sim w}[A]_{wx}\text{tr}_\text{rest}\big(M^l_{j\{w,x\}}(s)\big),}
\end{equation*}
where 
\begin{align*}
M_{j\{w,x\}}^l(s)=&\big[U_j^l(s)U_{j-1}^l(s)\dots U_1^1(s)\ \big(\big(\rho_w^\text{in}-\rho_x^\text{in}\big)\otimes\lvert 0\dots 0\rangle_1\langle 0\dots 0\rvert\big)\\
& {U_1^1}^\dagger(s)\dots{U_{j-1}^l}^\dagger(s){U_j^l}^\dagger(s),\\
&\hspace{15pt}{U_{j+1}^l}^\dagger(s)\dots {U_{m_\text{out}}^\text{out}}^\dagger(s)\big(\mathbb{1}_\text{in,hidden}\otimes\big(\rho^\text{out}_w-\rho^\text{out}_x\big)\big)U_{m_\text{out}}^\text{out}(s)\dots U_{j+1}^l(s)\big].
\end{align*}
\end{prop}
\begin{proof}

The derivation is analogous to the proof of \cref{prop:DQNN_K}. Given the graph-based loss
\begin{equation*}
\mathcal{L}_{G} \equiv \sum_{w,x\in V} [A]_{wx} d_{\text{HS}}(\mathcal{E}(\rho_w^\text{in}),\mathcal{E}(\rho_x^\text{in}),
\end{equation*}
we take the derivative with respect to the step parameter. We get
\begin{align*}\label{derusv}
\frac{d\mathcal{L}_\text{G}(s)}{ds}=&\lim\limits_{\epsilon \rightarrow 0}\frac{\mathcal{L}_\text{G}(s+\epsilon)-\mathcal{L}_\text{G}(s)}{\epsilon}\nonumber\\
=&2i\sum\limits_{w,x\in V}[A]_{wx}\tr\big(\big(\mathbbm{1}_\text{in, hidden}\otimes \big( \rho^\text{out}_w- \rho^\text{out}_x\big)\big)\\&\Big(\big[K^\text{out}_{m_\text{out}}, U^\text{out}_{m_\text{out}}\ldots U^1_1\big( \big){U^1_1}^\dagger\ldots {U^\text{out}_{m_\text{out}}}^\dagger\big]+\ldots \nonumber\\
&+\; U^\text{out}_{m_\text{out}}\ldots U^1_2 \big[K^1_1, U^1_1\big(\rho_w^\text{in}-\rho_x^\text{in}\big){U^1_1}^\dagger\big]{U^1_2}^\dagger\ldots {U^\text{out}_{m_\text{out}}}^\dagger\Big)\big)\nonumber\\
=&2i\sum\limits_{w,x\in V}[A]_{wx}\tr\big(\big[U^\text{out}_{m_\text{out}}\ldots U^1_1\big(\rho_w^\text{in}-\rho_x^\text{in}\big){U^1_1}^\dagger\ldots {U^\text{out}_{m_\text{out}}}^\dagger,\\&\big(\mathbbm{1}_\text{in, hidden}\otimes \big( \rho^\text{out}_w- \rho^\text{out}_x\big)\big)\big]K^\text{out}_{m_\text{out}}+\ldots+ \big[ U^1_1\big(\rho_w^\text{in}-\rho_x^\text{in}\big){U^1_1}^\dagger,\nonumber\\
&{U^1_2}^\dagger\ldots {U^\text{out}_{m_\text{out}}}^\dagger\big(\mathbbm{1}_\text{in, hidden}\otimes \big( \rho^\text{out}_w- \rho^\text{out}_x\big)\big)U^\text{out}_{m_\text{out}}\ldots U^1_2 \big] K^1_1\big)\nonumber\\
=&2i\sum\limits_{w,x\in V}[A]_{wx}\tr\big( M^\text{out}_{m_\text{out} \{ w,x\}}(s)K^\text{out}_{m_\text{out}}(s) +\ldots M^1_{1 \{ w,x\}}(s)K^1_1(s)\big),
\end{align*}
where
\begin{align*}
M_{j\{w,x\}}^l(s)=&\big[U_j^l(s)U_{j-1}^l(s)\dots U_1^1(s)\ \big(\big(\rho_w^\text{in}-\rho_x^\text{in}\big)\\
&\otimes\lvert 0\dots 0\rangle_1\langle 0\dots 0\rvert\big) {U_1^1}^\dagger(s)\dots{U_{j-1}^l}^\dagger(s){U_j^l}^\dagger(s),\\
&\hspace{15pt}{U_{j+1}^l}^\dagger(s)\dots {U_{m_\text{out}}^\text{out}}^\dagger(s)\big(\mathbbm{1}_\text{in,hidden}\otimes\big(\rho^\text{out}_w(s)-\rho^\text{out}_x(s)\big)\big)U_{m_\text{out}}^\text{out}(s)\dots U_{j+1}^l(s)\big].
\end{align*}
As a next step we expand $K^l_{j\text{,G}}(s)$ using Pauli matrices
\begin{equation*}
K^l_{j\text{,G}}(s)=\sum\limits_{\alpha_1,\ldots,\alpha_{m_{l-1}},\beta}K^l_{j,\alpha_1,\ldots,\alpha_{m_{l-1}},\beta}(s)\big( \sigma^{\alpha_1}\otimes\ldots\otimes\sigma^{\alpha_{m_{l-1}}}\otimes\sigma^\beta\big).
\end{equation*}
Since $d_s\mathcal{L}_\text{G}(s)$ is linear in the coefficients $K^l_{j,\alpha_1,\ldots,\alpha_{m_{l-1}},\beta}(s)$ we can solve 
\begin{eqnarray*}
\min\limits_{K^l_{j,\alpha_1,\ldots,\beta}}\big(\frac{d\mathcal{L}_\text{G}(s)}{ds}-\lambda\sum\limits_{\alpha_1,\ldots,\beta}K^l_{j,\alpha_1\ldots,\beta}(s)^2\big),
\end{eqnarray*} 
analogously to the proof in \cref{prop:DQNN_K} using the Lagrange multiplier $\lambda\in\mathbbm{R}$. Inserting the resulting coefficients
\begin{align*}
K^l_{j,\alpha_1\ldots,\beta}=&\frac{i}{\lambda}\sum\limits_{w,x \in V}[A]_{wx}\tr_{\alpha_1,\ldots,\alpha_{m_{l-1}},\beta}\Big(\tr_\text{rest}\big(M^l_{j\{w,x\}}\big)\big( \sigma^{\mu_1}\otimes\ldots\otimes\sigma^\eta\big)\Big)
\end{align*}
gives the desired expression for the update matrices, namely
\begin{align*}\label{Kusv}
K^l_{j\text{,G}}(s)=&{2^{m_{l-1}+1}i\eta}\sum\limits_{w,x\in V}[A]_{wx}\tr_\text{rest}\big(M^l_{j\{w,x\}}(s)\big),
\end{align*}
where $\eta=1/\lambda$ is the learning rate.
\end{proof}
Using the above-reasoned ansatz
\begin{equation*}
K^l_{j\text{,SV+G}}(s) = K^l_{j\text{,SV}}(s) + \gamma \cdot K^l_{j\text{,G}}(s)
\end{equation*}
we end up with the update matrix
\begin{equation*}
K^l_{j\text{,SV+G}}(s) = \frac{2^{m_{l-1}\eta}i}{S}\sum\limits_{x=1}^S\tr_\text{rest}\big(M^l_{j\{x\}}(s)\big) + \gamma 2^{m_{l-1}+1}i\eta\sum\limits_{w,x\in V}[A]_{wx}\tr_\text{rest}\big(M^l_{j\{w,x\}}(s)\big),
\end{equation*} 
for maximising the training loss $\mathcal{L}_\text{SV+G}=\mathcal{L}_\text{SV} + \gamma \mathcal{L}_{G}$, where $\gamma<0$ is the parameter to tune the influence of the graph structure.

\section{Classical simulation}
\label{sec:graph_classicalSim}

In the last section, we derived the update rule for the training loss, including the graph's information. Using these results and the algorithm described in \cref{sec:DQNN_trainingalgorithm} we can simulate the training on a quantum computer using QuTip \cite{Qutip}, a quantum toolbox in Python. The results are presented analogously to the training of the DQNN without graph structure in \cref{sec:DQNN_classical}. The code can be found at \cite{GithubKerstin}.

As already mentioned in \cref{sec:DQNN_classical} we are limited to small quantum systems due to the exponential scaling of the Hilbert space dimension with number of qubits. Despite this constraint, we will present three numerical studies following \cite{Beer2021}.

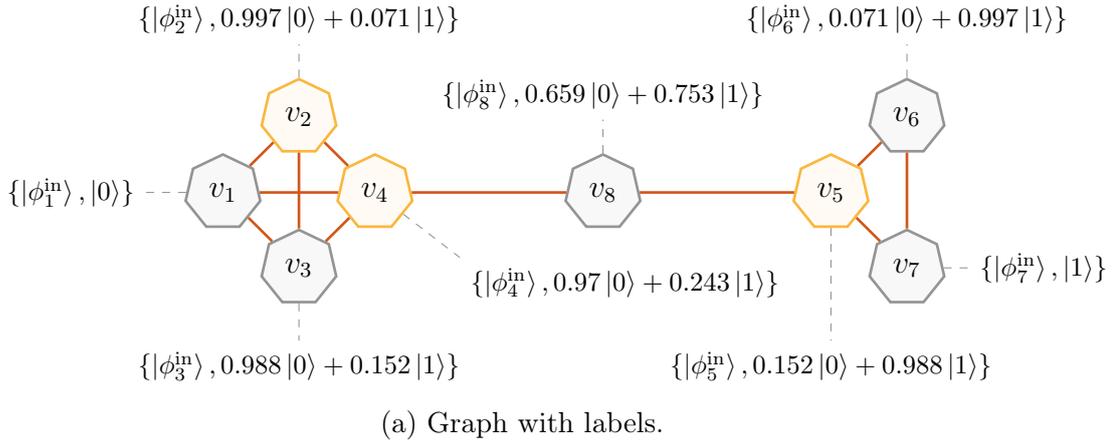
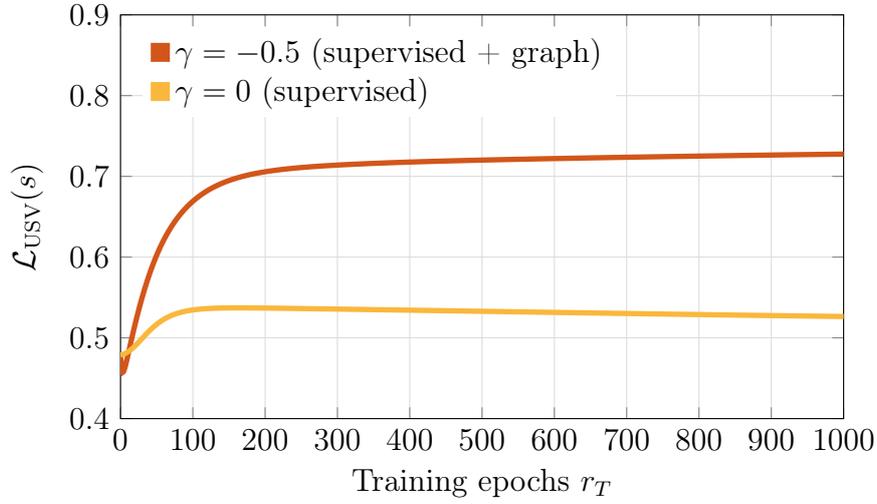
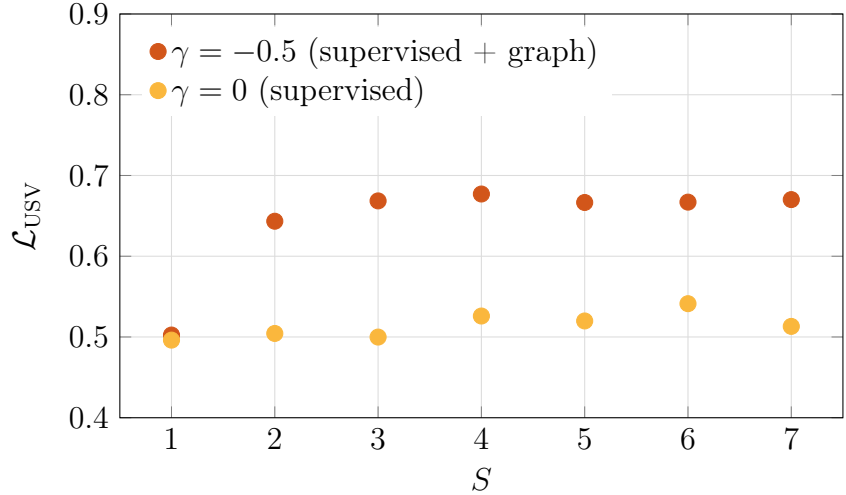
\begin{figure}
\centering
\begin{subfigure}{0.95\textwidth}\centering
\begin{tikzpicture}
\tikzstyle{place}=[circle,draw,fill=color0L,minimum size=6mm]
\node[vertex0] at (-1,0) (a1) {$v_1$};
\node[ font=\fontsize{10}{0}\selectfont] at ([shift={(180:2)}]a1) (a1l) {$\{\ket{\phi_1^\text{in}},\ket{0}\}$};
\draw[color0,dashed] (a1) -- (a1l.east);
\node[vertex2]   at (0,1)  (a2)  {$v_2$}; 
\node[ font=\fontsize{10}{0}\selectfont]  at ([shift={(90:1.3)}]a2) (a2l) {$\{\ket{\phi_2^\text{in}},0.997\ket{0}+0.071\ket{1}\}$};
\draw[color0,dashed] (a2) -- (a2l.south);
\node[vertex0]   at (0,-1)  (a3) {$v_3$};
\node[ font=\fontsize{10}{0}\selectfont]  at ([shift={(-90:1.3)}]a3) (a3l) {$\{\ket{\phi_3^\text{in}},0.988\ket{0}+0.152\ket{1}\}$};
\draw[color0,dashed] (a3) -- (a3l.north);
\node[vertex2] at (1,0) (a4) {$v_4$};
\node[ font=\fontsize{10}{0}\selectfont]  at ([shift={(-20:3.5)}]a4) (a4l) {$\{\ket{\phi_4^\text{in}},0.97\ket{0}+0.243\ket{1}\}$};
\draw[color0,dashed] (a4) -- (a4l.north west);
\node[vertex0]  at (4,0) (b1) {$v_8$};
\node[ font=\fontsize{10}{0}\selectfont]  at ([shift={(90:1.3)}]b1) (b1l) {$\{\ket{\phi_8^\text{in}},0.659\ket{0}+0.753\ket{1}\}$};
\draw[color0,dashed] (b1) -- (b1l.south);
\node[vertex2] at (7,0) (c1) {$v_5$};
\node[ font=\fontsize{10}{0}\selectfont]  at ([shift={(-90:2.3)}]c1) (c1l) {$\{\ket{\phi_5^\text{in}},0.152\ket{0}+0.988\ket{1}\}$};
\draw[color0,color0,dashed] (c1) -- (c1l.north);
\node[vertex0] at (8,1) (c2) {$v_6$};
\node[ font=\fontsize{10}{0}\selectfont]  at ([shift={(90:1.3)}]c2) (c2l) {$\{\ket{\phi_6^\text{in}},0.071\ket{0}+0.997\ket{1}\}$};
\draw[color0,dashed] (c2) -- (c2l.south);
\node[vertex0] at (8,-1) (c3) {$v_7$};
\node[ font=\fontsize{10}{0}\selectfont]  at ([shift={(0:1.8)}]c3) (c3l) {$\{\ket{\phi_7^\text{in}},\ket{1}\}$};
\draw[color0,dashed] (c3) -- (c3l.west);
\draw[line3] (a1) -- (a2);		
\draw[line3] (a1) -- (a3);
\draw[line3] (a1) -- (a4);
\draw[line3] (a2) -- (a3);
\draw[line3] (a2) -- (a4);
\draw[line3] (a3) -- (a4);
\draw[line3] (a4) --  (b1) -- (c1);
\draw[line3] (c1) -- (c2);		
\draw[line3] (c1) -- (c3);
\draw[line3] (c2) -- (c3);
\end{tikzpicture}
\subcaption{Graph with labels.}\label{fig:graph_connectedClustersA}
\end{subfigure}

\begin{subfigure}{0.95\textwidth}\centering
\begin{tikzpicture}
\begin{axis}[
xmin=0,   xmax=10,
ymin=0.4,   ymax=0.9,
width=.8\linewidth, 
height=.50\linewidth,
grid=major,
grid style={color0M},
xlabel= Training epochs $r_T$, 
xticklabels={0,0,100,200,300,400,500,600,700,800,900,1000},
ylabel=$\mathcal{L}_\text{USV}(s)$,legend pos=north west,legend cell align={left},legend style={draw=none,legend image code/.code={\filldraw[##1] (-.5ex,-.5ex) rectangle (0.5ex,0.5ex);}}]
\addplot[line width=2pt,  color=color3] table [x=step times epsilon, y=SsvGraphTestingUsv, col sep=comma] {numerics/connectedClustersRandomShuffled_8pairs3sv_3-1network_adjT0i65and1_g-0i5_delta0_lda1_ep0i01_plot.csv};
\addlegendentry{$\gamma=-0.5$ (supervised + graph)} 
\addplot[line width=2pt, color=color2] table [x=step times epsilon, y=SsvTestingUsv, col sep=comma] {numerics/connectedClustersRandomShuffled_8pairs3sv_3-1network_adjT0i65and1_g-0i5_delta0_lda1_ep0i01_plot.csv};
\addlegendentry{$\gamma=0$ (supervised)} 
\end{axis}
\end{tikzpicture}
\subcaption{Validation loss during training with $S=3$ supervised training pairs. }\label{fig:graph_connectedClustersB}
\end{subfigure}

\begin{subfigure}{0.95\textwidth}\centering
\begin{tikzpicture}
\begin{axis}[
xmin=0.5,   xmax=7.5,
ymin=0.4,   ymax=0.9,
width=.8\linewidth, 
height=.5\linewidth,
grid=major,
grid style={color0M},
xlabel= $S$, 
ylabel=$\mathcal{L}_\text{USV}$,legend pos=north west,legend cell align={left},legend style={draw=none}]
\addplot[color=color3, only marks, mark size=3 pt,mark phase=0] table [x=numberSupervisedPairsList, y=SsvGraphTestingUsvMeanList, col sep=comma] {numerics/connectedClustersRandomShuffled_8pairs_3-1network_adjT0i65and1_g-0i5_delta0_lda1_ep0i01_rounds1000_shots30_plotmean.csv};
\addlegendentry{$\gamma=-0.5$ (supervised + graph)}
\addplot[color=color2, only marks, mark size=3 pt,mark phase=0] table [x=numberSupervisedPairsList, y=SsvTestingUsvMeanList, col sep=comma] {numerics/connectedClustersRandomShuffled_8pairs_3-1network_adjT0i65and1_g-0i5_delta0_lda1_ep0i01_rounds1000_shots30_plotmean.csv};
\addlegendentry{$\gamma=0$ (supervised)} 
\end{axis}
\end{tikzpicture}
\subcaption{Validation loss after training with  different $S$ averaged over 30 sets. }\label{fig:graph_connectedClustersC}
\end{subfigure}
\caption{\textbf{Connected clusters.} This figure compares the training (b) and generalisation behaviour (c) of optimising a \protect\threeone DQNN ($r_T=1000$ epochs, $\epsilon=0.01$) with and without using the graph structure represented in (a).}
\end{figure}
\begin{figure}
\centering
\begin{subfigure}{0.95\textwidth}\centering


\begin{tikzpicture}[scale=1.4, rotate=15]
\node[vertex0] at (0,0) (a1) {$v_1$};
\node[ font=\fontsize{10}{0}\selectfont] at ([shift={(90:0.9)}]a1) (a1l) {$\{\ket{\phi_1^\text{in}},\ket{0}\}$};
\draw[color0,dashed] (a1) -- (a1l);
\node[vertex2]   at (1,0) (a2)  {$v_2$}; 
\node[ font=\fontsize{10}{0}\selectfont] at ([shift={(270:0.9)}]a2) (a2l) {$\{\ket{\phi_2^\text{in}},0.99\ket{0}+0.21\ket{1}\}$};
\draw[color0,dashed] (a2) -- (a2l);
\node[vertex2]   at (2,0) (a3) {$v_3$};
\node[ font=\fontsize{10}{0}\selectfont] at ([shift={(90:0.9)}]a3) (a3l) {$\{\ket{\phi_3^\text{in}},0.96\ket{0}+0.28\ket{1}\}$};					
\draw[color0,dashed] (a3) -- (a3l);
\node[vertex0] at (3,0) (a4) {$v_4$};
\node[ font=\fontsize{10}{0}\selectfont] at ([shift={(270:0.9)}]a4) (a4l) {$\{\ket{\phi_4^\text{in}},0.89\ket{0}+0.45\ket{1}\}$};					
\draw[color0,dashed] (a4) -- (a4l);
\node[vertex0]  at (4,0) (a5) {$v_5$};
\node[ font=\fontsize{10}{0}\selectfont] at ([shift={(90:0.9)}]a5) (a5l) {$\{\ket{\phi_5^\text{in}},0.78\ket{0}+0.62\ket{1}\}$};					
\draw[color0,dashed] (a5) -- (a5l);
\node[vertex0] at (5,0) (a6) {$v_6$};
\node[ font=\fontsize{10}{0}\selectfont] at ([shift={(270:0.9)}]a6) (a6l) {$\{\ket{\phi_6^\text{in}},0.62\ket{0}+0.78\ket{1}\}$};					
\draw[color0,dashed] (a6) -- (a6l);
\node[vertex0] at (6,0) (a7) {$v_7$};
\node[ font=\fontsize{10}{0}\selectfont] at ([shift={(90:0.9)}]a7) (a7l) {$\{\ket{\phi_7^\text{in}},0.45\ket{0}+0.89\ket{1}\}$};					
\draw[color0,dashed] (a7) -- (a7l);
\node[vertex2] at (7,0) (a8) {$v_8$};
\node[ font=\fontsize{10}{0}\selectfont] at ([shift={(270:0.9)}]a8) (a8l) {$\{\ket{\phi_8^\text{in}},0.27\ket{0}+0.96\ket{1}\}$};					
\draw[color0,dashed] (a8) -- (a8l);
\node[vertex0] at (8,0) (a9) {$v_9$};
\node[ font=\fontsize{10}{0}\selectfont] at ([shift={(90:0.9)}]a9) (a9l) {$\{\ket{\phi_9^\text{in}},0.12\ket{0}+0.99\ket{1}\}$};					
\draw[color0,dashed] (a9) -- (a9l);
\node[vertex0] at (9,0) (a10) {};
\node at (9,0) (a10t) {$v_{10}$};
\node[ font=\fontsize{10}{0}\selectfont] at ([shift={(270:0.9)}]a10) (a10l) {$\{\ket{\phi_{10}^\text{in}},\ket{1}\}$};					
\draw[color0,dashed] (a10) -- (a10l);
\draw[line3] (a1) -- (a2) -- (a3) -- (a4) -- (a5) -- (a6) -- (a7) -- (a8) -- (a9) -- (a10);

\end{tikzpicture}
\subcaption{Graph with labels.}\label{fig:graph_lineA}
\end{subfigure}

\begin{subfigure}{0.95\textwidth}\centering
\begin{tikzpicture}
\begin{axis}[
xmin=0,   xmax=10,
ymin=0.4,   ymax=0.9,
width=.8\linewidth, 
height=.5\linewidth,
grid=major,
grid style={color0M},
xlabel= Training epochs $r_T$, 
xticklabels={0,0,100,200,300,400,500,600,700,800,900,1000},
ylabel=$\mathcal{L}_\text{USV}(s)$,legend pos=south east,legend cell align={left},legend style={draw=none,legend image code/.code={\filldraw[##1] (-.5ex,-.5ex) rectangle (0.5ex,0.5ex);}}]
\addplot[line width=2pt, color=color3] table [x=step times epsilon, y=SsvGraphTestingUsv, col sep=comma] {numerics/lineOutputRandomShuffled_10pairs3sv_3-1network_adjT0i93and1_g-0i5_delta0_lda1_ep0i01_plot.csv};
\addlegendentry{$\gamma=-0.5$ (supervised + graph)}
\addplot[line width=2pt, color=color2] table [x=step times epsilon, y=SsvTestingUsv, col sep=comma] {numerics/lineOutputRandomShuffled_10pairs3sv_3-1network_adjT0i93and1_g-0i5_delta0_lda1_ep0i01_plot.csv};
\addlegendentry{$\gamma=0$ (supervised)} 
\end{axis}
\end{tikzpicture}
\subcaption{Validation loss during training with $S=3$ supervised training pairs. }\label{fig:graph_lineB}
\end{subfigure}

\begin{subfigure}{0.95\textwidth}\centering
\begin{tikzpicture}
\begin{axis}[
xmin=0.5,   xmax=7.5,
ymin=0.4,   ymax=0.9,
width=.95\linewidth, 
width=.8\linewidth, 
height=.5\linewidth,
grid=major,
grid style={color0M},
xlabel= $S$, 
ylabel=$\mathcal{L}_\text{USV}$,legend style={at={(0.97,0.5)},anchor=east},legend cell align={left},legend style={draw=none}]
\addplot[color=color3, only marks, mark size=3 pt,mark phase=0] table [x=numberSupervisedPairsList, y=SsvGraphTestingUsvMeanList, col sep=comma] {numerics/lineOutputRandomShuffled_10pairs_3-1network_adjT0i93and1_g-0i5_delta0_lda1_ep0i01_rounds1000_shots30_plotmean.csv};
\addlegendentry{$\gamma=-0.5$ (supervised + graph)}
\addplot[color=color2, only marks, mark size=3 pt,mark phase=0] table [x=numberSupervisedPairsList, y=SsvTestingUsvMeanList, col sep=comma] {numerics/lineOutputRandomShuffled_10pairs_3-1network_adjT0i93and1_g-0i5_delta0_lda1_ep0i01_rounds1000_shots30_plotmean.csv};
\addlegendentry{$\gamma=0$ (supervised)} 
\end{axis}
\end{tikzpicture}
\subcaption{Validation loss after training with  different $S$ averaged over 30 sets. }\label{fig:graph_lineC}
\end{subfigure}
\caption{\textbf{Line.} This illustration draws the comparison of the training (b) and generalisation behaviour (c) of optimising a \protect\threeone DQNN ($r_T=1000$ epochs, $\epsilon=0.01$) trained with and without using the graph structure (a).}
\end{figure}
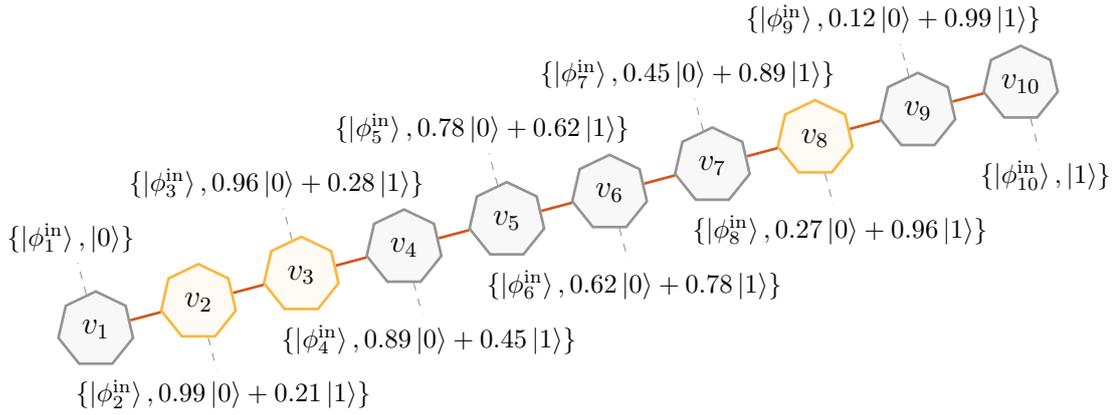
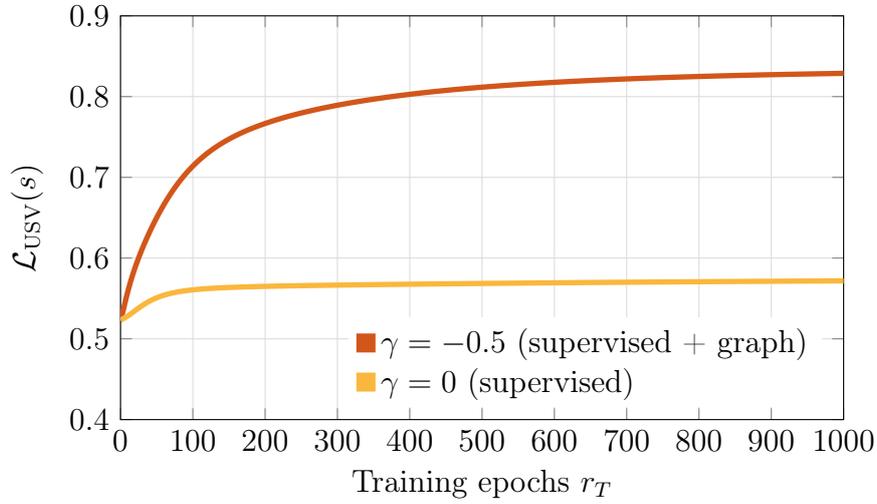
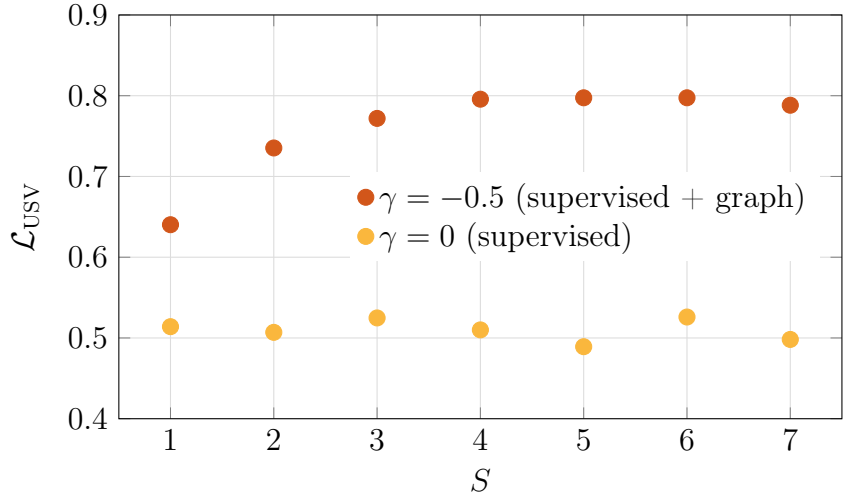

\subsection*{Example I: connected clusters}\label{sec:connectedclusters}

The first graph we study is a connected graph of degree $N=8$. The vertices have pairs of quantum states, including an input and output state, as labels. The first four vertices $v_1\hdots v_4$ form a cluster, as well as the vertices $v_5\hdots v_7$. Vertex $v_8$ connects the two clusters, see \cref{fig:graph_connectedClustersA}.  The output states are chosen in a way that connected vertices are associated to \emph{closer} output states in the sense of the Hilbert-Schmidt distance. The input states $\ket{\phi_x^\text{in}}$ are random $3$-qubit states built via a normal (Gaussian) distribution and are not linked to the graph structure in any way. Note that the coefficients depicted in \cref{fig:graph_connectedClustersA} are only recorded to three decimal places. 

The graph structure, saved in an adjacency matrix, is the foundation of the training loss $\mathcal{L}_{G}$. 
As introduced above, we also include a supervised learning part, packed in $\mathcal{L}_\text{SV}$. For this purpose, we assume that $S$ of the state pairs are used for training. In \cref{fig:graph_connectedClustersA} these $S=3$ supervised vertices are shaded. 

\cref{fig:graph_connectedClustersB} depicts the validation loss during $r_T=1000$ training epochs in the supervise case ($\gamma=0$) supervised and semi-supervised (i.e.\ supervised plus graph-based) case ($\gamma=-0.5$) with $S=3$. Although we do not reach a fidelity of $1$, the plot clarifies that the learning algorithm performs better when the graph structure is exploited.

We already observed that interpolation of the action on the supervised vertices on the unsupervised vertices is possible after training the network with three of the eight data pairs. We test how the number of supervised vertices $S$ affects the training process for a generalisation study. Therefore we randomly chose $S<N$ of the $N=8$ training pairs, trained the network in $r_T=1000$ steps and average the last values of the loss functions over $10$ completely independent training attempts. The results are depicted in \cref{fig:graph_connectedClustersC} and show that for $S\ge2$ the graph-based loss term optimises the training significantly.

\subsection*{Example II: Line}

In the second example, $N=10$ pairs of quantum states are the labels of vertices aligned in a line graph, see \cref{fig:graph_lineA}: the states were chosen to be -- according to the fidelity -- evenly spaced along a line between the endpoints associated with $\ket{0}$ and $\ket{1}$. In the same way, as in the connected-clusters data set, the input states $\ket{\phi_x^\text{in}}$ are random $3$-qubit states and the output states suit the graph structure. Again we assume three vertices to be supervised. 

In \cref{fig:graph_lineB} the validation loss is depicted. The training loss involving the graph structure ($\gamma=0$) yields a much higher validation loss function during the training. As in the example above the graph, structure supports the training. The maximum value of the fidelity is not reached, but after only $376$ training epochs, we exceed $\mathcal{L}_\text{USV}(s)=0.8$. 

In the generalisation study, see \cref{fig:graph_lineC}, it becomes clear that a validation loss of over $0.6$ with only $5$ of $10$ supervised vertices can be achieved when the graph structure is exploited.

\subsection*{Example III: Classic deep walk}

\begin{figure}
\centering
\begin{tikzpicture}
\begin{axis}[
xmin=0,   xmax=10.5,
ymin=0.0,   ymax=0.4,
width=.95\linewidth, 
width=.8\linewidth, 
height=.5\linewidth,
grid=major,
grid style={color0M},
xlabel= $S$, 
ylabel=$\mathcal{L}_\text{USV}(s)$,legend pos=north west,legend cell align={left},legend style={draw=none}]
\addplot[color=color3, only marks, mark size=3 pt,mark phase=0] table [x=numberSupervisedPairsList, y=SsvGraphTestingUsvMeanList, col sep=comma] {numerics/DeepWalk_32pairs_2-3network_g-0i5_delta0_lda1_ep0i01_rounds2000_shots5_plotmean.csv};
\addlegendentry{$\gamma=-0.5$ (supervised + graph)}
\addplot[color=color2, only marks, mark size=3 pt,mark phase=0] table [x=numberSupervisedPairsList, y=SsvTestingUsvMeanList, col sep=comma] {numerics/DeepWalk_32pairs_2-3network_g-0i5_delta0_lda1_ep0i01_rounds2000_shots5_plotmean.csv};
\addlegendentry{$\gamma=0$ (supervised)} 
\end{axis}
\end{tikzpicture}
\caption{\textbf{Deep walk.} The plot describes the generalisation behaviour of a \protect\twothree DQNN ($r_T=2000$ epochs, $\epsilon=0.01$) trained with and without using the graph structure of a graph with $32$ vertices produced by a classical deep walk. Each data point demonstrates an averaged over $5$ independent training attempts. }\label{fig:graph_walk}
\end{figure}
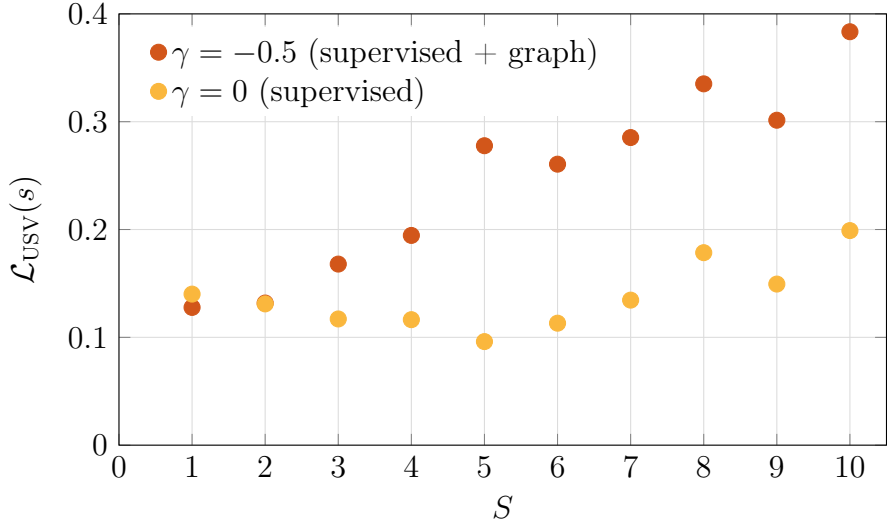

Using the quantum states forming connected clusters or a line graph we could, as described above, observe that including the graph structure of the output states in the training process can increase the training success. However, the input states, fed into the QNN, were random states. In the following we describe an ansatz, where both, the input and the desired output states, are related to a graph. 

We use a synthetic graph, motivated by the social distance attachment model in \cite{Talaga2019}. In contrast to the graphs used in this chapter before to every vertex not only labels are assigned, but also a embedding vector. The graph is build classically be randomly assigning labels to vertices such that the average number of labels per vertex is a constant. The number of possible labels is $8$, this means we can describe the labels of every vertex $x$ as a string of $8$ binary numbers. Our graph contains $32$ vertices. The average number of labels per vertex is $3$.

To create edges between the labelled vertices, let $d\left(\xi_{w}, \xi_{x}\right)$ denote the hamming distance between the label vectors $\xi_{w}$ and $\xi_{x}$ for vertices $w$ and $x$. We denote with $h$ the level of homophily and $b$ the characteristic distance. In graph theory, homophily describes how often linked vertices have the same labels or similar features \cite{Zhu2020}. Further we generate an edge between two vertices $w$ and $x$ with the probability
$$p_{wx}=\frac{1}{1+\left[b^{-1} d\left(\xi_{w}, \xi_{x}\right)\right]^{h}}.$$
As higher we choose the value of $h$, the chance of vertices with similar labels are more likely connected. 

As a last step we assign every vertex to an embedding vector $\vec{e}=\{\alpha,\beta,\gamma,\delta\}$. The embedding vector is computed using the method \emph{DeepWalk} \cite{Perozzi2014} using walks of length 1 and the number of walks per vertex as $10$. These embedding vectors will be used to construct input quantum states.

We now construct the corresponding quantum data, namely the quantum input and output states of the vertices. We adopt the graph structure and change the labels in the following way. Every vertex is assigned to a pair of quantum states: a $2$-qubit state $\ket{\phi_x^\text{E}}$ based on the embeddings and a $3$-qubit state $\ket{\phi_x^\text{L}}$ build based on the labels. $\ket{\phi_x^\text{E}}$ is the normalised version of the superposition $\alpha\ket{00}+\beta\ket{01}+\gamma\ket{10}+\delta\ket{11}$. To build $\ket{\phi_x^\text{L}}$ we link the states $\{\ket{000},\dots,\ket{111}\}$ to the $8$ possible labels. The output state assigned to a specific vertex is now built as the superposition of these basis states assigned to the vertex' labels. If for example a vertex has the labels $2$, $3$ and $8$ the output state would be a superposition of the states $\ket{001}$,$\ket{010}$ and $\ket{111}$. 

We use $\ket{\phi_x^\text{E}}$ as input states and $\ket{\phi_x^\text{L}}$ as desired output states to train DQNNs. The generalisation analysis in \cref{fig:graph_walk}, shows that with $3\le S\le 10$ supervised vertices, the validation loss is lower when ignoring the problem's graph structure. For larger values of $S$ both learning strategies are about equally good, which is discussed in the appendix, see \cref{fig:apdx_graph_walk} using a smaller amount of training rounds, since a study of the loss functions for all $S$ and $r_T=2000$ exceeds the computational power of the authors of this thesis. The evaluation of the loss functions is plotted in the appendix as well, see \cref{fig:apdx_graph_walk}.

\section{NISQ device implementation}
\label{sec:graph_quantumAlg}
\begin{figure}
\centering
\begin{subfigure}{0.95\textwidth}
\input{numerics/ClustersQC}
\subcaption{Training with connected clusters data set.}\label{fig:graph_QC_A}
\end{subfigure}	
\begin{subfigure}{0.95\textwidth}\centering
\input{numerics/LineQC}
\subcaption{Training with line data set.}\label{fig:graph_QC_B}
\end{subfigure}
\caption{\textbf{Training a \DQNNNISQ  using graph-structured quantum data.} The figures depict the validation loss during training a \protect\threeone \DQNNNISQ in $500$ epochs with the datasets connected clusters (a), see \cref{fig:graph_connectedClustersA}, and line (b), see \cref{fig:graph_lineA}. In both training attempts $S=3$ training pairs are supervised.}
\label{fig:graph_QC}
\end{figure}
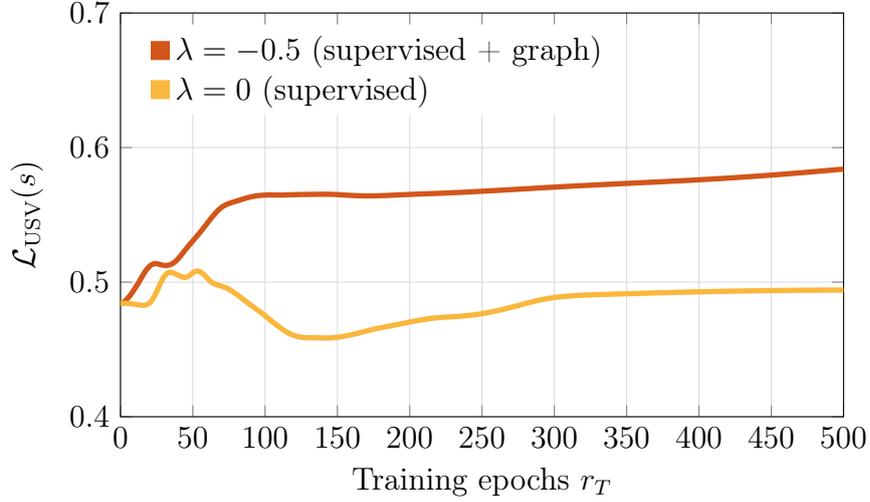
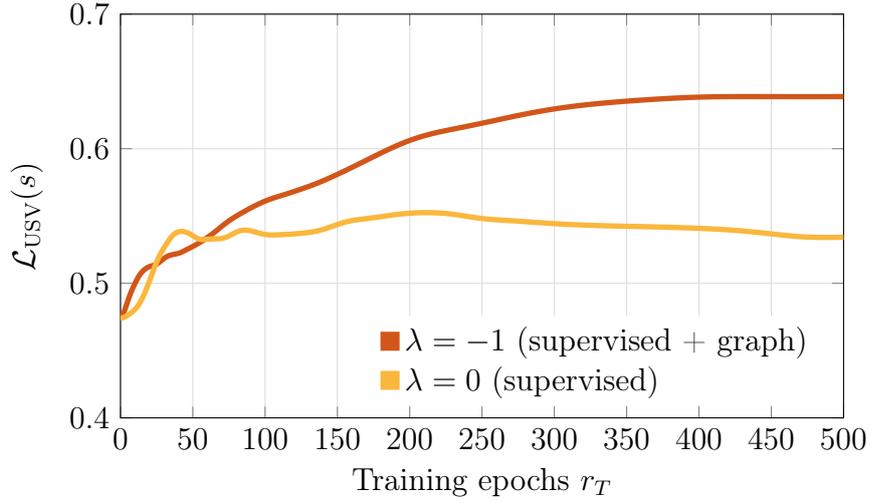

In the last section, we discussed training a DQNN using the newly introduced training loss function $\mathcal{L}_\text{SV+G}$. The results were done using the classical simulation presented in \cref{sec:DQNN_classical}. We will now test the behaviour of this new loss function using the NISQ device implementation, denoted with \DQNNNISQ and discussed in \cref{sec:DQNN_quantumalg}.

As in \cref{sec:DQNN_quantumalg} explained, the update of the parameter vector ${\omega}_{s}$ is done in classical manner using the gradient of the training loss, i.e.\ ${\omega}_{s+1} = {\omega}_{s} + {d \omega}$, where ${d\omega} = \eta {\nabla} \mathcal{L}_\text{SV+G} \left({\omega}_t\right)$ and
\begin{equation*}
\nabla _k \mathcal{L}_\text{SV+G} \left({\omega}_s\right) = \frac{\mathcal{L}_\text{SV+G}\left({\omega}_s + \epsilon{e}_k\right) - \mathcal{L}_\text{SV+G}\left({\omega}_s - \epsilon{e}_k\right)}{2\epsilon} + \mathcal{O}\left(\epsilon^2\right).
\end{equation*}
$\mathcal{L}_\text{SV+G}$ function includes both the fidelity and the Hilbert-Schmidt distance. Therefore, additionally to the implementation presented in \cref{sec:DQNN_quantumalg}, we have to evaluate the Hilbert-Schmidt distance. This can be done by three evaluations of the $\swap$ test, since $d_{\text{HS}}(\rho,\sigma) = \text{tr}(\rho^2) -2\text{tr}(\rho\sigma) +\text{tr}(\sigma^2)$
and $\text{tr}(\rho\sigma)={p}\cdot {c}$ can be measured with the $\swap$ test \cite{Cincio2018}, explained in \cref{subsec:DQNN_swap}.

The training of the \DQNNNISQ using the graph-structured quantum data is presented in \cref{fig:graph_QC_A} (connected clusters) and \cref{fig:graph_QC_A} (line). Despite the noise levels, we can observe that using the graph structure of the problem increases the validation loss reached after about $500$ epochs and seems to lead to a more stable training. 

The plots in \cref{fig:graph_QC} are based on numerics presented in the work of \cite{Struckmann2021}. We point the reader to this source for more experiments with graph-structured quantum data and NISQ devices and a comparison of the behaviour of \DQNNNISQ and QAOA with the semi-supervised loss function $\mathcal{L}_\text{SV+G}$. 

In this chapter we presented a variation of the DQNN training algorithm presented in \cref{chapter:DQNN} by including knowledge of the graph structure of the training data into the learning process. We could see that this extension can increase the reached validation loss. In the next chapter, we will extend the DQNN ansatz in another direction and follow a generative adversarial approach where two DQNNs, a generator and a discriminator model, are trained in a competitive manner.

%% file: numerics/ClustersQC.tex
\centering \begin{tikzpicture}
	\begin{axis}[
		xmin=0,   xmax=500,
		ymin=0.4,   ymax=0.7,
		width=.8\linewidth, 
		height=.50\linewidth,
		grid=major,
		grid style={color0M},
xlabel=Training epochs $r_T$, 
		ylabel=$\mathcal{L}_\text{USV}(s)$,legend pos=north west,legend cell align={left},legend style={draw=none,legend image code/.code={\filldraw[##1] (-.5ex,-.5ex) rectangle (0.5ex,0.5ex);}}]
		\addplot[line width=2pt, color=color3]table {%
			0 0.484223622154523
			1 0.484421241695101
			2 0.484756662628491
			3 0.485762331136139
			4 0.486943975082936
			5 0.488094933912217
			6 0.489291305539609
			7 0.490635650429775
			8 0.492125086630787
			9 0.493703162663613
			10 0.495327449330428
			11 0.496991830005565
			12 0.49872396730097
			13 0.500542717657639
			14 0.502425063974995
			15 0.504307881492237
			16 0.506116888259631
			17 0.507792659452144
			18 0.509298385898313
			19 0.510611108735085
			20 0.511711367065116
			21 0.512581151168806
			22 0.513208426508949
			23 0.513592201010195
			24 0.513745073893427
			25 0.513693983340103
			26 0.513482537002404
			27 0.513178743706485
			28 0.512870005884274
			29 0.512619970743567
			30 0.512455479269232
			31 0.51239024473137
			32 0.512433624137434
			33 0.512589956220094
			34 0.512866509048284
			35 0.513278725854187
			36 0.513840286897552
			37 0.514555551143934
			38 0.515421943313918
			39 0.516434257236113
			40 0.51758048301847
			41 0.518833931721938
			42 0.520157525867574
			43 0.521515861648883
			44 0.522882035464651
			45 0.524237776299723
			46 0.525571564391327
			47 0.526878055081873
			48 0.528158037875802
			49 0.529416879723156
			50 0.530662073176876
			51 0.531901771936599
			52 0.533144510297895
			53 0.534398921031698
			54 0.535672706823233
			55 0.536971220590711
			56 0.538296431170206
			57 0.539646729592171
			58 0.541017514448683
			59 0.542402133341285
			60 0.54379269101804
			61 0.545180435770545
			62 0.546555723070725
			63 0.547907753712847
			64 0.549224344341019
			65 0.55049197098081
			66 0.551696286074808
			67 0.552823209383003
			68 0.553860471197291
			69 0.55479922004272
			70 0.555635216189084
			71 0.556369317052567
			72 0.557007252025332
			73 0.557558872318922
			74 0.558037102122131
			75 0.558456762545267
			76 0.558833348117514
			77 0.55918177645835
			78 0.559515160953641
			79 0.559843754849069
			80 0.560174283179741
			81 0.560509822550762
			82 0.56085022347317
			83 0.561192913121572
			84 0.56153386130015
			85 0.561868521118678
			86 0.562192598160087
			87 0.562502537410025
			88 0.562795679992191
			89 0.56307014460017
			90 0.563324581664159
			91 0.563557966148765
			92 0.563769526250754
			93 0.563958802004559
			94 0.564125755924928
			95 0.564270848355347
			96 0.564395025483086
			97 0.564499608935617
			98 0.56458610339849
			99 0.56465596352009
			100 0.564710393148808
			101 0.564750269438317
			102 0.564776258566763
			103 0.564789113383789
			104 0.564790054092709
			105 0.564781083176694
			106 0.564765101436007
			107 0.564745760276811
			108 0.564727071328269
			109 0.564712866228328
			110 0.564706239779922
			111 0.564709112958315
			112 0.564722019832436
			113 0.564744163149112
			114 0.564773713707403
			115 0.564808269001299
			116 0.564845353822461
			117 0.564882846980824
			118 0.564919249784114
			119 0.564953760607553
			120 0.564986170888441
			121 0.565016639515851
			122 0.565045427234888
			123 0.565072675332836
			124 0.565098291893733
			125 0.565121969395869
			126 0.565143312302347
			127 0.565162018624778
			128 0.565178046595626
			129 0.56519170822823
			130 0.565203658638575
			131 0.56521478280866
			132 0.565226010194161
			133 0.565238105412277
			134 0.565251486732195
			135 0.565266113439668
			136 0.565281462482665
			137 0.565296591203529
			138 0.565310263931786
			139 0.565321110715088
			140 0.565327787161978
			141 0.565329112021606
			142 0.565324168847585
			143 0.565312366064225
			144 0.565293454804972
			145 0.565267507199092
			146 0.565234861182418
			147 0.565196042220946
			148 0.565151676735264
			149 0.565102414561008
			150 0.565048876547605
			151 0.564991637547764
			152 0.564931245488539
			153 0.564868266564798
			154 0.564803338389436
			155 0.564737210229147
			156 0.564670753491173
			157 0.564604935201166
			158 0.564540759070816
			159 0.564479188817557
			160 0.56442107330773
			161 0.56436709146603
			162 0.564317727792895
			163 0.564273279734886
			164 0.564233889631734
			165 0.564199589204541
			166 0.56417034440911
			167 0.564146091937718
			168 0.564126763569786
			169 0.564112298782195
			170 0.564102648262124
			171 0.564097771192724
			172 0.564097628326995
			173 0.564102172091545
			174 0.564111334973195
			175 0.564125018148732
			176 0.564143083004001
			177 0.564165348022035
			178 0.564191592122586
			179 0.564221563283404
			180 0.564254989162969
			181 0.564291585578718
			182 0.564331059670711
			183 0.564373107115643
			184 0.564417405770145
			185 0.564463610236029
			186 0.564511352035184
			187 0.564560248196241
			188 0.564609917870178
			189 0.564660003467903
			190 0.564710191053592
			191 0.564760224988878
			192 0.564809913852278
			193 0.56485912747679
			194 0.564907787359426
			195 0.56495585385028
			196 0.565003313264895
			197 0.565050166879268
			198 0.56509642246388
			199 0.565142088238807
			200 0.565187169064409
			201 0.56523166501492
			202 0.565275572685901
			203 0.565318889274497
			204 0.565361618656627
			205 0.565403777779875
			206 0.565445401259469
			207 0.565486542477718
			208 0.565527270680433
			209 0.565567665044039
			210 0.565607807781306
			211 0.565647778561099
			212 0.565687651770379
			213 0.565727496836666
			214 0.565767380592917
			215 0.565807370052213
			216 0.565847534157886
			217 0.565887943858571
			218 0.565928670744577
			219 0.56596978501703
			220 0.566011353555013
			221 0.566053438439349
			222 0.566096095827596
			223 0.566139374867808
			224 0.566183316486752
			225 0.56622795223595
			226 0.566273303640524
			227 0.566319382453434
			228 0.566366191865809
			229 0.566413728271739
			230 0.566461982921604
			231 0.566510942902712
			232 0.566560591316133
			233 0.56661090702797
			234 0.566661864661612
			235 0.566713435391513
			236 0.566765588655451
			237 0.566818294372887
			238 0.566871524938537
			239 0.566925256322268
			240 0.566979468000388
			241 0.567034141942483
			242 0.567089261216904
			243 0.567144808805869
			244 0.567200766974228
			245 0.567257117195352
			246 0.567313840405652
			247 0.567370917337021
			248 0.5674283288174
			249 0.567486056090749
			250 0.567544081256572
			251 0.567602387833799
			252 0.567660961296685
			253 0.567719789343229
			254 0.567778861724175
			255 0.567838169659147
			256 0.567897705078115
			257 0.567957460017528
			258 0.568017426410199
			259 0.568077596288314
			260 0.5681379622004
			261 0.568198517551371
			262 0.56825925665289
			263 0.568320174457219
			264 0.568381266120499
			265 0.568442526600138
			266 0.568503950417159
			267 0.568565531575047
			268 0.568627263523275
			269 0.56868913905211
			270 0.568751150095065
			271 0.568813287523186
			272 0.56887554106028
			273 0.568937899397749
			274 0.569000350477782
			275 0.569062881820479
			276 0.56912548075596
			277 0.569188134493067
			278 0.569250830061194
			279 0.569313554230971
			280 0.569376293511944
			281 0.569439034254653
			282 0.569501762805435
			283 0.569564465630212
			284 0.569627129355895
			285 0.569689740745579
			286 0.56975228667436
			287 0.569814754170175
			288 0.569877130533295
			289 0.569939403488311
			290 0.570001561297842
			291 0.570063592794087
			292 0.570125487341168
			293 0.570187234785589
			294 0.570248825452434
			295 0.570310250202823
			296 0.570371500515798
			297 0.57043256853379
			298 0.570493447032029
			299 0.570554129322792
			300 0.570614609149143
			301 0.570674880629974
			302 0.570734938285303
			303 0.570794777122103
			304 0.570854392729004
			305 0.5709137813315
			306 0.570972939790382
			307 0.571031865560186
			308 0.571090556638456
			309 0.571149011525521
			310 0.571207229192598
			311 0.571265209043546
			312 0.571322950862433
			313 0.571380454757828
			314 0.57143772112764
			315 0.571494750662641
			316 0.571551544384839
			317 0.571608103694522
			318 0.571664430394036
			319 0.571720526671939
			320 0.57177639505733
			321 0.571832038372447
			322 0.571887459709943
			323 0.571942662442133
			324 0.571997650247466
			325 0.572052427130017
			326 0.572106997415825
			327 0.572161365727687
			328 0.572215536953566
			329 0.572269516224183
			330 0.572323308904472
			331 0.572376920591513
			332 0.572430357107946
			333 0.57248362448639
			334 0.572536728950416
			335 0.57258967690204
			336 0.572642474920975
			337 0.572695129771138
			338 0.572747648403571
			339 0.57280003794727
			340 0.572852305688569
			341 0.572904459048265
			342 0.572956505566821
			343 0.573008452901079
			344 0.5730603088266
			345 0.573112081235315
			346 0.57316377812184
			347 0.573215407560312
			348 0.573266977680227
			349 0.573318496649696
			350 0.573369972668477
			351 0.573421413966301
			352 0.573472828799595
			353 0.573524225442947
			354 0.573575612177178
			355 0.573626997278901
			356 0.57367838901477
			357 0.573729795639225
			358 0.573781225391473
			359 0.57383268648842
			360 0.573884187114215
			361 0.573935735410569
			362 0.573987339472027
			363 0.574039007346981
			364 0.574090747041284
			365 0.574142566520113
			366 0.574194473705997
			367 0.574246476474566
			368 0.574298582651545
			369 0.574350800013253
			370 0.574403136289687
			371 0.574455599167007
			372 0.574508196286692
			373 0.574560935241424
			374 0.574613823570409
			375 0.574666868757276
			376 0.574720078231745
			377 0.574773459373589
			378 0.574827019516291
			379 0.57488076594874
			380 0.57493470591524
			381 0.574988846615324
			382 0.575043195204302
			383 0.575097758794059
			384 0.575152544452735
			385 0.575207559202499
			386 0.57526281001614
			387 0.575318303814252
			388 0.575374047464415
			389 0.575430047782217
			390 0.575486311532587
			391 0.5755428454298
			392 0.575599656135647
			393 0.575656750256722
			394 0.575714134342213
			395 0.57577181488285
			396 0.575829798310445
			397 0.575888090996836
			398 0.575946699251496
			399 0.576005629318064
			400 0.576064887370727
			401 0.576124479511122
			402 0.57618441176572
			403 0.57624469008304
			404 0.57630532033021
			405 0.576366308289035
			406 0.576427659652161
			407 0.576489380019813
			408 0.576551474896875
			409 0.576613949689567
			410 0.576676809701082
			411 0.576740060126122
			412 0.576803706044976
			413 0.576867752417865
			414 0.576932204079809
			415 0.57699706573574
			416 0.577062341955281
			417 0.577128037166953
			418 0.577194155651968
			419 0.577260701538009
			420 0.577327678793168
			421 0.577395091219913
			422 0.577462942448762
			423 0.577531235931582
			424 0.57759997493472
			425 0.57766916253233
			426 0.577738801600027
			427 0.577808894808722
			428 0.577879444618295
			429 0.577950453270964
			430 0.578021922784527
			431 0.578093854945868
			432 0.578166251304986
			433 0.578239113169566
			434 0.578312441599857
			435 0.578386237403699
			436 0.578460501131707
			437 0.578535233072833
			438 0.578610433250475
			439 0.578686101419153
			440 0.578762237061663
			441 0.578838839386632
			442 0.578915907326573
			443 0.578993439536619
			444 0.579071434394099
			445 0.579149889998935
			446 0.579228804174731
			447 0.579308174470428
			448 0.57938799816257
			449 0.579468272258284
			450 0.579548993499102
			451 0.579630158365602
			452 0.579711763082764
			453 0.579793803625912
			454 0.579876275727211
			455 0.579959174882765
			456 0.580042496360336
			457 0.580126235207613
			458 0.580210386260909
			459 0.580294944154121
			460 0.580379903327919
			461 0.5804652580391
			462 0.580551002370094
			463 0.58063713023847
			464 0.580723635406267
			465 0.580810511488971
			466 0.580897751964028
			467 0.580985350178838
			468 0.581073299358146
			469 0.581161592610728
			470 0.581250222935186
			471 0.581339183224718
			472 0.581428466270754
			473 0.581518064765417
			474 0.581607971302744
			475 0.581698178378599
			476 0.581788678389182
			477 0.581879463628071
			478 0.581970526281816
			479 0.582061858424114
			480 0.582153452008622
			481 0.58224529886046
			482 0.582337390666447
			483 0.582429718964161
			484 0.582522275129987
			485 0.582615050366313
			486 0.582708035688105
			487 0.582801221909019
			488 0.582894599627272
			489 0.582988159211506
			490 0.583081890786905
			491 0.583175784221857
			492 0.583269829115456
			493 0.583364014786105
			494 0.583458330261493
			495 0.583552764270214
			496 0.583647305235324
			497 0.583741941270094
			498 0.583836660176209
			499 0.58393144944459
			500 0.584026296259023
		};
		\addlegendentry{$\lambda=-0.5$ (supervised + graph)} 
		\addplot[line width=2pt, color=color2]  table {%
			0 0.484223622154523
			1 0.484270824795874
			2 0.484282247421946
			3 0.484270303888327
			4 0.484236454483699
			5 0.484180247875738
			6 0.484101618158856
			7 0.484001573072666
			8 0.483882129982793
			9 0.483745891328957
			10 0.483595640347811
			11 0.483434403924753
			12 0.483266429368539
			13 0.48309931784231
			14 0.482947125604645
			15 0.482833691014291
			16 0.482794896570317
			17 0.482878303422269
			18 0.48313917036516
			19 0.483633418755489
			20 0.48440933900192
			21 0.485499683608091
			22 0.486915600136625
			23 0.488643999338786
			24 0.490648296765139
			25 0.492870139965584
			26 0.495229946374741
			27 0.497627121676603
			28 0.499944742182984
			29 0.502063590009103
			30 0.5038829058654
			31 0.505335038102111
			32 0.506385628893404
			33 0.507030026838759
			34 0.507297345763132
			35 0.5072544385125
			36 0.506994132634044
			37 0.506606731810888
			38 0.506154004675943
			39 0.505665002932652
			40 0.505152076723056
			41 0.504632669935907
			42 0.5041458053838
			43 0.503755385569621
			44 0.503538244912803
			45 0.503561738050502
			46 0.503860646771315
			47 0.504424087735622
			48 0.505193582144721
			49 0.506066051902263
			50 0.506909555636215
			51 0.507599673556797
			52 0.508051005230105
			53 0.508219802496413
			54 0.508099085688884
			55 0.507715083846697
			56 0.507103760839027
			57 0.506291273521007
			58 0.505308260005977
			59 0.504202985834657
			60 0.503036191695445
			61 0.501885262736526
			62 0.500837770456353
			63 0.499955411694993
			64 0.499255727623707
			65 0.498720229983136
			66 0.49829953970777
			67 0.49793599753099
			68 0.497594618889661
			69 0.497260727179011
			70 0.496926878439287
			71 0.496585292411567
			72 0.496213427833282
			73 0.495779324421727
			74 0.495264214493159
			75 0.494670024660189
			76 0.494018694097729
			77 0.49333927491095
			78 0.492645677426265
			79 0.491932566005197
			80 0.491185639437244
			81 0.490397554166048
			82 0.4895793923463
			83 0.488753682999009
			84 0.487937690770817
			85 0.487132473679807
			86 0.486325594866018
			87 0.485505783065673
			88 0.484675016219937
			89 0.483846173091966
			90 0.483031355112244
			91 0.482231841559214
			92 0.481440092851166
			93 0.480650595208659
			94 0.479865389604331
			95 0.479090076278644
			96 0.478322838961094
			97 0.477549788004649
			98 0.476753308262361
			99 0.475923098224942
			100 0.475063524366475
			101 0.474187104797095
			102 0.47330429141442
			103 0.472419447109444
			104 0.471533057018762
			105 0.470649776467636
			106 0.469776921431328
			107 0.46892076547802
			108 0.468081784326018
			109 0.46725743201502
			110 0.466449117916977
			111 0.465663281258771
			112 0.464908973793373
			113 0.464190716159662
			114 0.46351017119852
			115 0.462869131242305
			116 0.462273412151021
			117 0.461729356671829
			118 0.461239165820465
			119 0.460800244881752
			120 0.460408218145649
			121 0.460061346171698
			122 0.459758373651543
			123 0.459497590424935
			124 0.459274922855961
			125 0.459088935587714
			126 0.458939770189428
			127 0.45882927534408
			128 0.458754166624535
			129 0.45870907233863
			130 0.458685077794044
			131 0.458677017043618
			132 0.458676898016514
			133 0.458680439751391
			134 0.45867832054973
			135 0.458670906164753
			136 0.458652687363821
			137 0.458632177960523
			138 0.458601452165522
			139 0.458575835053732
			140 0.458541269230513
			141 0.458528780391652
			142 0.458507302774662
			143 0.458531227631492
			144 0.458528920167832
			145 0.458606230032334
			146 0.45861962971602
			147 0.458760154852197
			148 0.458797058757785
			149 0.458963941777174
			150 0.459073295449708
			151 0.459207811258441
			152 0.459404489389493
			153 0.459529317918941
			154 0.459739155297692
			155 0.4599277376467
			156 0.460103174974497
			157 0.460350632737877
			158 0.46054586349325
			159 0.460779608677404
			160 0.461039984080425
			161 0.461258426104269
			162 0.461533871592702
			163 0.461792650878399
			164 0.462039041626186
			165 0.462331913097663
			166 0.462586785364388
			167 0.462858135073356
			168 0.463146872859398
			169 0.463398981500309
			170 0.463682962818962
			171 0.46395485098016
			172 0.464207928170137
			173 0.464488360502947
			174 0.464737281736271
			175 0.464986942391604
			176 0.465242562604829
			177 0.465463357096593
			178 0.465697695407224
			179 0.465918192868
			180 0.466121441527193
			181 0.466342610771489
			182 0.466543523643422
			183 0.466751522022658
			184 0.46696924002245
			185 0.467171206021695
			186 0.467391220901976
			187 0.467604291100407
			188 0.467812142323759
			189 0.468032089779527
			190 0.468235540876617
			191 0.468446243762
			192 0.468657508740453
			193 0.468859846533669
			194 0.469076569335714
			195 0.469286557997455
			196 0.469502005547352
			197 0.469725434882461
			198 0.469940354038683
			199 0.47016447224277
			200 0.470381158282795
			201 0.470592643679585
			202 0.470805428690835
			203 0.471004364378328
			204 0.471205224357622
			205 0.471399567547398
			206 0.471588163789701
			207 0.471779775688382
			208 0.471961283979881
			209 0.472143502693845
			210 0.472318690019582
			211 0.472484269004637
			212 0.472646490716858
			213 0.472794212771498
			214 0.472936790473254
			215 0.473069361632355
			216 0.473192067223753
			217 0.473311220702556
			218 0.473419135673499
			219 0.473524212258898
			220 0.473621413068452
			221 0.473711734394536
			222 0.473798480425481
			223 0.473875859150255
			224 0.473951266303789
			225 0.474020132111643
			226 0.474086373335846
			227 0.47415180267429
			228 0.474214172390205
			229 0.474279571031123
			230 0.474343412079011
			231 0.474410361070591
			232 0.474478768857732
			233 0.474548729041954
			234 0.474623270833598
			235 0.474699251173907
			236 0.474781700849843
			237 0.474867755586065
			238 0.474960585284208
			239 0.475059978515884
			240 0.475165403225012
			241 0.475279223033072
			242 0.475398401109223
			243 0.475526153099472
			244 0.475659502224103
			245 0.475800616509643
			246 0.475948465604488
			247 0.476103159555744
			248 0.476265786004739
			249 0.476434541556086
			250 0.476611632785859
			251 0.47679435740407
			252 0.476984813464065
			253 0.477180608842485
			254 0.47738298784376
			255 0.477590723462041
			256 0.477803971723669
			257 0.478022834105204
			258 0.478246442273744
			259 0.47847586807281
			260 0.478709412485046
			261 0.478948629861268
			262 0.479191324197223
			263 0.479439274504082
			264 0.479690142142228
			265 0.479945855806559
			266 0.480204111993532
			267 0.48046698357502
			268 0.480732109531303
			269 0.481001722992008
			270 0.481273122695474
			271 0.48154887370927
			272 0.481825547821787
			273 0.482106468600602
			274 0.48238692470576
			275 0.482671803636402
			276 0.482954106971282
			277 0.483241676228507
			278 0.483523268477778
			279 0.483812319845007
			280 0.484089525819723
			281 0.484379280423409
			282 0.484646532714168
			283 0.484937853802265
			284 0.485186360219128
			285 0.485484409039332
			286 0.485699500813913
			287 0.486017453851381
			288 0.486177315155297
			289 0.486530151409479
			290 0.486632881551415
			291 0.486979849024774
			292 0.487115989704577
			293 0.487337663113884
			294 0.487588574379931
			295 0.487689727393294
			296 0.487955735482603
			297 0.488091413713726
			298 0.488235088776238
			299 0.488457498951612
			300 0.488537287218995
			301 0.488715708347783
			302 0.488863744411513
			303 0.488938848250352
			304 0.489115454571263
			305 0.489199097434071
			306 0.489292674543337
			307 0.489435651349223
			308 0.489483472883161
			309 0.489593914537591
			310 0.489689176643101
			311 0.48973081903469
			312 0.489841912827278
			313 0.489894577958116
			314 0.489949174894956
			315 0.490042254787024
			316 0.490071299483953
			317 0.490142320311354
			318 0.490206499152837
			319 0.490234530756381
			320 0.49031146634294
			321 0.490348744886186
			322 0.490390684411263
			323 0.490457299567098
			324 0.490480988040826
			325 0.490537324184783
			326 0.49058284434308
			327 0.490609536838553
			328 0.490668243147735
			329 0.490694782204659
			330 0.490733796199855
			331 0.490780412162387
			332 0.490801224174423
			333 0.490848872784934
			334 0.490878314720541
			335 0.490907230098522
			336 0.490951322315236
			337 0.490972026177587
			338 0.491012046218549
			339 0.491043887134371
			340 0.491069721851842
			341 0.491111567563235
			342 0.491134726189647
			343 0.491171384745646
			344 0.491204584282214
			345 0.491230745678418
			346 0.491270687204919
			347 0.491295738030419
			348 0.491330936193465
			349 0.491363594574235
			350 0.491390549277652
			351 0.491428214160887
			352 0.491453752961709
			353 0.491488170976946
			354 0.491519079792873
			355 0.491547105420313
			356 0.491582580864045
			357 0.491608463715346
			358 0.491642849269271
			359 0.491672043014897
			360 0.491701872057536
			361 0.49173530572346
			362 0.491762192544837
			363 0.491796520292981
			364 0.491824371459297
			365 0.491856087945133
			366 0.491887025152
			367 0.491915508100792
			368 0.491948521971356
			369 0.491975750760158
			370 0.492008308378833
			371 0.492036612360081
			372 0.4920669365444
			373 0.492097193231316
			374 0.492125269605251
			375 0.492156703848856
			376 0.492183760201566
			377 0.492214891973048
			378 0.492242263412486
			379 0.492271954429574
			380 0.492300279857015
			381 0.492328207607092
			382 0.492357301734361
			383 0.492383828600635
			384 0.492413028633414
			385 0.492438794290218
			386 0.492467403241039
			387 0.492492964414615
			388 0.492520531981023
			389 0.492546194090738
			390 0.492572577986649
			391 0.492598401561795
			392 0.492623683354965
			393 0.492649578535095
			394 0.492673935284115
			395 0.492699764145296
			396 0.492723362966227
			397 0.492749010270956
			398 0.492771949359631
			399 0.492797359957778
			400 0.492819649142554
			401 0.492844847467477
			402 0.49286640587817
			403 0.492891514109175
			404 0.492912158737609
			405 0.492937429906998
			406 0.492956830364048
			407 0.49298271975295
			408 0.493000290287541
			409 0.493027606414596
			410 0.493042282434579
			411 0.493072496559414
			412 0.493082286265426
			413 0.493118158135003
			414 0.493119252411278
			415 0.493166063002318
			416 0.493151156088913
			417 0.493218853398595
			418 0.493174738338874
			419 0.493279619008127
			420 0.493188963612771
			421 0.493342591593254
			422 0.493213564858397
			423 0.493371306084978
			424 0.493292373398484
			425 0.493344425903027
			426 0.493397861449525
			427 0.493330288237044
			428 0.493450169285064
			429 0.493388729287748
			430 0.493434821227202
			431 0.493481203770551
			432 0.493429423912185
			433 0.493523101338365
			434 0.493488936223938
			435 0.493511569243986
			436 0.493565141530881
			437 0.493521212784694
			438 0.493591546781536
			439 0.493582306836507
			440 0.493585412195875
			441 0.493641612900081
			442 0.493608439403284
			443 0.493657267441469
			444 0.493666367515158
			445 0.493658693179109
			446 0.493710704760392
			447 0.493689573085022
			448 0.493721323391904
			449 0.493740639908289
			450 0.493729741926219
			451 0.49377405445613
			452 0.493763136054333
			453 0.49378366139529
			454 0.493806517706291
			455 0.493796652336623
			456 0.493832927169133
			457 0.493828691284436
			458 0.493842992646763
			459 0.493865119113319
			460 0.493857812994574
			461 0.493887410875924
			462 0.493886541311069
			463 0.493898111275966
			464 0.49391730894497
			465 0.49391265956056
			466 0.493937122197085
			467 0.4939371764449
			468 0.493948011259263
			469 0.493963241236014
			470 0.493960975547992
			471 0.493981362557879
			472 0.4939810119121
			473 0.493992100100818
			474 0.494002922544416
			475 0.494002939961374
			476 0.494019575047196
			477 0.494018471610405
			478 0.494029978466824
			479 0.494036304956526
			480 0.494038726798123
			481 0.494051293832496
			482 0.49404998458638
			483 0.494061308698641
			484 0.494063591783984
			485 0.494068454660595
			486 0.494076368970948
			487 0.494076034476087
			488 0.494085792751975
			489 0.494085338801574
			490 0.494091996921573
			491 0.494095125731996
			492 0.494096939610949
			493 0.494103351967225
			494 0.494102359707713
			495 0.494109001856366
			496 0.494108543649641
			497 0.494112557570138
			498 0.494114563394839
			499 0.494115238775865
			500 0.494119299064522
		};
		\addlegendentry{$\lambda=0$ (supervised)} 
	\end{axis}
	\draw[white] (-2.5,0) -- (-1,0); 
\end{tikzpicture}

%% file: numerics/LineQC.tex
\begin{tikzpicture}
	\begin{axis}[
		xmin=0,   xmax=500,
		ymin=0.4,   ymax=0.7,
		width=.8\linewidth, 
		height=.5\linewidth,
		grid=major,
		grid style={color0M},
xlabel=Training epochs $r_T$, 
		ylabel=$\mathcal{L}_\text{USV}(s)$,legend pos=south east,legend cell align={left},legend style={draw=none,legend image code/.code={\filldraw[##1] (-.5ex,-.5ex) rectangle (0.5ex,0.5ex);}}]
		\addplot[line width=2pt, color=color3] table {%
			0 0.474322957622956
			1 0.475638554349218
			2 0.477724922346339
			3 0.480364838492256
			4 0.483473361481981
			5 0.486730809353776
			6 0.489843022773478
			7 0.492659285328389
			8 0.4952030158853
			9 0.497577588174624
			10 0.499862251960663
			11 0.50204633511637
			12 0.504057454145159
			13 0.505820385127926
			14 0.507308070490651
			15 0.50855336998311
			16 0.509613110468779
			17 0.510527595660843
			18 0.51130355923087
			19 0.511929320704881
			20 0.512403754510045
			21 0.512756887596731
			22 0.513050661868223
			23 0.513361109954405
			24 0.513755280623977
			25 0.514275692182805
			26 0.514938134684025
			27 0.515733100710797
			28 0.516623694107457
			29 0.517548071764595
			30 0.518428910259335
			31 0.519197817125887
			32 0.519824932387951
			33 0.520319660993989
			34 0.520715161218502
			35 0.521043828528481
			36 0.521322318295205
			37 0.521559490459689
			38 0.521771579395357
			39 0.521991509125642
			40 0.522259459679125
			41 0.522605783241406
			42 0.52303819806265
			43 0.523541039984377
			44 0.524086046116446
			45 0.524644869032023
			46 0.525198928882893
			47 0.525743395324105
			48 0.526284493651708
			49 0.526832129420569
			50 0.527392398174683
			51 0.527964683234238
			52 0.528545487523271
			53 0.529137055492536
			54 0.529750484951287
			55 0.530398265562258
			56 0.531082997385959
			57 0.531793849116863
			58 0.532514882746602
			59 0.533238311836665
			60 0.533969687735517
			61 0.53472043534015
			62 0.535496737308271
			63 0.536295502496986
			64 0.537109569627783
			65 0.537935701444412
			66 0.538777374525284
			67 0.539639780400667
			68 0.540522088474359
			69 0.541414002879505
			70 0.542299653021678
			71 0.543165758392809
			72 0.544007286536608
			73 0.544825757871075
			74 0.545622617689483
			75 0.546394717541769
			76 0.547136378465792
			77 0.547845577662851
			78 0.548527606622775
			79 0.549192279344318
			80 0.549847473176245
			81 0.550495020173536
			82 0.551132165835537
			83 0.551756488962036
			84 0.552369502004101
			85 0.55297593342229
			86 0.553579830000376
			87 0.554181256612935
			88 0.554776506717088
			89 0.555361448274044
			90 0.55593486896752
			91 0.556498784079447
			92 0.557055598082687
			93 0.557604977120462
			94 0.558143431007542
			95 0.55866688262845
			96 0.559173651054054
			97 0.559665037801753
			98 0.560143191110193
			99 0.560608514447701
			100 0.561059094023666
			101 0.561492502287374
			102 0.561908099598543
			103 0.562307666800608
			104 0.562693950218254
			105 0.563068685424858
			106 0.563432015632322
			107 0.563783741563134
			108 0.56412505212904
			109 0.56445900149239
			110 0.564789313882484
			111 0.565118728098269
			112 0.565448422116086
			113 0.565778872327237
			114 0.566111086384367
			115 0.566446899568397
			116 0.566788081110381
			117 0.567135205503705
			118 0.567487414232436
			119 0.567843262006876
			120 0.568201803707005
			121 0.56856293358441
			122 0.568926802917064
			123 0.569293055649882
			124 0.569660738769903
			125 0.570029007237269
			126 0.570397928239507
			127 0.570768602649264
			128 0.571142512430588
			129 0.571520746229527
			130 0.571903813455062
			131 0.572292118565618
			132 0.572686508164265
			133 0.573088272576165
			134 0.573498567944307
			135 0.573917807656676
			136 0.574345583558621
			137 0.574781149269747
			138 0.575223978982215
			139 0.575673899625057
			140 0.576130754182966
			141 0.576594008523044
			142 0.577062730142708
			143 0.577535970562429
			144 0.57801318946775
			145 0.578494341189146
			146 0.57897958387782
			147 0.579468914245233
			148 0.579962056239926
			149 0.580458653475904
			150 0.580958526925706
			151 0.581461729259317
			152 0.581968351808297
			153 0.582478280768474
			154 0.582991131566438
			155 0.583506409033345
			156 0.584023736345264
			157 0.584542954992135
			158 0.585064037170798
			159 0.585586923569062
			160 0.586111446942816
			161 0.586637400453603
			162 0.587164669553769
			163 0.587693295638743
			164 0.58822341066223
			165 0.588755096872715
			166 0.589288281275482
			167 0.589822733042974
			168 0.590358142993749
			169 0.590894207197583
			170 0.591430651417136
			171 0.591967194469316
			172 0.592503499400101
			173 0.593039163765928
			174 0.593573761590255
			175 0.594106907139124
			176 0.594638297186336
			177 0.595167708334176
			178 0.595694958396523
			179 0.596219860645449
			180 0.596742196319382
			181 0.597261712103421
			182 0.597778131565798
			183 0.598291164146948
			184 0.598800502464177
			185 0.599305810381276
			186 0.599806711717825
			187 0.600302788915144
			188 0.600793594647895
			189 0.601278672349903
			190 0.601757577999784
			191 0.602229896345946
			192 0.602695248433662
			193 0.603153291279119
			194 0.603603712956651
			195 0.604046226737567
			196 0.604480566854117
			197 0.604906486960113
			198 0.60532376119025
			199 0.605732187165885
			200 0.606131590231055
			201 0.606521828338294
			202 0.606902797104871
			203 0.607274434551137
			204 0.6076367249745
			205 0.607989701412642
			206 0.608333446277623
			207 0.608668089995073
			208 0.608993807797416
			209 0.609310815112261
			210 0.60961936218265
			211 0.609919728610675
			212 0.610212218424351
			213 0.610497156061758
			214 0.610774883405693
			215 0.611045757753196
			216 0.611310150420203
			217 0.611568445589288
			218 0.611821039009744
			219 0.612068336238162
			220 0.612310750238191
			221 0.612548698309274
			222 0.61278259845497
			223 0.613012865407759
			224 0.613239906586287
			225 0.613464118271444
			226 0.613685882255673
			227 0.613905563154757
			228 0.614123506484346
			229 0.614340037508196
			230 0.614555460776001
			231 0.614770060198677
			232 0.614984099466875
			233 0.615197822608506
			234 0.615411454503227
			235 0.615625201221559
			236 0.615839250124495
			237 0.616053769732578
			238 0.616268909437519
			239 0.616484799172631
			240 0.616701549174434
			241 0.616919249955886
			242 0.617137972576503
			243 0.617357769244892
			244 0.61757867423626
			245 0.617800705062289
			246 0.618023863801296
			247 0.618248138486437
			248 0.618473504457742
			249 0.618699925605956
			250 0.618927355466362
			251 0.619155738152438
			252 0.619385009146614
			253 0.619615095984602
			254 0.619845918879358
			255 0.620077391330862
			256 0.620309420760248
			257 0.620541909193438
			258 0.620774754002931
			259 0.621007848699533
			260 0.621241083751264
			261 0.621474347396593
			262 0.621707526415272
			263 0.621940506822934
			264 0.622173174465059
			265 0.622405415499975
			266 0.622637116776591
			267 0.622868166127017
			268 0.623098452604202
			269 0.623327866698044
			270 0.623556300559877
			271 0.623783648255684
			272 0.624009806055623
			273 0.62423467275448
			274 0.624458150007413
			275 0.624680142659865
			276 0.624900559050257
			277 0.625119311268301
			278 0.625336315359015
			279 0.625551491470566
			280 0.625764763951383
			281 0.625976061407216
			282 0.626185316731573
			283 0.626392467123282
			284 0.626597454103156
			285 0.626800223538426
			286 0.627000725679212
			287 0.627198915206373
			288 0.627394751285328
			289 0.627588197616489
			290 0.62777922247065
			291 0.627967798697535
			292 0.628153903697958
			293 0.62833751935424
			294 0.628518631918867
			295 0.628697231866549
			296 0.628873313718745
			297 0.629046875851522
			298 0.629217920297083
			299 0.629386452546902
			300 0.629552481360995
			301 0.629716018584413
			302 0.629877078969472
			303 0.630035680000852
			304 0.630191841720583
			305 0.630345586550715
			306 0.630496939112625
			307 0.630645926043223
			308 0.630792575809353
			309 0.630936918522595
			310 0.631078985757348
			311 0.631218810375514
			312 0.631356426361305
			313 0.631491868669382
			314 0.631625173088717
			315 0.631756376123241
			316 0.631885514888775
			317 0.632012627024214
			318 0.632137750613946
			319 0.632260924118159
			320 0.632382186308157
			321 0.632501576204768
			322 0.632619133019081
			323 0.632734896095644
			324 0.632848904858748
			325 0.632961198762309
			326 0.633071817243457
			327 0.633180799679325
			328 0.633288185346073
			329 0.633394013378934
			330 0.633498322732053
			331 0.633601152137073
			332 0.63370254005956
			333 0.633802524652533
			334 0.633901143706496
			335 0.633998434595489
			336 0.634094434218974
			337 0.634189178939678
			338 0.634282704517973
			339 0.634375046043656
			340 0.63446623786619
			341 0.634556313524326
			342 0.634645305675767
			343 0.634733246027142
			344 0.634820165264218
			345 0.634906092982165
			346 0.634991057615737
			347 0.635075086369479
			348 0.635158205148384
			349 0.635240438489642
			350 0.635321809496265
			351 0.635402339773316
			352 0.635482049367364
			353 0.635560956709647
			354 0.635639078563306
			355 0.635716429975006
			356 0.63579302423123
			357 0.635868872819479
			358 0.635943985394575
			359 0.636018369750124
			360 0.636092031795154
			361 0.636164975535883
			362 0.636237203062654
			363 0.63630871454218
			364 0.63637950821544
			365 0.636449580401686
			366 0.636518925509106
			367 0.636587536052655
			368 0.636655402679487
			369 0.636722514202274
			370 0.636788857640627
			371 0.636854418270754
			372 0.636919179683513
			373 0.636983123851052
			374 0.63704623120225
			375 0.637108480707192
			376 0.637169849970902
			377 0.637230315336497
			378 0.637289851997912
			379 0.637348434122293
			380 0.637406034982128
			381 0.637462627097158
			382 0.637518182385982
			383 0.637572672327212
			384 0.637626068129796
			385 0.637678340911981
			386 0.637729461888191
			387 0.637779402562924
			388 0.637828134930629
			389 0.637875631680429
			390 0.637921866404425
			391 0.63796681380819
			392 0.638010449921952
			393 0.638052752310778
			394 0.63809370028199
			395 0.638133275087871
			396 0.638171460121686
			397 0.638208241104958
			398 0.638243606263933
			399 0.638277546493163
			400 0.638310055504168
			401 0.638341129957176
			402 0.638370769574066
			403 0.638398977230737
			404 0.638425759027368
			405 0.638451124335209
			406 0.63847508581886
			407 0.638497659433252
			408 0.638518864394838
			409 0.638538723126828
			410 0.638557261178603
			411 0.638574507119782
			412 0.638590492409721
			413 0.638605251243601
			414 0.638618820376531
			415 0.638631238927413
			416 0.638642548164602
			417 0.638652791275567
			418 0.638662013123003
			419 0.638670259989975
			420 0.638677579316779
			421 0.638684019432312
			422 0.63868962928274
			423 0.638694458160287
			424 0.638698555434867
			425 0.638701970291237
			426 0.638704751474198
			427 0.638706947044222
			428 0.638708604145731
			429 0.638709768790033
			430 0.638710485654744
			431 0.638710797901287
			432 0.638710747011807
			433 0.638710372646624
			434 0.638709712523058
			435 0.638708802316218
			436 0.638707675582098
			437 0.638706363703026
			438 0.63870489585533
			439 0.638703298998758
			440 0.638701597887018
			441 0.638699815098515
			442 0.638697971086162
			443 0.638696084244934
			444 0.63869417099561
			445 0.638692245883009
			446 0.638690321686832
			447 0.638688409543125
			448 0.638686519074247
			449 0.63868465852517
			450 0.638682834903911
			451 0.638681054123887
			452 0.638679321146059
			453 0.638677640118804
			454 0.63867601451362
			455 0.638674447254935
			456 0.638672940842506
			457 0.638671497465145
			458 0.638670119104778
			459 0.638668807630104
			460 0.63866756487944
			461 0.638666392732574
			462 0.638665293171741
			463 0.638664268332054
			464 0.638663320541934
			465 0.638662452354238
			466 0.638661666568918
			467 0.638660966248107
			468 0.638660354724582
			469 0.638659835604536
			470 0.638659412765568
			471 0.638659090350721
			472 0.638658872759299
			473 0.638658764635116
			474 0.638658770852684
			475 0.63865889650175
			476 0.638659146870491
			477 0.638659527427553
			478 0.638660043803081
			479 0.638660701768778
			480 0.638661507217027
			481 0.638662466139061
			482 0.638663584602176
			483 0.638664868725976
			484 0.638666324657683
			485 0.638667958546563
			486 0.638669776517565
			487 0.638671784644317
			488 0.638673988921659
			489 0.638676395237947
			490 0.638679009347378
			491 0.638681836842644
			492 0.638684883128203
			493 0.63868815339452
			494 0.638691652593578
			495 0.638695385415992
			496 0.638699356270021
			497 0.638703569262747
			498 0.638708028183672
			499 0.638712736490922
			500 0.63871769730023
		};
		\addlegendentry{$\lambda=-1$ (supervised + graph)}
		\addplot[line width=2pt, color=color2] table {%
			0 0.474322957622956
			1 0.474258087679962
			2 0.474519379809711
			3 0.474973634271036
			4 0.475557257239658
			5 0.476229877996292
			6 0.47696968539066
			7 0.477771488511363
			8 0.478640062835756
			9 0.479588056191923
			10 0.480640515871036
			11 0.481836014508482
			12 0.483217178208388
			13 0.484816950446707
			14 0.486650504946363
			15 0.488714541512477
			16 0.490990637742238
			17 0.493449842784931
			18 0.496057666662194
			19 0.498780104398585
			20 0.501590188679513
			21 0.504469048002246
			22 0.507393180910916
			23 0.510314679550728
			24 0.513156480789944
			25 0.515833494460797
			26 0.51828851002434
			27 0.520514262762963
			28 0.522544963602462
			29 0.524432896472554
			30 0.526227657743887
			31 0.527962565730273
			32 0.529648153857161
			33 0.531271978624286
			34 0.532803453205008
			35 0.53420222426062
			36 0.535428320915482
			37 0.536451534722505
			38 0.537257315211999
			39 0.537847891009149
			40 0.538239358353308
			41 0.538456422503633
			42 0.538526256174316
			43 0.538472737689876
			44 0.538312379412723
			45 0.538052877109023
			46 0.537694841962039
			47 0.537237035505202
			48 0.53668401245419
			49 0.536052866307129
			50 0.535375348239935
			51 0.534694249596634
			52 0.534055940929827
			53 0.533501774602291
			54 0.533061041815819
			55 0.532747581071465
			56 0.532560059284616
			57 0.532484437177848
			58 0.532497473216492
			59 0.532570777471639
			60 0.532675086793126
			61 0.532784287383385
			62 0.532878608650092
			63 0.532946560300148
			64 0.532985579233491
			65 0.533001754156106
			66 0.533008599501435
			67 0.533024498037133
			68 0.533069252889015
			69 0.533160876292336
			70 0.533313559217651
			71 0.533537039885207
			72 0.533836562598654
			73 0.534212413714592
			74 0.534659064586649
			75 0.535164897536044
			76 0.535713274850204
			77 0.536284750950013
			78 0.536859222019991
			79 0.537416888080504
			80 0.53793829064296
			81 0.538404679215572
			82 0.538799600117446
			83 0.53911127033814
			84 0.539334172187768
			85 0.539468627552563
			86 0.539518646952981
			87 0.53948967145073
			88 0.539387822873149
			89 0.539220792649731
			90 0.538998960819452
			91 0.5387352103985
			92 0.538443195460958
			93 0.53813535871739
			94 0.537822185896255
			95 0.537512896051322
			96 0.537216420905781
			97 0.536941321556764
			98 0.536694527337149
			99 0.536480155870677
			100 0.536299616588244
			101 0.536152956285988
			102 0.536040169659409
			103 0.535961209564226
			104 0.535914856513611
			105 0.535897764817215
			106 0.535904776611134
			107 0.535930333995484
			108 0.535969686861192
			109 0.536018960909558
			110 0.536074466713592
			111 0.536132375443421
			112 0.536189431324441
			113 0.536244080487611
			114 0.536296840297239
			115 0.536349502639171
			116 0.536403890428655
			117 0.536461279210338
			118 0.536522743002586
			119 0.536589662922254
			120 0.536663586274013
			121 0.536745459437064
			122 0.53683508066738
			123 0.536931400621321
			124 0.537033391912148
			125 0.537140658152192
			126 0.537253316480926
			127 0.537371552009849
			128 0.537495533453385
			129 0.537625862928656
			130 0.537764037677366
			131 0.537912343835733
			132 0.538073237624255
			133 0.538248798163795
			134 0.538440697060841
			135 0.538650499507246
			136 0.53887976085035
			137 0.539129688045741
			138 0.539400686007284
			139 0.53969225832651
			140 0.540003313858932
			141 0.54033248734767
			142 0.540678122507913
			143 0.541038014625847
			144 0.541409308428245
			145 0.541788748680611
			146 0.542173062713352
			147 0.542559120853777
			148 0.54294381725803
			149 0.543323949217561
			150 0.543696345820548
			151 0.54405816690769
			152 0.544407080730293
			153 0.544741184682725
			154 0.545058839972418
			155 0.545358668772417
			156 0.545639732717635
			157 0.545901679383196
			158 0.54614469011477
			159 0.546369308462724
			160 0.54657635593399
			161 0.546767002143644
			162 0.546942843858898
			163 0.547105830173925
			164 0.547258058028405
			165 0.547401607029945
			166 0.54753851072667
			167 0.547670781435997
			168 0.547800352208833
			169 0.547928929529337
			170 0.548057883497243
			171 0.548188267442009
			172 0.548320910234566
			173 0.548456457603163
			174 0.548595330255174
			175 0.548737685884048
			176 0.548883463810771
			177 0.549032477159985
			178 0.549184455754081
			179 0.549339011360242
			180 0.549495597931725
			181 0.549653539355767
			182 0.549812104368087
			183 0.54997055166388
			184 0.550128118721188
			185 0.550284008631811
			186 0.55043742899895
			187 0.550587661431265
			188 0.550734096271026
			189 0.550876212080963
			190 0.551013546552098
			191 0.551145703456707
			192 0.551272378549959
			193 0.551393356005218
			194 0.551508467301777
			195 0.551617555562236
			196 0.551720478449269
			197 0.551817129606543
			198 0.551907438579218
			199 0.551991346638257
			200 0.55206879326023
			201 0.55213973232198
			202 0.552204154448439
			203 0.552262084782943
			204 0.552313561334678
			205 0.552358623959245
			206 0.552397323735524
			207 0.552429729979323
			208 0.552455915278096
			209 0.552475930653878
			210 0.552489795509285
			211 0.552497503954482
			212 0.552499026735546
			213 0.55249429862619
			214 0.552483206452396
			215 0.55246559449243
			216 0.552441281074182
			217 0.552410067941501
			218 0.552371739144243
			219 0.552326064013203
			220 0.552272812681695
			221 0.552211773710245
			222 0.552142760982455
			223 0.55206561369927
			224 0.551980202985055
			225 0.551886447690693
			226 0.551784328514695
			227 0.551673893786391
			228 0.551555264058006
			229 0.551428644077647
			230 0.551294338210183
			231 0.551152758618563
			232 0.551004424068734
			233 0.550849956871438
			234 0.550690081715046
			235 0.550525620230528
			236 0.550357475233579
			237 0.550186608304944
			238 0.550014018573511
			239 0.549840723718434
			240 0.549667738062195
			241 0.549496046735015
			242 0.549326582030243
			243 0.549160206765457
			244 0.548997702136859
			245 0.548839755695975
			246 0.548686950690103
			247 0.548539761285071
			248 0.548398553820385
			249 0.548263589411278
			250 0.548135025347771
			251 0.548012917846832
			252 0.547897228558167
			253 0.54778783264679
			254 0.54768452509512
			255 0.547587025891604
			256 0.547494987367166
			257 0.547408004222042
			258 0.547325623570017
			259 0.547247353744221
			260 0.547172673891051
			261 0.547101045964135
			262 0.54703192744987
			263 0.546964782410946
			264 0.546899091139302
			265 0.546834360184846
			266 0.546770132511222
			267 0.546705995638575
			268 0.546641586994795
			269 0.546576597895967
			270 0.546510777036538
			271 0.546443932286893
			272 0.546375929600954
			273 0.54630668985062
			274 0.546236185069936
			275 0.54616443405287
			276 0.546091496341405
			277 0.546017464897908
			278 0.545942458966336
			279 0.545866617796302
			280 0.545790094545867
			281 0.545713050020844
			282 0.545635647139825
			283 0.545558046901933
			284 0.545480405380918
			285 0.545402870980642
			286 0.545325582189338
			287 0.54524866650773
			288 0.545172240359565
			289 0.54509640915118
			290 0.545021267262255
			291 0.544946898459925
			292 0.544873376827978
			293 0.544800767593239
			294 0.544729127490279
			295 0.544658505068098
			296 0.544588941326576
			297 0.544520470420822
			298 0.544453120084244
			299 0.544386912003972
			300 0.544321862562307
			301 0.544257983828979
			302 0.544195284384334
			303 0.544133769945065
			304 0.544073444094076
			305 0.544014309112949
			306 0.543956366566315
			307 0.54389961751912
			308 0.543844062649956
			309 0.543789702400823
			310 0.543736536953583
			311 0.543684565893206
			312 0.543633787763793
			313 0.543584199728767
			314 0.543535797243062
			315 0.543488573600265
			316 0.543442519490387
			317 0.543397622780616
			318 0.543353868479775
			319 0.543311238718719
			320 0.543269712764894
			321 0.54322926720821
			322 0.543189876283098
			323 0.543151512149529
			324 0.543114145093358
			325 0.543077743761015
			326 0.543042275449093
			327 0.543007706328978
			328 0.542974001566142
			329 0.542941125438184
			330 0.542909041503681
			331 0.54287771273824
			332 0.542847101588517
			333 0.54281717002507
			334 0.542787879660082
			335 0.54275919187812
			336 0.542731067929846
			337 0.54270346904305
			338 0.542676356606785
			339 0.542649692380448
			340 0.542623438660451
			341 0.542597558423824
			342 0.542572015490288
			343 0.542546774668912
			344 0.542521801835145
			345 0.542497063958948
			346 0.542472529136239
			347 0.542448166615591
			348 0.54242394678298
			349 0.542399841121471
			350 0.542375822191836
			351 0.542351863630683
			352 0.542327940131439
			353 0.54230402741495
			354 0.54228010222462
			355 0.542256142341771
			356 0.542232126586269
			357 0.542208034798611
			358 0.542183847825866
			359 0.542159547504836
			360 0.542135116611391
			361 0.54211053877316
			362 0.542085798369835
			363 0.542060880424946
			364 0.542035770472296
			365 0.542010454401759
			366 0.541984918309938
			367 0.541959148362112
			368 0.541933130652006
			369 0.541906851062082
			370 0.541880295144121
			371 0.541853448024141
			372 0.541826294319147
			373 0.54179881806632
			374 0.541771002678331
			375 0.541742830924071
			376 0.541714284919802
			377 0.541685346126927
			378 0.541655995363904
			379 0.541626212828876
			380 0.541595978120107
			381 0.541565270252632
			382 0.541534067679179
			383 0.541502348313811
			384 0.541470089548955
			385 0.541437268265706
			386 0.541403860843968
			387 0.541369843170286
			388 0.541335190635774
			389 0.541299878125359
			390 0.541263880004742
			391 0.541227170103918
			392 0.541189721692436
			393 0.541151507449023
			394 0.541112499431284
			395 0.54107266904423
			396 0.541031987004392
			397 0.540990423303016
			398 0.540947947173665
			399 0.540904527063664
			400 0.540860130607913
			401 0.540814724609019
			402 0.540768275027763
			403 0.540720746982457
			404 0.540672104755702
			405 0.540622311811595
			406 0.54057133082575
			407 0.540519123726301
			408 0.540465651745106
			409 0.54041087548199
			410 0.540354754983396
			411 0.540297249833525
			412 0.540238319257638
			413 0.540177922239997
			414 0.540116017657329
			415 0.540052564426716
			416 0.539987521669031
			417 0.539920848891093
			418 0.539852506187793
			419 0.539782454464141
			420 0.539710655679167
			421 0.539637073114496
			422 0.539561671668334
			423 0.539484418175025
			424 0.539405281752161
			425 0.539324234177297
			426 0.5392412502944
			427 0.539156308449967
			428 0.539069390959917
			429 0.538980484607593
			430 0.53888958117126
			431 0.538796677979507
			432 0.538701778493387
			433 0.538604892912683
			434 0.538506038801856
			435 0.538405241730957
			436 0.538302535926285
			437 0.538197964923376
			438 0.538091582213049
			439 0.537983451870662
			440 0.537873649157572
			441 0.537762261081609
			442 0.537649386902038
			443 0.537535138564279
			444 0.537419641048705
			445 0.537303032616743
			446 0.537185464937767
			447 0.537067103081398
			448 0.536948125360618
			449 0.536828723012456
			450 0.53670909970581
			451 0.536589470869251
			452 0.536470062834675
			453 0.536351111796309
			454 0.536232862589086
			455 0.536115567294498
			456 0.535999483685465
			457 0.535884873525289
			458 0.535772000738897
			459 0.535661129476519
			460 0.535552522090776
			461 0.535446437048824
			462 0.535343126801071
			463 0.535242835627184
			464 0.535145797479366
			465 0.535052233843011
			466 0.534962351635678
			467 0.534876341167049
			468 0.534794374186013
			469 0.534716602045965
			470 0.534643154025184
			471 0.534574135845239
			472 0.534509628436246
			473 0.534449687002117
			474 0.534394340440274
			475 0.534343591167786
			476 0.534297415398877
			477 0.534255763906515
			478 0.534218563283319
			479 0.534185717695528
			480 0.534157111099533
			481 0.534132609865207
			482 0.53411206572636
			483 0.534095318958633
			484 0.534082201671039
			485 0.534072541090536
			486 0.534066162720593
			487 0.534062893264696
			488 0.534062563223026
			489 0.534065009093314
			490 0.534070075133281
			491 0.534077614669282
			492 0.534087490961469
			493 0.534099577657882
			494 0.534113758887216
			495 0.534129929051596
			496 0.534147992386593
			497 0.534167862356614
			498 0.534189460950495
			499 0.534212717935813
			500 0.534237570122047
		};
		\addlegendentry{$\lambda=0$ (supervised)} 
	\end{axis}
	\draw[white] (-2.5,0) -- (-1,0); 
\end{tikzpicture}

%% file: text/gan.tex
\chapter{Quantum generative adversarial networks}\hypertarget{QGAN}{}
\label{chapter:QGAN}

In the preceding chapters, we worked with training data in the form of pairs of input and output quantum states in order to learn the relation between them. We defined loss functions that compared the output of a \emph{quantum neural network} (QNN) with the desired output and demonstrated that is possible to update the network based on this and achieve good training results. The whole training algorithm rests upon the fact that we have access to related data pairs. In contrast, in the following we undertake the task of extending a set of quantum states to states which have similar properties.

We present a QNN training algorithm \cite{Beer2021b} which can not only learn the common characteristics of such data sets but also generates states with suiting requirements. The proposal is based on the \emph{dissipative quantum neural networks} (DQNNs) presented \cref{chapter:DQNN}. More precisely, we will use two DQNNs acting in competition, whereof one acts in a generative and the other in a discriminative manner, to learn from unlabelled quantum data.

In classical \emph{machine learning} (ML) we generally distinguish between \emph{discriminative} and \emph{generative} models. The in \cref{chapter:ML} discussed example of classifying handwritten digits based on the MNIST dataset \cite{MNIST} is a typical discriminative task: the \emph{neural network} (NN) learns in a supervised way to decide between the ten options. Such problems have in common that it is easy to find a proper training loss function which is aimed to be optimised during training \cite{Nielsen2015}. In contrast, generative  models produce data. Speaking in the example of handwritten digits, we would train a generative model to produce different \enquote{handwritten} digits from a random input. The downside of generative models is that these, due to the difficulty of approximating many probabilistic computations, are much more complicated to train than discriminative models. 

This changed with the arrival of \emph{generative adversarial networks} (GANs) \cite{Goodfellow2014}, consisting of a generative and discriminative part. The generative model generates samples by passing random noise through, for example, a multi-layer perceptron. The discriminative part, which can also be realised by a multi-layer perceptron, is trained to distinguish the data produced by the generator from the training data. On the other hand, the generative model aims to \enquote{fool} the discriminator. In this opposing way, it is possible to train the generator \cite{Goodfellow2014, Radford2015, Creswell2018} and GANs have since found a lot of applications \cite{Creswell2018}, ranging from classification or regression tasks \cite{Creswell2018, Zhu2016, Salimans2016} to the generation \cite{Reed2016} and improvement \cite{Ledig2017} of images.

In the field of quantum machine learning (QML), such generative adversarial processes, referred to as \emph{quantum generative adversarial networks} (QGANs) \cite{DallaireDemers2018,Lloyd2018a,Benedetti2019a,Chakrabarti2019,Hu2019,Zoufal2019, Zoufal2021,Niu2021}, were studied, as well. Some of these models are defined as a quantum-classical hybrid, include a quantum generator and a classical discriminator, and are used to learn from classical input data \cite{Zoufal2019, Zoufal2021}. Moreover, the authors of \cite{Lloyd2018a} present the usage of quantum or classical data and two quantum processors in an adversarial learning setting. Furthermore, the work of \cite{DallaireDemers2018} designs quantum circuits for the generator and discriminator NNs and trains them using data which has two possible labels assigned. A further recent proposal is the \emph{Entangling} QGAN \cite{Niu2021}: here, the discriminative model has access to a state produced by the generator a training data set and a parametrised $\swap$ test is used to \enquote{entangle} the states.

Since we could observe excellent learning behaviour using the in \cref{chapter:DQNN} presented DQNNs in a supervised ansatz, in this chapter, we introduce a QGAN based on these types of QNNs. This so called \emph{discriminative quantum generative adversarial network} (DQGAN) \cite{Beer2021b} is, in contrast to some of the above named proposals, a fully quantum architecture and is trained with quantum data. Different to the approach in \cite{DallaireDemers2018} in the DQGAN algorithm only the discriminator has access to training data, and we only work with unlabelled quantum states as training data. In fact, the generative model gets a random quantum state as input, whereas the discriminator receives either the output of the generative model or a training data state as input. Since quantum data is always rare, this model aims to produce states with similar features as the training data. In that way, we generated more quantum data of the same kind, which could be, for example, useful to train other QNNs.

We start this chapter with a discussion of classical GAN methods in \cref{sec:QGAN_classicalGAN} since the quantum analogues are based on these. It follows a presentation of the already advertised DQGAN architecture, see \cref{sec:QGAN_network}. We discuss the data sets and loss functions used for training and validation of DQGAN in \cref{sec:QGAN_loss}. Based on this, we formulate the update rules for training the DQGAN training algorithm in \cref{sec:QGAN_training}. We conclude with a classical simulation of the algorithm and show results of an implementation suitable for a NISQ device in \cref{sec:QGAN_classical} and \cref{sec:QGAN_QC}.

\section{Classical adversarial networks}
\label{sec:QGAN_classicalGAN}

In this section, we summarise the work of \cite{Goodfellow2014}, who propose the first GAN model and explore the particular case where the generative model generates samples by passing random noise through a multi-layer perceptron. The discriminative model is built as a multi-layer perceptron as well. It has access to generator's synthetic samples and samples from the stack of training data. However, the generator part has no direct admission to the training data. In general, GANs are usually realised with multi-layer NNs consisting of convolutional or fully connected layers \cite{Creswell2018}.

\FloatBarrier \subsection*{Network architecture}

For describing the proposal of \cite{Goodfellow2014} we call the generative model, capturing the data distribution and producing data, $G$. The other model, the discriminator, is denoted by $D$ and estimates the probability $D({x})$ that a given sample ${x}$ is based on the training data set rather than produced by the generator $G$. 

\begin{figure}[h!]
\centering
\begin{tikzpicture}[scale=1.85]
\node[] (pz) at (-1,0) {${z}$};
\node[perceptron1] (a) at (0,0) {};
\node[perceptron1] (b) at (1,-.5) {};
\node[perceptron1] (c) at (1,.5) {};
\node[] (pg) at (2,0) {$x$};
\draw[-stealth,shorten <=2pt, shorten >=2pt] (a) -- (b);
\draw[-stealth,shorten <=2pt, shorten >=2pt] (a) -- (c);
\node[color1] (G) at (.5,0) {$G$};
\draw[-stealth,shorten <=4pt, shorten >=4pt,color0] (-1.75,0) --node[above]{$p_\text{in}$} (pz);
\draw[-stealth,shorten <=4pt, shorten >=4pt,color0] (pz) --(a) ;
\draw[-stealth,shorten <=4pt, shorten >=4pt,color0] (1.3,0) --node[above]{$p_G$} (pg);

\node[perceptron0,ellipse,draw=color0] (ellipse) at (-.25,-1.5) {training data};
\draw[brace0](2.45,.75)--  (2.45,-1.75) ;
\node[] (pd) at (2,-1.5) {$x$};
\draw[-stealth,color0,shorten <=10pt] (ellipse) --node[above]{$p_\text{data}$} (pd);

\begin{scope}[shift={(4,-.55)}]
\node[] (pd) at (2,0) {$D(x)$};
\node[perceptron2] (d) at (0,-.5) {};
\node[perceptron2] (e) at (0,.5) {};
\node[perceptron2] (f) at (1,0) {};
\draw[-stealth,shorten <=4pt, shorten >=2pt] (d) -- (f);
\draw[-stealth,shorten <=4pt, shorten >=2pt] (e) -- (f);
\node[color2] (D) at (.5,0) {$D$};
\draw[-stealth,shorten <=4pt, shorten >=4pt,color0] (-1,0) -- (-.2,0);
\draw[-stealth,shorten <=4pt, shorten >=4pt,color0] (1.2,0) -- (pd);

\end{scope}
\end{tikzpicture}
\caption{\textbf{GAN.} The discriminative model $D$ gets the input $x$ either of the generator $G$ or from the training data set.}
\label{fig:QGAN_adver}
\end{figure}
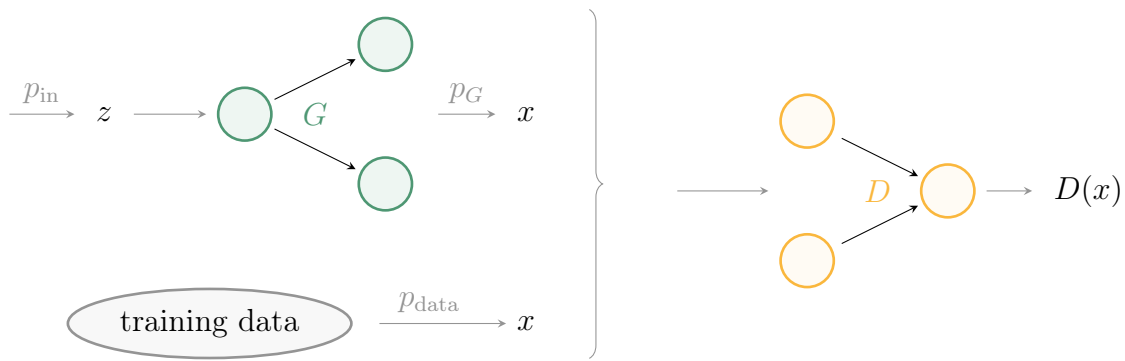

Here $G$ takes a random input ${z}$ with distribution $p_\text{in}$ and outputs ${x}$ with probability distribution $p_G$. On the other hand the probability distribution $p_\text{data}$ describes sampling ${x}$ from the training data.  The second network $D$ gets ${x}$ either based on $p_G$ or $p_\text{data}$ as an input. These relations are depicted in \cref{fig:QGAN_adver} for the case $G(x):[0,1]\rightarrow[0,1]\times [0,1]$, $D(x):[0,1]\times [0,1]\rightarrow[0,1]$, and $D(G(x)):[0,1]\rightarrow[0,1]$.

\FloatBarrier \subsection*{Training}

For the training we assume that both models, $G({x},\theta_G)$ and $D({x},\theta_D)$ are implemented by multilayer perceptrons and parametrised by $\theta_G$ and $\theta_D$, respectively. Both models are trained simultaneously but with a different aim. The training goal of model $G$ is to maximise the probability of $D$ making a mistake. The aim of $D$ is to always make the correct distinction. Therefore the whole problem corresponds to a \emph{minimax problem}: We train $D$ to maximise the probability of assigning the correct label to both training examples and samples from $G$. Simultaneously we train $G$ to minimise $\log(1-D(G({z}))$, i.e.\ we train the generator to produce fake data which cannot be distinguished by the discriminator $D$. 
\begin{equation*}
\mathcolorbox{\min_G 
\max_D ( \mathcal{L}(D,G))=\min_G 
\max_D \Big(\mathbbm{E}_{x \sim p_\text{data}} [\log{}  D({x})]  + \mathbbm{E}_{x \sim p_\text{in}(z)} [\log (1- D(G({z})))] \Big) }
\end{equation*}

During the training, we alternate between $r_D$ epochs of optimising $D$ and $r_G$ of optimising $G$. Here, it is essential to choose $r_D$ in a way that $G$ changes slowly enough \cite{Goodfellow2014}. $r_T$ describes how often alterations are repeated. The authors of \cite{Goodfellow2014} choose an algorithm based on \emph{minibatch stochastic gradient}, as shown in \cref{alg:QGAN_algorithm}. Further, they recommend using a momentum method for the update. Both of these techniques were explained in \cref{subsec:ML_alternativesGradient}. 

\begin{algorithm}
\caption{Minibatch stochastic gradient descent for GAN training. }
\label{alg:QGAN_algorithm}
\begin{algorithmic}
\For{for $r_T$ epochs}
\For{$r_D$ epochs}
\State Sample minibatch $\{{z^i}\}_{i=1}^S$ from noise prior $p_\text{in}({z})$.
\State Sample minibatch $\{{x^i}\}_{i=1}^S$ from data generating distribution $p_\text{data}({x})$.
\State Update the discriminator by ascending its stochastic gradient 
\begin{equation*}
\nabla_{\theta D} \frac{1}{S} \sum_{i=1}^{S}[\log D({x^i}) + \log (1- D(G({z^i})))].
\end{equation*}
\EndFor
\For{$r_G$ epochs}
\State Sample minibatch of $S$ noise samples $\{{z^i}, \hdots, {z^S}\}$ from noise prior $p_\text{in}({z})$.
\State Update the generator by descending its stochastic gradient
\begin{equation*}
\nabla_{\theta G} \frac{1}{S} \sum_{i=1}^{S}\log (1- D(G({z^i})))].
\end{equation*}
\EndFor
\EndFor
\end{algorithmic}
\end{algorithm}

At this point, we want to mention one of the main difficulties occurring when training a GAN, called \emph{mode collapse} \cite{Arjovsky2017}. This describes the situation where the generative model only produces a selection of the aimed training data. When we think in the example of producing \enquote{handwritten} digits, the generator would, for example, produce some of the digits never or only rarely.

Mode collapse often occurs when the discriminator is not trained well enough. On the other hand, when the discriminator is trained optimal, the gradient used in \cref{alg:QGAN_algorithm} vanishes. Different solutions were proposed since this is a common problem in generative ML. One of them is the in \cite{Arjovsky2017} discussed \emph{Wasserstein} GAN. Here, instead of the discriminative, a \emph{critic} network is applied. Using the Wasserstein distance \cite{Vaserstein1969} the two different kinds of outputs of the critic are compared and aimed to be maximised by the critic: the output of the critic getting training data as input and the output of it getting the generated data. In the same way as the original GAN, this results in a minimax problem, i.e.\
\begin{equation*}
\min_G \max_D ( \mathcal{L}(D,C))=
\min_G \max_D \Big(\frac{1}{S}\sum_{i=1}^S C(x_i)-C(G(z_i)) \Big). 
\end{equation*}

\FloatBarrier \subsection*{Theoretical results}

Before discussing GAN applications, we shortly want to mention some theoretical results of the original GAN, proposed in \cite{Goodfellow2014}, with the aim of getting a better intuition of the training algorithm.

Whereas the generator's and discriminator's data distribution changes during the training, the training samples generating distribution $p_\text{data}$ does not change. In the first phase of \cref{alg:QGAN_algorithm}, $G$ is fixed and only the discriminator is trained. We can formulate this phase as
\begin{align*}
\mathcal{L}(G,D)=&\int_x p_\text{data}({x}) \log (D({x}))dx + \int_z p_\text{in} ({z})\log(1-D(G({z}))dz\\
=&\int_x p_\text{data}({x}) \log (D({x})) +  p_G ({x})\log(1-D({x}))dx.
\end{align*}
For $(a,b)\in \mathbbm{R}^2\backslash \{0,0\}$, the function $y\rightarrow a \log (y)+b\log(1-y)$ achieves its maximum in $[0,1]$ at $\frac{a}{a+b}$. Hence for a fixed generator $G$, the optimal discriminator, denoted by $D^*$, is 
\begin{equation*}
D^*({x})=\frac{p_\text{data}({x})}{p_\text{data}({x}) + p_G({x})}.
\end{equation*}

After training the discriminator, the generator $G$ gets trained. The authors of \cite{Goodfellow2014} observed that the discriminator is more likely to classify the data at first. Experimental results show that after several training iterations, including updating $D$ and $G$, we can reach the state of $p_G=p_\text{data}$. The discriminator is unable to differentiate between the two distribution, i.e.\ $D({x})=\frac{1}{2}$.

Indeed the authors of \cite{Goodfellow2014} showed that the global minimum of the training criterion  
\begin{align*}
C(G)=&\max_D \mathcal{L}(G,D)\\
=&\mathbbm{E}_{x \sim p_\text{data}} [\log  D_G({x})]  + \mathbbm{E}_{z \sim p_\text{in}} [\log (1- D_G(G({z})))]
\end{align*}
is achieved if and only if $p_G=p_\text{data}$.

Moreover it was demonstrated that $p_G$ actually converges to $p_\text{data}$, given $G$ and $D$ have enough capacity at each step of \cref{alg:QGAN_algorithm} so that the discriminator is allowed to reach its optimum given $G$, and $p_G$ is updated so as to improve the criterion
\begin{equation*}
\mathbbm{E}_{{x}\sim p_\text{data}}[log D^*_G({x})] + \mathbbm{E}_{{x}\sim p_G} [log(1-D^*_G({x}))].
\end{equation*}

\FloatBarrier \subsection*{Applications}

\begin{figure}[h!]
\centering
\begin{subfigure}{0.49\linewidth}
\includegraphics[width=1\linewidth]{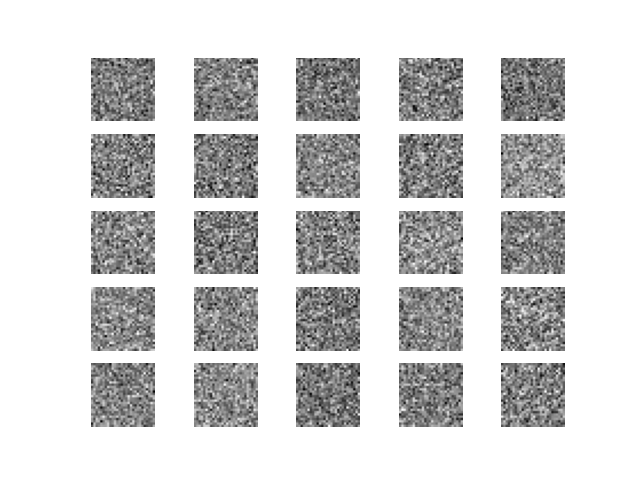}
\subcaption{Input $z$}
\end{subfigure}
\begin{subfigure}{0.49\linewidth}
\includegraphics[width=1\linewidth]{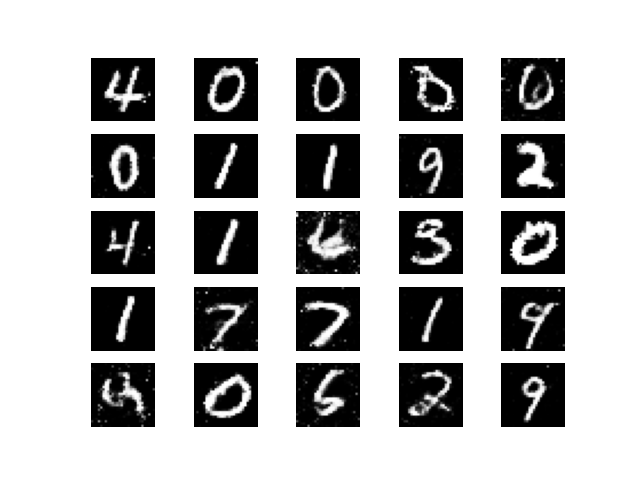}
\subcaption{Output $G(z)$}
\end{subfigure}
\caption{\textbf{Generating \enquote{handwritten} digits.} Using the code proposed in \cite{Goodfellow2014} and the MNIST data set \cite{MNIST} we trained a generator $G$ in a GAN context. After $90000$ epochs the $G$ generated the output depicted in (b) from the input in (a).}
\label{fig:QGAN_MNIST}
\end{figure}

The authors of \cite{Goodfellow2014} trained their proposed GANs at a range of datasets, including the already in \cref{sec:ML_data} discussed MNIST dataset \cite{MNIST}. Therefore we want to take the opportunity to continue the discussion of the demonstrative MNIST dataset and present in \cref{fig:QGAN_MNIST} the input and output of a, in the context of a GAN, trained generator. Although the authors do not claim that the resulting samples are better than samples generated by previously existing methods, the figure shows that after $90000$ training epochs, it is possible to recognise digits in the outputs. The output is quite diverse, so no mode collapse accrued.

Although generating pictures of digits is indeed a good demonstration example, we will discuss more practicable applications of GANs in the following lines. Based on the original proposal discussed above, generative NNs are applied in many areas today \cite{Creswell2018}. These network architectures allow training in semi-supervised and unsupervised learning manner, for example, when producing data pairs is expensive or impossible. 

So far, we have focused on the output of the generator. However, also the in GAN training acquired discriminator can be of use, for example, for classification or regression tasks \cite{Creswell2018, Zhu2016, Salimans2016}. Moreover, generators can be used to simply generate more labelled training samples for training further NNs \cite{Shrivastava2017, Liu2016}. 

Also, \emph{image synthesis}, i.e.\ the generation of images, plays a significant role in the usage of GANs. Many different techniques arose, for example, algorithms where both, the generator and the discriminator, are fed with additional labels. Using the latter ansatz, for example, \enquote{reverse captioning}, i.e.\ the generation of pictures due to their description is possible \cite{Reed2016}.

In contrast, the MNIST example displayed in \cref{fig:QGAN_MNIST} belongs in the category of GANs used for \emph{image-to-image translation}. Here an input image gets processed to an output image. The model \emph{pix2pix} \cite{Isola2017} can, for example, learn to generate maps from aerial photos or colourise greyscale images. Furthermore, \emph{super-resolution} is a scope of application. Using GANs, low-resolution images can be upscaled in a realistic and natural-looking way \cite{Ledig2017}.

For a more detailed discussion of different kinds of GANs and their applications, we point to \cite{Radford2015, Creswell2018}.

\section{Dissipative quantum generative adversarial network}
\label{sec:QGAN_network}

In analogy to the in \cref{fig:QGAN_adver} depicted classical GAN, the in the following presented DQGAN, proposed in \cite{Beer2021b}, is constructed of two DQNNs, the generative model, and the discriminative model, described through the completely positive maps $\mathcal{E}_G$ and $\mathcal{E}_D$, respectively. The number of qubits in the last layer of the generator is equal to the number of qubits in the first layer of the discriminator, so that the generator's output can be used as input for the discriminator. We are given a set of training states $\{\ket{\phi_x^T}\}_{x=1}^N$. Moreover, several sets of random states $\{\ket{\psi ^\text{in}_x}\}$ are needed during the training algorithm. The aim is that the generative DQNN produces states similar to the training states $\ket{\phi_x^T}$. Note that we assume the states of these sets to be pure. 

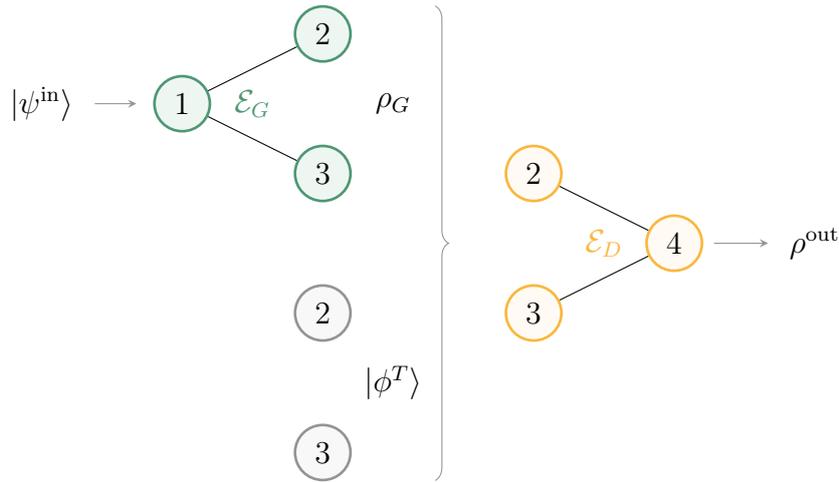
\begin{figure}[h!]
\centering
\begin{tikzpicture}[scale=1.85]
\node[] (pz) at (-1,0) {$\ket{\psi^\text{in}}$};
\node[perceptron1] (a) at (0,0) {1};
\node[perceptron1] (b) at (1,-.5) {3};
\node[perceptron1] (c) at (1,.5) {2};
\draw (a) -- (b);
\draw (a) -- (c);
\node[color1] (Ug) at (.5,0) {$\mathcal{E}_G$};
\node[] (Rg) at (1.5,0) {$\rho_G$};
\draw[-stealth,shorten <=4pt, shorten >=4pt,color0] (pz) -- (-.25,0);
\draw[brace0](1.8,0.7)-- (1.8,-2.7) ;
\begin{scope}[shift={(0,-2)}]
\node[] (Rt) at (1.5,0) {$\ket{\phi^T}$};
\node[perceptron0] (b) at (1,-.5) {3};
\node[perceptron0] (c) at (1,.5) {2};
\end{scope}
\begin{scope}[shift={(2.5,-1)}]
\node[] (pd) at (2,0) {$\rho^\text{out}$};
\node[perceptron2] (d) at (0,-.5) {3};
\node[perceptron2] (e) at (0,.5) {2};
\node[perceptron2] (f) at (1,0) {4};
\draw (d) -- (f);
\draw (e) -- (f);
\node[color2] (Ud) at (.5,0) {$\mathcal{E}_D$};
\draw[-stealth,shorten <=4pt, shorten >=4pt,color0] (f) -- (pd);
\end{scope}
\end{tikzpicture}
\caption{\textbf{DQGAN.} The here depicted DQGAN consists of four qubits. Qubits $2$ and $3$ are shared by the generative and the discriminative DQNN. The state of these qubits is either the generator applied on the input state, i.e.\ $\mathcal{E}_G (\ket{\psi^\text{in}}\bra{\psi^\text{in}})$ or a given training state $\ket{\phi^T}$. }
\label{fig:QGAN_qgan}
\end{figure}

\cref{fig:QGAN_qgan} presents a minimalistic DQGAN: both, the generator and the discriminator consist of two layers and involve three qubits, respectively. Since the qubits of the second layer of the generator, qubits $2$ and $3$, are shared with the discriminator, the whole DQNN includes four qubits, if the discriminator gets a generated state $\mathcal{E}_G (\ket{\psi^\text{in}}\bra{\psi^\text{in}})$ as an input. In the other case, these qubits are directly initialised by a training data state and the DQNN consists only of qubits $2$, $3$ and $4$. In equations we can describe this as
\begin{singlespace}
\begin{equation*}
\mathcolorbox{	\rho^\text{out}=
\begin{cases} 
\mathcal{E}_D (\mathcal{E}_G (\ket{\psi^\text{in}}\bra{\psi^\text{in}})) &\mbox{for generated data}\\
\mathcal{E}_D (\ket{\phi^T}\bra{\phi^T}) & \mbox{for training data.} 
\end{cases}}
\end{equation*}
\end{singlespace}

Note that in DQNN manner, the generator depicted in \cref{fig:QGAN_qgan} consists of two two-qubit unitaries $U_{G1}$ and $U_{G2}$, acting on qubits $1$ and $2$, and on qubits $1$ and $3$. On the contrary, the discriminator is described by a single three-qubit unitary $U_D$. To clarify the network structure of the DQGAN, we explain in the following how the states migrate through the network, assuming the depicted example.

The generator gets a random state $\ket{\psi_x^\text{in}}$ as an input. Next, we initialise the output of the generator as $\ket{00}$, tensor this state to the input state, apply the network unitaries and trace out the input layer. This can be described as
\begin{equation*}
\rho_{G,x}=\mathcal{E}_G (\ket{\psi_x^\text{in}}\bra{\psi_x^\text{in}}) = \tr_{\{1\}}\left( U_{G2} U_{G1} ( \ket{\psi_x^\text{in}}\bra{\psi_x^\text{in}} \otimes \ket{00}\bra{00}) U_{G1}^\dagger U_{G2}^\dagger \right).
\end{equation*}

The discriminator's input is either the output of the generator $\rho_{G,x}$ leading to 
\begin{align*}
	\rho_{G+D,x}=&\tr_{\{2,3\}}(U_D(\rho_{G,x}\otimes \ket{0}\bra{0})U_D^\dagger)
\end{align*}
or in the other case the training state $\ket{\phi_x^T}$ leading to
\begin{align*}
\rho_{D,x}=&\tr_{\{2,3\}}(U_D(\ket{\phi_x^T}\bra{\phi_x^T}\otimes \ket{0}\bra{0})U_D^\dagger).
\end{align*}

\section{Loss functions}
\label{sec:QGAN_loss}

In the same way as the in \cref{subsec:DQNN_trainingloss} described loss function for the supervised DQNN, the loss functions to train the DQGAN are based on the fidelity. The discriminative model $\mathcal{E}_D$ aims to output the state $\ket{1}$ when fed with a state $\ket{\psi^\text{in}}$ and identify \enquote{false} data, i.e.\ states generated by $\mathcal{E}_G$, with outputting the state $\ket{0}$. 

During the training only $S<N$ of the states $\{\ket{\phi_x^T}\}_{x=1}^N$ are used. However, we aim that at the end of the training that every of the generator's output states is close to a state of the full data set $\{\ket{\phi_x^T}\}_{x=1}^N$. In other words we desire that the generator does not reproduces only the $S$ given states but extends this data set.

\FloatBarrier\subsection*{Training loss}

In analogy to the classical case we can describe the training process through
\begin{equation}
\label{eq:dqgan_minimax}
\mathcolorbox{\min_G 
\max_D \left(\frac{1}{S}\sum_{x=1}^S \bra{0} \mathcal{E}_D (\mathcal{E}_G (\ket{\psi_x^\text{in}}\bra{\psi_x^\text{in}}))\ket{0} + \frac{1}{S}\sum_{x=1}^S \bra{1} \mathcal{E}_D (\ket{\phi_x^T}\bra{\phi_x^T})\ket{1} \right),
}
\end{equation}
and the updates of the discriminator and the generator take place consecutively. For updating the generator we maximise the loss function
\begin{equation*}
\mathcal{L}_{D}(\mathcal{E}_D,\mathcal{E}_G)=\frac{1}{S}\sum_{x=1}^S \bra{0} \mathcal{E}_D (\mathcal{E}_G (\ket{\psi_x^\text{in}}\bra{\psi_x^\text{in}}))\ket{0} + \frac{1}{S}\sum_{x=1}^S \bra{1} \mathcal{E}_D (\ket{\phi_x^T}\bra{\phi_x^T})\ket{1} .
\end{equation*}
The generator is trained through maximising
\begin{equation*}
\mathcal{L}_{G}(\mathcal{E}_D,\mathcal{E}_G)=\frac{1}{S}\sum_{x=1}^S \bra{1} \mathcal{E}_D (\mathcal{E}_G (\ket{\psi_x^\text{in}}\bra{\psi_x^\text{in}}))\ket{1}.
\end{equation*}
As shown in \cref{alg:QGAN_algorithmQ}, in every training round two new sets of random states $\{\ket{\psi ^\text{in}_x}\}_{x=1}^S$ are used to evaluate $\mathcal{L}_{D}$ and $\mathcal{L}_{G}$, respectively. Note further that $\mathcal{L}_G$ differs from the corresponding term in \cref{eq:dqgan_minimax}: the fidelity is calculated with respect to $\ket{1}$ instead of $\ket{0}$ and hence the generator is trained by maximising $\mathcal{L}_G$ rather than minimising. Using this arrangement it will be more convenient to compare the loss functions in \cref{sec:QGAN_classical} and \cref{sec:QGAN_QC}.

\FloatBarrier\subsection*{Validation loss}

In the final stages of the training, we aim that every output of the generator is close to one of the given states $\{\ket{\phi_x^T}\}_{x=1}^N$. Therefore we produce $V$ additional random states $\{\ket{\psi ^\text{in}_x}\}_{x=1}^V$. For every of these validation states, we search for the best suiting state of the data set through $\max_{x=1}^T \left( \bra{\phi_x^T} \mathcal{E}_G (\ket{\psi_i^\text{in}}\bra{\psi_i^\text{in}})\ket{\phi_x^T}\right)$ and build the mean value over all $V$ generated states, i.e.\
\begin{equation*}
\mathcal{L}_{V}(\mathcal{E}_D,\mathcal{E}_G)=\frac{1}{V}\sum_{i=v}^V \max_{x=1}^N \left( \bra{\phi_x^T} \mathcal{E}_G (\ket{\psi_i^\text{in}}\bra{\psi_i^\text{in}})\ket{\phi_x^T}\right).
\end{equation*}
This \emph{validation loss} function is used to test the training success. 

\section{Training algorithm}
\label{sec:QGAN_training}

Since the presented algorithm consists of two parts, we define the training epochs with three parameters: the overall repetition of alternating between training the discriminator and the generator $r_T$, a parameter $r_D$ describing the repetition of training discriminator and a parameter $r_G$ describing the repetition of training generator. These parameters are used in the \emph{for} loops of the pseudo-code in \cref{alg:QGAN_algorithmQ}, which gives an overview of the algorithm. The last three steps describe the validation process after the training. 

\begin{algorithm}
\caption{Training of a DQGAN.}
\label{alg:QGAN_algorithmQ}
\begin{algorithmic}
\State initialise network unitaries 
\For{$r_T$ epochs}
\State make a list of $S$ randomly chosen states of the training data list $\{\ket{\phi_x^T}\}_{x=1}^N$
\For{$r_D$ epochs}
\State make a list of $S$ random states $\ket{\psi _x^\text{in}}$
\State update the discriminator unitaries with maximising $\mathcal{L}_{D}$
\EndFor
\For{$r_G$ epochs}
\State make a list of $S$ random states $\ket{\psi_x^\text{in}}$
\State update the generator unitaries with maximising $\mathcal{L}_{G}$
\EndFor
\EndFor
\State make a list of $V$ random states $\ket{\psi_x^\text{in}}$
\State propagate each $\ket{\psi_x^\text{in}}$ through the generator to produce $V$ new states
\State compute $\mathcal{L}_{V}$
\end{algorithmic}
\end{algorithm}

\FloatBarrier\subsection*{Derivation of the update matrices}

Analogous to the DQNN update rule presented in \cref{sec:DQNN_trainingalgorithm} the unitaries will be updated through
\begin{equation*}
U_j^l(t+\epsilon)=e^{i\epsilon K_{j}^l(t)} U_j^l(t).
\end{equation*}
Before we derive the form of the update matrices $K_{j}^l$ in general we take a smaller step first and discuss the update of the minimal example which is depicted in \cref{fig:QGAN_qgan} and consists of three unitaries. These unitaries can be updated via
\begin{align*}
U_D(t+\epsilon)=&e^{i\epsilon K_{D}(t)} U_D(t)\\ 
U_{G1}(t+\epsilon)=&e^{i\epsilon K_{G1}(t)} U_{G1}(t)\\  
U_{G2}(t+\epsilon)=&e^{i\epsilon K_{G2}(t)} U_{G2}(t).  
\end{align*}
We want to use this minimal example and explain how the update changes the output states. Note that in the following the unitaries act on the current layers, e.g. is $U_{G1}$ is actually $U_{G1}\otimes \mathbbm{1}$ and $U_{G2}$ represents $\mathbbm{1} \otimes U_{G2}$. 

In the first part of the algorithm the generator is fixed and the discriminator is updated. When the discriminator is fed with training data we get the output state 
\begin{align*}
\rho_{\mathrm{out}}^{D}(t+\epsilon)
=&\tr_{\{2,3\}}\Big(e^{i\epsilon K_{D}}U_D \ \left(\ket{\phi^T_x} \bra{\phi^T_x} \otimes \ket{0}\bra{0}\right) \ U_D^\dagger e^{-i\epsilon K_{D}}\Big)\\
=&\tr_{\{2,3\}}\Big(U_D\ \left(\ket{\phi^T_x} \bra{\phi^T_x} \otimes \ket{0}\bra{0}\right)\ U_D^\dagger +i\epsilon\ \left[K_{D}, U_D\ \left(\ket{\phi^T_x} \bra{\phi^T_x} \otimes \ket{0}\bra{0}\right)U_D^\dagger\right]  \\
&+\mathcal{O}(\epsilon^2)\Big)\\	
=&\rho_{\mathrm{out}}^{D}(t)+i\epsilon\ \tr_{\{2,3\}}\Big(\left[K_{D}, U_D\ \left(\ket{\phi^T_x} \bra{\phi^T_x} \otimes \ket{0}\bra{0}\right) \ U_D^\dagger\right]  \Big)+\mathcal{O}(\epsilon^2).
\end{align*}
In the case the discriminator gets input of the generator we get the output state
\begin{align*}
\rho_{\mathrm{out}}^{G+D}(t+\epsilon)
=&\tr_{\{1,2,3\}}\Big(e^{i\epsilon K_{D}}U_D  U_{G2} U_{G1} ( \ket{\psi_x^\text{in}}\bra{\psi_x^\text{in}} \otimes \ket{000}\bra{000}) U_{G1}^\dagger U_{G2}^\dagger U_D^\dagger e^{-i\epsilon K_{D}}\Big)\\
=&\tr_{\{1,2,3\}}\Big(U_D  U_{G2} U_{G1} ( \ket{\psi_x^\text{in}}\bra{\psi_x^\text{in}} \otimes \ket{000}\bra{000}) U_{G1}^\dagger U_{G2}^\dagger  U_D^\dagger \\
& +i\epsilon\ \left[K_{D}, U_D  U_{G2} U_{G1} ( \ket{\psi_x^\text{in}}\bra{\psi_x^\text{in}} \otimes \ket{000}\bra{000}) U_{G1}^\dagger U_{G2}^\dagger  U_D^\dagger\right] +\mathcal{O}(\epsilon^2)\Big)\\	
=&\rho_{\mathrm{out}}^{G+D}(t)\\
&+i\epsilon\ \tr_{\{1,2,3\}}\Big(\left[K_{D}, U_D U_{G2} U_{G1} ( \ket{\psi_x^\text{in}}\bra{\psi_x^\text{in}} \otimes \ket{000}\bra{000}) U_{G1}^\dagger U_{G2}^\dagger  U_D^\dagger\right]  \Big)\\
& +\mathcal{O}(\epsilon^2).
\end{align*}
The update of the generator, assuming a fixed discriminator, is
\begin{align*}
\rho_{\mathrm{out 2}}^{G+D}(t+\epsilon)
=&\tr_{\{1,2,3\}}\Big(U_D e^{i\epsilon K_{G2}}U_{G2} e^{i\epsilon K_{G1}}U_{G1} ( \ket{\psi_x^\text{in}}\bra{\psi_x^\text{in}} \otimes \ket{000}\bra{000})\\
& U_{G1}^\dagger e^{-i\epsilon K_{G1}} U_{G2}^\dagger e^{-i\epsilon K_{G2}}  U_D^\dagger \Big)\\
=&\tr_{\{1,2,3\}}\Big(U_D \Big(   U_{G2} U_{G1} ( \ket{\psi_x^\text{in}}\bra{\psi_x^\text{in}} \otimes \ket{000}\bra{000}) U_{G1}^\dagger U_{G2}^\dagger	\\
&+ i\epsilon \ U_{G2} \left[K_{G1}, U_{G1} ( \ket{\psi_x^\text{in}}\bra{\psi_x^\text{in}} \otimes \ket{000}\bra{000}) U_{G1}^\dagger  \right] U_{G2}^\dagger \\
&+ i\epsilon \left[K_{G2}, U_{G2} U_{G1} ( \ket{\psi_x^\text{in}}\bra{\psi_x^\text{in}} \otimes \ket{000}\bra{000}) U_{G1}^\dagger U_{G2}^\dagger  \right]
\Big) U_D^\dagger +\mathcal{O}(\epsilon^2) \Big) \\		
=&\rho_{\mathrm{out 2}}^{G+D}(t)\\
&+i\epsilon\ \tr_{\{1,2,3\}}\Big(U_D \Big(  
U_{G2} \left[K_{G1}, U_{G1} ( \ket{\psi_x^\text{in}}\bra{\psi_x^\text{in}} \otimes \ket{000}\bra{000}) U_{G1}^\dagger  \right] U_{G2}^\dagger \\
&+ \left[K_{G2}, U_{G2} U_{G1} ( \ket{\psi_x^\text{in}}\bra{\psi_x^\text{in}} \otimes \ket{000}\bra{000}) U_{G1}^\dagger U_{G2}^\dagger  \right]
\Big) U_D^\dagger  \Big)+\mathcal{O}(\epsilon^2).
\end{align*}

To derive the update matrices in general, i.e.\ beyond this minimal example, we assume in the following a generator consisting of unitaries $U_1^1 \dots U_{m_g}^g$ and a discriminator built of unitaries $U_1^{g+1}\dots U_{m_{L+1}}^{L+1}$. The update matrices $K_j^l$ update the generator if $l\le g$, where $g$ is the generator's number of perceptron layers. Otherwise, the matrices describe discriminator updates. Note that in the sections describing the numerical results, \cref{sec:QGAN_classical} and \cref{sec:QGAN_QC}, we will always use $g=1$, i.e.\ a generator with just one perceptron layer connecting two layers of qubits.

\begin{prop}
The update matrix for a DQGAN trained with pure states $\ket{\phi^\text{T}_x}$  has to be of the form
\begin{equation*}
K^l_j(t) = \frac{\eta 2^{m_{l-1}}i}{S}\sum_x\tr_\text{rest}\big(M^l_{j}(x,t)\big),
\end{equation*}
where 
\begin{align*}
M_j^l =& \Big[ U_{j}^{l} \dots U_{1}^{1} \ ( \ket{\psi_x^\text{in}}\bra{\psi_x^\text{in}} \otimes \ket{0...0}\bra{0...0})  U_{1}^{1 \dagger} \dots U_{j}^{l \dagger}, \\
&U_{j+1}^{l\dagger}\dots U_{m_{L+1}}^{L+1 \dagger} \left(\mathbbm{1}_\mathrm{in+hid}\otimes \ket{1}\bra{1}\right)U_{m_{L+1}}^{L+1 } \dots U_{l+1}^{l}\Big]
\end{align*}
for $l\le g$ and 
\begin{align*}
M_j^l =& \Big[  U_{j}^{l} \dots U_{1}^{g+1} \left(\ket{\phi^T_x} \bra{\phi^T_x} \otimes \ket{0...0}\bra{0...0}\right) U_{1}^{g+1 \dagger}\dots U_{j}^{l \dagger}   \\
&- U_{j}^{l} \dots U_{1}^{g+1} U_{m_g}^{g} \dots U_{1}^{1} \ ( \ket{\psi_x^\text{in}}\bra{\psi_x^\text{in}} \otimes \ket{0...0}\bra{0...0})   U_{1}^{1 \dagger} \dots U_{m_{g}}^{g\dagger}  U_{1}^{g+1 \dagger}\dots U_{j}^{l \dagger} ,\\
&U_{j+1}^{l\dagger}\dots U_{m_{L+1}}^{L+1 \dagger} \left(\mathbbm{1}_\mathrm{in+hid}\otimes \ket{1}\bra{1}\right)U_{m_{L+1}}^{L+1 } \dots U_{l+1}^{l}\Big]
\end{align*}
else. Here $U_j^l$ is assigned to the $j$th perceptron acting on layers $l-1$and $l$, $g$ is the number of perceptron layers of the generator, and $\eta$ is the learning rate.
\end{prop}
\begin{proof}
To study the training of the discriminator, we need to compute the output state after an update with $K_D$. Note that in the following the unitaries act on the current layers, e.g. $U_1^l$ denotes actually $U_1^l\otimes \mathbbm{1}^l_{2,3,\dots,m_l}$. First we fix the generator. To derive the update for the discriminator, we compute the state when it is fed with the training data, i.e.\
\begin{align*}
\rho_{\mathrm{out}}^{D}(t+\epsilon)
=&\tr_\mathrm{in(D)+hid}\Big(e^{i\epsilon K_{m_{L+1}}^{L+1}}U_{m_{L+1}}^{L+1} \dots e^{i\epsilon K_{1}^{g+1}}U_{1}^{g+1} \ \left(\ket{\phi^T_x} \bra{\phi^T_x} \otimes \ket{0...0}\bra{0...0}\right)  \\
&  U_{1}^{g+1 \dagger}e^{-i\epsilon K_{1}^{g+1}} \dots U_{m_{L+1}}^{L+1 \dagger}e^{-i\epsilon K_{m_{L+1}}^{L+1}}\Big)\\
=&\rho_{\mathrm{out}}^{D}(t)+i\epsilon\ \tr_\mathrm{in(D)+hid}\Big(
\big[K_{m_{L+1}}^{L+1},U_{m_{L+1}}^{L+1} \dots U_{1}^{g+1} \left(\ket{\phi^T_x} \bra{\phi^T_x} \otimes \ket{0...0}\bra{0...0}\right)\\
& U_{1}^{g+1 \dagger} \dots U_{m_{L+1}}^{L+1 \dagger} \big]+\dots + U_{m_{L+1}}^{L+1} \dots U_{2}^{g+1} \big[K_{1}^{g+1}, U_{1}^{g+1} \ \ket{\phi^T_x} \bra{\phi^T_x} \\
&\otimes \ket{0...0}\bra{0...0} \ U_{1}^{g+1 \dagger} \big]  U_{2}^{g+1 \dagger} \dots U_{m_{L+1}}^{L+1 \dagger}\Big)+\mathcal{O}(\epsilon^2).
\end{align*}
Analogously we formulate the state when the discriminator gets the generator's output as input as
\begin{align*}
\rho_{\mathrm{out}}^{G+D}(t+\epsilon)
=&\tr_\mathrm{in(G)+hid}\Big(e^{i\epsilon K_{m_{L+1}}^{L+1}}U_{m_{L+1}}^{L+1} \dots e^{i\epsilon K_{1}^{g+1}}U_{1}^{g+1 } U_{m_g}^{g}\dots U_1^1  ( \ket{\psi_x^\text{in}}\bra{\psi_x^\text{in}} \\
&\otimes \ket{0...0}\bra{0...0})   U_1^{1\dagger} \dots U_{m_g}^{g\dagger}   U_{1}^{g+1\dagger}e^{-i\epsilon K_{1}^{g+1}} \dots U_{m_{L+1}}^{L+1\dagger}e^{-i\epsilon K_{m_{L+1}}^{L+1}}\Big)\\
=&\rho_{\mathrm{out}}^{D}(t)+i\epsilon\ \tr_\mathrm{in(G)+hid}\Big(
\big[K_{m_{L+1}}^{L+1},U_{m_{L+1}}^{L+1} \dots U_{1}^{1} \ ( \ket{\psi_x^\text{in}}\bra{\psi_x^\text{in}}\\
& \otimes \ket{0...0}\bra{0...0})  U_{1}^{1 \dagger} \dots U_{m_{L+1}}^{L+1 \dagger} \big]+\dots \\
&+ U_{m_{L+1}}^{L+1} \dots U_{2}^{g+1} \big[K_{1}^{g+1}, U_{1}^{g+1} \ U_{m_g}^{g}\dots U_1^1  ( \ket{\psi_x^\text{in}}\bra{\psi_x^\text{in}} \otimes \ket{0...0}\bra{0...0}) \\
&  U_1^{1\dagger} \dots U_{m_g}^{g\dagger} \ U_{1}^{g+1 \dagger} \big] U_{2}^{g+1 \dagger} \dots U_{m_{L+1}}^{L+1 \dagger} \Big)   +\mathcal{O}(\epsilon^2).	
\end{align*}
Further, the derivative of the discriminator loss function is of the form
\begin{align*}
\frac{d\mathcal{L}_D}{dt}=&\lim_{\epsilon\rightarrow 0}\frac{\mathcal{L}_D(t)+i\epsilon\frac{1}{S} \sum_{x=1}^S\bra{1}\tr_\mathrm{in+hid}(\dots)\ket{1}-\mathcal{L}_D(t)}{\epsilon}\\
=&\frac{i}{S}\ \sum_{x=1}^S\tr_\mathrm{in+hid}\Big(\mathbbm{1}_\mathrm{in+hid}\otimes \ket{1}\bra{1}\Big(\Big(
\big[K_{m_{L+1}}^{L+1},U_{m_{L+1}}^{L+1} \dots U_{1}^{g+1} \ket{\phi^T_x} \bra{\phi^T_x} \\
&\otimes \ket{0...0}\bra{0...0} U_{1}^{g+1 \dagger} \dots U_{m_{L+1}}^{L+1 \dagger} \big]  +\hdots \\
& + U_{m_{L+1}}^{L+1} \dots U_{2}^{g+1} \left[K_{1}^{g+1}, U_{1}^{g+1} \ \left(\ket{\phi^T_x} \bra{\phi^T_x} \otimes \ket{0...0}\bra{0...0}\right) \ U_{1}^{g+1 \dagger} \right]\\
&  U_{2}^{g+1 \dagger} \dots U_{m_{L+1}}^{L+1 \dagger}\Big)  \\
&- \Big(
\big[K_{m_{L+1}}^{L+1},U_{m_{L+1}}^{L+1} \dots U_{1}^{1} \ ( \ket{\psi_x^\text{in}}\bra{\psi_x^\text{in}}  \otimes \ket{0...0}\bra{0...0}) U_{1}^{1 \dagger} \dots U_{m_{L+1}}^{L+1 \dagger} \big]+\dots \\
& + U_{m_{L+1}}^{L+1} \dots U_{2}^{g+1} \big[K_{1}^{g+1}, U_{1}^{g+1} \ U_{m_g}^{g}\dots U_1^1  ( \ket{\psi_x^\text{in}}\bra{\psi_x^\text{in}} \otimes \ket{0...0}\bra{0...0}) \\
&  U_1^{1\dagger} \dots U_{m_g}^{g\dagger} \ U_{1}^{g+1 \dagger} \big] U_{2}^{g+1 \dagger} \dots U_{m_{L+1}}^{L+1 \dagger} \Big)\Big)\Big) \\
=&\frac{i}{S}\ \sum_{x=1}^S\tr_\mathrm{in+hid}\Big( \Big[  U_{m_{L+1}}^{L+1} \dots U_{1}^{g+1} \left(\ket{\phi^T_x} \bra{\phi^T_x} \otimes \ket{0...0}\bra{0...0}\right) U_{1}^{g+1 \dagger} \dots U_{m_{L+1}}^{L+1 \dagger}\\
&-U_{m_{L+1}}^{L+1} \dots U_{1}^{1} \ ( \ket{\psi_x^\text{in}}\bra{\psi_x^\text{in}} \otimes \ket{0...0}\bra{0...0})  U_{1}^{1 \dagger} \dots U_{m_{L+1}}^{L+1 \dagger} ,\\
&\mathbbm{1}_\mathrm{in+hid}\otimes \ket{1}\bra{1}\Big]K_{m_{L+1}}^{L+1}+\dots +\Big[ U_{1}^{g+1} \left(\ket{\phi^T_x} \bra{\phi^T_x} \otimes \ket{0...0}\bra{0...0}\right) U_{1}^{g+1 \dagger}\\
& - U_{1}^{g+1} U_{m_g}^{g} \dots U_{1}^{1} \ ( \ket{\psi_x^\text{in}}\bra{\psi_x^\text{in}} \otimes \ket{0...0}\bra{0...0}) \\
&   U_{1}^{1 \dagger} \dots U_{m_{g}}^{g\dagger}  U_{1}^{g+1\dagger} , U_{m_{L+1}}^{L+1\dagger}\dots U_{2}^{g+1\dagger} \mathbbm{1}_\mathrm{in+hid}\otimes \ket{1}\bra{1}U_{2}^{g+1} \dots U_{m_{L+1}}^{L+1}\Big]K_{1}^{g+1}\Big)\\
=&\frac{i}{S}\ \sum_{x=1}^S\tr_\mathrm{in+hid}\left(M_{m_{L+1}}^{L+1}K_{m_{L+1}}^{L+1}+\dots+M_{1}^{g+1}K_{1}^{g+1}\right).
\end{align*}
Note that at this point $ \ket{\phi^T_x} \bra{\phi^T_x} \otimes \ket{0...0}\bra{0...0}$ denotes $\mathbbm{1}_{in(G)+hid(G)} \otimes  \ket{\phi^T_x} \bra{\phi^T_x} \otimes \ket{0...0}\bra{0...0}$, to match the dimension of the other summand. To this point the generator was fixed. As a next step we fix the discriminator. Using the state
\begin{align*}
\rho_{\mathrm{out 2}}^{G+D}(s+\epsilon)
=&\tr_\mathrm{in(G)+hid}\Big(U_{m_{L+1}}^{L+1 } \dots U_{1}^{g+1}  \ e^{i\epsilon K_{m_{g}}^{g}}U_{m_{g}}^{g} \dots e^{i\epsilon K_{1}^{1}}U_{1}^{1 } \ ( \ket{\psi_x^\text{in}}\bra{\psi_x^\text{in}} \\
&\otimes \ket{0...0}\bra{0...0})   U_{1}^{1\dagger} e^{-i\epsilon K_{1}^{1}}\dots U_{m_{g}}^{g\dagger} e^{-i\epsilon K_{m_{g}}^{g}}U_{1}^{g+1\dagger}\dots U_{m_{L+1}}^{L+1 \dagger}\Big)\\
=&\rho_{\mathrm{out}}^{D}(t)+i\epsilon\ \tr_\mathrm{in()+hid}\Big(
U_{m_{L+1}}^{L+1 } \dots U_{1}^{g+1}\big[K_{m_{g}}^{g},U_{m_{g}}^{g} \dots U_{1}^{1} \ ( \ket{\psi_x^\text{in}}\bra{\psi_x^\text{in}} \\
& \otimes \ket{0...0}\bra{0...0})  U_{1}^{1 \dagger} \dots U_{m_{g}}^{g \dagger} \big] U_{1}^{g+1\dagger}\dots U_{m_{L+1}}^{L+1 \dagger} +\dots \\
&+ U_{m_{L+1}}^{L+1} \dots U_{2}^{1} \left[K_{1}^{1}, U_{1}^{1} \ ( \ket{\psi_x^\text{in}}\bra{\psi_x^\text{in}} \otimes \ket{0...0}\bra{0...0}) \ U_{1}^{1 \dagger} \right] U_{2}^{1 \dagger} \dots U_{m_{L+1}}^{L+1 \dagger} \Big) \\
& +\mathcal{O}(\epsilon^2)	
\end{align*}
the derivative of the loss function for training the generator can be written as
\begin{align*}
\frac{d\mathcal{L}_G}{dt}=&\lim_{\epsilon\rightarrow 0}\frac{\mathcal{L}_G(t)+i\epsilon\frac{1}{S} \sum_x\bra{1}\tr_\mathrm{in+hid}(\dots)\ket{1}-\mathcal{L}_G(t)}{\epsilon}\\
=&\frac{i}{S}\ \sum_{x=1}^S\tr\Big(\mathbbm{1}_\mathrm{in+hid}\otimes \ket{1}\bra{1}\Big(\Big(
U_{m_{l+1}}^{l+1 } \dots U_{1}^{g+1}\big[K_{m_{g}}^{g},U_{m_{g}}^{g} \dots U_{1}^{1} \ ( \ket{\psi_x^\text{in}}\bra{\psi_x^\text{in}} \\
&\otimes \ket{0...0}\bra{0...0})  U_{1}^{1 \dagger} \dots U_{m_{g}}^{g \dagger} \big] U_{1}^{g+1\dagger}\dots U_{m_{l+1}}^{l+1 \dagger} +\dots \\
&+ U_{m_{l+1}}^{l+1} \dots U_{2}^{1} \left[K_{1}^{1}, U_{1}^{1} \ ( \ket{\psi_x^\text{in}}\bra{\psi_x^\text{in}} \otimes \ket{0...0}\bra{0...0}) \ U_{1}^{1 \dagger} \right] U_{2}^{1 \dagger} \dots U_{m_{l+1}}^{l+1 \dagger} \Big)   \Big)\Big) \\
=&\frac{i}{S}\ \sum_{x=1}^S\tr\Big( \Big[ U_{m_{g}}^{g} \dots U_{1}^{1} \ ( \ket{\psi_x^\text{in}}\bra{\psi_x^\text{in}} \otimes \ket{0...0}\bra{0...0})  U_{1}^{1 \dagger} \dots U_{m_{g}}^{g \dagger}, \\
&U_{1}^{g+1\dagger}\dots U_{m_{l+1}}^{l+1 \dagger} \left(\mathbbm{1}_\mathrm{in+hid}\otimes \ket{1}\bra{1}\right) U_{m_{l+1}}^{l+1 } \dots U_{1}^{g+1}\Big]K_{m_{g}}^{g}+\dots \\
&+\Big[ U_{1}^{1} \ ( \ket{\psi_x^\text{in}}\bra{\psi_x^\text{in}} \otimes \ket{0...0}\bra{0...0})  U_{1}^{1 \dagger} , \\
&U_{2}^{1 \dagger} \dots U_{m_{l+1}}^{l+1 \dagger}  \left(\mathbbm{1}_\mathrm{in+hid}\otimes \ket{1}\bra{1}\right) U_{m_{l+1}}^{l+1} \dots U_{2}^{1}\Big]K_{1}^{1}\Big)\\
\equiv&\frac{i}{S}\ \sum_{x=1}^S\tr\left(M_{m_{g}}^{g}K_{m_{g}}^{g}+\dots+M_{1}^{1}K_{1}^{1}\right).
\end{align*}
In both updates, we parametrise the parameter matrices analogously to the proof of \cref{prop:DQNN_K} as
\begin{equation*}
K_j^l(t)=\sum_{\alpha_1,\alpha_2,\dots,\alpha_{m_{l-1}},\beta}K^l_{j,\alpha_1,\dots,\alpha_{m_{l-1}},\beta}(t)\left(\sigma^{\alpha_1}\otimes\ \dots\ \otimes\sigma^{\alpha_{m_{l-1}}}\otimes\sigma^\beta\right),
\end{equation*}
where the $\alpha_i$ denote the qubits in the previous layer and $\beta$ denotes the current qubit in layer $l$. To reach the maximum of the loss function as a function of the parameters \emph{fastest}, we maximise $\frac{d\mathcal{L}}{dt}$. Since this is a linear function, the extrema are at $\pm\infty$. To ensure that we get a finite solution, we introduce a Lagrange multiplier $\lambda\in\mathbbm{R}$. Hence, to find $K_j^l$ we have to solve a following maximisation problem for the discriminator update and for the generator update. Since both are solved analogously we only formulate the maximisation for the discriminator update here, namely
\begin{align*}
\max_{K^l_{j,\alpha_1,\dots,\beta}}&\left(\frac{dC(t)}{dt}-\lambda\sum_{\alpha_i,\beta}{K^l_{j,\alpha_1,\dots,\beta}}(t)^2\right)\\
=&\max_{K^l_{j,\alpha_1,\dots,\beta}}\left(\frac{i}{S}\sum_{x=1}^S \tr\left(M_{m_{L+1}}^{L+1}K_{m_{L+1}}^{L+1}+\dots+M_{1}^{g+1}K_{1}^{g+1}\right)-\lambda\sum_{\alpha_1,\dots,\beta}{K^l_{j,\alpha_1,\dots,\beta}}(t)^2\right)\\
=&\max_{K^l_{j,\alpha_1,\dots,\beta}}\Big(\frac{i}{S}\sum_{x=1}^S\tr_{\alpha_1,\dots,\beta}\left(\tr_\mathrm{rest}\left(M_{m_{L+1}}^{L+1}K_{m_{L+1}}^{L+1}+\dots+M_{1}^{g+1}K_{1}^{g+1}\right)\right)\\
&-\lambda\sum_{\alpha_1,\dots,\beta}{K^l_{j,\alpha_1,\dots,\beta}}(t)^2\Big).
\end{align*}
Taking the derivative with respect to $K^l_{j,\alpha_1,\dots,\beta}$ leads to
\begin{align*}
\frac{i}{S}\sum_{x=1}^S\tr_{\alpha_1,\dots,\beta}\left(\tr_\mathrm{rest}\left(M_j^l(t)\right)\left(\sigma^{\alpha_1}\otimes\ \dots\ \otimes\sigma^\beta\right)\right)-2\lambda K^l_{j,\alpha_1,\dots,\beta}(t)=0,
\end{align*}
hence,
\begin{align*}
K^l_{j,\alpha_1,\dots,\beta}(t)=\frac{i}{2S\lambda}\sum_{x=1}^S\tr_{\alpha_1,\dots,\beta}\left(\tr_\mathrm{rest}\left(M_j^l(t)\right)\left(\sigma^{\alpha_1}\otimes\ \dots\ \otimes\sigma^\beta\right)\right)
\end{align*}
This produces the matrix 
\begin{align*}
K_j^l(t)=&\sum_{\alpha_1,\dots,\beta}K^l_{j,\alpha_1,\dots,\beta}(t)\left(\sigma^{\alpha_1}\otimes\ \dots\ \otimes\sigma^\beta\right)\\
=&\frac{i}{2S\lambda}\sum_{\alpha_1,\dots,\beta}\sum_{x=1}^S\tr_{\alpha_1,\dots,\beta}\left(\tr_\mathrm{rest}\left(M_j^l(t)\right)\left(\sigma^{\alpha_1}\otimes\ \dots\ \otimes\sigma^\beta\right)\right)\left(\sigma^{\alpha_1}\otimes\ \dots\ \otimes\sigma^\beta\right)\\
=&\frac{\eta2^{m_{l-1}}i}{2S}\sum_{x=1}^S\tr_\mathrm{rest}\left(M_j^l(t)\right),
\end{align*}
where $\eta=1/\lambda$ is the learning rate and $\tr_\text{rest}$ traces out all qubits that the perceptron unitary $U_j^l$ does not act on.

Notice again that $K_j^l$ updates the generator if $j\le g$ for the generator's number of perceptron layers $g$. The definition of $M_j^l$ is
\begin{align*}
M_j^l =& \Big[ U_{j}^{l} \dots U_{1}^{1} \ ( \ket{\psi_x^\text{in}}\bra{\psi_x^\text{in}} \otimes \ket{0...0}\bra{0...0})  U_{1}^{1 \dagger} \dots U_{j}^{l \dagger}, \\
&U_{j+1}^{l\dagger}\dots U_{m_{L+1}}^{L+1 \dagger} \left(\mathbbm{1}_\mathrm{in+hid}\otimes \ket{1}\bra{1}\right)U_{m_{L+1}}^{L+1 } \dots U_{l+1}^{l}\Big]
\end{align*}
for $l\le g$ and 
\begin{align*}
M_j^l =& \Big[  U_{j}^{l} \dots U_{1}^{g+1} \left(\ket{\phi^T_x} \bra{\phi^T_x} \otimes \ket{0...0}\bra{0...0}\right) U_{1}^{g+1 \dagger}\dots U_{j}^{l \dagger}   \\
&- U_{j}^{l} \dots U_{1}^{g+1} U_{m_g}^{g} \dots U_{1}^{1} \ ( \ket{\psi_x^\text{in}}\bra{\psi_x^\text{in}} \otimes \ket{0...0}\bra{0...0})   U_{1}^{1 \dagger} \dots U_{m_{g}}^{g\dagger}  U_{1}^{g+1 \dagger}\dots U_{j}^{l \dagger} ,\\
&U_{j+1}^{l\dagger}\dots U_{m_{L+1}}^{L+1 \dagger} \left(\mathbbm{1}_\mathrm{in+hid}\otimes \ket{1}\bra{1}\right)U_{m_{L+1}}^{L+1 } \dots U_{l+1}^{l}\Big]
\end{align*}
else.
\end{proof}

\section{Classical simulation}
\label{sec:QGAN_classical}
Just like in \cref{chapter:DQNN} and \cref{chapter:graphs}, where we discussed the initially proposed DQNNs \cite{Beer2020} and the extension to graph-structured quantum data \cite{Beer2021}, we use classical simulation to study the above-described training methods for DQGANs.

In contrary to the corresponding sections in the chapters mentioned above, we structure this section into two parts. The first part describes how the data set used for training the DQGAN. Next, we use the classical simulation to demonstrate how the generator's and discriminator's training losses and the validation loss evolve during the training process. In the last part, we explain how to study the generated states' diversity. The used code can be found at \cite{GithubKerstin}.

\subsection*{Training set}
As an exemplary data set we use a set of pure one-qubit states which build a line on the Bloch sphere, namely
\begin{equation}
\text{data}_\text{line}=\left\{\frac{(N-x)\ket{0}+(x-1)\ket{1}}{||(N-x)\ket{0}+(x-1)\ket{1}||}\right\}_{x=1}^{N},
\label{eqn:GAN_lineN}
\end{equation}
see also \cref{fig:QGAN_blocha}. To use $\text{data}_\text{line}$ for the training it is shuffled. The first $S$ of the resulting set, denoted as $\{\ket{\phi^T_x}\}_{x=1}^{S}$, will be used for the training process and can be accessed by the generative model. The full data set $\{\ket{\phi^T_x}\}_{x=1}^{N}$ with $N>S$ is used for computing the validation loss. In the following we choose $N=50$ and $S=10$.

\subsection*{Evaluation of the loss functions}
\begin{figure}
\centering
\begin{subfigure}{\textwidth}\centering
\begin{tikzpicture}
\begin{axis}[
xmin=0,   xmax=20,
ymin=0.2,   ymax=1.5,
width=0.8\linewidth, 
height=0.5\linewidth,
grid=major,grid style={color0M},
xlabel= Training epochs $r_T$, 
xticklabels={0,0,100,200,300,400,500,600,700,800,900,1000},
ylabel=$\mathcal{L}(t)$,legend pos=north east,legend cell align={left},legend style={draw=none,legend image code/.code={\filldraw[##1] (-.5ex,-.5ex) rectangle (0.5ex,0.5ex);}}]
\coordinate (0,0) ;
\addplot[mark size=1.5 pt,  color=color2] table [x=step times epsilon, y=costFunctionDis, col sep=comma] {numerics/QGAN_50data10sv_100statData_100statData_1-1networkGen_1-1networkDis_lda1_ep0i01_rounds1000_roundsGen1_roundsDis1_line_plot1_training.csv};
\addlegendentry[scale=1]{Training loss $\mathcal{L}_\text{D}$} 
\addplot[mark size=1.5 pt,  color=color1] table [x=step times epsilon, y=costFunctionGen, col sep=comma] {numerics/QGAN_50data10sv_100statData_100statData_1-1networkGen_1-1networkDis_lda1_ep0i01_rounds1000_roundsGen1_roundsDis1_line_plot1_training.csv};
\addlegendentry[scale=1]{Training loss $\mathcal{L}_\text{G}$} 
\addplot[mark size=1.5 pt,  color=color3] table [x=step times epsilon, y=costFunctionTest, col sep=comma] {numerics/QGAN_50data10sv_100statData_100statData_1-1networkGen_1-1networkDis_lda1_ep0i01_rounds1000_roundsGen1_roundsDis1_line_plot1_training.csv};
\addlegendentry[scale=1]{Validation loss $\mathcal{L}_\text{V}$} 
\draw [line width=0.5mm,dashed] (60,0) -- (60,200);
\node at (70,10) {(b)};
\draw [line width=0.5mm,dashed] (100,0) -- (100,200);
\node at (110,10) {(c)};
\draw [line width=0.5mm,dashed] (160,0) -- (160,200);
\node at (170,10) {(d)};
\end{axis}
\end{tikzpicture}
\subcaption{Loss functions.}
\label{fig:GAN_line}
\end{subfigure}
\begin{subfigure}{\textwidth}\centering
\begin{tikzpicture}[scale=1]
\begin{axis}[
ybar,
bar width=1.5pt,
xmin=0,   xmax=51,
ymin=0,   ymax=12,
width=.8\linewidth, 
height=.28\linewidth,
grid=major,
grid style={color0M},
xlabel= State index $x$, 
ylabel=Counter,legend pos=north east,legend cell align={left},legend style={draw=none,legend image code/.code={\filldraw[##1] (-.5ex,-.5ex) rectangle (0.5ex,0.5ex);}}]
\addplot[color=color2, fill=color2] table [x=indexDataTest, y=countOutTest, col sep=comma] {numerics/QGAN_50data10sv_100statData_100statData_1-1networkGen_1-1networkDis_lda1_ep0i01_rounds300_roundsGen1_roundsDis1_line_plot1_statisticsUSV.csv};
\addlegendentry[scale=1]{Validation states} 
\addplot[color=color1,fill=color1] table [x=indexDataTrain, y=countOutTrain, col sep=comma] {numerics/QGAN_50data10sv_100statData_100statData_1-1networkGen_1-1networkDis_lda1_ep0i01_rounds300_roundsGen1_roundsDis1_line_plot1_statisticsSV.csv};
\addlegendentry[scale=1]{Supervised states} 
\end{axis}
\end{tikzpicture}
\subcaption{Diversity of the generator's output after $r_T=300$ training epochs.} \label{fig:GAN_line300}
\end{subfigure}
\begin{subfigure}{\textwidth}\centering
\begin{tikzpicture}[scale=1]
\begin{axis}[
ybar,
bar width=1.5pt,
xmin=0,   xmax=51,
ymin=0,   ymax=25,
width=.8\linewidth, 
height=.28\linewidth,
grid=major,
grid style={color0M},
xlabel= State index $x$, 
ylabel=Counter,legend pos=north east,legend cell align={left},legend style={draw=none,legend image code/.code={\filldraw[##1] (-.5ex,-.5ex) rectangle (0.5ex,0.5ex);}}]
\addplot[color=color2, fill=color2] table [x=indexDataTest, y=countOutTest, col sep=comma] {numerics/QGAN_50data10sv_100statData_100statData_1-1networkGen_1-1networkDis_lda1_ep0i01_rounds500_roundsGen1_roundsDis1_line_plot1_statisticsUSV.csv};
\addlegendentry[scale=1]{Validation states} 
\addplot[color=color1,fill=color1] table [x=indexDataTrain, y=countOutTrain, col sep=comma] {numerics/QGAN_50data10sv_100statData_100statData_1-1networkGen_1-1networkDis_lda1_ep0i01_rounds500_roundsGen1_roundsDis1_line_plot1_statisticsSV.csv};
\addlegendentry[scale=1]{Supervised states} 
\end{axis}
\end{tikzpicture}
\subcaption{Diversity of the generator's output after $r_T=500$ training epochs.} \label{fig:GAN_line500}
\end{subfigure}
\begin{subfigure}{\textwidth}\centering
\begin{tikzpicture}[scale=1]
\begin{axis}[
ybar,
bar width=1.5pt,
xmin=0,   xmax=51,
ymin=0,   ymax=110,
width=.8\linewidth, 
height=.28\linewidth,
grid=major,
grid style={color0M},
xlabel= State index $x$, 
ylabel=Counter,legend pos=north east,legend cell align={left},legend style={draw=none,legend image code/.code={\filldraw[##1] (-.5ex,-.5ex) rectangle (0.5ex,0.5ex);}}]
\addplot[color=color2, fill=color2] table [x=indexDataTest, y=countOutTest, col sep=comma] {numerics/QGAN_50data10sv_100statData_100statData_1-1networkGen_1-1networkDis_lda1_ep0i01_rounds800_roundsGen1_roundsDis1_line_plot1_statisticsUSV.csv};
\addlegendentry[scale=1]{Validation states} 
\addplot[color=color1,fill=color1] table [x=indexDataTrain, y=countOutTrain, col sep=comma] {numerics/QGAN_50data10sv_100statData_100statData_1-1networkGen_1-1networkDis_lda1_ep0i01_rounds800_roundsGen1_roundsDis1_line_plot1_statisticsSV.csv};
\addlegendentry[scale=1]{Supervised states} 
\end{axis}
\end{tikzpicture}
\subcaption{Diversity of the generator's output after $r_T=800$ training epochs.} \label{fig:GAN_line800}
\end{subfigure}
\caption{\textbf{Training a DQGAN.} (a) depicts the evolution of the loss functions during the training of a \protect\oneoneone DQGAN in $r_T=1000$ epochs with $\eta=1$ and $\epsilon=0.01$ using $50$ data pairs where $10$ are used as training pairs. The dashed lines mark the diversity checks after 300 (b), 500 (c) and 800 (d) training epochs $r_T$ for the generator's output.}
\end{figure}
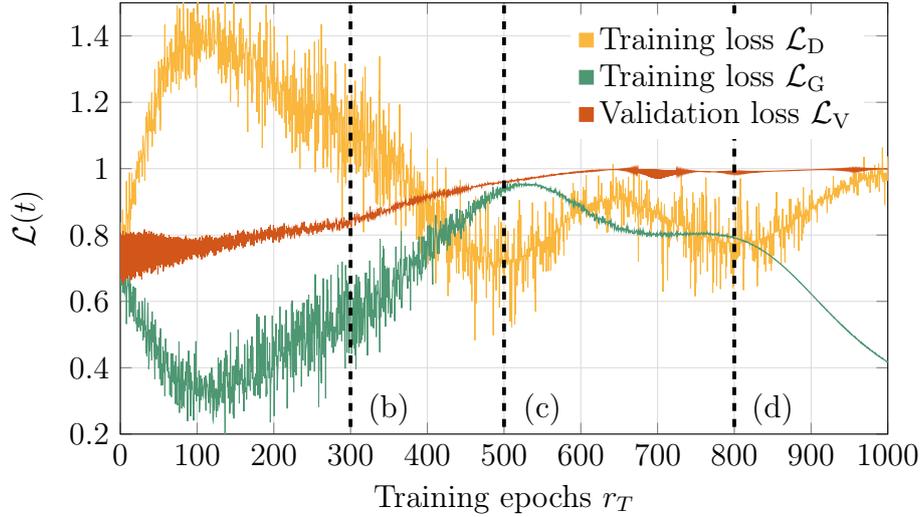
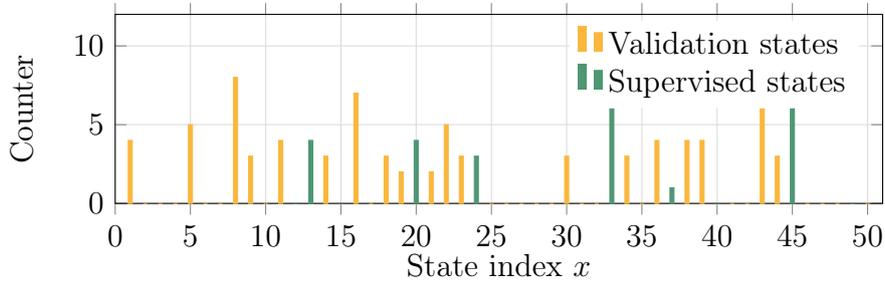
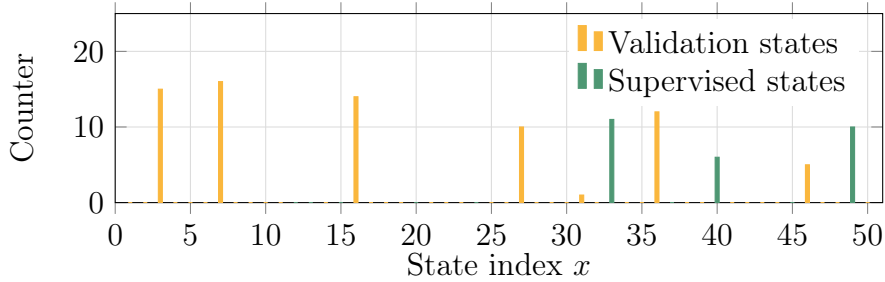
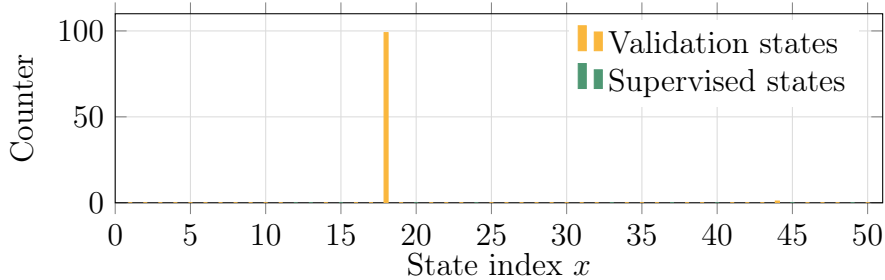

In \cref{fig:GAN_line}, both training losses and the validation loss during training of a 1-1-1 DQGAN are plotted. For all in the following presented examples we choose $r_D=r_G=1$, $\epsilon=0.01$ and $\eta=1$ and a network architecture of 1-1-1. Note that in \cref{fig:apdx_line} in the appendix, the training of a 1-3-1 DQGAN is documented as well. 

Compared to the prior loss function plots in this work, the evaluation of the losses is much more eventful. The validation loss reaches values over $0.95$ after $r_T=475$ training epochs. In the first training epochs, the training loss of the generator shrinks and the discriminator training loss increases. This behaviour is inverted after $r_T=100$. Following the remaining training, we find that this process is repeated. The generator's and discriminator's opposing goals, formulated in the training loss functions, explain this behaviour. 

Furthermore, we can observe the saturation of the validation loss at nearly 1. This means that every output of the generator is close to one state of $\text{data}_\text{line}$. Note that here we define the closeness through the fidelity of two quantum states: the larger the fidelity is, the closer the states are. However, from this plot, it is not clear if the generator produces the same state time after time or, in the other extreme, reproduces all states of the set $\text{data}_\text{line}$. In both cases, the validation loss would be maximal. Hence, we have to study the \emph{diversity} of the generator's output in the following. 

\subsection*{Diversity of the generator's output}
Since our aim is to train the generator for producing extended data sets, we want its output to be diverse. We build a set of $100$ by the generator produced states to test the diversity. For each of these, we find the element with the index $x$ in $\text{data}_\text{line}$, which is the closest concerning the fidelity of two states.  

\begin{figure}
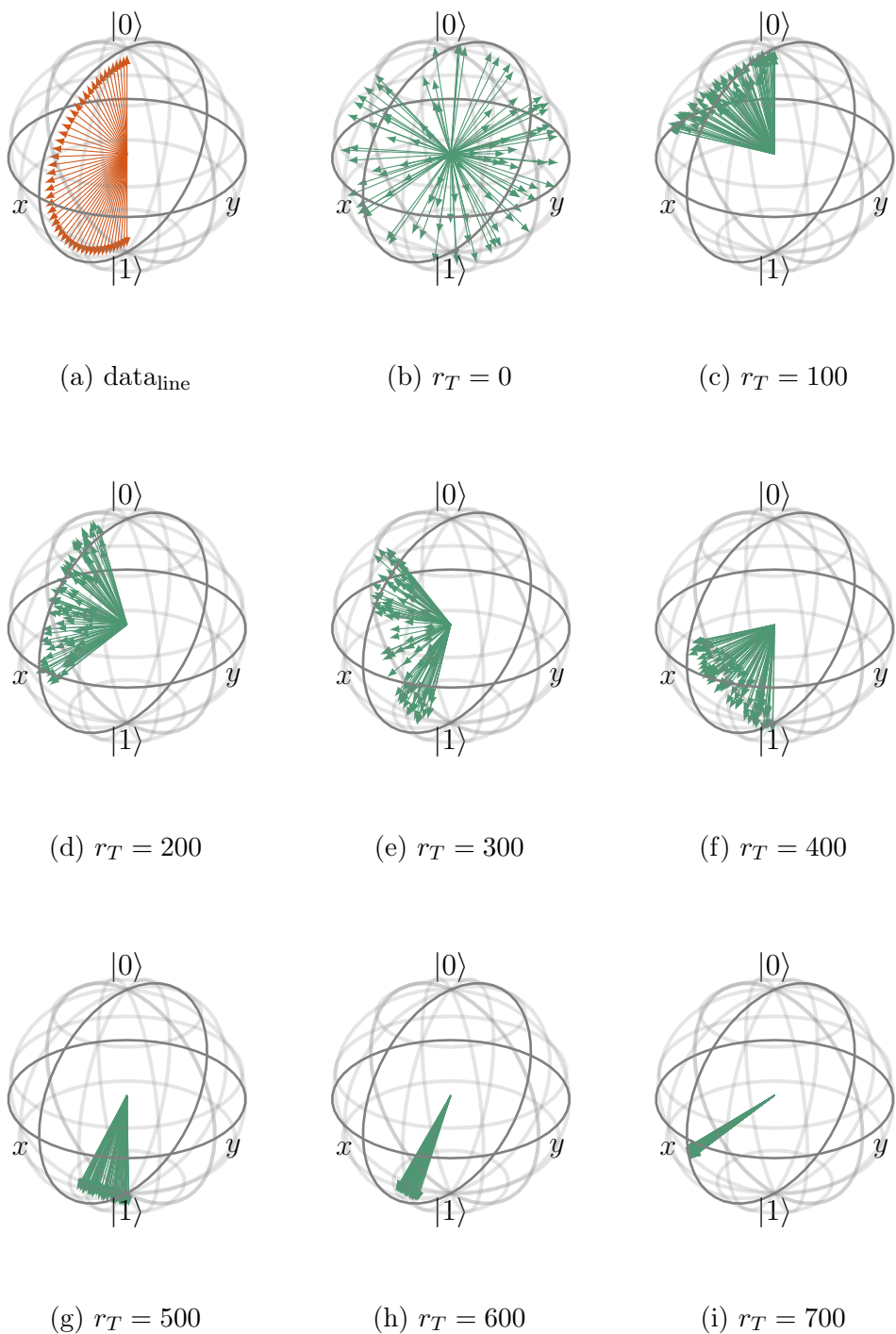

	\centering
	\begin{subfigure}{0.3\linewidth}
		\input{numerics/T}
		\subcaption{$\text{data}_\text{line}$}
		\label{fig:QGAN_blocha}
	\end{subfigure}
	\begin{subfigure}{0.3\linewidth}
		\input{numerics/0}
		\subcaption{$r_T=0$}
	\end{subfigure}
	\begin{subfigure}{0.3\linewidth}
		\input{numerics/100}
		\subcaption{$r_T=100$}
	\end{subfigure}
	
	\vspace{1cm}
	
	\begin{subfigure}{0.3\linewidth}
		\input{numerics/200}
		\subcaption{$r_T=200$}
	\end{subfigure}
	\begin{subfigure}{0.3\linewidth}
		\input{numerics/300}
		\subcaption{$r_T=300$}
	\end{subfigure}
	\begin{subfigure}{0.3\linewidth}
		\input{numerics/400}
		\subcaption{$r_T=400$}
	\end{subfigure}
	
	\vspace{1cm}
	
	\begin{subfigure}{0.3\linewidth}
		\input{numerics/500}
		\subcaption{$r_T=500$}
	\end{subfigure}
	\begin{subfigure}{0.3\linewidth}
		\input{numerics/600}
		\subcaption{$r_T=600$}
	\end{subfigure}
	\begin{subfigure}{0.3\linewidth}
		\input{numerics/700}
		\subcaption{$r_T=700$}
	\end{subfigure}
	\caption{\textbf{Output of the generator.} To compare the output of the generator (b-i), during the training of a \protect\oneoneone DQGAN, to the data set $\text{data}_\text{line}$ (a) we plot the states in Bloch spheres.}
	\label{fig:QGAN_bloch}
\end{figure}

Following this procedure, we count how many times every element in $\text{data}_\text{line}$ was (approximately) reproduced by the generator. In \cref{fig:GAN_line300,fig:GAN_line500,fig:GAN_line800} these numbers are depicted in the form of histograms. According to the definition of $\text{data}_\text{line}$ in \cref{eqn:GAN_lineN}, $x=1$ corresponds to the state $\ket{0}$ and $x=50$ to  $\ket{1}$, respectively. The number in the histogram related to, for example, $x=1$ shows how often the output of the generator was approximately $\ket{0}$.  The different colours describe whether an element of $\text{data}_\text{line}$ was used as a training state or not.  We can see that the DQGAN can generalise, and not only the training states get reproduced. Specifically, we show this diversity checks after $300$, $500$, and $800$ training epochs. Comparing \cref{fig:GAN_line300,fig:GAN_line500,fig:GAN_line800} with the plots of the loss functions in \cref{fig:GAN_line} (see the dashed lines), let us resume that the training of the DQGAN has to be stopped, in this case, after about $300$ training epochs to have a good balance between a large diversity and a significant validation loss.

\cref{fig:QGAN_bloch}, where the output of the generator is plotted for different training epochs in Bloch spheres, confirms this statement. At every of these training steps we build a set of $100$ states produced by the generator  and plot these states in a Bloch sphere. For the discussion of other training data sets we point to \cref{fig:apnx_Div} in the appendix.

\section{NISQ device implementation}
\label{sec:QGAN_QC} 

The last section described results of the classical simulation of the DQGAN. In the following we want to discuss \DQGANNISQ, i.e.\ the implementation of the DQGAN algorithm for a NISQ device. In \cref{sec:DQNN_quantumalg} we described \DQNNNISQ in full detail. 

Two main things are different in the implementation of \DQGANNISQ compared to the \DQGANNISQ. Firstly, we have two instead of one training functions. Depending on the phase of the algorithm, see \cref{alg:QGAN_algorithmQ}, we update the parameters ${\omega^G}_{s}$ of the generator or the discriminator's parameters ${\omega^D}_{s}$ using the gradient of the training loss $\mathcal{L}_\text{G}$ or $\mathcal{L}_\text{D}$, respectively. Whereas one measurement per input is needed for computing $\mathcal{L}_\text{G}$, two are needed for executing $\mathcal{L}_\text{D}$ for one input.

Secondly, we have to implement the quantum circuit, in order that the discriminator can be used in two ways, i.e.\ 
\begin{equation*}
\rho^\text{out}=
\begin{cases} 
\mathcal{E}_D (\mathcal{E}_G (\ket{\psi^\text{in}}\bra{\psi^\text{in}})) &\mbox{for generated data}\\
\mathcal{E}_D (\ket{\phi^T} \bra{\phi^T}) & \mbox{for training data.} 
\end{cases}
\end{equation*}

\begin{sloppypar}
Moreover, we could observe that a small change compared to the \DQNNNISQ implementation presented in \cref{sec:DQNN_quantumalg} is helpful for the training the \DQGANNISQ. We describe the modification, which increases the computational power of the \DQGANNISQ without using additional qubits, in the appendix in  \cref{fig:GAN_circuit_implementation}. In the following we denote this modification with a $^+$ when describing the DQNN architecture, for example in 2-3-2$^+$. Further we use different learning rates for the generator and discriminator, denoted by $\eta_D$ and $\eta_G$, respectively.
\end{sloppypar}

\begin{figure}[H]
	\centering
	\begin{subfigure}{\textwidth}\centering
		\begin{tikzpicture}[scale=1]
			\begin{axis}[
				ybar,
				bar width=1.5pt,
				xmin=0,   xmax=50,
				ymin=0,   ymax=20,
				width=.8\linewidth, 
				height=.28\linewidth,
				grid=major,
				grid style={color0M},
				xlabel= State index $x$, 
				ylabel=Counts,legend pos=north east,legend cell align={left},legend style={draw=none,legend image code/.code={\filldraw[##1] (-.5ex,-.5ex) rectangle (0.5ex,0.5ex);}}]
				\addplot[color=color2, fill=color2] table [x=indexDataTest, y=countOutTest, col sep=comma] {numerics/dqnn_q_eq_line_v2_epoch_100_vs.csv};
				\addlegendentry[scale=1]{Validation states} 
				\addplot[color=color1,fill=color1] table [x=indexDataTrain, y=countOutTrain, col sep=comma] {numerics/dqnn_q_eq_line_v2_epoch_100_ts.csv};
				\addlegendentry[scale=1]{Training states} 
			\end{axis}
		\end{tikzpicture}
		\subcaption{Diversity of the generator's output after $r_T=100$ training epochs.} \label{fig:QGANQ_eq_line_a}
	\end{subfigure}
	\begin{subfigure}{\textwidth}\centering
		\begin{tikzpicture}[scale=1]
			\begin{axis}[
				ybar,
				bar width=1.5pt,
				xmin=0,   xmax=50,
				ymin=0,   ymax=20,
				width=.8\linewidth, 
				height=.28\linewidth,
				grid=major,
				grid style={color0M},
				xlabel= State index $x$, 
				ylabel=Counts,legend pos=north east,legend cell align={left},legend style={draw=none,legend image code/.code={\filldraw[##1] (-.5ex,-.5ex) rectangle (0.5ex,0.5ex);}}]
				\addplot[color=color2, fill=color2] table [x=indexDataTest, y=countOutTest, col sep=comma] {numerics/dqnn_q_eq_line_v2_epoch_440_vs.csv};
				\addlegendentry[scale=1]{Validation states} 
				\addplot[color=color1,fill=color1] table [x=indexDataTrain, y=countOutTrain, col sep=comma] {numerics/dqnn_q_eq_line_v2_epoch_440_ts.csv};
				\addlegendentry[scale=1]{Training states} 
			\end{axis}
		\end{tikzpicture}
		\subcaption{Diversity of the generator's output after $r_T=440$ training epochs.} \label{fig:QGANQ_eq_line_b}
	\end{subfigure}
	\caption{\textbf{Training a \DQGANNISQ.} The training set features $S=10$ equally spaced quantum states from $\text{data}_\text{line}$. The remaining states from $\text{data}_\text{line}$ are used as validation states. The \DQGANNISQ features a 1-1$^+$ generator and a 1-1$^+$ discriminator, and employs the hyper-parameters $r_D=4$, $\eta_D=0.5$, $r_G=1$ and $\eta_G=0.1$.}
	\label{fig:QGANQ_eq_line}
\end{figure}
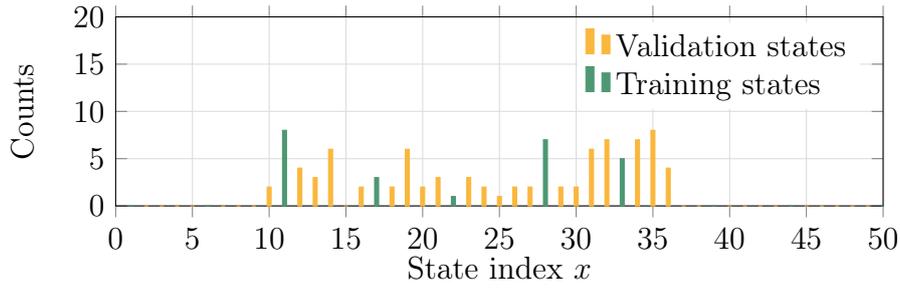
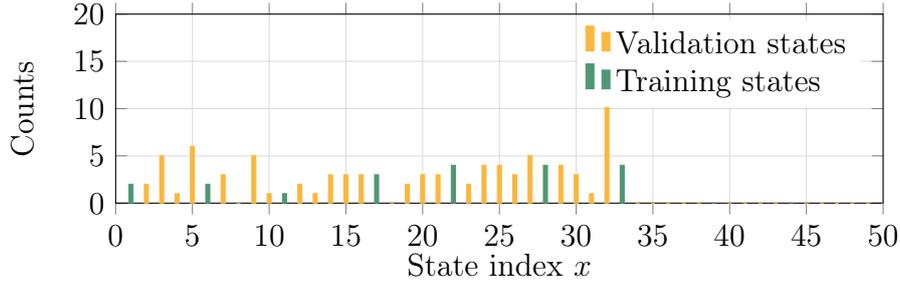

For training a \DQGANNISQ we use the same loss functions $\mathcal{L}_G$ and $\mathcal{L}_D$ as well as the same training data, $\text{data}_\text{line}$. We choose $S=10$ equally spaced training states of the set for the training. Further, we select that for each of the $r_T$ epochs, the discriminator is trained $r_D=4$ times with a learning rate $\eta_D=0.5$ and the discriminator $r_G=1$ times with a learning rate $\eta_G=0.1$. The results of the training are depicted in \cref{fig:QGANQ_eq_line}. 

After $r_T=100$ epochs, see \cref{fig:QGANQ_eq_line_a}, we see that the generator produces states in a little more than half of the training data range. The generator's diversity is improved to two-thirds of the training data range after $r_T=440$ training epochs which is depicted in \cref{fig:QGANQ_eq_line_b}. Although not all states of the aimed data sets get produced by the generator, a majority of the produced states is closer to a validation state than a training state. We see this as a training success, since the generator does not only learn to reproduce the training states but instead learns to extend the given training data. For more numerical results using \DQGANNISQ we point to \cref{apdx:QGAN} and \cite{Mueller2021}.

The numerical experiments with DQGANs and \DQGANsNISQ have shown that DQNNs can be trained in a generative adversarial context. Moreover, we could observe that, in contrast to the DQNN and graph-DQNN, the choice of the number of training rounds is crucial: a DQGAN trained in to many training epochs leads to low diversity in the generator's output states. This result builds almost the end of this thesis: in the ensuing chapter, we want to conclude the results of this work and give an outlook.

%% file: numerics/T.tex
	\centering
	\begin{tikzpicture}		
		\begin{axis}[
			hide x axis,
			hide y axis,
			minor xtick={},
			minor ytick={},
			scaled x ticks=manual:{}{\pgfmathparse{#1}},
			scaled y ticks=manual:{}{\pgfmathparse{#1}},
			tick align=outside,
			x grid style={white!69.0196078431373!black},
			xmajorticks=false,
			xmin=-0.1, xmax=0.1,
			xtick style={color=black},
			xtick={-0.8,-0.6,-0.4,-0.2,0,0.2,0.4,0.6,0.8},
			xticklabels={},
			y grid style={white!69.0196078431373!black},
			ymajorticks=false,
			ymin=-0.1, ymax=0.1,
			ytick style={color=black},
			ytick={-0.8,-0.6,-0.4,-0.2,0,0.2,0.4,0.6,0.8},
			yticklabels={},
			scale only axis,
			height=1\textwidth,
			width=1\textwidth
			]
			\addplot [line width=1pt, gray]
			table {%
				0.055932384721948 -0.027966192360974
				0.0623345722215435 -0.023915499774857
				0.067544617782594 -0.0194984516295455
				0.0714903843556341 -0.0148061433896845
				0.0741286411994976 -0.00993135477271091
				0.0754437860468987 -0.00496618143026167
				0.0754458267525361 0
				0.0741678500363723 0.00488219134898792
				0.0716631963388365 0.00960104779301211
				0.0680025353232887 0.0140837862033308
				0.0632710021271912 0.0182647650550156
				0.05756551555759 0.0220857868321307
				0.0509923604825966 0.0254961802412983
				0.0436650808240095 0.0284527206185055
				0.0357026986628325 0.0309194440256736
				0.0272282497987915 0.0328674049719624
				0.0183676066173988 0.0342744205547852
				0.00924854475533992 0.0351248359310476
				5.73172703936822e-17 0.0354093379612954
				-0.00924854475533978 0.0351248359310476
				-0.0183676066173987 0.0342744205547852
				-0.0272282497987914 0.0328674049719624
				-0.0357026986628324 0.0309194440256736
				-0.0436650808240094 0.0284527206185056
				-0.0509923604825964 0.0254961802412983
			};

		\draw[-latex,draw=color3] (axis cs:2.22044604925031e-17,-5.55111512312578e-17) -- (axis cs:2.30268479181514e-17,0.0641500299099584);
		\draw[-latex,draw=color3] (axis cs:2.22044604925031e-17,-5.55111512312578e-17) -- (axis cs:-0.00218554024119369,0.0631208481224792);
		\draw[-latex,draw=color3] (axis cs:2.22044604925031e-17,-5.55111512312578e-17) -- (axis cs:-0.004466281566784,0.0619236265147808);
		\draw[-latex,draw=color3] (axis cs:2.22044604925031e-17,-5.55111512312578e-17) -- (axis cs:-0.0068411733521636,0.0605428336003828);
		\draw[-latex,draw=color3] (axis cs:2.22044604925031e-17,-5.55111512312578e-17) -- (axis cs:-0.00930776312791541,0.0589624157318192);
		\draw[-latex,draw=color3] (axis cs:2.22044604925031e-17,-5.55111512312578e-17) -- (axis cs:-0.011861920695898,0.0571660317259214);
		\draw[-latex,draw=color3] (axis cs:2.22044604925031e-17,-5.55111512312578e-17) -- (axis cs:-0.0144975412613505,0.0551373650567816);
		\draw[-latex,draw=color3] (axis cs:2.22044604925031e-17,-5.55111512312578e-17) -- (axis cs:-0.0172062354971531,0.0528605242379481);
		\draw[-latex,draw=color3] (axis cs:2.22044604925031e-17,-5.55111512312578e-17) -- (axis cs:-0.0199770185117742,0.0503205400051501);
		\draw[-latex,draw=color3] (axis cs:2.22044604925031e-17,-5.55111512312578e-17) -- (axis cs:-0.0227960144502543,0.0475039641819735);
		\draw[-latex,draw=color3] (axis cs:2.22044604925031e-17,-5.55111512312578e-17) -- (axis cs:-0.0256461986921964,0.0443995692477969);
		\draw[-latex,draw=color3] (axis cs:2.22044604925031e-17,-5.55111512312578e-17) -- (axis cs:-0.0285072048955575,0.040999139311142);
		\draw[-latex,draw=color3] (axis cs:2.22044604925031e-17,-5.55111512312578e-17) -- (axis cs:-0.0313552288003455,0.0372983323217337);
		\draw[-latex,draw=color3] (axis cs:2.22044604925031e-17,-5.55111512312578e-17) -- (axis cs:-0.0341630638260246,0.0332975802000907);
		\draw[-latex,draw=color3] (axis cs:2.22044604925031e-17,-5.55111512312578e-17) -- (axis cs:-0.0369003039282427,0.0290029789242301);
		\draw[-latex,draw=color3] (axis cs:2.22044604925031e-17,-5.55111512312578e-17) -- (axis cs:-0.0395337456857073,0.0244271059497561);
		\draw[-latex,draw=color3] (axis cs:2.22044604925031e-17,-5.55111512312578e-17) -- (axis cs:-0.0420280130484217,0.0195896898129568);
		\draw[-latex,draw=color3] (axis cs:2.22044604925031e-17,-5.55111512312578e-17) -- (axis cs:-0.0443464139066422,0.014518049115268);
		\draw[-latex,draw=color3] (axis cs:2.22044604925031e-17,-5.55111512312578e-17) -- (axis cs:-0.0464520177427543,0.00924721828462473);
		\draw[-latex,draw=color3] (axis cs:2.22044604925031e-17,-5.55111512312578e-17) -- (axis cs:-0.0483089193383694,0.00381968817989624);
		\draw[-latex,draw=color3] (axis cs:2.22044604925031e-17,-5.55111512312578e-17) -- (axis cs:-0.0498836273591769,-0.00171528773339761);
		\draw[-latex,draw=color3] (axis cs:2.22044604925031e-17,-5.55111512312578e-17) -- (axis cs:-0.0511464923486321,-0.007302836617833);
		\draw[-latex,draw=color3] (axis cs:2.22044604925031e-17,-5.55111512312578e-17) -- (axis cs:-0.052073070637501,-0.0128840307571189);
		\draw[-latex,draw=color3] (axis cs:2.22044604925031e-17,-5.55111512312578e-17) -- (axis cs:-0.0526453131618606,-0.0183977984180745);
		\draw[-latex,draw=color3] (axis cs:2.22044604925031e-17,-5.55111512312578e-17) -- (axis cs:-0.052852474192092,-0.023783062897172);
		\draw[-latex,draw=color3] (axis cs:2.22044604925031e-17,-5.55111512312578e-17) -- (axis cs:-0.0526916553255279,-0.0289809592432878);
		\draw[-latex,draw=color3] (axis cs:2.22044604925031e-17,-5.55111512312578e-17) -- (axis cs:-0.0521679328601289,-0.0339369630927459);
		\draw[-latex,draw=color3] (axis cs:2.22044604925031e-17,-5.55111512312578e-17) -- (axis cs:-0.0512940573199503,-0.0386027717076953);
		\draw[-latex,draw=color3] (axis cs:2.22044604925031e-17,-5.55111512312578e-17) -- (axis cs:-0.0500897561880745,-0.0429378032608788);
		\draw[-latex,draw=color3] (axis cs:2.22044604925031e-17,-5.55111512312578e-17) -- (axis cs:-0.0485807082128569,-0.046910222378351);
		\draw[-latex,draw=color3] (axis cs:2.22044604925031e-17,-5.55111512312578e-17) -- (axis cs:-0.0467972845176006,-0.0504974508009966);
		\draw[-latex,draw=color3] (axis cs:2.22044604925031e-17,-5.55111512312578e-17) -- (axis cs:-0.0447731649631611,-0.0536861734787588);
		\draw[-latex,draw=color3] (axis cs:2.22044604925031e-17,-5.55111512312578e-17) -- (axis cs:-0.0425439373370426,-0.0564718949192805);
		\draw[-latex,draw=color3] (axis cs:2.22044604925031e-17,-5.55111512312578e-17) -- (axis cs:-0.0401457740167783,-0.0588581328281185);
		\draw[-latex,draw=color3] (axis cs:2.22044604925031e-17,-5.55111512312578e-17) -- (axis cs:-0.0376142595136155,-0.060855353582091);
		\draw[-latex,draw=color3] (axis cs:2.22044604925031e-17,-5.55111512312578e-17) -- (axis cs:-0.0349834171206572,-0.0624797572977202);
		\draw[-latex,draw=color3] (axis cs:2.22044604925031e-17,-5.55111512312578e-17) -- (axis cs:-0.0322849578228774,-0.0637520117685745);
		\draw[-latex,draw=color3] (axis cs:2.22044604925031e-17,-5.55111512312578e-17) -- (axis cs:-0.029547752750974,-0.0646960180844455);
		\draw[-latex,draw=color3] (axis cs:2.22044604925031e-17,-5.55111512312578e-17) -- (axis cs:-0.0267975135955062,-0.065337770247101);
		\draw[-latex,draw=color3] (axis cs:2.22044604925031e-17,-5.55111512312578e-17) -- (axis cs:-0.0240566541565915,-0.0657043499775278);
		\draw[-latex,draw=color3] (axis cs:2.22044604925031e-17,-5.55111512312578e-17) -- (axis cs:-0.0213443002645144,-0.0658230786697224);
		\draw[-latex,draw=color3] (axis cs:2.22044604925031e-17,-5.55111512312578e-17) -- (axis cs:-0.0186764137576768,-0.0657208325778324);
		\draw[-latex,draw=color3] (axis cs:2.22044604925031e-17,-5.55111512312578e-17) -- (axis cs:-0.0160659978810271,-0.0654235154238606);
		\draw[-latex,draw=color3] (axis cs:2.22044604925031e-17,-5.55111512312578e-17) -- (axis cs:-0.0135233552050466,-0.0649556746095114);
		\draw[-latex,draw=color3] (axis cs:2.22044604925031e-17,-5.55111512312578e-17) -- (axis cs:-0.0110563739568825,-0.064340242614963);
		\draw[-latex,draw=color3] (axis cs:2.22044604925031e-17,-5.55111512312578e-17) -- (axis cs:-0.00867082372898774,-0.0635983832746936);
		\draw[-latex,draw=color3] (axis cs:2.22044604925031e-17,-5.55111512312578e-17) -- (axis cs:-0.00637064635765862,-0.0627494227026011);
		\draw[-latex,draw=color3] (axis cs:2.22044604925031e-17,-5.55111512312578e-17) -- (axis cs:-0.00415823202974032,-0.0618108460266788);
		\draw[-latex,draw=color3] (axis cs:2.22044604925031e-17,-5.55111512312578e-17) -- (axis cs:-0.00203467423430762,-0.0607983432355738);
		\draw[-latex,draw=color3] (axis cs:2.22044604925031e-17,-5.55111512312578e-17) -- (axis cs:2.14387894410375e-17,-0.0597258899161682);
			\addplot [line width=1pt, gray]
			table {%
				0.0509923604825966 0.0254961802412983
				0.0508144641284987 0.0336005929090441
				0.0497944599148205 0.0412382479523173
				0.0479333324664563 0.0482835303342209
				0.0452440451226282 0.0546143936758346
				0.0417524500760907 0.0601143339364639
				0.037498116911366 0.0646746848297206
				0.0325349964548976 0.0681972109485291
				0.0269318258092109 0.070596942882889
				0.0207721745142527 0.0718051631034812
				0.01415403323771 0.0717724145042109
				0.0071888573755637 0.0704713688388215
				4.95445232744598e-17 0.0678993644195169
				-0.00727949779649591 0.0640804064800908
				-0.0145097470086934 0.0590664244732719
				-0.0215474174463378 0.0529376020749203
				-0.0282495927738551 0.0458016394615269
				-0.0344779045327252 0.0377918721596146
				-0.0401027351683544 0.02906425164204
				-0.045007256248083 0.0197932818456229
				-0.0490910657483456 0.0101670926121571
				-0.0522732094178073 0.000381904731788546
				-0.0544944142980677 -0.0093638079477985
				-0.0557184226814535 -0.0188751326819909
				-0.0559323847219479 -0.027966192360974
			};
			\addplot [line width=1pt, gray]
			table {%
				-0.0509923604825965 0.0254961802412983
				-0.0575655155575899 0.0220857868321308
				-0.0632710021271911 0.0182647650550156
				-0.0680025353232886 0.0140837862033308
				-0.0716631963388364 0.00960104779301212
				-0.0741678500363722 0.00488219134898792
				-0.075445826752536 5.55111512312578e-18
				-0.0754437860468986 -0.00496618143026165
				-0.0741286411994975 -0.00993135477271089
				-0.071490384355634 -0.0148061433896845
				-0.0675446177825939 -0.0194984516295455
				-0.0623345722215434 -0.0239154997748569
				-0.0559323847219479 -0.027966192360974
				-0.0484394210758185 -0.0315637412958728
				-0.0399854657685109 -0.0346284291376835
				-0.0307266660277852 -0.0370903669253938
				-0.0208422029492727 -0.0388920801741739
				-0.010529764374659 -0.0399907505274216
				4.15741797238211e-17 -0.0403599520257423
				0.0105297643746591 -0.0399907505274216
				0.0208422029492728 -0.0388920801741739
				0.0307266660277853 -0.0370903669253938
				0.039985465768511 -0.0346284291376835
				0.0484394210758185 -0.0315637412958729
				0.055932384721948 -0.027966192360974
			};
			\addplot [line width=1pt, gray]
			table {%
				-0.0559323847219479 -0.027966192360974
				-0.0551463389473285 -0.0364650049385081
				-0.0533918736718876 -0.0442175159422661
				-0.0507201120554238 -0.0510906699570057
				-0.0471991951659225 -0.0569744685513325
				-0.0429114496861602 -0.0617830381553776
				-0.0379504213490275 -0.0654547945888852
				-0.0324179352791624 -0.0679518368417097
				-0.0264213147396773 -0.0692587298307084
				-0.0200708551160697 -0.0693808452382716
				-0.0134776145915092 -0.0683424240104661
				-0.00675155020197458 -0.0661845074482186
				4.59424754914535e-17 -0.0629628600612049
				0.00667350998943558 -0.0587459801111312
				0.0131701712563796 -0.0536132659893323
				0.0193965581943491 -0.047653380358582
				0.0252651061814176 -0.0409628306341268
				0.0306944565538663 -0.033644764489429
				0.035609721098378 -0.0258079627377343
				0.0399427149438239 -0.0175659989181445
				0.043632201434514 -0.00903652479518527
				0.0466241861150618 -0.000340633328074654
				0.0488722895495046 0.00839775487461268
				0.0503382203579635 0.0170525392232275
				0.0509923604825966 0.0254961802412983
			};
			
		\draw (axis cs:-0.067775454902823,-0.0338877274514115) node[
		text=black,
		rotate=0.0
		]{$x$};
		\draw (axis cs:0.0677754549028231,-0.0338877274514115) node[
		text=black,
		rotate=0.0
		]{$y$};
		\draw (axis cs:4.65145166854011e-17,0.0821231106811569) node[
		text=black,
		rotate=0.0
		]{$\left|0\right>$};
		\draw (axis cs:4.24856997397892e-17,-0.0750100843935497) node[
		text=black,
		rotate=0.0
		]{$\left|1\right>$};
			\path [draw=gray, draw opacity=0.2, line width=1.5pt]
			(axis cs:4.61498269077781e-17,0.0678993644195169)
			--(axis cs:0.00727949779649602,0.0640804064800908)
			--(axis cs:0.0145097470086935,0.0590664244732719)
			--(axis cs:0.0215474174463379,0.0529376020749203)
			--(axis cs:0.0282495927738552,0.0458016394615269)
			--(axis cs:0.0344779045327253,0.0377918721596145)
			--(axis cs:0.0401027351683544,0.02906425164204)
			--(axis cs:0.0450072562480832,0.0197932818456229)
			--(axis cs:0.0490910657483457,0.0101670926121571)
			--(axis cs:0.0522732094178074,0.000381904731788534)
			--(axis cs:0.0544944142980677,-0.00936380794779851)
			--(axis cs:0.0557184226814536,-0.0188751326819909)
			--(axis cs:0.055932384721948,-0.027966192360974)
			--(axis cs:0.0551463389473286,-0.0364650049385081)
			--(axis cs:0.0533918736718877,-0.0442175159422661)
			--(axis cs:0.0507201120554239,-0.0510906699570057)
			--(axis cs:0.0471991951659226,-0.0569744685513324)
			--(axis cs:0.0429114496861603,-0.0617830381553776)
			--(axis cs:0.0379504213490276,-0.0654547945888852)
			--(axis cs:0.0324179352791625,-0.0679518368417097)
			--(axis cs:0.0264213147396774,-0.0692587298307084)
			--(axis cs:0.0200708551160698,-0.0693808452382716)
			--(axis cs:0.0134776145915093,-0.0683424240104661)
			--(axis cs:0.00675155020197468,-0.0661845074482186)
			--(axis cs:4.90903664049098e-17,-0.0629628600612049);
			
			\path [draw=gray, draw opacity=0.2, line width=1.5pt]
			(axis cs:4.61498269077781e-17,0.0678993644195169)
			--(axis cs:0.0101544920027974,0.0667053867877091)
			--(axis cs:0.020137991428914,0.0643230362828052)
			--(axis cs:0.0297591368270551,0.0607974761223566)
			--(axis cs:0.0388344683181779,0.0561980267483865)
			--(axis cs:0.0471929691513353,0.0506162953449908)
			--(axis cs:0.0546801441075774,0.0441636243077408)
			--(axis cs:0.0611614735973195,0.0369679868512813)
			--(axis cs:0.0665251251082729,0.0291704844588019)
			--(axis cs:0.070683853036046,0.0209216143464113)
			--(axis cs:0.0735760688446164,0.0123774751565389)
			--(axis cs:0.0751661111789884,0.00369606684746908)
			--(axis cs:0.0754437860468987,-0.00496618143026166)
			--(axis cs:0.0744232777973233,-0.0134575506566607)
			--(axis cs:0.0721415510444264,-0.0216337660258267)
			--(axis cs:0.0686563719396976,-0.029360285906548)
			--(axis cs:0.0640440753942115,-0.0365141358387519)
			--(axis cs:0.058397194857189,-0.0429853032951052)
			--(axis cs:0.0518220553094327,-0.0486777288690523)
			--(axis cs:0.0444364105288039,-0.0535099434149068)
			--(axis cs:0.0363671845306746,-0.0574154086354438)
			--(axis cs:0.0277483561230086,-0.0603426212897374)
			--(axis cs:0.0187190060587047,-0.0622550394329442)
			--(axis cs:0.00942152920549162,-0.0631308838522766)
			--(axis cs:5.16219932618017e-17,-0.0629628600612049);
			
			\path [draw=gray, draw opacity=0.2, line width=1.5pt]
			(axis cs:4.61498269077781e-17,0.0678993644195169)
			--(axis cs:0.00882060034619592,0.0695453993484673)
			--(axis cs:0.0173975404854276,0.0699507423596018)
			--(axis cs:0.0255753688716571,0.0691273754439867)
			--(axis cs:0.0332114944663671,0.0671113284462138)
			--(axis cs:0.0401786482982443,0.0639604106140683)
			--(axis cs:0.0463666431010718,0.0597515400389833)
			--(axis cs:0.0516834596461562,0.0545778430882969)
			--(axis cs:0.0560557237614775,0.0485456808049625)
			--(axis cs:0.0594286624613889,0.0417717351200313)
			--(axis cs:0.0617656411096943,0.034380258416009)
			--(axis cs:0.0630473872169115,0.0265005589841592)
			--(axis cs:0.0632710021271912,0.0182647650550156)
			--(axis cs:0.0624488515704235,0.00980588335167267)
			--(axis cs:0.0606074119001638,0.00125614571373374)
			--(axis cs:0.0577861326190897,-0.0072543802363878)
			--(axis cs:0.0540363589543951,-0.0155989531941736)
			--(axis cs:0.0494203418289704,-0.0236551382113147)
			--(axis cs:0.044010347252373,-0.0313056543357297)
			--(axis cs:0.0378878632976223,-0.0384391407352494)
			--(axis cs:0.0311428905894132,-0.0449508906628518)
			--(axis cs:0.0238732916360776,-0.0507435987700406)
			--(axis cs:0.0161841653827142,-0.0557281613662269)
			--(axis cs:0.00818720607280406,-0.0598245613842071)
			--(axis cs:5.05053110819088e-17,-0.0629628600612049);
			
			\path [draw=gray, draw opacity=0.2, line width=1.5pt]
			(axis cs:4.61498269077781e-17,0.0678993644195169)
			--(axis cs:0.00388317023821108,0.0714369444879401)
			--(axis cs:0.0076314304071923,0.0736651488012985)
			--(axis cs:0.0111801631180523,0.0745778683961119)
			--(axis cs:0.0144719224518979,0.0741945894471299)
			--(axis cs:0.017457023636514,0.0725574893292975)
			--(axis cs:0.0200937924043597,0.0697282771874704)
			--(axis cs:0.0223485316263501,0.0657849904902597)
			--(axis cs:0.0241952695908673,0.0608189171683992)
			--(axis cs:0.0256153545799167,0.0549317673220386)
			--(axis cs:0.0265969558472004,0.048233174183445)
			--(axis cs:0.027134523418005,0.0408385644387348)
			--(axis cs:0.0272282497987915,0.0328674049719624)
			--(axis cs:0.026883566871143,0.0244418071449537)
			--(axis cs:0.0261107017560061,0.015685450504537)
			--(axis cs:0.0249243067405071,0.00672277440771821)
			--(axis cs:0.0233431706556289,-0.00232162266862923)
			--(axis cs:0.0213900123699502,-0.0113234415578231)
			--(axis cs:0.0190913511828616,-0.0201590666372326)
			--(axis cs:0.0164774436550527,-0.0287059447376105)
			--(axis cs:0.0135822715900816,-0.0368430269751744)
			--(axis cs:0.0104435613023838,-0.0444513397561309)
			--(axis cs:0.00710280987803661,-0.0514147475522974)
			--(axis cs:0.0036052898759796,-0.0576209641481942)
			--(axis cs:4.62018371500936e-17,-0.0629628600612049);
			
			\path [draw=gray, draw opacity=0.2, line width=1.5pt]
			(axis cs:4.61498269077781e-17,0.0678993644195169)
			--(axis cs:-0.00262532190184761,0.0716238064054855)
			--(axis cs:-0.00515759541198561,0.0740306329228037)
			--(axis cs:-0.00755340803198377,0.0751121577206319)
			--(axis cs:-0.00977428885608032,0.074886499547558)
			--(axis cs:-0.0117870730016795,0.0733945968222206)
			--(axis cs:-0.0135640354694092,0.0706969820223057)
			--(axis cs:-0.0150828361172965,0.0668705313972792)
			--(axis cs:-0.0163263212528452,0.0620053609009858)
			--(axis cs:-0.0172822267662812,0.0562019914862611)
			--(axis cs:-0.0179428239715013,0.0495688612207312)
			--(axis cs:-0.0183045435883866,0.0422202213926501)
			--(axis cs:-0.0183676066173987,0.0342744205547852)
			--(axis cs:-0.0181356840038013,0.0258525546739698)
			--(axis cs:-0.0176156004828913,0.0170774426835206)
			--(axis cs:-0.0168170921233428,0.00807287372726411)
			--(axis cs:-0.0157526219395293,-0.0010369360109445)
			--(axis cs:-0.0144372534873764,-0.0101277439931996)
			--(axis cs:-0.0128885784702847,-0.0190756092453687)
			--(axis cs:-0.0111266909015634,-0.027757218189607)
			--(axis cs:-0.00917419713630969,-0.0360502787808923)
			--(axis cs:-0.007056247968563,-0.0438340518089831)
			--(axis cs:-0.00480057591808647,-0.0509900851285851)
			--(axis cs:-0.00243751781971411,-0.0574032112362675)
			--(axis cs:4.04901684926626e-17,-0.0629628600612049);
			
			\path [draw=gray, draw opacity=0.2, line width=1.5pt]
			(axis cs:4.61498269077781e-17,0.0678993644195169)
			--(axis cs:-0.0071888573755636,0.0704713688388215)
			--(axis cs:-0.0141540332377099,0.0717724145042109)
			--(axis cs:-0.0207721745142526,0.0718051631034812)
			--(axis cs:-0.0269318258092108,0.070596942882889)
			--(axis cs:-0.0325349964548975,0.0681972109485291)
			--(axis cs:-0.0374981169113659,0.0646746848297206)
			--(axis cs:-0.0417524500760906,0.0601143339364639)
			--(axis cs:-0.0452440451226281,0.0546143936758346)
			--(axis cs:-0.0479333324664562,0.0482835303342209)
			--(axis cs:-0.0497944599148204,0.0412382479523173)
			--(axis cs:-0.0508144641284986,0.0336005929090441)
			--(axis cs:-0.0509923604825964,0.0254961802412983)
			--(axis cs:-0.0503382203579634,0.0170525392232276)
			--(axis cs:-0.0488722895495045,0.00839775487461271)
			--(axis cs:-0.0466241861150617,-0.000340633328074623)
			--(axis cs:-0.0436322014345139,-0.00903652479518522)
			--(axis cs:-0.0399427149438238,-0.0175659989181444)
			--(axis cs:-0.0356097210983779,-0.0258079627377343)
			--(axis cs:-0.0306944565538663,-0.033644764489429)
			--(axis cs:-0.0252651061814176,-0.0409628306341268)
			--(axis cs:-0.019396558194349,-0.047653380358582)
			--(axis cs:-0.0131701712563795,-0.0536132659893323)
			--(axis cs:-0.00667350998943552,-0.0587459801111312)
			--(axis cs:3.64988027510844e-17,-0.0629628600612049);
			
			\path [draw=gray, draw opacity=0.2, line width=1.5pt]
			(axis cs:4.61498269077781e-17,0.0678993644195169)
			--(axis cs:4.61498269077781e-17,0.0678993644195169)
			--(axis cs:4.61498269077781e-17,0.0678993644195169)
			--(axis cs:4.61498269077781e-17,0.0678993644195169)
			--(axis cs:4.61498269077781e-17,0.0678993644195169)
			--(axis cs:4.61498269077781e-17,0.0678993644195169)
			--(axis cs:4.61498269077781e-17,0.0678993644195169)
			--(axis cs:4.61498269077781e-17,0.0678993644195169)
			--(axis cs:4.61498269077781e-17,0.0678993644195169)
			--(axis cs:4.61498269077781e-17,0.0678993644195169)
			--(axis cs:4.61498269077781e-17,0.0678993644195169)
			--(axis cs:4.61498269077781e-17,0.0678993644195169)
			--(axis cs:4.61498269077781e-17,0.0678993644195169)
			--(axis cs:4.61498269077781e-17,0.0678993644195169)
			--(axis cs:4.61498269077781e-17,0.0678993644195169)
			--(axis cs:4.61498269077781e-17,0.0678993644195169)
			--(axis cs:4.61498269077781e-17,0.0678993644195169)
			--(axis cs:4.61498269077781e-17,0.0678993644195169)
			--(axis cs:4.61498269077781e-17,0.0678993644195169)
			--(axis cs:4.61498269077781e-17,0.0678993644195169)
			--(axis cs:4.61498269077781e-17,0.0678993644195169)
			--(axis cs:4.61498269077781e-17,0.0678993644195169)
			--(axis cs:4.61498269077781e-17,0.0678993644195169)
			--(axis cs:4.61498269077781e-17,0.0678993644195169)
			--(axis cs:4.61498269077781e-17,0.0678993644195169);
			
			\path [draw=gray, draw opacity=0.2, line width=1.5pt]
			(axis cs:0.0344779045327253,0.0377918721596145)
			--(axis cs:0.0385232318489047,0.0400232801669803)
			--(axis cs:0.0418605485562577,0.0424696295108759)
			--(axis cs:0.0444387727821861,0.0450835572023401)
			--(axis cs:0.04622271215434,0.0478157921552087)
			--(axis cs:0.0471929691513353,0.0506162953449908)
			--(axis cs:0.0473454448090874,0.0534353132373704)
			--(axis cs:0.0466905158501246,0.0562243176580048)
			--(axis cs:0.0452519653923232,0.0589368158900558)
			--(axis cs:0.0430657461072718,0.0615290247067986)
			--(axis cs:0.0401786482982443,0.0639604106140683)
			--(axis cs:0.0366469353134886,0.0661941053937697)
			--(axis cs:0.0325349964548976,0.0681972109485291)
			--(axis cs:0.0279140543807352,0.0699410104865683)
			--(axis cs:0.0228609509846968,0.0714011044296315)
			--(axis cs:0.017457023636514,0.0725574893292975)
			--(axis cs:0.0117870730016796,0.0733945968222206)
			--(axis cs:0.00593841470658748,0.0739013075195291)
			--(axis cs:5.22213358926415e-17,0.0740709519555328)
			--(axis cs:-0.00593841470658735,0.0739013075195291)
			--(axis cs:-0.0117870730016795,0.0733945968222206)
			--(axis cs:-0.0174570236365139,0.0725574893292975)
			--(axis cs:-0.0228609509846967,0.0714011044296315)
			--(axis cs:-0.0279140543807351,0.0699410104865684)
			--(axis cs:-0.0325349964548975,0.0681972109485291);
			
			\path [draw=gray, draw opacity=0.2, line width=1.5pt]
			(axis cs:0.0544944142980677,-0.00936380794779851)
			--(axis cs:0.0607424576443773,-0.00553788089785755)
			--(axis cs:0.0658317834178951,-0.00136442573565185)
			--(axis cs:0.0696913849833833,0.0030708256437037)
			--(axis cs:0.0722782252137269,0.00768043436961534)
			--(axis cs:0.0735760688446164,0.0123774751565389)
			--(axis cs:0.0735935968894255,0.0170774535900896)
			--(axis cs:0.0723620177138319,0.0216998794413739)
			--(axis cs:0.0699323825297053,0.0261694856761292)
			--(axis cs:0.06637279102771,0.0304171079808011)
			--(axis cs:0.0617656411096943,0.034380258416009)
			--(axis cs:0.0562050403305504,0.0380034390749989)
			--(axis cs:0.0497944599148205,0.0412382479523173)
			--(axis cs:0.0426446780951416,0.0440433306367063)
			--(axis cs:0.0348720298633531,0.0463842291259075)
			--(axis cs:0.0265969558472004,0.048233174183445)
			--(axis cs:0.0179428239715015,0.0495688612207312)
			--(axis cs:0.00903498337370289,0.0503762424661978)
			--(axis cs:5.79696847869999e-17,0.0506463607072081)
			--(axis cs:-0.00903498337370275,0.0503762424661978)
			--(axis cs:-0.0179428239715013,0.0495688612207312)
			--(axis cs:-0.0265969558472003,0.0482331741834451)
			--(axis cs:-0.034872029863353,0.0463842291259075)
			--(axis cs:-0.0426446780951415,0.0440433306367064)
			--(axis cs:-0.0497944599148204,0.0412382479523173);
			
			\path [draw=gray, draw opacity=0.2, line width=1.5pt]
			(axis cs:0.0507201120554239,-0.0510906699570057)
			--(axis cs:0.0565609590651723,-0.0472747518057331)
			--(axis cs:0.0613301657082578,-0.0431082802595031)
			--(axis cs:0.0649598886791089,-0.0386759442202936)
			--(axis cs:0.0674078370267178,-0.0340644369589177)
			--(axis cs:0.0686563719396976,-0.029360285906548)
			--(axis cs:0.068710948435455,-0.0246479548415487)
			--(axis cs:0.0675980813142587,-0.020008261110116)
			--(axis cs:0.0653630148466158,-0.0155171224696908)
			--(axis cs:0.0620672593643557,-0.0112446243735427)
			--(axis cs:0.0577861326190897,-0.0072543802363878)
			--(axis cs:0.0526064137047538,-0.00360314475373164)
			--(axis cs:0.0466241861150618,-0.00034063332807463)
			--(axis cs:0.0399429168214412,0.00249050171630844)
			--(axis cs:0.0326717918632217,0.00485458617481469)
			--(axis cs:0.0249243067405071,0.00672277440771821)
			--(axis cs:0.016817092123343,0.0080728737272641)
			--(axis cs:0.00846894181056969,0.00888917467416798)
			--(axis cs:5.16478709967531e-17,0.00916231208453441)
			--(axis cs:-0.00846894181056956,0.00888917467416798)
			--(axis cs:-0.0168170921233428,0.00807287372726411)
			--(axis cs:-0.024924306740507,0.00672277440771822)
			--(axis cs:-0.0326717918632216,0.0048545861748147)
			--(axis cs:-0.039942916821441,0.00249050171630846)
			--(axis cs:-0.0466241861150617,-0.000340633328074623);
			
			\path [draw=gray, draw opacity=0.2, line width=1.5pt]
			(axis cs:0.0264213147396774,-0.0692587298307084)
			--(axis cs:0.0295499108881756,-0.0672074933249455)
			--(axis cs:0.0321439459822504,-0.0649541055298417)
			--(axis cs:0.0341624691064748,-0.062541064169394)
			--(axis cs:0.0355761091183335,-0.0600129423968703)
			--(axis cs:0.0363671845306746,-0.0574154086354438)
			--(axis cs:0.0365295301100312,-0.0547942951650467)
			--(axis cs:0.0360680803805556,-0.0521947392039001)
			--(axis cs:0.034998255383991,-0.0496604130991129)
			--(axis cs:0.0333451955216741,-0.0472328532122973)
			--(axis cs:0.0311428905894132,-0.0449508906628518)
			--(axis cs:0.0284332438772451,-0.0428501816307415)
			--(axis cs:0.0252651061814177,-0.0409628306341268)
			--(axis cs:0.0216933074897345,-0.0393170971658028)
			--(axis cs:0.0177777066035202,-0.0379371742708856)
			--(axis cs:0.0135822715900816,-0.0368430269751744)
			--(axis cs:0.0091741971363098,-0.0360502787808923)
			--(axis cs:0.00462305889704288,-0.03557013556395)
			--(axis cs:4.68959485039218e-17,-0.0354093379612954)
			--(axis cs:-0.00462305889704276,-0.03557013556395)
			--(axis cs:-0.00917419713630969,-0.0360502787808923)
			--(axis cs:-0.0135822715900815,-0.0368430269751744)
			--(axis cs:-0.0177777066035201,-0.0379371742708856)
			--(axis cs:-0.0216933074897344,-0.0393170971658028)
			--(axis cs:-0.0252651061814176,-0.0409628306341268);
			
			\path [draw=gray, draw opacity=0.2, line width=1.5pt]
			(axis cs:4.90903664049098e-17,-0.0629628600612049)
			--(axis cs:4.98582695475355e-17,-0.0629628600612049)
			--(axis cs:5.05053110819088e-17,-0.0629628600612049)
			--(axis cs:5.1020419947983e-17,-0.0629628600612049)
			--(axis cs:5.13947824902443e-17,-0.0629628600612049)
			--(axis cs:5.16219932618017e-17,-0.0629628600612049)
			--(axis cs:5.16981646233591e-17,-0.0629628600612049)
			--(axis cs:5.16219932618017e-17,-0.0629628600612049)
			--(axis cs:5.13947824902443e-17,-0.0629628600612049)
			--(axis cs:5.1020419947983e-17,-0.0629628600612049)
			--(axis cs:5.05053110819088e-17,-0.0629628600612049)
			--(axis cs:4.98582695475355e-17,-0.0629628600612049)
			--(axis cs:4.90903664049098e-17,-0.0629628600612049)
			--(axis cs:4.82147406897052e-17,-0.0629628600612049)
			--(axis cs:4.72463746006781e-17,-0.0629628600612049)
			--(axis cs:4.62018371500936e-17,-0.0629628600612049)
			--(axis cs:4.50990006633316e-17,-0.0629628600612049)
			--(axis cs:4.39567349784446e-17,-0.0629628600612049)
			--(axis cs:4.27945845779971e-17,-0.0629628600612049)
			--(axis cs:4.16324341775496e-17,-0.0629628600612049)
			--(axis cs:4.04901684926626e-17,-0.0629628600612049)
			--(axis cs:3.93873320059006e-17,-0.0629628600612049)
			--(axis cs:3.83427945553161e-17,-0.0629628600612049)
			--(axis cs:3.7374428466289e-17,-0.0629628600612049)
			--(axis cs:3.64988027510844e-17,-0.0629628600612049);
			
			\path [draw=gray, draw opacity=0.2, line width=1.5pt]
			(axis cs:4.61498269077781e-17,0.0678993644195169)
			--(axis cs:-0.0071888573755636,0.0704713688388215)
			--(axis cs:-0.0141540332377099,0.0717724145042109)
			--(axis cs:-0.0207721745142526,0.0718051631034812)
			--(axis cs:-0.0269318258092108,0.070596942882889)
			--(axis cs:-0.0325349964548975,0.0681972109485291)
			--(axis cs:-0.0374981169113659,0.0646746848297206)
			--(axis cs:-0.0417524500760906,0.0601143339364639)
			--(axis cs:-0.0452440451226281,0.0546143936758346)
			--(axis cs:-0.0479333324664562,0.0482835303342209)
			--(axis cs:-0.0497944599148204,0.0412382479523173)
			--(axis cs:-0.0508144641284987,0.0336005929090441)
			--(axis cs:-0.0509923604825965,0.0254961802412983)
			--(axis cs:-0.0503382203579634,0.0170525392232276)
			--(axis cs:-0.0488722895495045,0.0083977548746127)
			--(axis cs:-0.0466241861150617,-0.000340633328074628)
			--(axis cs:-0.0436322014345139,-0.00903652479518523)
			--(axis cs:-0.0399427149438238,-0.0175659989181444)
			--(axis cs:-0.0356097210983779,-0.0258079627377343)
			--(axis cs:-0.0306944565538663,-0.033644764489429)
			--(axis cs:-0.0252651061814176,-0.0409628306341268)
			--(axis cs:-0.019396558194349,-0.047653380358582)
			--(axis cs:-0.0131701712563796,-0.0536132659893323)
			--(axis cs:-0.00667350998943552,-0.0587459801111312)
			--(axis cs:3.64988027510844e-17,-0.0629628600612049);
			
			\path [draw=gray, draw opacity=0.2, line width=1.5pt]
			(axis cs:4.61498269077781e-17,0.0678993644195169)
			--(axis cs:-0.0101310334249443,0.067885062021662)
			--(axis cs:-0.0200459396258624,0.0666681103667981)
			--(axis cs:-0.0295585542308932,0.0642791463364559)
			--(axis cs:-0.0384935930098653,0.0607725178902347)
			--(axis cs:-0.0466905158501245,0.0562243176580048)
			--(axis cs:-0.0540067528917355,0.0507298689460372)
			--(axis cs:-0.0603202130213146,0.0444008082560728)
			--(axis cs:-0.065531042429919,0.0373619163086138)
			--(axis cs:-0.0695626454312482,0.0297478456926204)
			--(axis cs:-0.0723620177138318,0.0216998794413739)
			--(axis cs:-0.0738994713402084,0.0133628336253831)
			--(axis cs:-0.0741678500363722,0.00488219134898792)
			--(axis cs:-0.0731813427099581,-0.00359847184324147)
			--(axis cs:-0.0709740036741705,-0.0119397377575633)
			--(axis cs:-0.0675980813142587,-0.020008261110116)
			--(axis cs:-0.0631222447841624,-0.0276783614834889)
			--(axis cs:-0.0576297826598268,-0.0348333178111007)
			--(axis cs:-0.0512168300315005,-0.041366402314056)
			--(axis cs:-0.0439906627321908,-0.047181697132346)
			--(axis cs:-0.0360680803805555,-0.0521947392039001)
			--(axis cs:-0.027573884443893,-0.05633303784383)
			--(axis cs:-0.0186394440992098,-0.0595365055803608)
			--(axis cs:-0.00940133157147264,-0.0617578366915671)
			--(axis cs:3.39671758941925e-17,-0.0629628600612049);
			
			\path [draw=gray, draw opacity=0.2, line width=1.5pt]
			(axis cs:4.61498269077781e-17,0.0678993644195169)
			--(axis cs:-0.00889909571292206,0.0650263952082629)
			--(axis cs:-0.0177055739245985,0.0609669425280065)
			--(axis cs:-0.0262466362760574,0.0557882340046795)
			--(axis cs:-0.034352388631302,0.0495834020589722)
			--(axis cs:-0.0418605485562576,0.0424696295108759)
			--(axis cs:-0.0486210373888861,0.03458535287651)
			--(axis cs:-0.0545002195436469,0.0260866310944095)
			--(axis cs:-0.0593845687426075,0.0171428483746272)
			--(axis cs:-0.0631835779909137,0.0079319670423858)
			--(axis cs:-0.065831783417895,-0.00136442573565184)
			--(axis cs:-0.0672898355335579,-0.0105660274192746)
			--(axis cs:-0.0675446177825939,-0.0194984516295455)
			--(axis cs:-0.0666084742950045,-0.0279973822836888)
			--(axis cs:-0.0645176603136265,-0.0359120992533603)
			--(axis cs:-0.0613301657082577,-0.0431082802595031)
			--(axis cs:-0.0571230823360894,-0.0494700404455236)
			--(axis cs:-0.05198969008453,-0.0549012230812163)
			--(axis cs:-0.0460364263628607,-0.0593259979394237)
			--(axis cs:-0.0393798829183074,-0.0626888556004247)
			--(axis cs:-0.0321439459822504,-0.0649541055298416)
			--(axis cs:-0.0244571646606119,-0.0661049939210508)
			--(axis cs:-0.0164504013995769,-0.0661425556662849)
			--(axis cs:-0.00825478973747233,-0.0650843056366037)
			--(axis cs:3.50838580740854e-17,-0.0629628600612049);
			
			\path [draw=gray, draw opacity=0.2, line width=1.5pt]
			(axis cs:4.61498269077781e-17,0.0678993644195169)
			--(axis cs:-0.00394726447329938,0.0630865180119895)
			--(axis cs:-0.00788298542484386,0.0570621644864902)
			--(axis cs:-0.0117284728039547,0.0499203589976665)
			--(axis cs:-0.0154041005257724,0.0417848875598817)
			--(axis cs:-0.0188316848881039,0.0328073527077875)
			--(axis cs:-0.0219370055955046,0.0231638689889124)
			--(axis cs:-0.0246523230470764,0.0130504430974669)
			--(axis cs:-0.0269187319417674,0.00267723434869771)
			--(axis cs:-0.0286881942233449,-0.00773799889145597)
			--(axis cs:-0.0299251148875929,-0.017976878323615)
			--(axis cs:-0.0306073603152185,-0.0278273787710699)
			--(axis cs:-0.0307266660277852,-0.0370903669253938)
			--(axis cs:-0.0302884327046089,-0.0455853265487866)
			--(axis cs:-0.0293109589629978,-0.0531549774819399)
			--(axis cs:-0.0278242005516113,-0.0596686064155129)
			--(axis cs:-0.0258681738299319,-0.0650240459421965)
			--(axis cs:-0.0234911347278989,-0.0691483494538933)
			--(axis cs:-0.0207476632653872,-0.0719973007607729)
			--(axis cs:-0.01769677059439,-0.0735539614484872)
			--(axis cs:-0.0144001240342285,-0.0738264936298922)
			--(axis cs:-0.0109204596889088,-0.0728455029595643)
			--(axis cs:-0.00732022561154824,-0.0706611316831989)
			--(axis cs:-0.00366047395808582,-0.0673401007172423)
			--(axis cs:3.93873320059006e-17,-0.0629628600612049);
			
			\path [draw=gray, draw opacity=0.2, line width=1.5pt]
			(axis cs:4.61498269077781e-17,0.0678993644195169)
			--(axis cs:0.00267064365280734,0.0628932927172908)
			--(axis cs:0.00533547602544641,0.0566716121942877)
			--(axis cs:0.00794114221859689,0.0493310950781913)
			--(axis cs:0.0104334980949618,0.0409987391152328)
			--(axis cs:0.0127592353488724,0.0318298479435099)
			--(axis cs:0.0148676178483273,0.0220046663127384)
			--(axis cs:0.0167122279612792,0.01172364026827)
			--(axis cs:0.0182526104936192,0.00120150087948746)
			--(axis cs:0.0194557024733789,-0.00933951174143812)
			--(axis cs:0.0202969502704566,-0.0196768770517153)
			--(axis cs:0.0207610403241808,-0.0295950199593726)
			--(axis cs:0.0208422029492728,-0.0388920801741739)
			--(axis cs:0.0205440857539435,-0.0473858222496948)
			--(axis cs:0.0198792290414676,-0.0549183755632186)
			--(axis cs:0.0188682055467781,-0.0613596119204972)
			--(axis cs:0.0175385076464042,-0.066609095793118)
			--(axis cs:0.0159232752101634,-0.0705966614374114)
			--(axis cs:0.01405995678768,-0.0732817688728959)
			--(axis cs:0.0119889875704479,-0.0746518587705577)
			--(axis cs:0.00975255217758281,-0.074719962238456)
			--(axis cs:0.00739348170279928,-0.0735218276854307)
			--(axis cs:0.0049543153104582,-0.0711128091358271)
			--(axis cs:0.00247653906794367,-0.0675647259230112)
			--(axis cs:4.50990006633316e-17,-0.0629628600612049);
			
			\path [draw=gray, draw opacity=0.2, line width=1.5pt]
			(axis cs:4.61498269077781e-17,0.0678993644195169)
			--(axis cs:0.00727949779649602,0.0640804064800908)
			--(axis cs:0.0145097470086935,0.0590664244732719)
			--(axis cs:0.0215474174463379,0.0529376020749203)
			--(axis cs:0.0282495927738552,0.0458016394615269)
			--(axis cs:0.0344779045327253,0.0377918721596145)
			--(axis cs:0.0401027351683544,0.02906425164204)
			--(axis cs:0.0450072562480832,0.0197932818456229)
			--(axis cs:0.0490910657483457,0.0101670926121571)
			--(axis cs:0.0522732094178074,0.000381904731788534)
			--(axis cs:0.0544944142980677,-0.00936380794779851)
			--(axis cs:0.0557184226814536,-0.0188751326819909)
			--(axis cs:0.055932384721948,-0.027966192360974)
			--(axis cs:0.0551463389473286,-0.0364650049385081)
			--(axis cs:0.0533918736718877,-0.0442175159422661)
			--(axis cs:0.0507201120554239,-0.0510906699570057)
			--(axis cs:0.0471991951659226,-0.0569744685513324)
			--(axis cs:0.0429114496861603,-0.0617830381553776)
			--(axis cs:0.0379504213490276,-0.0654547945888852)
			--(axis cs:0.0324179352791625,-0.0679518368417097)
			--(axis cs:0.0264213147396774,-0.0692587298307084)
			--(axis cs:0.0200708551160698,-0.0693808452382716)
			--(axis cs:0.0134776145915093,-0.0683424240104661)
			--(axis cs:0.00675155020197468,-0.0661845074482186)
			--(axis cs:4.90903664049098e-17,-0.0629628600612049);
			
			\path [draw=gray, draw opacity=0.2, line width=1.5pt]
			(axis cs:4.61498269077781e-17,0.0678993644195169)
			--(axis cs:4.61498269077781e-17,0.0678993644195169)
			--(axis cs:4.61498269077781e-17,0.0678993644195169)
			--(axis cs:4.61498269077781e-17,0.0678993644195169)
			--(axis cs:4.61498269077781e-17,0.0678993644195169)
			--(axis cs:4.61498269077781e-17,0.0678993644195169)
			--(axis cs:4.61498269077781e-17,0.0678993644195169)
			--(axis cs:4.61498269077781e-17,0.0678993644195169)
			--(axis cs:4.61498269077781e-17,0.0678993644195169)
			--(axis cs:4.61498269077781e-17,0.0678993644195169)
			--(axis cs:4.61498269077781e-17,0.0678993644195169)
			--(axis cs:4.61498269077781e-17,0.0678993644195169)
			--(axis cs:4.61498269077781e-17,0.0678993644195169)
			--(axis cs:4.61498269077781e-17,0.0678993644195169)
			--(axis cs:4.61498269077781e-17,0.0678993644195169)
			--(axis cs:4.61498269077781e-17,0.0678993644195169)
			--(axis cs:4.61498269077781e-17,0.0678993644195169)
			--(axis cs:4.61498269077781e-17,0.0678993644195169)
			--(axis cs:4.61498269077781e-17,0.0678993644195169)
			--(axis cs:4.61498269077781e-17,0.0678993644195169)
			--(axis cs:4.61498269077781e-17,0.0678993644195169)
			--(axis cs:4.61498269077781e-17,0.0678993644195169)
			--(axis cs:4.61498269077781e-17,0.0678993644195169)
			--(axis cs:4.61498269077781e-17,0.0678993644195169)
			--(axis cs:4.61498269077781e-17,0.0678993644195169);
			
			\path [draw=gray, draw opacity=0.2, line width=1.5pt]
			(axis cs:-0.0325349964548975,0.0681972109485291)
			--(axis cs:-0.0366469353134885,0.0661941053937697)
			--(axis cs:-0.0401786482982442,0.0639604106140683)
			--(axis cs:-0.0430657461072717,0.0615290247067986)
			--(axis cs:-0.0452519653923232,0.0589368158900558)
			--(axis cs:-0.0466905158501245,0.0562243176580048)
			--(axis cs:-0.0473454448090873,0.0534353132373704)
			--(axis cs:-0.0471929691513352,0.0506162953449908)
			--(axis cs:-0.04622271215434,0.0478157921552087)
			--(axis cs:-0.044438772782186,0.0450835572023401)
			--(axis cs:-0.0418605485562576,0.0424696295108759)
			--(axis cs:-0.0385232318489046,0.0400232801669803)
			--(axis cs:-0.0344779045327252,0.0377918721596146)
			--(axis cs:-0.0297911682042386,0.0358196707236675)
			--(axis cs:-0.0245442668273852,0.0341466505267455)
			--(axis cs:-0.0188316848881039,0.0328073527077875)
			--(axis cs:-0.0127592353488723,0.0318298479435099)
			--(axis cs:-0.0064416852963864,0.0312348605985429)
			--(axis cs:4.47527584642421e-17,0.031035103245361)
			--(axis cs:0.00644168529638647,0.0312348605985429)
			--(axis cs:0.0127592353488724,0.0318298479435099)
			--(axis cs:0.018831684888104,0.0328073527077875)
			--(axis cs:0.0245442668273853,0.0341466505267455)
			--(axis cs:0.0297911682042387,0.0358196707236675)
			--(axis cs:0.0344779045327253,0.0377918721596145);
			
			\path [draw=gray, draw opacity=0.2, line width=1.5pt]
			(axis cs:-0.0497944599148204,0.0412382479523173)
			--(axis cs:-0.0562050403305503,0.0380034390749989)
			--(axis cs:-0.0617656411096942,0.034380258416009)
			--(axis cs:-0.0663727910277099,0.0304171079808011)
			--(axis cs:-0.0699323825297052,0.0261694856761292)
			--(axis cs:-0.0723620177138318,0.0216998794413739)
			--(axis cs:-0.0735935968894254,0.0170774535900896)
			--(axis cs:-0.0735760688446163,0.0123774751565389)
			--(axis cs:-0.0722782252137268,0.00768043436961536)
			--(axis cs:-0.0696913849833832,0.0030708256437037)
			--(axis cs:-0.065831783417895,-0.00136442573565184)
			--(axis cs:-0.0607424576443772,-0.00553788089785754)
			--(axis cs:-0.0544944142980677,-0.0093638079477985)
			--(axis cs:-0.0471868779427508,-0.0127606318558619)
			--(axis cs:-0.0389464557693328,-0.0156535070571707)
			--(axis cs:-0.0299251148875929,-0.017976878323615)
			--(axis cs:-0.0202969502704566,-0.0196768770517153)
			--(axis cs:-0.0102538170554242,-0.0207133950678727)
			--(axis cs:4.191215737084e-17,-0.0210616889470168)
			--(axis cs:0.0102538170554243,-0.0207133950678728)
			--(axis cs:0.0202969502704566,-0.0196768770517153)
			--(axis cs:0.029925114887593,-0.017976878323615)
			--(axis cs:0.0389464557693329,-0.0156535070571707)
			--(axis cs:0.0471868779427509,-0.0127606318558619)
			--(axis cs:0.0544944142980677,-0.00936380794779851);
			
			\path [draw=gray, draw opacity=0.2, line width=1.5pt]
			(axis cs:-0.0466241861150617,-0.000340633328074628)
			--(axis cs:-0.0526064137047537,-0.00360314475373163)
			--(axis cs:-0.0577861326190896,-0.0072543802363878)
			--(axis cs:-0.0620672593643556,-0.0112446243735427)
			--(axis cs:-0.0653630148466157,-0.0155171224696908)
			--(axis cs:-0.0675980813142587,-0.020008261110116)
			--(axis cs:-0.0687109484354549,-0.0246479548415487)
			--(axis cs:-0.0686563719396976,-0.029360285906548)
			--(axis cs:-0.0674078370267177,-0.0340644369589176)
			--(axis cs:-0.0649598886791088,-0.0386759442202936)
			--(axis cs:-0.0613301657082577,-0.0431082802595031)
			--(axis cs:-0.0565609590651723,-0.0472747518057331)
			--(axis cs:-0.0507201120554239,-0.0510906699570057)
			--(axis cs:-0.0439010944250877,-0.0544757200301463)
			--(axis cs:-0.036222116185828,-0.0573564292633277)
			--(axis cs:-0.0278242005516113,-0.0596686064155129)
			--(axis cs:-0.0188682055467781,-0.0613596119204972)
			--(axis cs:-0.00953086468913152,-0.0623903139964142)
			--(axis cs:4.65467435283404e-17,-0.0627365970464728)
			--(axis cs:0.00953086468913158,-0.0623903139964142)
			--(axis cs:0.0188682055467781,-0.0613596119204972)
			--(axis cs:0.0278242005516114,-0.0596686064155129)
			--(axis cs:0.0362221161858281,-0.0573564292633277)
			--(axis cs:0.0439010944250878,-0.0544757200301463)
			--(axis cs:0.0507201120554239,-0.0510906699570057);
			
			\path [draw=gray, draw opacity=0.2, line width=1.5pt]
			(axis cs:-0.0252651061814176,-0.0409628306341268)
			--(axis cs:-0.028433243877245,-0.0428501816307415)
			--(axis cs:-0.0311428905894131,-0.0449508906628518)
			--(axis cs:-0.0333451955216741,-0.0472328532122973)
			--(axis cs:-0.0349982553839909,-0.0496604130991129)
			--(axis cs:-0.0360680803805555,-0.0521947392039001)
			--(axis cs:-0.0365295301100312,-0.0547942951650467)
			--(axis cs:-0.0363671845306745,-0.0574154086354438)
			--(axis cs:-0.0355761091183334,-0.0600129423968703)
			--(axis cs:-0.0341624691064747,-0.062541064169394)
			--(axis cs:-0.0321439459822504,-0.0649541055298416)
			--(axis cs:-0.0295499108881756,-0.0672074933249455)
			--(axis cs:-0.0264213147396773,-0.0692587298307084)
			--(axis cs:-0.0228102639303068,-0.0710683912857712)
			--(axis cs:-0.0187792633032251,-0.0726011089946252)
			--(axis cs:-0.0144001240342285,-0.0738264936298922)
			--(axis cs:-0.00975255217758274,-0.074719962238456)
			--(axis cs:-0.00492245247686641,-0.0752634291630205)
			--(axis cs:4.44089209850063e-17,-0.075445826752536)
			--(axis cs:0.00492245247686647,-0.0752634291630205)
			--(axis cs:0.00975255217758281,-0.074719962238456)
			--(axis cs:0.0144001240342286,-0.0738264936298922)
			--(axis cs:0.0187792633032252,-0.0726011089946252)
			--(axis cs:0.0228102639303069,-0.0710683912857712)
			--(axis cs:0.0264213147396774,-0.0692587298307084);
			
			\path [draw=gray, draw opacity=0.2, line width=1.5pt]
			(axis cs:3.64988027510844e-17,-0.0629628600612049)
			--(axis cs:3.57308996084587e-17,-0.0629628600612049)
			--(axis cs:3.50838580740854e-17,-0.0629628600612049)
			--(axis cs:3.45687492080112e-17,-0.0629628600612049)
			--(axis cs:3.41943866657499e-17,-0.0629628600612049)
			--(axis cs:3.39671758941925e-17,-0.0629628600612049)
			--(axis cs:3.38910045326351e-17,-0.0629628600612049)
			--(axis cs:3.39671758941925e-17,-0.0629628600612049)
			--(axis cs:3.41943866657499e-17,-0.0629628600612049)
			--(axis cs:3.45687492080112e-17,-0.0629628600612049)
			--(axis cs:3.50838580740854e-17,-0.0629628600612049)
			--(axis cs:3.57308996084587e-17,-0.0629628600612049)
			--(axis cs:3.64988027510844e-17,-0.0629628600612049)
			--(axis cs:3.7374428466289e-17,-0.0629628600612049)
			--(axis cs:3.83427945553161e-17,-0.0629628600612049)
			--(axis cs:3.93873320059006e-17,-0.0629628600612049)
			--(axis cs:4.04901684926626e-17,-0.0629628600612049)
			--(axis cs:4.16324341775496e-17,-0.0629628600612049)
			--(axis cs:4.27945845779971e-17,-0.0629628600612049)
			--(axis cs:4.39567349784446e-17,-0.0629628600612049)
			--(axis cs:4.50990006633316e-17,-0.0629628600612049)
			--(axis cs:4.62018371500936e-17,-0.0629628600612049)
			--(axis cs:4.72463746006781e-17,-0.0629628600612049)
			--(axis cs:4.82147406897052e-17,-0.0629628600612049)
			--(axis cs:4.90903664049098e-17,-0.0629628600612049);

		\end{axis}
		
	\end{tikzpicture}

%% file: text/conclusion.tex
\chapter{Conclusion and outlook}\hypertarget{Con}{}
\label{chapter:concl}

In this thesis, we investigated \emph{quantum neural networks} (QNNs). These architectures combine two of the most exciting research areas of the 21st century: machine learning and quantum computation. We summarise our results and name ideas for future research questions in the following. 

We entered the topic of QNNs in two steps. Firstly, we introduced the reader to their classical counterparts. There, we began with explaining the functionality of the building blocks of \emph{neural networks} (NNs) and their activation functions. It followed an overview of some popular network architectures and methods. Our focus was on supervised feed-forward NNs. Next, the general approach of optimising NNs was clarified. With a view to the following chapters, we focused on the gradient descent and back-propagation methods. 

Secondly, we gave an overview of the field of \emph{quantum information}. We introduced \emph{qubits}, the quantum analogous to the classical binary bits, and explained their characteristics, including the phenomenons of superposition, mixed states and entanglement. To prepare for the definitions of quantum loss functions, we explained how to compare two quantum states. Moreover, we introduced quantum circuits, which are used to exploit quantum mechanics for quantum computing, and gave an intuition on how the characteristics of quantum mechanics can be used to outperform classical algorithms. Since quantum computation became experimentally possible in the last years, we described the state of the art of these devices. Moreover, we introduced the topic of QNNs with the focus on the implementation, challenges and opportunities. 

\subsection*{Dissipative quantum neural networks}
We presented \emph{dissipative quantum neural networks} (DQNNs) \cite{Beer2020, Beer2021, Beer2021a, Beer2021b}, which are designed for fully quantum learning tasks, and are capable of universal quantum computation. These types of QNNs are built of layers of qubits, which are connected via perceptrons. Such a perceptron is engineered as an arbitrary unitary operation and acts on qubits of two consecutive qubit layers. 

We described the propagation of an input state $\rho^{\text{in}}$ through the network using a completely positive map $\mathcal{E}$. This map is defined as a composition of layer-to-layer transition maps, i.e.\ $\rho^\text{out}=\mathcal{E}\left(\rho^{\text{in}}\right)$. Such a transition map not only contains tensoring the current layer's state to the state of the next layer qubits and applying the perceptron unitaries operations, but also tracing out the qubits from the first of the two layers. For this reason, these QNNs are called \emph{dissipative}.

Further, we presented training and validation loss functions, which are based on the fidelity of two states and compare the output state of the network with the desired state. Moreover, we discussed how the perceptron unitaries get updated in every training epoche based on the knowledge gained from the computation of the training loss. We formulated the update of the unitary $U_j^l$, assigned to the $j$th perceptron acting on layers $l-1$ and $l$, as $U_j^l\rightarrow e^{i\epsilon K_j^l} U_j^l$. Next, we made clear that for such an update, only two states are needed: the output state of the previous layer, $\rho^{l-1}$, obtained by feed-forward propagation through the network, and the state of the following layer $\sigma^l$, obtained by back-propagation of the desired output up to the current layer. This allows us to train deep DQNNs since the memory requirements scale only with the width, not the length of the QNN.

After explaining why DQNNs are capable of universal quantum computing, we presented examples of classical simulations. We could observe that using $S$ training data pairs of the form $\{\ket{\phi^{\text{in}}}, Y\ket{\phi^{\text{in}}}\}$, these QNNs can learn the unknown unitary $Y$ in a feasible number of training epochs, which could be demonstrated by using validation data pairs of the same form. Varying the number $S$ has shown that for only a small number of training pairs, nearly perfect training results can be reached, i.e.\ the fidelity between the desired state and the network's output is nearly one. Further, we could observe that the training algorithm is robust to noise in the training data. Next, we presented an implementation of the DQNN training algorithm on a NISQ device, called \DQNNNISQ \cite{Beer2021a}, and tested it with the same learning task. Despite the high noise levels, we observed that the \DQNNNISQ could generalise the information provided through the training data pairs. 

Using the task of learning an unknown unitary, we compared the training success of \DQNNsNISQ with an implementation of the \emph{quantum approximate optimisation algorithm} (QAOA)\cite{Farhi2014, Farhi2016, Hadfield2019}.  Whereas in the \DQNNNISQ context, a perceptron acts on layers of different qubits, in the QAOA setting, a perceptron is defined as a sequence of operations, and all perceptrons act on the identical qubits. We could observe that both QNN architectures succeed in the learning task. However, the results indicate that the \DQNNNISQ is more suitable for learning an unknown unitary operation compared to the QAOA.

As described in \cref{sec:QI_QNN}, many different models for a quantum perceptron and QNNs were proposed in the last years. An open research task is to perform a more exhaustive comparison, including more versions of QNNs and find out which QNN architectures best suits the training tasks. Further, in the following years, it will be essential to optimise the implementations of QNNs in general and the implementation of \DQNNsNISQ in particular for quantum devices of coming generations.

\subsection*{No free lunch theorem}
To understand the ultimate limits for QNNs, we presented the \emph{quantum no free lunch} (QNFL) theorem \cite{Poland2020}. This result describes a bound on the probability that a quantum device, which can be modelled as a unitary process and is optimised with quantum data pairs, gives an incorrect output for a random input. The theorem enabled us to review the learning behaviour of QNNs.

For describing the proof of the QNFL theorem, we motivated a quantum risk by comparing the unknown unitary and the device approximated unitary using an average over all pure states. We compared the resulting bound with the classical \emph{no free lunch} theorem for invertible functions. Further, we found that the results gained with the DQNN algorithm are close to achieving the QNFL bound. However, a slight discrepancy was expected since the process of evaluating the quantum risk includes empirical averages.

The here presented work on the QNFL theorem was already generalised by \cite{Sharma2020a} to the case where these quantum states can be entangled to a reference system. Beyond that, the understanding of ultimate limits for quantum learning devices in different variations is key for further progress in the field of quantum machine learning. 

\subsection*{Quantum machine learning with graph-structured data}
The above mentioned QNN architectures were trained with quantum data pairs. Moreover, we studied two extensions of this ansatz. The first one includes a possible underlying graph structure of the training data \cite{Beer2021a}. The structure of quantum devices leads to structured quantum data. Hence, we assumed to have knowledge about the underlying graph and found a training loss function including such a structure in the training process of the DQNNs and the \DQNNsNISQ. 

In the first two examples, we chose the desired output states to have a structure of a line graph and a graph describing a connected cluster. The input states were chosen to be random. In both examples, we could observe that including the graph structure in the training process leads to better generalisation results than a simple supervised ansatz.

Moreover, to also make the input states assigned to the graph structure, we used a synthetic classical graph, where each vertex was assigned to a label and an embedding vector. The graph was constructed via a classical deep walk. We used the embedding of a vertex to construct an input state and the labels for constructing desired output states. Thus, both parts of a training data pair relied on the graph. Using a so constructed data set, we noticed that the training process leads to better results when including the knowledge on the graph structure, especially if only a few of the desired output states were supervised.  

The in this work studied graphs were synthetically constructed. An exciting research direction would be to test the in \cite{Beer2021a} proposed graph-based loss function, using output data generated from actual quantum devices of which the build-in structure is known.

\subsection*{Quantum generative adversarial networks}
In the second extension, we discussed a different training goal. Whereas the original DQNN and the DQNN making use of graph structure were trained with data pairs in order to learn an underlying relation, we aimed for characterising a set of quantum states in order to extend it to quantum states having similar properties with the \emph{discriminative quantum adversarial network} (DQGAN).

After describing the original classical \emph{generative adversarial networks} (GANs), we explained how to build the DQGANs. Such an architecture is built of two DQNNs, where one of them plays a generative and one a discriminative role. The discriminator gets either the generator's output or a training data state as an input and is trained to distinguish well between these categories. The generator's goal is to trick the discriminator. These training goals can be realised in two different training losses, which are optimised successively in one training epoch. 

Using the DQGAN, it is possible to extent a given data set to states with similar characteristics. We described a way to study the diversity of such a generated data set. In this regard, we found that, different to the training DQNN and DQNN with graph structure, more training epochs do not give better results. When training DQGANs, we could observe that a more extended training can lead to a maximum validation loss. However, after some training epochs, our examples showed that the generator only produces a small set of different states. We could assert that a carefully chosen number of training rounds can lead to both, a good validation loss and the generator's output diversity.

In this thesis, we only discussed a limited amount of examples for training data used to train DQGANs. The study of other data sets is of interest. One example could be a set of states with similar degrees of entanglement (concerning a chosen entanglement measure) \cite{Schatzki2021}. Moreover, the application of the data sets produced by the generator is to study. In classical machine learning, the output of GANs is used, for example, for the training of other NN architectures, when training data is rare. Whether such an ansatz can be executed successfully using DQGANs, is left open as a future research topic. Further, GANs are not the only adaption of simple feed-forward NNs. It is also of interest to study the applications and behaviour of DQNNs in a recurrent or convolutional neural network settings. We leave these topics open for future work.

\subsection*{Conclusion}
This work introduced DQNNs, which are designed for fully quantum learning tasks, are capable of universal quantum computation and have low memory requirements while training. We showed via the QNFL theorem that these QNNs can be nearly optimally trained when learning an unknown unitary operation. Moreover, we demonstrated that DQNNs can make use of the graph structure of quantum data and can also be trained in a generative adversarial setting in order to extend quantum data sets.

Since we are still at the beginning of the age of quantum computers, this work leads to many interesting further research questions. We believe that QNNs will be essential for analysing and processing huge amounts of quantum data produced by the quantum devices of future generations. Today, we can already look back on many successes in the field of quantum computation in the last few years. Therefore we presume that quantum computing, including more qubits and less noise, compared to nowadays' NISQ devices, will be experimentally possible in the following decades. Quantum computers will not solve all of our world's problems. However, if applied wisely, we believe that they can lead to more exciting and fruitful applications in the coming time.

%% file: text/appendixDQNN.tex
\chapter{Dissipative quantum neural networks}\label{apdx:DQNN}
\markboth{Appendix}{Appendix} 

In \cref{sec:DQNN_classical} the classical simulation of a \emph{dissipative quantum neural network} (DQNN) was explained. The numerical simulation, including a generalisation analysis and noise robustness testing, was done with 2-3-2 DQNNs. In the following, we want to embrace the opportunity of an appendix and show some more examples. We study the following DQNN architectures: 1-1-1 in \cref{fig:1-1-1}, 1-2-1 in \cref{fig:1-2-1}, 2-3-2 in \cref{fig:2-3-2}, 3-4-3 in \cref{fig:3-4-3}, and 2-3-4-3-2 in \cref{fig:2-3-4-3-2}. Note that the results in \cref{fig:2-3-2} were already discussed in \cref{sec:DQNN_classical}. We replotted them here here again for better comparison. All results are obtained with the code available at \cite{GithubKerstin}. The training data pairs of all examples are of the form $\{\ket{\phi^{\text{in}}_x}, \ket{\phi^{\text{SV}}_x}\} $ with $\ket{\phi^\text{SV}_x} = Y\ket{\phi^\text{in}_x}$, where $Y$ is a unitary. In the plots values of the the training loss 
\begin{equation*}
\mathcal{L}_\text{SV}=\frac{1}{S}\sum_{x=1}^S F(\ket{\phi^{\text{SV}}_x}\bra{\phi^{\text{SV}}_x},\rho_x^{\text{out}}) = \frac{1}{S}\sum_{x=1}^S \braket{\phi^{\text{SV}}_x|\rho_x^{\text{out}}|\phi^{\text{SV}}_x},
\end{equation*}
and the validation loss
\begin{equation*}
\mathcal{L}_\text{USV}=\frac{1}{N-S}\sum_{x=S+1}^{N} \langle\phi^{\text{USV}}_x\rvert\rho_x^{\text{out}}\lvert\phi^{\text{USV}}_x\rangle,
\end{equation*} 
with $\ket{\phi^\text{USV}_x} = Y\ket{\phi^\text{in}_x}$ are depicted.

\subsection*{Training}
At first we comment on the evaluation of the loss function during a single training session. In all examples the validation and the training loss reach values near $1$. However, we can notice differences in the process: For training a 1-1-1 DQNN with $S=20$ training pairs and $N-S=80$ validation pairs the validation loss $\mathcal{L}_\text{USV}$ reaches a value $0.95$ at $r_T=382$. For the other architectures the same value is reached at $r_T=164$ (1-2-1), $r_T=420$ (2-3-2), and $r_T=552$ (3-4-3). We can see as well that the distance between the validation and training loss is bigger. 

But, especially in the first $500$ rounds of training the 2-3-4-3-2, the validation loss is increasing rather slowly and a validation loss value of $0.95$ is not reached until $r_T=795$. Despite that, a fidelity of nearly $1$ is reached at the end of the training. Note though that these values are not averaged over many training attempts and therefore are probably not very comparable, since every training process is unique due to the randomly assigned unitary initialisation and the generation of the training data pairs.

\subsection*{Generalisation}

A better comparison can be made for the different generalisation analysis plots. Here, every data point depicts an average over 10 independent training attempts. 

For training a 1-1-1 DQNN in $1000$ training epochs, the validation loss $\mathcal{L}_\text{USV}$ reaches a value higher than $0.8$ with only $S=2$ supervised training pairs. For the other architectures, the same value is reached at $S=2$ (1-2-1), $S=5$ (2-3-2), $S=9$ (3-4-3), and $S=6$ (2-3-4-3-2). As expected via the \emph{quantum no free lunch} theorem, explained in \cref{chapter:NFL}, the generalisation of states of bigger dimensions general needs more training data pairs. Further, we want to point out that the values $S=5$ (2-3-2) and $S=6$ (2-3-4-3-2) lead to the assumption that additional layers do not improve the generalisation behaviour.  

\subsection*{Noise robustness}
Analogously to the generalisation analysis, every data point in in the noise robustness plots is generated via an average over 10 independent training attempts. 

Comparing the training of the 1-1-1 and the 1-2-1 DQNN, we can see that the additional qubit does not improve the resistance against noise. Further, we notice that the noise robustness shrinks with the number of qubits in the input layer. For $\delta=0.3$, the 1-1-1 DQNN reaches a validation loss of $0.79$ and the 1-2-1 a validation loss of $0.83$. Using the same $\delta$, we reach validation loss values of  $0.73$ (2-3-2), $0.50$ (2-3-4-3-2), and $0.62$ (3-4-3) for the other DQNN architectures.

\newpage
\begin{figure}
\centering
\begin{subfigure}{0.95\textwidth}\centering
\begin{tikzpicture}
\begin{axis}[
xmin=0,   xmax=10,
ymin=0,   ymax=1,
width=.8\linewidth, 
height=.48\linewidth,
grid=major,
grid style={color0M},
xlabel= Training epochs $r_T$, 
xticklabels={0,0,100,200,300,400,500,600,700,800,900,1000},
ylabel=$\mathcal{L}(t)$,legend style={draw=none},legend pos=south east,legend cell align={left},legend style={draw=none,legend image code/.code={\filldraw[##1] (-.5ex,-.5ex) rectangle (0.5ex,0.5ex);}}]
\addplot[line width=2pt, color=color1] table [x=step times epsilon, y=SsvTestingUsv, col sep=comma] {numerics/randomUnitary_100pairs20sv_1-1-1network_delta0_lda1_ep0i01_plot.csv};
\addlegendentry[mark size=10 pt,scale=1]{Validation loss $\mathcal{L}_\text{USV}$} 
\addplot[line width=2pt, color=color2] table [x=step times epsilon, y=SsvTraining, col sep=comma] {numerics/randomUnitary_100pairs20sv_1-1-1network_delta0_lda1_ep0i01_plot.csv};
\addlegendentry[scale=1]{Training loss $\mathcal{L}_\text{SV}$} 
\end{axis}
\end{tikzpicture}
\subcaption{Training with $S=20$.}
\end{subfigure}
\begin{subfigure}{0.95\textwidth}\centering
\begin{tikzpicture}[scale=1]
\begin{axis}[
xmin=0.5,   xmax=14.5,
ymin=0,   ymax=1,
width=.8\linewidth, 
height=.48\linewidth,
grid=major,
grid style={color0M},
xlabel= $S$, 
ylabel=$\mathcal{L}$,legend style={draw=none},legend pos=south east,legend cell align={left}]
\addplot[color=color1, only marks, mark size=3 pt,mark phase=0] table [x=numberSupervisedPairsList, y=SsvTestingUsvMeanList, col sep=comma] {numerics/randomUnitary_100pairs_1-1-1network_delta0_lda1_ep0i01_rounds1000_shots10_plotmean.csv};
\addlegendentry[mark size=3 pt,scale=1]{Validation loss $\mathcal{L}_\text{USV}$} 
\addplot[color=color2, only marks, mark size=3 pt,mark phase=0] table [x=numberSupervisedPairsList, y=SsvTrainingMeanList, col sep=comma] {numerics/randomUnitary_100pairs_1-1-1network_delta0_lda1_ep0i01_rounds1000_shots10_plotmean.csv};
\addlegendentry[scale=1]{Training loss $\mathcal{L}_\text{SV}$} 
\end{axis}
\end{tikzpicture}
\subcaption{Generalisation analysis.}
\end{subfigure}
\begin{subfigure}{0.95\textwidth}\centering
\begin{tikzpicture}[scale=1]
\begin{axis}[
xmin=-0.05,   xmax=1.05,
ymin=0,   ymax=1,
width=.8\linewidth, 
height=.48\linewidth,
grid=major,
grid style={color0M},
xlabel= $\delta$, 
ylabel=$\mathcal{L}$,legend style={draw=none},legend pos=south west,legend cell align={left}]
\addplot[color=color1, only marks, mark size=3 pt,mark phase=0] table [x=deltaList, y=SsvTestingUsvMeanList, col sep=comma] {numerics/randomUnitary_100pairs_20sv_1-1-1network_deltaVar_lda1_ep0i01_rounds1000_shots10_plotmean.csv};
\addlegendentry[mark size=3 pt,scale=1]{Validation loss $\mathcal{L}_\text{USV}$} 
\addplot[color=color2, only marks, mark size=3 pt,mark phase=0] table [x=deltaList, y=SsvTrainingMeanList, col sep=comma] {numerics/randomUnitary_100pairs_20sv_1-1-1network_deltaVar_lda1_ep0i01_rounds1000_shots10_plotmean.csv};
\addlegendentry[scale=1]{Training loss $\mathcal{L}_\text{SV}$} 
\end{axis}
\end{tikzpicture}
\subcaption{Testing the noise robustness for $S=20$.}
\end{subfigure}
\caption{\textbf{1-1-1 DQNN.} The plots describe the training (a) and generalisation behaviour (b) as well as the noise robustness (c) of training the DQNN in $k=1000$ training epochs with $\eta=1$, $\epsilon=0.01$ and $N=100$ data pairs (based on a unitary $Y \in \mathcal{U}(2)$).}
\label{fig:1-1-1}
\end{figure}
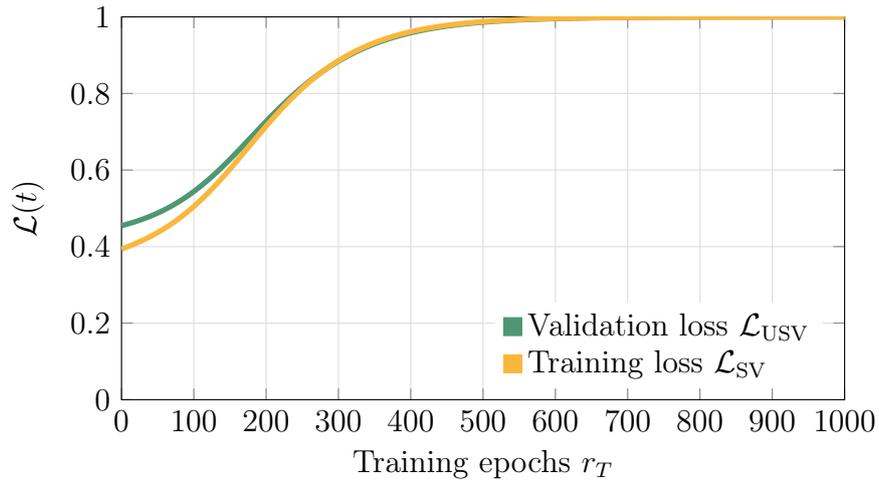
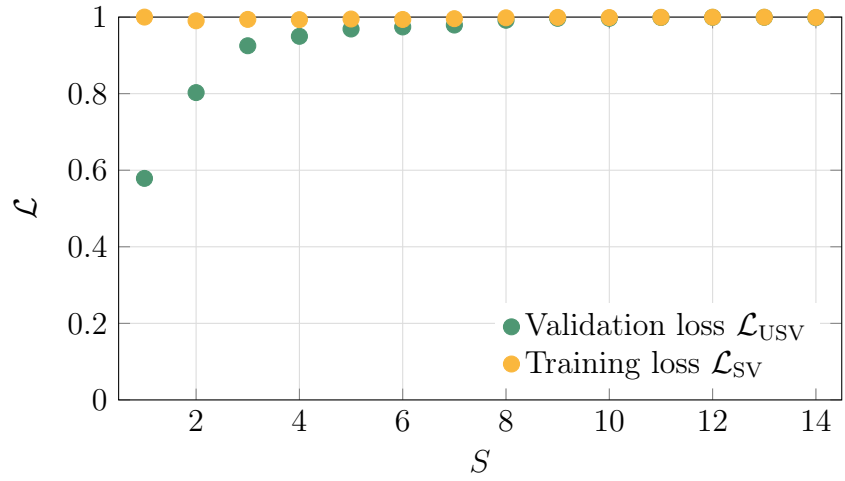
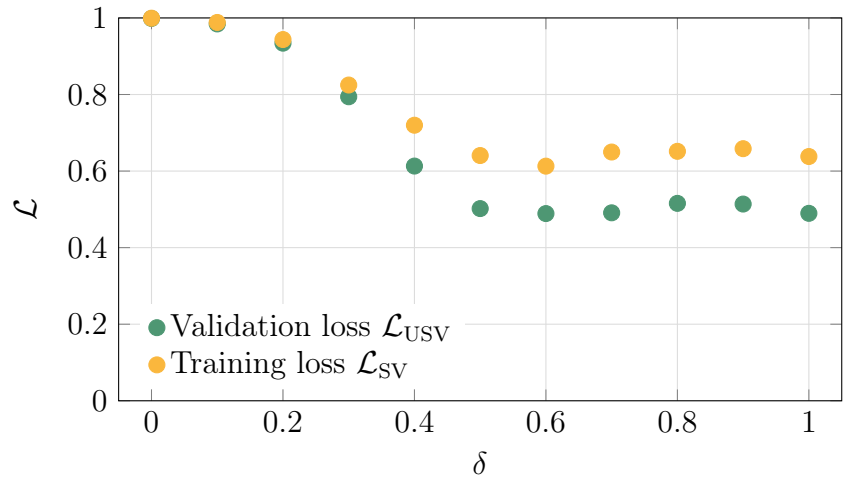

\begin{figure}
\centering
\begin{subfigure}{0.95\textwidth}\centering
\begin{tikzpicture}
\begin{axis}[
xmin=0,   xmax=10,
ymin=0,   ymax=1,
width=.8\linewidth, 
height=.48\linewidth,
grid=major,
grid style={color0M},
xlabel= Training epochs $r_T$, 
xticklabels={0,0,100,200,300,400,500,600,700,800,900,1000},
ylabel=$\mathcal{L}(t)$,legend style={draw=none},legend pos=south east,legend cell align={left},legend style={draw=none,legend image code/.code={\filldraw[##1] (-.5ex,-.5ex) rectangle (0.5ex,0.5ex);}}]
\addplot[line width=2pt, color=color1] table [x=step times epsilon, y=SsvTestingUsv, col sep=comma] {numerics/randomUnitary_100pairs20sv_1-2-1network_delta0_lda1_ep0i01_plot.csv};
\addlegendentry[mark size=10 pt,scale=1]{Validation loss $\mathcal{L}_\text{USV}$} 
\addplot[line width=2pt, color=color2] table [x=step times epsilon, y=SsvTraining, col sep=comma] {numerics/randomUnitary_100pairs20sv_1-2-1network_delta0_lda1_ep0i01_plot.csv};
\addlegendentry[scale=1]{Training loss $\mathcal{L}_\text{SV}$} 
\end{axis}
\end{tikzpicture}
\subcaption{Training with $S=20$.}
\end{subfigure}
\begin{subfigure}{0.95\textwidth}\centering
\begin{tikzpicture}[scale=1]
\begin{axis}[
xmin=0.5,   xmax=14.5,
ymin=0,   ymax=1,
width=.8\linewidth, 
height=.48\linewidth,
grid=major,
grid style={color0M},
xlabel= $S$, 
ylabel=$\mathcal{L}$,legend style={draw=none},legend pos=south east,legend cell align={left}]
\addplot[color=color1, only marks, mark size=3 pt,mark phase=0] table [x=numberSupervisedPairsList, y=SsvTestingUsvMeanList, col sep=comma] {numerics/randomUnitary_100pairs_1-2-1network_delta0_lda1_ep0i01_rounds1000_shots10_plotmean.csv};
\addlegendentry[mark size=3 pt,scale=1]{Validation loss $\mathcal{L}_\text{USV}$} 
\addplot[color=color2, only marks, mark size=3 pt,mark phase=0] table [x=numberSupervisedPairsList, y=SsvTrainingMeanList, col sep=comma] {numerics/randomUnitary_100pairs_1-2-1network_delta0_lda1_ep0i01_rounds1000_shots10_plotmean.csv};
\addlegendentry[scale=1]{Training loss $\mathcal{L}_\text{SV}$} 
\end{axis}
\end{tikzpicture}
\subcaption{Generalisation analysis.}
\end{subfigure}
\begin{subfigure}{0.95\textwidth}\centering
\begin{tikzpicture}[scale=1]
\begin{axis}[
xmin=-0.05,   xmax=1.05,
ymin=0,   ymax=1,
width=.8\linewidth, 
height=.48\linewidth,
grid=major,
grid style={color0M},
xlabel= $\delta$, 
ylabel=$\mathcal{L}$,legend style={draw=none},legend pos=south west,legend cell align={left}]
\addplot[color=color1, only marks, mark size=3 pt,mark phase=0] table [x=deltaList, y=SsvTestingUsvMeanList, col sep=comma] {numerics/randomUnitary_100pairs_20sv_1-2-1network_deltaVar_lda1_ep0i01_rounds1000_shots10_plotmean.csv};
\addlegendentry[mark size=3 pt,scale=1]{Validation loss $\mathcal{L}_\text{USV}$} 
\addplot[color=color2, only marks, mark size=3 pt,mark phase=0] table [x=deltaList, y=SsvTrainingMeanList, col sep=comma] {numerics/randomUnitary_100pairs_20sv_1-2-1network_deltaVar_lda1_ep0i01_rounds1000_shots10_plotmean.csv};
\addlegendentry[scale=1]{Training loss $\mathcal{L}_\text{SV}$} 
\end{axis}
\end{tikzpicture}
\subcaption{Testing the noise robustness for $S=20$.}
\end{subfigure}
\caption{\textbf{1-2-1 DQNN.} The plots describe the training (a) and generalisation behaviour (b) as well as the noise robustness (c) of training the DQNN in $k=1000$ training epochs with $\eta=1$, $\epsilon=0.01$ and $N=100$ data pairs (based on a unitary $Y \in \mathcal{U}(4)$).}
\label{fig:1-2-1}
\end{figure}

\begin{figure}
\centering
\begin{subfigure}{0.95\textwidth}\centering
\begin{tikzpicture}
\begin{axis}[
xmin=0,   xmax=10,
ymin=0,   ymax=1,
width=.8\linewidth, 
height=.48\linewidth,
grid=major,
grid style={color0M},
xlabel= Training epochs $r_T$, 
xticklabels={0,0,100,200,300,400,500,600,700,800,900,1000},
ylabel=$\mathcal{L}(t)$,legend style={draw=none},legend pos=south east,legend cell align={left},legend style={draw=none,legend image code/.code={\filldraw[##1] (-.5ex,-.5ex) rectangle (0.5ex,0.5ex);}}]
\addplot[line width=2pt, color=color1] table [x=step times epsilon, y=SsvTestingUsv, col sep=comma] {numerics/randomUnitary_100pairs20sv_2-3-2network_delta0_lda1_ep0i01_plot.csv};
\addlegendentry[mark size=10 pt,scale=1]{Validation loss $\mathcal{L}_\text{USV}$} 
\addplot[line width=2pt, color=color2] table [x=step times epsilon, y=SsvTraining, col sep=comma] {numerics/randomUnitary_100pairs20sv_2-3-2network_delta0_lda1_ep0i01_plot.csv};
\addlegendentry[scale=1]{Training loss $\mathcal{L}_\text{SV}$} 
\end{axis}
\end{tikzpicture}
\subcaption{Training with $S=20$.}
\end{subfigure}
\begin{subfigure}{0.95\textwidth}\centering
\begin{tikzpicture}[scale=1]
\begin{axis}[
xmin=0.5,   xmax=14.5,
ymin=0,   ymax=1,
width=.8\linewidth, 
height=.48\linewidth,
grid=major,
grid style={color0M},
xlabel= $S$, 
ylabel=$\mathcal{L}$,legend style={draw=none},legend pos=south east,legend cell align={left}]
\addplot[color=color1, only marks, mark size=3 pt,mark phase=0] table [x=numberSupervisedPairsList, y=SsvTestingUsvMeanList, col sep=comma] {numerics/randomUnitary_100pairs_2-3-2network_delta0_lda1_ep0i01_rounds1000_shots10_plotmean.csv};
\addlegendentry[mark size=3 pt,scale=1]{Validation loss $\mathcal{L}_\text{USV}$} 
\addplot[color=color2, only marks, mark size=3 pt,mark phase=0] table [x=numberSupervisedPairsList, y=SsvTrainingMeanList, col sep=comma] {numerics/randomUnitary_100pairs_2-3-2network_delta0_lda1_ep0i01_rounds1000_shots10_plotmean.csv};
\addlegendentry[scale=1]{Training loss $\mathcal{L}_\text{SV}$} 
\end{axis}
\end{tikzpicture}
\subcaption{Generalisation analysis.}
\end{subfigure}
\begin{subfigure}{0.95\textwidth}\centering
\begin{tikzpicture}[scale=1]
\begin{axis}[
xmin=-0.05,   xmax=1.05,
ymin=0,   ymax=1,
width=.8\linewidth, 
height=.48\linewidth,
grid=major,
grid style={color0M},
xlabel= $\delta$, 
ylabel=$\mathcal{L}$,legend style={draw=none},legend pos=south west,legend cell align={left}]
\addplot[color=color1, only marks, mark size=3 pt,mark phase=0] table [x=deltaList, y=SsvTestingUsvMeanList, col sep=comma] {numerics/randomUnitary_100pairs_20sv_2-3-2network_deltaVar_lda1_ep0i01_rounds1000_shots10_plotmean.csv};
\addlegendentry[mark size=3 pt,scale=1]{Validation loss $\mathcal{L}_\text{USV}$} 
\addplot[color=color2, only marks, mark size=3 pt,mark phase=0] table [x=deltaList, y=SsvTrainingMeanList, col sep=comma] {numerics/randomUnitary_100pairs_20sv_2-3-2network_deltaVar_lda1_ep0i01_rounds1000_shots10_plotmean.csv};
\addlegendentry[scale=1]{Training loss $\mathcal{L}_\text{SV}$} 
\end{axis}
\end{tikzpicture}
\subcaption{Testing the noise robustness for $S=20$.}
\end{subfigure}
\caption{\textbf{2-3-2 DQNN.} The plots describe the training (a) and generalisation behaviour (b) as well as the noise robustness (c) of training the DQNN in $k=1000$ training epochs with $\eta=1$, $\epsilon=0.01$ and $N=100$ data pairs (based on a unitary $Y \in \mathcal{U}(4)$).}
\label{fig:2-3-2}
\end{figure}

\begin{figure}
\centering
\begin{subfigure}{0.95\textwidth}\centering
\begin{tikzpicture}
\begin{axis}[
xmin=0,   xmax=10,
ymin=0,   ymax=1,
width=.8\linewidth, 
height=.48\linewidth,
grid=major,
xlabel= Training epochs $r_T$, 
xticklabels={0,0,100,200,300,400,500,600,700,800,900,1000},
ylabel=$\mathcal{L}(t)$,legend style={draw=none},legend pos=south east,legend cell align={left},legend style={draw=none,legend image code/.code={\filldraw[##1] (-.5ex,-.5ex) rectangle (0.5ex,0.5ex);}}]
\addplot[line width=2pt, color=color1] table [x=step times epsilon, y=SsvTestingUsv, col sep=comma] {numerics/randomUnitary_100pairs20sv_3-4-3network_delta0_lda1_ep0i01_plot.csv};
\addlegendentry[mark size=10 pt,scale=1]{Validation loss $\mathcal{L}_\text{USV}$} 
\addplot[line width=2pt, color=color2] table [x=step times epsilon, y=SsvTraining, col sep=comma] {numerics/randomUnitary_100pairs20sv_3-4-3network_delta0_lda1_ep0i01_plot.csv};
\addlegendentry[scale=1]{Training loss $\mathcal{L}_\text{SV}$} 
\end{axis}
\end{tikzpicture}
\subcaption{Training with $S=20$.}
\end{subfigure}
\begin{subfigure}{0.95\textwidth}\centering
\begin{tikzpicture}[scale=1]
\begin{axis}[
xmin=0.5,   xmax=14.5,
ymin=0,   ymax=1,
width=.8\linewidth, 
height=.48\linewidth,
grid=major,
grid style={color0M},
xlabel= $S$, 
ylabel=$\mathcal{L}$,legend style={draw=none},legend pos=south east,legend cell align={left}]
\addplot[color=color1, only marks, mark size=3 pt,mark phase=0] table [x=numberSupervisedPairsList, y=SsvTestingUsvMeanList, col sep=comma] {numerics/randomUnitary_100pairs_3-4-3network_delta0_lda1_ep0i01_rounds1000_shots10_plotmean.csv};
\addlegendentry[mark size=3 pt,scale=1]{Validation loss $\mathcal{L}_\text{USV}$} 
\addplot[color=color2, only marks, mark size=3 pt,mark phase=0] table [x=numberSupervisedPairsList, y=SsvTrainingMeanList, col sep=comma] {numerics/randomUnitary_100pairs_3-4-3network_delta0_lda1_ep0i01_rounds1000_shots10_plotmean.csv};
\addlegendentry[scale=1]{Training loss $\mathcal{L}_\text{SV}$} 
\end{axis}
\end{tikzpicture}
\subcaption{Generalisation analysis.}
\end{subfigure}
\begin{subfigure}{0.95\textwidth}\centering
\begin{tikzpicture}[scale=1]
\begin{axis}[
xmin=-0.05,   xmax=1.05,
ymin=0,   ymax=1,
width=.8\linewidth, 
height=.48\linewidth,
grid=major,
grid style={color0M},
xlabel= $\delta$, 
ylabel=$\mathcal{L}$,legend style={draw=none},legend pos=south west,legend cell align={left}]
\addplot[color=color1, only marks, mark size=3 pt,mark phase=0] table [x=deltaList, y=SsvTestingUsvMeanList, col sep=comma] {numerics/randomUnitary_100pairs_20sv_3-4-3network_deltaVar_lda1_ep0i01_rounds1000_shots10_plotmean.csv};
\addlegendentry[mark size=3 pt,scale=1]{Validation loss $\mathcal{L}_\text{USV}$} 
\addplot[color=color2, only marks, mark size=3 pt,mark phase=0] table [x=deltaList, y=SsvTrainingMeanList, col sep=comma] {numerics/randomUnitary_100pairs_20sv_3-4-3network_deltaVar_lda1_ep0i01_rounds1000_shots10_plotmean.csv};
\addlegendentry[scale=1]{Training loss $\mathcal{L}_\text{SV}$} 
\end{axis}
\end{tikzpicture}
\subcaption{Testing the noise robustness for $S=20$.}
\end{subfigure}
\caption{\textbf{3-4-3 DQNN.} The plots describe the training (a) and generalisation behaviour (b) as well as the noise robustness (c) of training the DQNN in $k=1000$ training epochs with $\eta=1$, $\epsilon=0.01$ and $N=100$ data pairs (based on a unitary $Y \in \mathcal{U}(9)$).}
\label{fig:3-4-3}
\end{figure}

\begin{figure}
\centering
\begin{subfigure}{0.95\textwidth}\centering
\begin{tikzpicture}
\begin{axis}[
xmin=0,   xmax=10,
ymin=0,   ymax=1,
width=.8\linewidth, 
height=.48\linewidth,
grid=major,
grid style={color0M},
xlabel= Training epochs $r_T$, 
xticklabels={0,0,100,200,300,400,500,600,700,800,900,1000},
ylabel=$\mathcal{L}(t)$,legend style={draw=none},legend pos=south east,legend cell align={left},legend style={draw=none,legend image code/.code={\filldraw[##1] (-.5ex,-.5ex) rectangle (0.5ex,0.5ex);}}]
\addplot[line width=2pt, color=color1] table [x=step times epsilon, y=SsvTestingUsv, col sep=comma] {numerics/randomUnitary_100pairs20sv_2-3-4-3-2network_delta0_lda1_ep0i01_plot.csv};
\addlegendentry[mark size=10 pt,scale=1]{Validation loss $\mathcal{L}_\text{USV}$} 
\addplot[line width=2pt, color=color2] table [x=step times epsilon, y=SsvTraining, col sep=comma] {numerics/randomUnitary_100pairs20sv_2-3-4-3-2network_delta0_lda1_ep0i01_plot.csv};
\addlegendentry[scale=1]{Training loss $\mathcal{L}_\text{SV}$} 
\end{axis}
\end{tikzpicture}
\subcaption{Training with $S=20$.}
\end{subfigure}
\begin{subfigure}{0.95\textwidth}\centering
\begin{tikzpicture}[scale=1]
\begin{axis}[
xmin=0.5,   xmax=14.5,
ymin=0,   ymax=1,
width=.8\linewidth, 
height=.48\linewidth,
grid=major,
grid style={color0M},
xlabel= $S$, 
ylabel=$\mathcal{L}$,legend style={draw=none},legend pos=south east,legend cell align={left}]
\addplot[color=color1, only marks, mark size=3 pt,mark phase=0] table [x=numberSupervisedPairsList, y=SsvTestingUsvMeanList, col sep=comma] {numerics/randomUnitary_100pairs_2-3-4-3-2network_delta0_lda1_ep0i01_rounds1000_shots10_plotmean.csv};
\addlegendentry[mark size=3 pt,scale=1]{Validation loss $\mathcal{L}_\text{USV}$} 
\addplot[color=color2, only marks, mark size=3 pt,mark phase=0] table [x=numberSupervisedPairsList, y=SsvTrainingMeanList, col sep=comma] {numerics/randomUnitary_100pairs_2-3-4-3-2network_delta0_lda1_ep0i01_rounds1000_shots10_plotmean.csv};
\addlegendentry[scale=1]{Training loss $\mathcal{L}_\text{SV}$} 
\end{axis}
\end{tikzpicture}
\subcaption{Generalisation analysis.}
\end{subfigure}
\begin{subfigure}{0.95\textwidth}\centering
\begin{tikzpicture}[scale=1]
\begin{axis}[
xmin=-0.05,   xmax=1.05,
ymin=0,   ymax=1,
width=.8\linewidth, 
height=.48\linewidth,
grid=major,
grid style={color0M},
xlabel= $\delta$, 
ylabel=$\mathcal{L}$,legend style={draw=none},legend pos=south west,legend cell align={left}]
\addplot[color=color1, only marks, mark size=3 pt,mark phase=0] table [x=deltaList, y=SsvTestingUsvMeanList, col sep=comma] {numerics/randomUnitary_100pairs_20sv_2-3-4-3-2network_deltaVar_lda1_ep0i01_rounds1000_shots10_plotmean.csv};
\addlegendentry[mark size=3 pt,scale=1]{Validation loss $\mathcal{L}_\text{USV}$} 
\addplot[color=color2, only marks, mark size=3 pt,mark phase=0] table [x=deltaList, y=SsvTrainingMeanList, col sep=comma] {numerics/randomUnitary_100pairs_20sv_2-3-4-3-2network_deltaVar_lda1_ep0i01_rounds1000_shots10_plotmean.csv};
\addlegendentry[scale=1]{Training loss $\mathcal{L}_\text{SV}$} 
\end{axis}
\end{tikzpicture}
\subcaption{Testing the noise robustness for $S=20$.}
\end{subfigure}
\caption{\textbf{2-3-4-3-2 DQNN.} The plots describe the training (a) and generalisation behaviour (b) as well as the noise robustness (c) of training the DQNN in $k=1000$ training epochs with $\eta=1$, $\epsilon=0.01$ and $N=100$ data pairs (based on a unitary $Y \in \mathcal{U}(4)$).}
\label{fig:2-3-4-3-2}
\end{figure}

%% file: text/appendixgraph.tex
\chapter{Training with graph-structured quantum data}
\markboth{Appendix}{Appendix} 
\label{apdx:graphs}

In \cref{sec:graph_classicalSim}, we described how to build a set of quantum data pairs with an underlying graph-structure using a classical walk. In the following, we plot the evolution of the validation loss after training a DQNN with
\begin{equation*}
		\mathcal{L}_\text{SV+G}=\mathcal{L}_\text{SV} + \gamma \mathcal{L}_{G},
\end{equation*}
where
\begin{equation*}
	\mathcal{L}_\text{SV} \equiv \frac{1}{S}\sum\limits_{x=1}^S\bra{\phi^\text{SV}_x}\mathcal{E}\big(\rho^\text{in}_x\big)\ket{\phi^\text{SV}_x},
\end{equation*}
and
\begin{equation*}
		\mathcal{L}_{G} \equiv \sum_{w,x\in V} [A]_{wx} d_{\text{HS}}(\mathcal{E}(\rho_w),\mathcal{E}(\rho_x)).
\end{equation*}
All results are generated with the code available at \cite{GithubKerstin}. In \cref{fig:apdx_graph_walk},
\begin{equation*}
	\mathcal{L}_\text{USV}=\frac{1}{N-S}\sum_{x=S+1}^{N} \langle\phi^{\text{USV}}_x\rvert\mathcal{E}\big(\rho^\text{in}_x\big)\lvert\phi^{\text{USV}}_x\rangle
\end{equation*}
is plotted for a training with $\gamma=0$ and $\gamma=-0.5$ using the graph-structured data built based on a classical deep walk. Since we discussed the generalisation behaviour of this training example only for $1\le S\le 10$ in \cref{sec:graph_classicalSim},  \cref{fig:apdx_graph_walk} plots the values of the loss functions after $r_T=1000$ training rounds for $1\le S\le 32$. The figure shows that for cases with less than $15$ of $32$ supervised vertices, the validation loss is lower when ignoring the problem's graph structure. After that threshold both learning strategies are about equally good.

\begin{figure}[H]
\centering
\begin{tikzpicture}
\begin{axis}[
	xmin=0,   xmax=20,
	ymin=0.0,   ymax=0.5,
	width=.8\linewidth, 
	height=.50\linewidth,
	grid=major,
	grid style={color0M},
	xlabel= Training epochs $r_T$, 
	xticklabels={0,0,200,400,600,800,1000,1200,1400,1600,1800,2000},
	ylabel=$\mathcal{L}_\text{USV}(s)$,legend pos=south east,legend cell align={left},legend style={draw=none,legend image code/.code={\filldraw[##1] (-.5ex,-.5ex) rectangle (0.5ex,0.5ex);}}]
	\addplot[mark size=1.5 pt,  color=color3] table [x=step times epsilon, y=SsvGraphTestingUsv, col sep=comma] {numerics/DeepWalk_32pairs10sv_2-3network_g-0i5_delta0_lda1_ep0i01_plot.csv};
	\addlegendentry{$\gamma=-0.5$ (supervised + graph)} 
	\addplot[line width=2pt, color=color2] table [x=step times epsilon, y=SsvTestingUsv, col sep=comma] {numerics/DeepWalk_32pairs10sv_2-3network_g-0i5_delta0_lda1_ep0i01_plot.csv};
	\addlegendentry{$\gamma=0$ (supervised)} 
\end{axis} 
\end{tikzpicture}
\caption{\textbf{Deep walk training.} The figure describes the training of a \protect\twothree DQNN ($r_T=2000$ epochs, $\epsilon=0.01$) trained with and without using the graph structure of a graph with $32$ vertices produced by a classical deep walk. $S=10$ supervised data pairs are used. }\label{fig:apdx_graph_walk}
\end{figure}
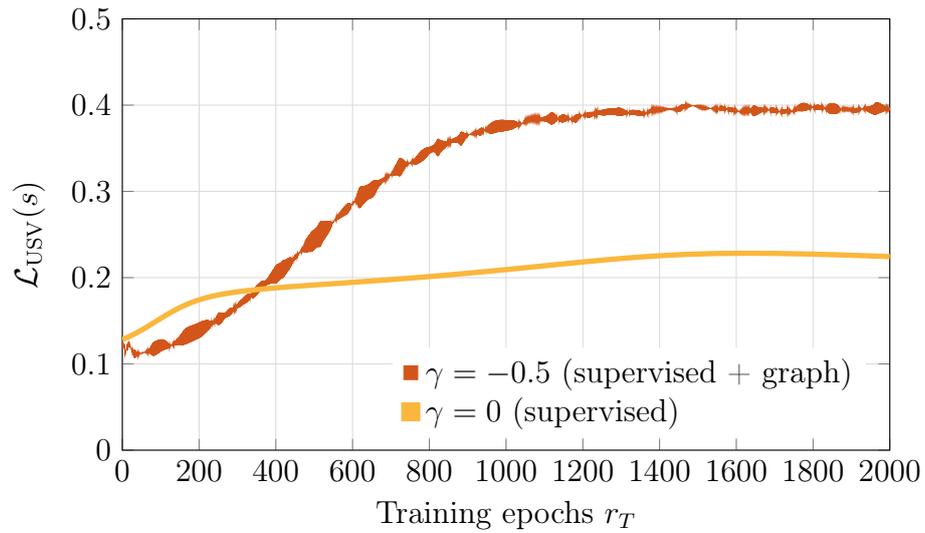

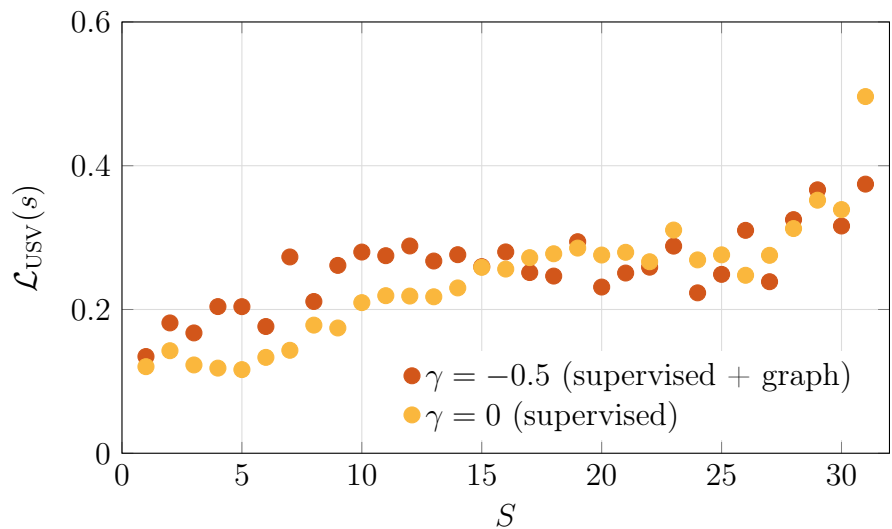
\begin{figure}[H]
	\centering
	\begin{tikzpicture}
		\begin{axis}[
			xmin=0,   xmax=32,
			ymin=0.0,   ymax=0.6,
			width=.95\linewidth, 
			width=.8\linewidth, 
			height=.5\linewidth,
			grid=major,
			grid style={color0M},
			xlabel= $S$, 
			ylabel=$\mathcal{L}_\text{USV}(s)$,legend pos=south east,,legend cell align={left},legend style={draw=none}]
			\addplot[color=color3, only marks, mark size=3 pt,mark phase=0] table [x=numberSupervisedPairsList, y=SsvGraphTestingUsvMeanList, col sep=comma] {numerics/DeepWalk_32pairs_2-3network_g-0i5_delta0_lda1_ep0i01_rounds1000_shots10_plotmean.csv};
			\addlegendentry{$\gamma=-0.5$ (supervised + graph)}
			\addplot[color=color2, only marks, mark size=3 pt,mark phase=0] table [x=numberSupervisedPairsList, y=SsvTestingUsvMeanList, col sep=comma] {numerics/DeepWalk_32pairs_2-3network_g-0i5_delta0_lda1_ep0i01_rounds1000_shots10_plotmean.csv};
			\addlegendentry{$\gamma=0$ (supervised)} 
		\end{axis}
	\end{tikzpicture}
	\caption{\textbf{Deep walk generalisation behaviour.} The plot describes the generalisation behaviour of a \protect\twothree DQNN ($r_T=1000$ epochs, $\epsilon=0.01$) trained with and without using the graph structure of a graph with $32$ vertices produced by a classical deep walk. Each data point demonstrates an averaged over $10$ independent training attempts.}\label{fig:apdx_graph_walk}
\end{figure}

%% file: text/appendixGAN.tex
\chapter{Quantum generative adversial networks}
\markboth{Appendix}{Appendix} 
\label{apdx:QGAN}

In \cref{sec:QGAN_classical} we discussed the classical simulation of the DQGAN algorithm. In the following we extend the numerical examples of this section. All results are generated with the code available at \cite{GithubKerstin}.

First, in \cref{fig:apdx_line} we compare the behaviour of the loss functions,
\begin{align*}
\mathcal{L}_{D}(\mathcal{E}_D,\mathcal{E}_G)&=\frac{1}{S}\sum_{x=1}^S \bra{0} \mathcal{E}_D (\mathcal{E}_G (\ket{\psi_x^\text{in}}\bra{\psi_x^\text{in}}))\ket{0} + \frac{1}{S}\sum_{x=1}^S \bra{1} \mathcal{E}_D (\ket{\phi_x^T}\bra{\phi_x^T})\ket{1} ,\\
\mathcal{L}_{G}(\mathcal{E}_D,\mathcal{E}_G)&=\frac{1}{S}\sum_{x=1}^S \bra{1} \mathcal{E}_D (\mathcal{E}_G (\ket{\psi_x^\text{in}}\bra{\psi_x^\text{in}}))\ket{1}, \\
\mathcal{L}_{V}(\mathcal{E}_D,\mathcal{E}_G)
&=\frac{1}{V}\sum_{i=v}^V \max_{x=1}^N \left( \bra{\phi_x^T} \mathcal{E}_G (\ket{\psi_i^\text{in}}\bra{\psi_i^\text{in}})\ket{\phi_x^T}\right),
\end{align*}
for two different DQGAN architectures: 1-1-1 and 1-3-1. For the training we use $S=10$ states of the data sets
\begin{align*}
\text{data}_\text{line}&=\left\{\frac{(N-x)\ket{0}+(x-1)\ket{1}}{||(N-x)\ket{0}+(x-1)\ket{1}||}\right\}_{x=1}^{N},\\
\text{data}_\text{line'}&=\left\{\frac{(N-x)\ket{000}+(x-1)\ket{001}}{||(N-x)\ket{000}+(x-1)\ket{001}||}\right\}_{x=1}^{N},
\end{align*} for $N=50$, respectively. In both plots the opposed behaviour of $\mathcal{L}_{G}$ and $\mathcal{L}_{D}$ can be observed. Further, we notice that the validation loss does not reach values as high as in the case of the 1-1-1 network.

\begin{figure}[H]
\begin{subfigure}{0.99\linewidth}
\centering
\begin{tikzpicture}
\begin{axis}[
xmin=0,   xmax=20,
ymin=0.2,   ymax=1.5,
width=0.9\linewidth, 
height=0.7\linewidth,
grid=major,grid style={color0M},
xlabel= Training epochs $r_T$, 
xticklabels={0,0,100,200,300,400,500,600,700,800,900,1000},
ylabel=$\mathcal{L}(t)$,legend pos=north east,legend cell align={left},legend style={draw=none,legend image code/.code={\filldraw[##1] (-.5ex,-.5ex) rectangle (0.5ex,0.5ex);}}]
\coordinate (0,0) ;
\addplot[mark size=1.5 pt, color=color2] table [x=step times epsilon, y=costFunctionDis, col sep=comma] {numerics/QGAN_50data10sv_100statData_100statData_1-1networkGen_1-1networkDis_lda1_ep0i01_rounds1000_roundsGen1_roundsDis1_line_plot1_training.csv};
\addlegendentry[scale=1]{Training loss $\mathcal{L}_\text{D}$} 
\addplot[mark size=1.5 pt, color=color1] table [x=step times epsilon, y=costFunctionGen, col sep=comma] {numerics/QGAN_50data10sv_100statData_100statData_1-1networkGen_1-1networkDis_lda1_ep0i01_rounds1000_roundsGen1_roundsDis1_line_plot1_training.csv};
\addlegendentry[scale=1]{Training loss $\mathcal{L}_\text{G}$} 
\addplot[mark size=1.5 pt, color=color3] table [x=step times epsilon, y=costFunctionTest, col sep=comma] {numerics/QGAN_50data10sv_100statData_100statData_1-1networkGen_1-1networkDis_lda1_ep0i01_rounds1000_roundsGen1_roundsDis1_line_plot1_training.csv};
\addlegendentry[scale=1]{Validation loss $\mathcal{L}_\text{V}$} 
\end{axis}
\end{tikzpicture}
\subcaption{Training a \protect\oneoneone DQGAN.}
\label{fig:apdx_line3}
\end{subfigure}
\begin{subfigure}{0.99\linewidth}
\centering
\begin{tikzpicture}
\begin{axis}[
xmin=0,   xmax=20,
ymin=0,   ymax=2,
width=0.9\linewidth, 
height=0.7\linewidth,
grid=major,grid style={color0M},
xlabel= Training epochs $r_T$, 
xticklabels={0,0,100,200,300,400,500,600,700,800,900,1000},
ylabel=$\mathcal{L}(t)$,legend pos=north east,legend cell align={left},legend style={draw=none,legend image code/.code={\filldraw[##1] (-.5ex,-.5ex) rectangle (0.5ex,0.5ex);}}]
\coordinate (0,0) ;
\addplot[mark size=1.5 pt,color=color2] table [x=step times epsilon, y=costFunctionDis, col sep=comma] {numerics/QGAN_50data10sv_100statData_100statData_1-3networkGen_3-1networkDis_lda1_ep0i01_rounds1000_roundsGen1_roundsDis1_connectedLine_training.csv};
\addlegendentry[mark size=10 pt,scale=1]{Training loss $\mathcal{L}_\text{D}$} 
\addplot[mark size=1.5 pt,color=color1] table [x=step times epsilon, y=costFunctionGen, col sep=comma] {numerics/QGAN_50data10sv_100statData_100statData_1-3networkGen_3-1networkDis_lda1_ep0i01_rounds1000_roundsGen1_roundsDis1_connectedLine_training.csv};
\addlegendentry[scale=1]{Training loss $\mathcal{L}_\text{G}$} 
\addplot[mark size=1.5 pt,color=color3] table [x=step times epsilon, y=costFunctionTest, col sep=comma] {numerics/QGAN_50data10sv_100statData_100statData_1-3networkGen_3-1networkDis_lda1_ep0i01_rounds1000_roundsGen1_roundsDis1_connectedLine_training.csv};
\addlegendentry[scale=1]{Validation loss $\mathcal{L}_\text{V}$} 
\end{axis}
\end{tikzpicture}
\subcaption{Training a \protect\oneothreeone DQGAN.}
\label{fig:apdx_line3}
\end{subfigure}
\caption{\textbf{Training a DQGAN.} The evolution of the training losses and validation loss during the training of a \protect\oneoneone (a) and a \protect\oneothreeone (b) DQGAN in $r_T=1000$ epochs with $\eta=1$ and $\epsilon=0.01$ using $50$ data pairs of the data sets $\text{data}_\text{line}$ or $\text{data}_\text{line'}$ whereof $10$ are used for training.}
\label{fig:apdx_line}
\end{figure}
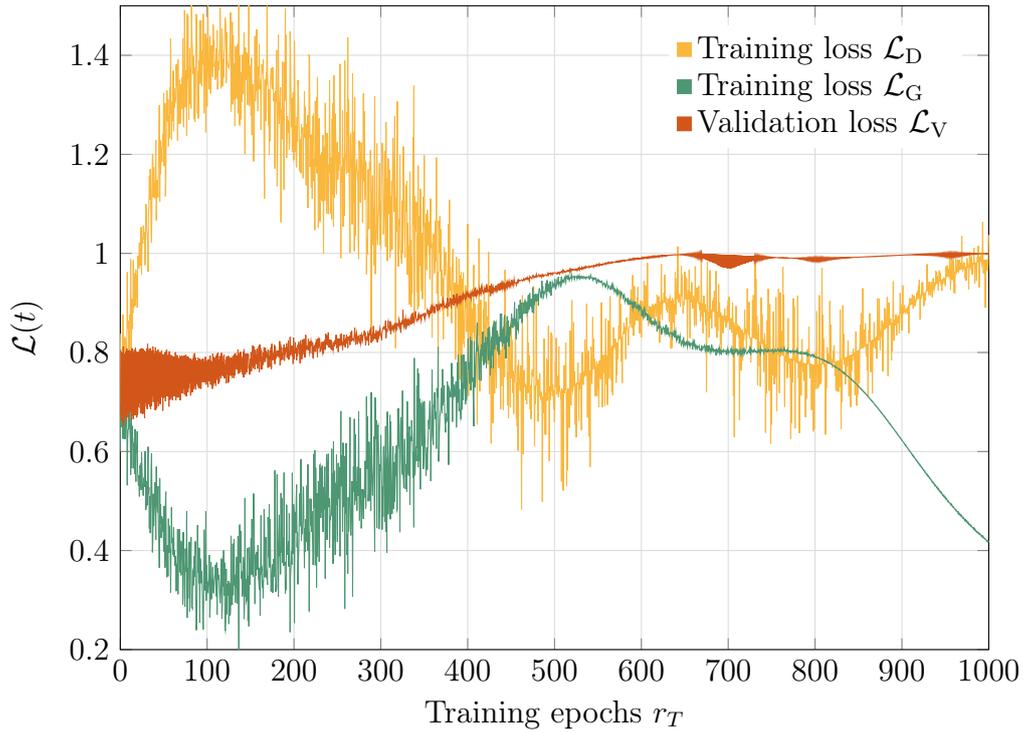
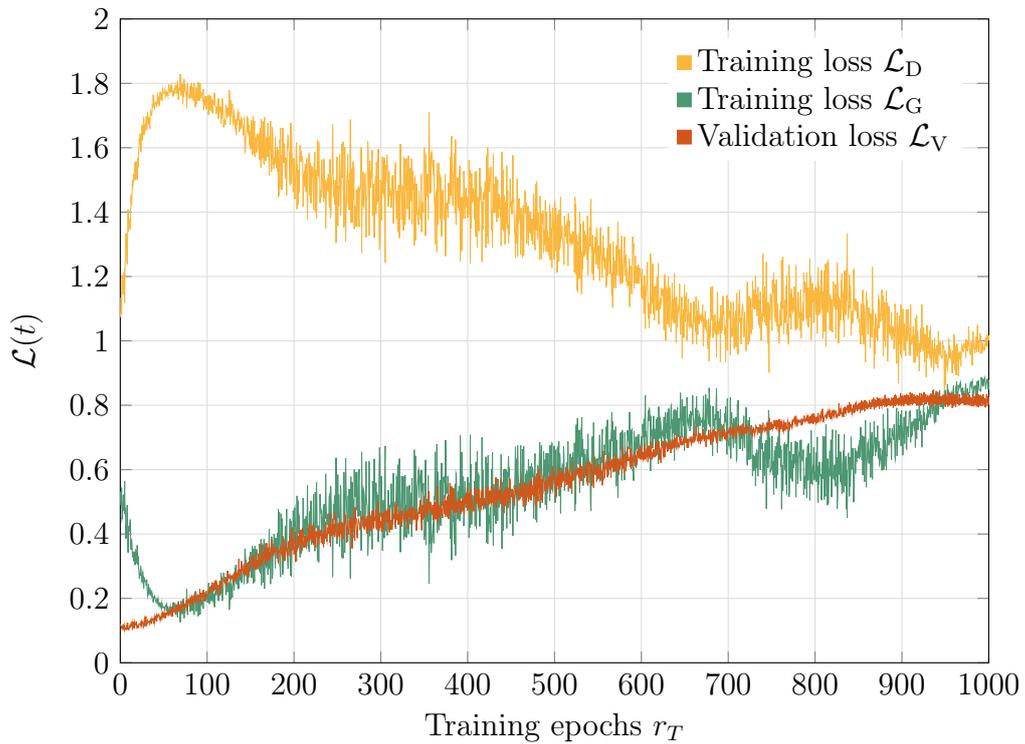

\begin{figure}[H]
\centering
\begin{subfigure}{\textwidth}\centering
\begin{tikzpicture}[scale=1]
\begin{axis}[
ybar,
bar width=1.5pt,
xmin=0,   xmax=51,
ymin=0,   ymax=9,
width=.8\linewidth, 
height=.28\linewidth,
grid=major,
grid style={color0M},
xlabel= State index $x$, 
ylabel=Counts,legend pos=north east,legend cell align={left}]
\addplot[color=color3, fill=color3] table [x=index,y=countTTMean, col sep=comma] {numerics/QGAN_50data10sv_100statData_100statData_1-1networkGen_1-1networkDis_lda1_ep0i01_rounds200_roundsGen1_roundsDis1_line_statMean.csv};
\end{axis}
\end{tikzpicture}
\caption{Line trained with DQNN.} \label{fig:GAN_lineComp}
\end{subfigure}
\begin{subfigure}{\textwidth}\centering
\begin{tikzpicture}[scale=1]
\begin{axis}[
ybar,
bar width=1.5pt,
xmin=0,   xmax=51,
ymin=0,   ymax=30,
width=.8\linewidth, 
height=.28\linewidth,
grid=major,
grid style={color0M},
xlabel= State index $x$, 
ylabel=Counts,legend pos=north east,legend cell align={left}]
\addplot[color=color3, fill=color3] table [x=index,y=countTTMean, col sep=comma] {numerics/QGAN_50data10sv_100statData_100statData_1-1networkGen_1-1networkDis_lda1_ep0i01_rounds200_roundsGen1_roundsDis1_CvsLi_statMean.csv};
\end{axis}
\end{tikzpicture}
\caption{Two clusters trained with DQNN.} \label{fig:GAN_ClusComp}
\end{subfigure}
\begin{subfigure}{\textwidth}\centering
\begin{tikzpicture}[scale=1]
\begin{axis}[
ybar,
bar width=1.5pt,
xmin=0,   xmax=51,
ymin=0,   ymax=15,
width=.8\linewidth, 
height=.28\linewidth,
grid=major,
grid style={color0M},
xlabel= State index $x$, 
ylabel=Counts,legend pos=north east,legend cell align={left}]
\addplot[color=color3, fill=color3] table [x=index,y=countTTMean, col sep=comma] {numerics/QGAN_50data10sv_100statData_100statData_1-1networkGen_1-1networkDis_lda1_ep0i01_rounds200_roundsGen1_roundsDis1_conCvsLi_statMean.csv};
\end{axis}
\end{tikzpicture}
\caption{Two clusters plus $\frac{1}{\sqrt{2}}(\ket{0}+\ket{1})$ trained with DQNN.} \label{fig:GAN_Clus+Comp}
\end{subfigure}
\begin{subfigure}{\textwidth}\centering
\begin{tikzpicture}[scale=1]
\begin{axis}[
ybar,
bar width=1.5pt,
xmin=0,   xmax=51,
ymin=0,   ymax=12,
width=.8\linewidth, 
height=.28\linewidth,
grid=major,
grid style={color0M},
xlabel= State index $x$, 
ylabel=Counts,legend pos=north east,legend cell align={left}]
\addplot[color=color3, fill=color3] table [x=indexDataTest,y=countOutTest, col sep=comma] {numerics/dqnn_q_eq_cluster_epoch_200_vs.csv};
\end{axis}
\end{tikzpicture}
\caption{Two clusters trained with DQNN\textsubscript{Q}.} \label{fig:qgan_q_cluster}
\end{subfigure}
\caption{\textbf{Diversity analysis of a DQGAN.} This plot describes the output's diversity of a \protect\oneoneone DQGAN (DQGAN\textsubscript{Q}) trained in 200 epochs with $\eta=1$ ($\eta_D=0.5,\eta_G=0.1$) and $\epsilon=0.01$ ($\epsilon = 0.25$) using $10$ quantum states of the data sets $\text{data}_\text{line}$ (a), $\text{data}_\text{cl}$ (b,d) and  $\text{data}_\text{cl+}$ (c). The states from $\text{data}_\text{line}$ are used as validation states.}
\label{fig:apnx_Div}
\end{figure}
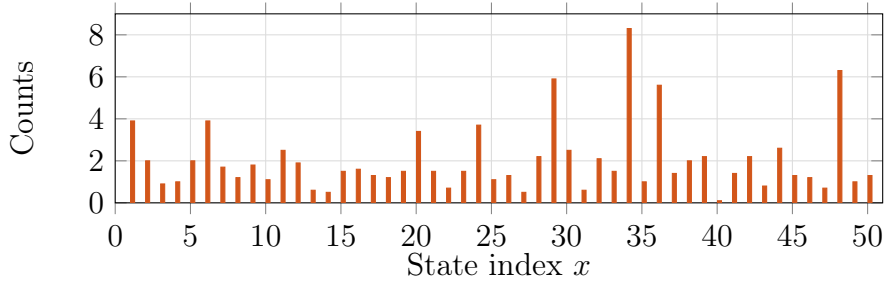
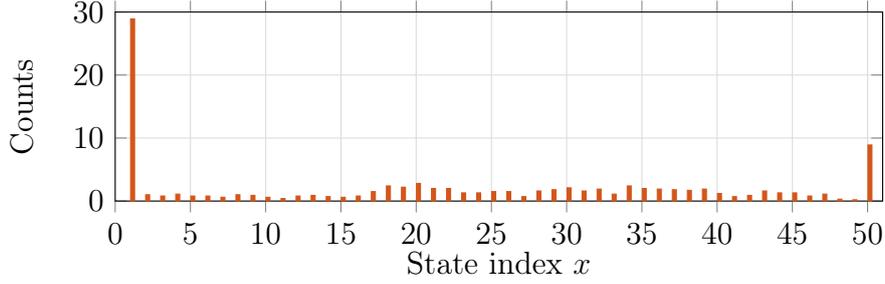
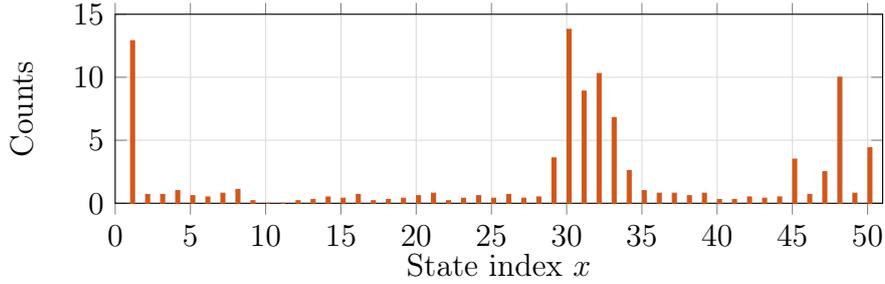
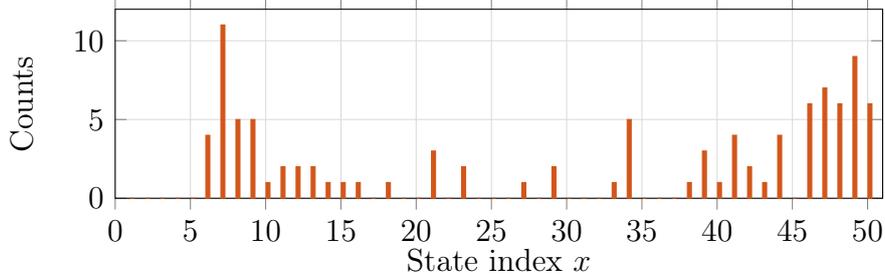

So far, we have only described single runs of the DQGAN algorithm. In the following, we average in ten independent training attempts the data points of the bar diagrams introduced in \cref{sec:QGAN_classical}.

\cref{fig:GAN_lineComp} depicts such an averaged bar diagram resulting after 200 training epochs of training a DQGAN with ten randomly chosen training states of the data set $\text{data}_\text{line}$. We can see that all elements of $\text{data}_\text{line}$ get produced relatively equally.

Further, we extend our study with a second data set,
\begin{equation*}
\text{data}_\text{cl}= \left\{\frac{(2N-1)\ket{0}+(x-1)\ket{1}}{||(2N-1)\ket{0}+(x-1)\ket{1}||}\right\}_{x=1}^{\tfrac{N}{2}}\cup\left\{\frac{(2N-1)\ket{0}+(x-1)\ket{1}}{||(2N-1)\ket{0}+(x-1)\ket{1}||}\right\}_{x=\tfrac{3N}{2}}^{2N}. 
\end{equation*}
If we train a DQGAN and compare the generator's output with the $\text{data}_\text{cl}$ data set, we expect that mainly some of the first and last states get produced. To study this, we train the DQGAN with $S=10$ randomly chosen training states of this set. \cref{fig:GAN_ClusComp} depicts the distribution of the generator's output after 200 training epochs. As expected, the generator does not produce all elements in $\text{data}_\text{line}$ equally often. Due to the average of ten independent training attempts, the states $\ket{0}$ and $\ket{1}$ are very prominent in this plot. Since the state $\ket{0}$ is produced more often, we assume that the training states randomly chosen in every training attempt the $S=10$ training data states were more often chosen of the first part of the cluster.

Further, by removing an arbitrary state of the data set $\text{data}_\text{cl}$ and replacing it by $\frac{1}{\sqrt{2}}(\ket{0}+\ket{1})$ we obtain the connected cluster data set $\text{data}_\text{cl+}$. \cref{fig:GAN_Clus+Comp} shows the diversity of a generator resulting by training a DQGAN with this data set. We can see that some states in the middle are generated more often compared to the plot in \cref{fig:GAN_ClusComp}. However, the state $\frac{1}{\sqrt{2}}(\ket{0}+\ket{1})$ is not produced very often ($x=25$) and the resulting peak in the histogram is rather shifted more in the direction of the  $\ket{1}$ state ($x=50$).

\begin{sloppypar}
In \cref{fig:qgan_q_cluster} the generator's diversity is depicted after a single training run of \DQGANNISQ using the data set $\text{data}_\text{cl}$ in $r_T=200$ training epochs and compared the output to the data set $\text{data}_\text{line}$. We can see that the generator is able to extend the clustered training data. However, similar to the training results of a DQNN shown in \cref{fig:GAN_ClusComp}, the \DQGANNISQ does not achieve to produce the full range of training data. Note that we here use a slightly different implementation, compared to the one presented in \cref{fig:DQNN_qnncircuit}. The differences are marked with orange dashed lines in \cref{fig:GAN_circuit_implementation}.
\end{sloppypar}

\begin{figure}
\centering
\begin{subfigure}[t]{1\linewidth}
\centering
\begin{tikzpicture}[scale=1.6]
\foreach \x in {-.5,.5} {
\draw[line0] (0,\x) -- (2,-1);
\draw[line1] (0,\x) -- (2,-1);
\draw[line0] (0,\x) -- (2,0);
\draw[line1] (0,\x) -- (2,0);
\draw[line0] (0,\x) -- (2,1);
}
\draw[line0] (0,-.5) -- (2,1);
\draw[line1, dash pattern=on 1pt off 5pt] (0,-.5) -- (2,1);
\draw[line1, dash pattern=on 6pt off 2pt] (0,.5) -- (2,1);
\foreach \x in {-1,0,1} {
\draw[line0] (2,\x) -- (4,-0.5);
\draw[line2] (2,\x) -- (4,-0.5);
\draw[line0] (2,\x) -- (4,0.5);
\draw[line2] (2,\x) -- (4,0.5);
}
\node[perceptron0] at (0,-0.5) {};
\node[perceptron0] at (0,0.5) {};
\node[perceptron0] at (2,-1) {};
\node[perceptron0] at (2,0) {};
\node[perceptron0] at (2,1) {};
\node[perceptron0] at (4,-0.5) {};
\node[perceptron0] at (4,0.5) {};
\end{tikzpicture}
\subcaption{Network. }
\end{subfigure}
\vspace*{10mm}

\begin{subfigure}[t]{1\linewidth}
\centering
\begin{tikzpicture}[scale=1.3]
\matrix[row sep=0.3cm, column sep=0.5cm] (circuit) {
\node(start3){$\ket{\phi^\text{in}}$};  
& \node[halfcross,label={\small 2}] (c13){};
& \node[operator0] (c23){$u^{\otimes 2}$};
& \node[]{}; 
& \node[dcross](end3){}; 
& \node[]{}; 
& \node[]{}; 
& \node[]{}; \\
\node(start2){$\ket{000}$};
& \node[halfcross,label={\small 3}] (c12){};
& \node[]{}; 
& \node[]{}; 
& \node[operator0] (c32){$u^{\otimes 3}$};
& \node[]{}; 
& \node[dcross](end2){}; 
& \node[]{};  \\
\node(start1){$\ket{00}$};
& \node[halfcross,label={\small 2}] (c11){};
& \node[]{}; 
& \node[]{}; 
& \node[]{}; 
& \node[]{}; 
& \node[operator0] (c41){$u^{\otimes 3}$};
& \node (end1){$\rho ^\text{out}$}; \\
};
\begin{pgfonlayer}{background}
\draw[] (start1) -- (end1)  
(start2) -- (end2)
(start3) -- (end3);
\node[operator1, minimum height=2cm] at (-.4,0.5) {$U^1$};
\node[operator2,minimum height=2cm] at (1.3,-.7) {$U^2$};
\end{pgfonlayer}
\end{tikzpicture}
\subcaption{Implementation. }
\end{subfigure}
\vspace*{10mm}
\begin{subfigure}[t]{1\linewidth}
\centering
\begin{tikzpicture}[]
\draw[white] (0,2.2)-- (1,2.2);
\draw (0.2,1)-- (1.8,1);
\draw (0.2,0)-- (1.8,0);
\node[operator1,minimum height=1.5cm] at (1,0.5){$U^1$};
\node[] at (2.3,0.5){$=$};
\begin{scope}[xshift=2.8cm]
\draw (0,1)-- (4,1);
\draw (0,0)-- (4,0);
\node[operator1,minimum height=1.5cm] at (0.7,0.5){$U^1_a$};
\node[operator0]at (2,1) {$u^{\otimes 2}$};
\node[operator0]at (2,0) {$u^{\otimes 3}$};
\node[operator1,minimum height=1.5cm] at (3.3,0.5){$U^1_b$};
\node[draw=color3,line width=1pt,dashed,rounded corners=.35cm,minimum height=2.5cm,minimum width=2.8cm] at (2.7,0.5){};
\end{scope}
\end{tikzpicture}
\subcaption{$U^1$ expressed by an unitary $U^1_a$, an additional unitary $U^1_b$ and  $u$-gates. }
\end{subfigure}
\begin{subfigure}[t]{1\linewidth}
\centering
\begin{tikzpicture}[yscale=.6]
\draw (0,4)-- (2,4);
\draw (0,3)-- (2,3);
\draw (0,2)-- (2,2);
\draw (0,1)-- (2,1);
\draw (0,0)-- (2,0);
\node[operator1,minimum height=2.8cm] at (1,2){$U^1_{a,b}$};
\node[] at (2.5,2){$=$};
\begin{scope}[xshift=3cm]
\draw[white] (0,4.5)-- (7,4.5);
\draw (0,4)-- (7,4);
\draw (0,3)-- (7,3);
\draw (0,2)-- (7,2);
\draw (0,1)-- (7,1);
\draw (0,0)-- (7,0);
\node[operator1,minimum height=1.5cm,line width=1pt, dash pattern=on 6pt off 2pt] at (1,3){};
\node[operator1,minimum height=0.9cm, line width=1pt, dash pattern=on 1pt off 5pt] at (2,2.5){};
\node[operator1,minimum height=2.1cm] at (3,2.5){};
\node[operator1,minimum height=1.5cm] at (4,2){};
\node[operator1,minimum height=2.8cm] at (5,2){};
\node[operator1,minimum height=2.1cm] at (6,1.5){};
\draw[line0] (0,3)-- (1.5,3);
\draw (0,3)-- (1.5,3);
\draw[line0] (2.5,3)-- (3.5,3);
\draw (2.5,3)-- (3.5,3);
\draw[line0] (4.5,3)-- (5.5,3);
\draw (4.5,3)-- (5.5,3);
\draw[line0] (2.5,2)-- (7,2);
\draw (2.5,2)-- (7,2);
\draw[line0] (4.5,1)-- (7,1);
\draw (4.5,1)-- (7,1);
\end{scope}
\end{tikzpicture}
\subcaption{$U^1_a$ and $U^1_b$ expressed by two-qubit unitaries. }
\end{subfigure}
\vspace*{10mm}
\caption{\textbf{Implementation of a \DQNNNISQP.} A \DQNNNISQP of the architecture 2-3-2$^+$ (a) can be implemented as quantum circuit using $u$-gates and unitary operations representing the layers of the network (b). Different to the \DQNNNISQ, $U^l$ is decomposed into two unitaries, $U^1_a$ and $U^1_b$, and  $u$-gates. The difference of a \DQNNNISQP compared to a \DQGANNISQ is marked in orange dashed lines (c). $U^1_a$ and $U^1_b$ are expressed via two-qubit unitaries (d).}
\label{fig:GAN_circuit_implementation}
\end{figure}

%% file: text/CV.tex
\chapter*{Curriculum vitae}\addcontentsline{toc}{section}{Curriculum vitae}
\begin{tabular}{ l l }
\textcolor{color1}{\textbf{Name:}} & Kerstin Beer\\ 
\textcolor{color1}{\textbf{Date of birth:}} & 17.08.1993\\ 
\textcolor{color1}{\textbf{Place of birth:}} & Hanover, Germany
\end{tabular}

\subsection*{Academic career}
\begin{flushleft}
\cventry{2017 - 2021}{PhD student}{Leibniz Universität Hannover}{Faculty of Mathematics and Physics, Institute for Theoretical Physics}{Quantum information group, Prof. Dr. Tobias J. Osborne}{Research, teaching, IT administration.}{}
\cventry{2015 - 2017}{Master of Science in Physics}{Leibniz Universität Hannover}{}{}{Thesis: Contextuality and Cohomology}
\cventry{2012 - 2015}{Bachelor of Science in Physics}{Leibniz Universität Hannover}{}{}{ Thesis: Quantum Compiling - Zerlegung unitärer Quantenoperationen}
\cventry{2012 - 2017}{Scholarship}{Heinrich Böll Stiftung}{}{}{}
\cventry{2004 - 2012}{Abitur}{Albert-Einstein-Schule}{Laatzen}{}{}
\end{flushleft}

\vspace{1.5cm}

\vspace*{\fill} 
\newpage